\documentclass[aps,12pt,a4paper,english,preprint,sort&compress,numbers,sort]{myrevtex4} 
\usepackage[latin1]{inputenc}
\usepackage{babel}
\usepackage{natbib}
\usepackage{amsmath}
\usepackage{amssymb}
\usepackage{bm}
\usepackage[cspex,bbgreekl]{mathbbol}
\usepackage{bbm}
\usepackage{wasysym}
\usepackage{graphicx}  % if you want to include PostScript figures
\usepackage[figuresright]{rotating}  % if you have landscape tables

\makeatletter

\providecommand{\LyX}{L\kern-.1667em\lower.25em\hbox{Y}\kern-.125emX\@}
\renewcommand{\bibnumfmt}[1]{#1. }
\def\v#1{{\bf #1}}

\newcommand{\LSM}{LaSr$_2$Mn$_2$O$_7$}
\newcommand{\LSXM}{La$_{2-2x}$Sr$_{1+2x}$Mn$_2$O$_7$}
\newcommand{\NAV}{$\alpha$'-NaV$_2$O$_5$}
\newcommand{\NAXV}{$\alpha$'-Na$_x$V$_2$O$_5$}
\newcommand{\YBA}{Yb$_4$As$_3$}

\newcommand{\UPD}{UPd$_2$Al$_3~$~}

\newcommand{\la}{\langle}
\newcommand{\ra}{\rangle}

\newcommand{\ua}{\uparrow}
\newcommand{\da}{\downarrow}

\newcommand{\vk}{\bf k\rm}

\newcommand{\vq}{\bf q\rm}

\newcommand{\boldsigma}{{\bm \sigma}}
\newcommand{\boldtau}{{\bm \tau}}

\renewcommand{\theequation}{\arabic{section}.\arabic{equation}}
\newcommand{\resetdoublenumb}{\setcounter{equation}{0}} 
\renewcommand{\thefigure}{\arabic{section}.\arabic{figure}}
\newcommand{\resetdoublenumbf}{\setcounter{figure}{0}} 
\usepackage{babel}
\makeatother
\begin{document}

\title{Strongly correlated electrons}

\author{PETER FULDE}

\affiliation{Max-Planck-Institut für Physik komplexer Systeme, 01187 Dresden}

\author{PETER THALMEIER}

\affiliation{Max-Planck-Institut für Chemische Physik fester Stoffe, 01187
  Dresden} 

\author{GERTRUD ZWICKNAGL}

\affiliation{Institut für Mathematische Physik, Technische Universität
  Braunschweig, 38106 Braunschweig}

\maketitle

\tableofcontents

\newpage
\section{Introduction}
\resetdoublenumb 
\resetdoublenumbf

\label{Sect:Introduction}

The field of strongly correlated electron systems has been constantly growing
for almost three decades. A milestone in its development was the discovery by
Andres, Graebner and Ott \cite{Andres75} of heavy-quasiparticle excitations in
CeAl$_3$. Additional verve came from the discovery of superconductivity in
the related compounds CeCu$_2$Si$_2$ \cite{Steglich79}, UBe$_{13}$ \cite{Ott83}
and UPt$_3$ \cite{Stewart84}. But a real great push for the
field was provided by the discovery of high-temperature superconductivity in
the copper-oxide based perovskites \cite{Bednorz86}. Were it not for strong
electron correlations La$_2$CuO$_4$, one of the key compounds of that class of
materials and the basis of the hole doped superconductors
La$_{2-x}$Ba$_x$CuO$_4$ and La$_{2-x}$Sr$_x$CuO$_4$ would be metallic. Instead
it is an antiferromagnet which remains insulating even above the
N\'eel temperature where 
the unit cell is not doubled anymore. Therefore, electron correlations are
apparently so strong that the metallic character of the material is suppressed
in favor of an insulating state. That electron correlation may induce a metal
to insulator transition had been suggested long before the discovery of heavy
quasiparticles and high-T$_c$ cuprates. The names of Mott \cite{MottBook90} and
Hubbard \cite{Hubbard63} stand  for that phenomenon. At their time the
interests in the effects of strong correlations resulted from the transition
metal oxides and their various phase transitions. It is worth recalling that
the famous Verwey \cite{Verwey41} transition in magnetite Fe$_3$O$_4$ falls
into the same category. One may even go back to Wigner \cite{Wigner34} or
Heitler-London \cite{Heitler27} who dealt with strongly correlated electrons
long before corresponding experiments were available. While Wigner pointed out
that electrons may form a lattice when their correlations become sufficiently
strong, Heitler and London developed a theory for chemical bonding based on
strongly correlated electrons. It is the opposite limit of H\"uckel's theory
\cite{Hueckel31,HueckelE31,Hueckel32} based on molecular orbitals in which
electron correlations are completely neglected. This raises the question of how
to quantify the strength of electronic correlations. For example, one would
like to know by how much electrons are stronger correlated in LaCu$_2$O$_4$
than, e.g., in iron or nickel or in transition metal oxides.

The differences between systems with strongly and with weakly correlated
electrons may be seen by considering the ground state of the simplest possible
example, i.e., of a H$_2$ molecule in the Heitler-London- and in the molecular
orbital limit. The Heitler-London form of the ground-state wavefunction is

%1
\begin{equation}
\psi_{\rm HL} \left( {\bf r}_1, {\bf r}_2 \right) =  \frac{1}{2} \left[ \phi_1
({\bf r}_1) \phi_2 ({\bf r}_2) + \phi_2 ({\bf r}_1) \phi_1 ({\bf r}_2) \right]
\left( \alpha_1 \beta_2 - \beta_1 \alpha_2\right)
\label{psiHL}
\end{equation}

\noindent where the single-electron wavefunctions $\phi_{1,2}({\bf r})$ are
centered on atoms 1 and 2 of the molecule and $\alpha$ and $\beta$ denote
spinors for up and down spins. In distinction to Eq. (\ref{psiHL}) the
molecular-orbital form of the ground-state wavefunction is 

%2
\begin{eqnarray}
\psi_{\rm MO} \left( {\bf r}_1, {\bf r}_2 \right) 
= && \frac{1}{2^{3/2}} \left[ \phi_1 ({\bf r}_1) \phi_1 ({\bf r}_2) 
+ \phi_1 ({\bf r}_1) \phi_2 ({\bf r}_2) 
+ \phi_2 ({\bf r}_1) \phi_1 ({\bf r}_2) \right. \nonumber \\
&& + \left. \phi_2 ({\bf r}_1) \phi_2 ({\bf r}_2) \right] \left( \alpha_1
\beta_2 - \beta_1 \alpha_2 \right)~~~.
\label{psiMO}
\end{eqnarray}

\noindent It is seen that $\psi_{MO}({\bf r}_1, {\bf r}_2)$ but not $\psi_{\rm
HL}({\bf r}_1, {\bf r}_2)$ contains ionic configurations $\phi_1 ({\bf r}_1)
\phi_1 ({\bf r}_2)$ and $\phi_2 ({\bf r}_1) \phi_2 ({\bf r}_2)$. In
Eq. (\ref{psiMO}) they have equal weight like the nonionic configurations. But
ionic configurations cost additional Coulomb repulsion energy of the 
electrons. Therefore they are completely suppressed in the Heitler-London- or
strong correlation limit. This demonstrates an important feature of electron
correlations, namely a partial suppression of electronic charge fluctuations on
an atomic site. The former are called interatomic correlations because charge
fluctuations at an atomic site are caused by an overlap of wavefunctions
of different atoms. They are favored by a kinetic energy gain due to
electron delocalization. Reducing them compared with uncorrelated electrons
keeps the Coulomb repulsions small. 

In addition to interatomic correlations we must also consider
intra-atomic correlations. Consider an atom of a solid in a configuration with
a given number of electrons, for example, a C atom in diamond with, e.g., 4 or
5 valence electrons. Those electrons will optimize their on-site Coulomb
repulsions by arranging according to Hund's rules and by in-out
correlations. Hund's rules ensure that electrons on an atom are optimally
distributed over the angular segments of the atom, so that their repulsions are
as small as possible. In-out correlations achieve the same by proper radial
distribution of the electrons. The intra-atomic correlations are strongest for
4$f$ electrons, i.e., for atoms or ions of the lanthanide series. But also in
actinides or transition-metal ions they play a big role. Large overlaps with
atomic wavefunctions of the chemical environment will weaken them. This is
understandable: before the electrons can fully establish intra-atomic
correlations they leave for the neighboring sites by hopping off the
site. Interatomic correlations can be strong even when intra-atomic
correlations are moderate or weak. Let us make a gedanken experiment and
consider a Si crystal with artificially enlarged lattice parameter. The
intra-atomic correlations on a Si site are fairly moderate, but the interatomic
correlations are becoming strong when the lattice constant is increased, i.e.,
when the limit of separate atoms is approached. In that case fluctuations in
the electron number at a site reduce to zero. 

From the above considerations it follows that a suitable measure of the
{\it interatomic} correlation strength is the reduction of electron number
fluctuations on a given atom. An independent-electron or Hartree-Fock
description implies too large fluctuations. Let $| \psi_0 \rangle$ denote the
exact ground state of an electronic system and $| \Phi_{\rm SCF} \rangle$ the
corresponding self-consistent field (SCF) or Hartree-Fock (HF) state. The
normalized mean-square deviation of the electron number n$_i$ on atom $i$ is
given by 

%3
\begin{eqnarray}
\Sigma (i) =\frac{\langle \Phi_{\rm SCF} \mid \left( \Delta n_i \right)^2 \mid
\Phi_{\rm SCF} \rangle - \langle \psi_0 \mid \left( \Delta n_i \right)^2 \mid
\psi_0 \rangle}{\langle \Phi_{\rm SCF} \mid \left( \Delta n_i \right)^2 \mid
\Phi_{\rm SCF}\rangle}  
\label{sumi}
\end{eqnarray}

\noindent where $\Delta n_i = n_i - \Bar{n}_i$ and $\Bar{n}_i$ denotes the
average value. One notices that $0 \leq \Sigma (i) \leq 1$. When $\Sigma (i) =
0$ the interatomic correlations vanish, i.e., the Coulomb repulsions between
the electrons can be treated in mean-field approximation. In a solid atoms or
ions with 
strongly correlated electrons have $\Sigma (i)$ values near unity. One can also
define a correlation strength for different bonds instead of atoms. In that
case the denominator is modified when heteropolar bonds are considered. Then we
must subtract from $\langle \Phi_{\rm SCF} \mid (\Delta n_i)^2 \mid \Phi_{\rm
  SCF} \rangle$ a term $(\Delta n)^2_{\rm pc}$. It takes into account that some
number fluctuations are required even when the electrons are perfectly
correlated in order to ensure a heteropolar charge distribution within the
bond. Let $\alpha_{\rm p}$ denote the bond polarity. It is defined by the
difference in the average occupation numbers of the two half-bonds 1 and 2
which form the heteropolar bond, i.e., $\Bar{n}_{1(2)} = (1 \pm \alpha_{\rm
  p})$. In that case $(\Delta n)^2_{\rm pc} = \alpha_{\rm p} (1 - \alpha_{\rm
  p})$. Those considerations apply to a solid as well as to a molecule. 

For the H$_2$ molecule one checks immediately that approximating $| \psi_0
\rangle$ by $\psi_{MO}({\bf r}_1, {\bf r}_2)$ gives $\Sigma = 0$ while a 
replacement by $\psi_{\rm HL} ({\bf r}_1, {\bf r}_2)$ yields $\Sigma = 1$ since
$\langle \psi_0 \mid (\Delta n)^2 \mid \psi_0 \rangle = 0$ in that case. For a
C=C or N=N $\pi$ bond one finds $\Sigma \approx 0.5$ while for a C-C or N-N
$\sigma$ bond $\Sigma$ = 0.30 and 0.35, respectively. Let us consider the
ground state of 
La$_2$CuO$_4$ and let P(d$^\nu$) denote the probability of finding $\nu$ 3$d$
electrons on a given Cu site. Within the independent electron or Hartree-Fock
approximation the average $d$ count is found to be $\Bar{n}_d \simeq 9.5$ and
the probabilities of different configurations are P(d$^{10}$) = 0.56, P(d$^9$)
= 0.38 and P(d$^8$) = 0.06. When correlations are included, i.e., the
correlated ground state $| \psi_0 \rangle$ is used the average $d$ electron
number changes to $\Bar{n}_d \simeq 9.3$ and P(d$^{10}$) = 0.29, P(d$^9$) =
0.70 while P(d$^8$) = 0.0. One notices that the d$^8$ configurations are almost
completely suppressed in agreement with photoemission experiments. The
fluctuations between the d$^9$ and d$^{10}$ configurations are fixed by the
value of $\Bar{n}_d$. A similar analysis for the oxygen atoms reveals that
there the 2p$^4$ configurations are {\it not} completely suppressed because the
Coulomb integrals are not as large as for Cu. Indeed, these configurations are
important for superexchange to occur, which determines the antiferromagnetic
coupling between Cu ions. In accordance with the above consideration one finds
$\Sigma({\rm Cu}) \simeq$ 0.8 and $\Sigma({\rm O}) \simeq$ 0.7
\cite{Oles87}. So indeed, correlations are quite strong in La$_2$CuO$_4$. On
the other hand, they are still smaller than those of 4$f$ electrons in a system
like CeAl$_3$.  

A measure for the strength of intra-atomic correlations is more difficult to
define. One way is by finding out to which extent Hund's rule correlations are
building up on a given atomic site $i$. A possible measure for that is the
degree of spin alignment at a given atomic site $i$

%4
\begin{equation}
S^2_i = \langle \psi_0 \mid {\bf S}^2(i) \mid \psi_0 \rangle  
\label{S2i}
\end{equation}

\noindent where ${\bf S}(i) = \sum\limits_\nu {\bf s}_\nu (i)$ and ${\bf s}_\nu
(i)$ is the spin operator for orbital $\nu$. The quantity $S^2_i$
should be compared with the values when the SCF ground-state wavefunction $|
\Phi_{\rm SCF} \rangle$ is used and when instead the ground state $| \Phi_{\rm
  loc} \rangle$ in the limit of complete suppression of interatomic charge
fluctuations is taken, i.e., for large atomic distances. Therefore we may
define  

%5
\begin{eqnarray}
\Delta S^2 _i = \frac{\langle \psi_0 \mid {\bf S}^2 (i) \mid \psi_0 \rangle -
\langle  \Phi_{\rm SCF} \mid {\bf S}^2 (i) \mid \Phi_{\rm SCF} \rangle}
{\langle \Phi_{\rm loc} \mid {\bf S}^2 (i) \mid \Phi_{\rm loc} \rangle -
\langle \Phi_{\rm SCF} \mid {\bf S}^2 (i) \mid \Phi_{\rm SCF} \rangle}
\label{deltaS}
\end{eqnarray}

\noindent as a possible measure of the strength of intra-atomic
correlations. Note that $0 \leq \Delta S^2_i \leq 1$. For example, for the
transition metals Fe, Co and Ni $\Delta S^2_i$ is approximately 0.5.

Those findings show that the much discussed transition metals are just in the
middle between the limits of uncorrelated and strongly correlated
electrons. Hund's rule correlations are important in them but relatively large
overlaps of atomic wavefunctions on neighboring sites prevent their complete
establishment. Starting from the work of Slater \cite{Slater36} and Van Vleck
\cite{vanVleck53} in particular Friedel \cite{FriedelBook69}, Gutzwiller
\cite{Gutzwiller64,Gutzwiller65}, Hubbard \cite{Hubbard64} and Kanamori
\cite{Kanamori63} have discussed their effects in detail. One of the outcome of
the studies of transition metals was the Hubbard Hamiltonian. It was in fact
used independently also by Gutzwiller and in a slightly modified version by
Kanamori. This Hamiltonian was extensively treated in various
approximations. The extended Hubbard model has remained until present times the
working horse of many studies of strongly correlated electrons
\cite{Imada98,Mancini04}. The shortcomings of that model are known. For
example, it considers $d$ electrons only, i.e., $s$ electrons are
neglected. Also it cannot provide for orbital relaxations when electrons hop on
or off a site because only one basis function per atomic orbital is
used. Nevertheless, it is believed that it covers the most important generic
effects of strongly correlated electrons. 

The valence electrons which are most strongly correlated are the 4$f$ ones
because their atomic wavefunction is close to the nucleus and the tendency to
delocalize is very small. In fact, in intermetallic rare-earth compounds only
$f$-electrons in Ce or Yb ions show a noticeable degree of itineracy. The
consequence are new low-energy scales which may appear in those compounds and
as a result heavy-quasiparticle excitations. Not always do quasiparticles show
conventional Fermi liquid behavior which governs the low-temperature
thermodynamic properties of many metals. In a number of cases one observes what
is called non-Fermi liquid behavior, i.e., quantities like the temperature
dependence of the specific heat or of the susceptibility deviate from normal
metallic behavior. In particular this holds true near a quantum critical point
where apparently no characteristic energy scale is prevailing. Fermi liquid
behavior requires that at low temperatures all thermodynamic quantities scale
with $k_BT^*$, a characteristic energy which in strongly correlated electron
systems takes the role of the Fermi energy. When such a characteristic scale
does not exist deviations from Fermi liquid behavior do occur. In a way it is
more astonishing that to good approximation Fermi liquid behavior is observed
in a number of strongly correlated electron systems than that it is not. A
study of the Hubbard model shows ways for obtaining deviations from standard
features of a metal.  

One interesting aspect of strong electron correlations is the possible
occurrence of charge order. A charge ordered state minimizes the repulsive
energy between electrons at the expense of the kinetic energy. Wigner was the
first to study this subject by considering a homogeneous electron gas and
specifying the conditions under which the formation of an electronic lattice is
possible. Chances for charge ordering are larger for inhomogeneous systems,
i.e., lattices than for homogenous ones, the reason being that the kinetic
energy gain of electrons due to delocalization may become very much reduced as
compared with homogeneous electron systems. A prototype example is Yb$_4$As$_3$
where charge order occurs close to room temperature and there are many other
cases.

While 4$f$ electrons are localized in most cases and are very strongly
correlated, 5$f$ electrons are more delocalized but still more strongly correlated
as, e.g., 3$d$ electrons in transition metals. It turns out that in this case a
dual picture applies: while 5$f$ electrons become itinerant in some of the
orbitals they remain localized in others. Such a model explains very well a
number of experiments on U compounds.

Heavy quasiparticles have also been observed in LiV$_2$O$_4$, a metal with 3$d$
electrons. A special feature of that material is that the 3$d$
electrons are placed on a pyrochlore or geometrically frustrated lattice. Model
calculations show that charge degrees of freedom of strongly correlated
electrons in frustrated lattice structures can give rise to new phenomena
at special band fillings. There may exist large numbers of low-energy
excitations for which Landau's Fermi liquid approach fails and there may be
even excitations with fractional electron charges. Although phenomena of this
kind have not been observed yet, the theoretical results may stimulate further
thinking.

Strongly correlated electrons show in addition to the quasiparticle bands also
satellite structures in photoemission experiments. They are contained in the
incoherent part of the one-particle Green function. It appears that detailed
studies of the incoherent part of Green's function have not been done to the
extent they deserve. The reason for their importance is the following. A
quasiparticle in a solid can be considered as a bare particle (electron or
hole) surrounded by a correlation hole. The whole object, i.e., particle plus
correlation hole moves in form of a Bloch wave through the system. The internal
degrees of freedom of the correlation hole give rise to excitations which are
contained in the incoherent part of Green's function. Therefore it is very
instructive to study general features of that incoherent part. Hubbard's upper
band can be considered a satellite feature for filling factors $n < 1/2$, i.e.,
for less than one electron per site. Other examples will be presented.

It is impossible to cover all aspects of strongly correlated electrons in a
review of reasonable size. Therefore selections have to be made. Naturally,
authors select topics for reviews for which they feel particularly
competent. These are usually areas in which they have actively worked. This
holds also true here and the selection we made here may do injustice to other
interesting developments in the field not covered here. So we apologize for an
incomplete covering of topics as well as for incomplete lists of contributions
of authors to the subject discussed here.

\newpage
 
\section{Special Features of Strong Correlations}
\resetdoublenumb 
\resetdoublenumbf

\label{Sect:SignStrongCorr}

Metals with strongly correlated electrons exhibit characteristic deviations
from the behavior of independent electrons. The latter are reflected in
thermodynamic and transport properties as well as in the high energy spectra. 

Traditional electron theory of metals proceeds from the electron gas model
formulated by Sommerfeld and Bethe \cite{SommerfeldBook33}. The electrons are
described as a system of non-interacting fermions. The eigenstates are formed
by filling single-particle levels in a manner consistent with the Pauli
principle which permits at most one electron per spin direction to occupy any
single-electron level. The ground state of an $N$ electron state is obtained by
filling the $N/2$ single particle levels with the lowest energies. It is
non-degenerate and characterized by a surface in ${\bf k}$-space separating the
occupied levels from their unoccupied counterparts. The existence of this
surface, the Fermi surface, follows directly for a system of independent
electrons. But note that the observation of a Fermi surface does not imply
that the independent electron approximation is a valid description of a
system. The low-temperature properties which are dominated by the low-energy
excitations are universal, the detailed character of the system under
consideration being reflected in a characteristic energy - the Fermi energy
$E_F$. The energy scale is set by the variation of the single particle levels
with wave number ${\bf k}$. A measure of it is the Fermi velocity $v_F$. The
linear variation with temperature of the specific heat, $C(T) \simeq \gamma T$,
and the temperature-independent magnetic susceptibility, $\chi_s (T) \to$ const
are also characteristic features of free electrons. Finally, the spectrum for
adding or removing a particle in a single-particle level ${\bf k}, \sigma$ to
the ground state $A ({\bf k}, \omega)$, exhibits a well-defined peak 

%2.1
\begin{equation}
A \left( {\bf k}, \omega \right) = \delta \left( \omega - \epsilon_k \right)
\label{Akome}
\end{equation} 

\noindent of weight unity centered at the single-particle energy $\epsilon_{\bf
  k}$. The independent electron model has proven to be very successful in
  explaining experimentally observed properties of simple metals. That was a
  surprise for some time since electron-electron repulsions are not weak in any
  metal and one might therefore expect that they modify strongly the properties
  of a system of independent electrons. That this is not necessarily the case
  was shown by Landau \cite{Landau56,Landau57,Landau58,AbrikosovBook63}.

The Landau theory assumes that there exists a one-to-one correspondence between
the excitations of the complex interacting electron system and those of 
independent electron. The former are called quasiparticles and their orbitals
and energies $E({\bf k})$ are determined from an effective Hamiltonian. It
contains an effective, not necessarily local potential. The many-body aspects
are contained in the construction of the effective potential which must be
determined specifically for the problem under consideration.

The quasiparticle energies may be altered when the overall configuration is
changed. A characteristic feature of interacting Fermi liquids is that the
energy dispersion $\tilde{E}_\sigma ({\bf k})$ of a quasiparticle depends on
how many other quasiparticles are present, 

%2.2
\begin{equation}
\tilde{E}_\sigma ({\bf k}) = E ({\bf k}) + \sum_{{\bf k}' \sigma'} f_{\sigma
  \sigma'} ({\bf k}, {\bf k}') \delta n_\sigma' ({\bf k}')~~~. 
\label{tilEsig}
\end{equation} 

Here $E ({\bf k})$ denotes the energy dispersion of a quasiparticle when there
are no other quasiparticles around (dilute gas limit). In systems with strong
correlations it reflects the electron interactions and hence cannot be
calculated from the overlap of single-electron wave functions. Interactions
among quasiparticles are characterized by the matrix $f_{\sigma \sigma'} ({\bf
  k}, {\bf k}')$. The deviations from a step-function-like Fermi distribution
$f(E({\bf k}), T = 0)$ are given by $\delta n_\sigma ({\bf k})$.

The scattering amplitudes $f_{\sigma \sigma'} ({\bf k}, {\bf k}')$ are
parameterized and the parameters are adjusted to experiments. Their form is
strictly applicable only to a homogeneous translationaly invariant electron
system. Therefore applying it to an inhomogeneous periodic solid requires some
modifications (see, e.g., Ref. \cite{Fulde88}) which are usually not
discussed. From this point of view Landau's theory is more of a useful
theoretical concept rather than a quantitative computational scheme.

The assumed one-to-one correspondence of the excitations implies that the low
temperature thermodynamic properties resemble those of independent electrons
but with renormalized parameters such as the effective electron or hole
mass. Also the weight of the peak in the spectral density $A({\bf k}, \omega)$
is modified to $Z \cdot \delta (\omega - E({\bf k}))$ where the renormalization
factor $0 < Z \leq 1$ describes the weight of the bare electron in the
quasiparticle. The latter contains in addition to the bare electron also the
correlation hole around it.

An interacting electron system to which Landau's theory applies has also a
Fermi surface. Luttinger has proven \cite{Luttinger60} that in case that
perturbation theory is applicable the volume enclosed by the Fermi surface is
independent of the electron interactions. Another important property of
quasiparticles is that they can be considered as 'rigid' with respect to
low-energy and long-wavelength perturbations. That is to say that excitations
involving degrees of freedom of the correlation hole show up at high energies
only and are neglected as regards low-temperature properties. They are
discussed in Sec. \ref{Sect:HighEnergyExcitations}.

When electron correlations are strong the quasiparticle concept is still
applicable for a number of substances. In that case the renormalization factor
$Z$ may become very small. This results in heavy quasiparticles because the
Fermi velocity is reduced by the same factor. Probably in many cases the
one-to-one correspondence between the excitations of a strongly correlated
electron system and a corresponding system of independent electrons is only
approximately fulfilled. But then the 
low-temperature properties of the system may still look very similar to those
of independent electrons with renormalized parameters. For example, the
specific heat will still be nearly linear in $T$ at low temperatures etc. 

However, we want to stress that from the observation of a specific heat linear
in $T$ or a temperature independent spin susceptibility in the low temperature
regime one may not conclude that the quasiparticle picture is applicable. In
fact, Luttinger liquids in quasi- one dimensional systems show many properties
as quasiparticles do. This is so despite the fact that the key assumption of a
one-to-one correspondence of excitations to those of independent electrons is
unjustified here. There are strongly correlated systems where the quasiparticle
picture seems totally inappropriate. This is outlined in the following
subsection \ref{Sect:DevFermiLiquBeha} and discussed in more detail in various
sections of this article.   

\subsection{Low-Energy Scales: a Signature of Strong Correlations}

\label{Sect:LowEnergyScales}

As mentioned above, the characteristic energy scale of a free electron gas is
the Fermi energy $E_F$ or, alternatively the Fermi temperature $T_F$. A typical
value for $E_F$ is 5 eV corresponding to a $T_F$ of 5 $\cdot 10^4$ K. A
special feature of strongly correlated electrons is that they introduce new
low-energy scales. It is customary to associate a temperature $T^*$ with
them. In metals with heavy quasiparticles, i.e., with very strong electron
correlations $T^*$ ranges from a few Kelvin to a few hundred Kelvin. As
correlations become weaker $T^*$ increases until it is no longer justified to
speak of a separate low-energy scale. The microscopic origin of the low-energy
scales can be quite different. A widely recognized case is the Kondo
effect. Here $T^*$ is the Kondo temperature, i.e., it is given by the binding
energy of the spins of the conduction electrons to local spins. Local spins
imply incomplete inner shells of an atom or ion. The fact that they remain
partially filled only, when surrounded by conduction electrons is due to strong
correlations. Any conduction electron which tries to enter the incomplete inner
shell is expelled by strong on-site Coulomb repulsion. An example are
Ce$^{3+}$ ions immersed in a sea of (generally) weakly correlated
conduction electron. Due to a weak hybridization the number of 4$f$ electrons
is nearly one. It forms a singlet with the conduction electrons. The
aforementioned CeAl$_3$ falls into that category. Breaking those singlets
results in low-energy excitations and fixes the low-energy scale $T^*$. The
low-energy excitations make it plausible that there will be a large
low-temperature specific heat. To explain heavy quasiparticles the
singlet-triplet excitations on different sites must lock together and form
coherent Bloch-like excitations. That takes place at a somewhat lower energy
scale $T_{\rm coh}$. One expects that $T_{\rm coh}$ is of order of $T^*$ but no
detailed theory for a relation between the two temperatures is available. There
is also no theory existing which tells us that the coherent excitations are in
one-to-one correspondence to excitations of (nearly) free
electrons. Nevertheless this assumption has worked remarkably well.

The origin of a low $T^*$ is quite different in the strongly correlated
semimetal Yb$_4$As$_3$. Here the Coulomb repulsion of the 4$f$ holes in
neighboring Yb ions leads to charge order in the form of well
separated chains of Yb$^{3+}$ ions. Spin excitations in those chains by which
light mobile 4$p$ holes of As are scattered, lead to low temperature properties
which resemble very much those of other systems with heavy quasiparticles
\cite{Fulde95,Kohgi97}. Despite of this the system is not really a heavy Landau
Fermi liquid any more as is explained in the next subsection and discussed in
more detail in Sec. \ref{subsect:YBA}.

A third mechanism is found to be responsible for a low energy scale $T^*$ in U
compounds like UPd$_2$Al$_3$ or UPt$_3$. Strong intra-atomic or Hund's rule
correlations lead here to pronounced anisotropies of the effective
hybridization of different 5$f$ orbitals. As a result some of the 5$f$
electrons remain localized while others delocalize. The crystalline environment
lifts degeneracies of the localized electrons on a low energy scale. The
delocalized or itinerant electrons couple to the excitations of the local
system and in this way generate a low $T^*$ \cite{Zwicknagl03}. Heavy
quasiparticles may also appear near a quantum critical point like in YMn$_2$
\cite{Pinettes94}. 

Finally, in Ce doped Nd$_2$CuO$_4$ a fourth origin of a low $T^*$ is
observed. Here it is essentially a fluctuating internal molecular field
originating from the Nd ions which causes a low energy scale in the strongly
correlated $d$-electron system of the Cu-O planes to which is couples
\cite{Brugger93,Fulde93}. 

It seems obvious that there will be other physical processes identified in the
future resulting in low-energy scales of strongly correlated electron systems.

\subsection{Deviations from Fermi-Liquid Behavior}

\label{Sect:DevFermiLiquBeha}

There is no obvious reason why strongly correlated metallic electron systems
should be Fermi liquids. But as pointed out above a large number of them behave
very nearly like ordinary metals, i.e., Fermi liquids with renormalized
parameters like the effective mass. Even in these cases, high energy
excitations show characteristic satellite structures which reflect strong
correlations in partially filled inner shells. This topic is discussed in
Sec. \ref{Sect:HighEnergyExcitations}. However, there are also numerous
examples where the Fermi liquid concept for low-energy excitations is not
applicable. 

One much discussed item is the separation of charge and spin degrees of freedom
and moreover the appearance of fractional charges. Separate spin and charge
excitations occur always when electron correlations are so strong that the
electrons remain localized. In that case the coupling of spins on different
sites leads to magnetic excitations with energies of order $J$, the intersite
coupling constant. In contrast, charge excitations from the partially filled
inner shells as observed, e.g., by photoelectron spectroscopy have much higher
energies. But this kind of spin-charge separation is trivial and does not
require further consideration. It is well known that in one dimension (1D) spin
and charge degrees of freedom lead to different kinds of excitations even when
the correlations are weak (Luttinger liquid). For a review see, e.g., Ref.
\cite{GogolinBook98}. Spin-charge separation is also found for kink excitations
(solitons) in polyacethylene \cite{Su79}. Those excitations exist even within
the independent electron approximation, but require inclusion of lattice
degrees of freedom. Doped polyacethylene can have also excitations with
fractional charges, again within the one-electron picture but requiring lattice
(chain) deformations \cite{Su81}. In 2D electron correlations, e.g., in
semiconducting inversion layers may become strong when a magnetic field is
applied perpendicular to the plane. The kinetic energy of the electrons is
strongly reduced in a high field and therefore the Coulomb repulsions become
dominant. This results in the fractional quantum Hall effect (FQHE) and
quasiparticles with fractional charges \cite{Laughlin83}. Thus in two
dimensions electron correlations are essential for the appearance of fractional
charges. The same holds true for 3D systems. There it turns out that
excitations with fractional charges may exist in certain geometrically
frustrated lattice structures like the pyrochlore lattice
\cite{Fulde02,Runge04}. There is also spin-charge separation. A Fermi liquid
description is inapplicable here. This intriguing possibility is discussed
in Sec. \ref{Sect:GeometricFrustration}. 

Another interesting case of a breakdown of Landau's Fermi liquid description is
found in Yb$_4$As$_3$. This system is metallic in a high temperature phase and
semimetallic in the low temperature phase \cite{Ochiai90,Kohgi97}. The change
is related to a partial electronic charge order in form of well separated
Yb$^{3+}$ chains with an effective spin 1/2 per site. It is well known that a
Heisenberg spin chain has a specific heat of the form $C = \gamma T$ like a
metal. It is due to spinons which obey Fermi statistics. The reader should note
that in one dimension one can convert fermions into bosons and vice versa
\cite{Luther75}. The coefficient $\gamma$ is large here because of a weak
coupling of the spins in a chain and therefore the specific heat resembles that
of heavy quasiparticles \cite{Fulde95}. But the charge carriers, which are
mainly 4$f$ holes in the high temperature phase consist of a small number of As
4$p$ holes in the low temperature phase \cite{Antonov98}. Therefore one may
speak of spin-charge separation and a breakdown of the conventional Fermi
liquid picture. The one-to-one correspondence between the excitations in the
low temperature phase and those of an independent electron system is no longer
given. Nevertheless the system shows many properties of an ordinary metal with
heavy quasiparticles at low temperature. A detailed discussion of that
interesting material is found in Sec. \ref{sect:ChargeOrdering}.

Another form of deviation from classical Fermi liquid behavior is found in the
cuprates perovskite structures. In the underdoped regime many of their physical
properties show marginal Fermi liquid behavior
\cite{Varma89,Littlewood91}. This implies that they can be described by
assuming a frequency dependence of the electron self-energy $\Sigma(\omega)$
for small values of $\omega$ of the form 

%2.3
\begin{equation}
Re \Sigma (\omega) \sim \omega~{\rm ln} \omega~~~;~~~~~{\rm Im} \Sigma
(\omega) \sim \mid \omega \mid
\label{ReSigom}
\end{equation} 

\noindent instead of the Fermi-liquid form

%2.4
\begin{equation}
Re \Sigma (\omega) \sim \omega~~~;~~~~~{\rm Im} \Sigma (\omega) \sim
\omega^2 ~~~.
\label{ReeSigom}
\end{equation}

\noindent The latter would be required for the one-to-one correspondence of the
excitations. The relations (\ref{ReSigom}) hold only for $T <
\omega$. Otherwise $T$ replaces $\omega$. They yield, e.g., a resistivity $\rho
(T) \sim T$ as is observed in a number of the strongly correlated cuprates. The
microscopic origin of marginal Fermi liquid behavior in the presence of strong
electron correlations has been an open problem. It is also unclear down to
which small $\omega$ (or $T$) values the relations (\ref{ReSigom})  must hold
in order to explain the relevant experiments. It is shown in Sec.
\ref{Sect:HighEnergyExcitations} that marginal Fermi liquid behavior is
obtained for a certain parameter range of the Hubbard Hamiltonian on a square
lattice near half filling when the one-site Coulomb repulsions dominate.

Last but not least, non-Fermi liquid behavior is also found near a quantum
critical point (QCP). It is a point in parameter space at which the system
would undergo a phase transition at $T = 0$, if we were able to reach the limit
of zero temperature. In that case quantum fluctuations instead of thermal
fluctuations determine the critical behavior of the system. It is intuitively
obvious that near a QCP the conventional Fermi liquid description breaks down
since the self-energy is no longer expected to be of the form
(\ref{ReeSigom}). Instead, quantum fluctuations down to arbitrary low wave
numbers will modify this form. The scattering length of
electrons diverges at a QCP while it must remain finite for a Fermi liquid. It
should be emphasized that those features do not require strong electron
correlations but appear also at QCP's of weakly correlated systems. An example
of the latter case is the theory of Moriya \cite{Moriya73} (see also
Ref. \cite{Murata72}) for the resistivity near a QCP of a weak
ferromagnet. Quantum critical points are discussed in Sec.
\ref{Sect:QuantumPhaseTransitions}.

\newpage

\section{Kondo Lattice Systems}
\resetdoublenumb 
\resetdoublenumbf

\label{Sect:KondoLattice}

The Kondo Hamiltonian describes magnetic impurities with free spins
embedded in a metal and interacting with metal electrons via exchange
scattering. The key ingredient is an antiferromagnetic interaction
term
%3.1
\begin{equation}
H_{\textrm{int}} = J{\bf s}(0)\cdot {\bf S}\quad ;\quad J>0
\label{eq:KondoHamiltonianImpurity}
\end{equation}
 where ${\bf S}$ and ${\bf s}(0)$ are the $S = 1/2$ impurity
spin and the conduction electron spin density at the impurity site
which is taken here to be the origin. The model explains the characteristic
Kondo behavior in dilute magnetic alloys which is determined by the
phenomena of asymptotic freedom and confinement. They give rise to
anomalies in the variation with temperature of equilibrium and transport
properties and the {}``quenching'' of the magnetic moment at low
temperatures. 

The presence of a highly complex many-body ground state is highlighted
by the breakdown of conventional perturbation theory which starts
from free electrons and magnetic moments. The divergence of the conduction
electron scattering matrix sets the low-energy characteristic scale
k$_{B}$T$_{K}$ where T$_{K}$ is usually referred to as the Kondo
temperature. Microscopically it arises because the local degeneracy
associated with the magnetic ion is removed through the exchange coupling
between the conduction electrons and the impurity spin. The coupling
leads to the formation of a singlet ground state and low-energy excitations
which can be described in terms of a local Fermi liquid. In close
analogy to confinement the local quasiparticles are composite objects
formed by conduction electrons and magnetic degrees of freedom.

The problem of magnetic impurities is well understood theoretically.
There is a wide variety of techniques available which allow for an
accurate description of the impurity contributions to physical properties.
For detailed discussion, we refer to \cite{Hewson93,KuramotoBook,CoxBook}
and references therein.

Challenging problems are posed by work on concentrated systems, in
particular on Ce-based compounds with heavy quasiparticles (heavy-fermion
systems).  At first glance these systems share many properties with dilute
magnetic alloys. Those materials differ from ordinary metals in that there
exists a characteristic temperature scale $T^{*}\simeq 10 - 100K$, that is much
smaller than the usual Fermi temperatures in ordinary metals, on which the
electronic behavior of the compounds changes drastically. In the
high-temperature regime for $T\gg T^{*}$ the systems behave like ordinary
magnetic rare-earth systems which have itinerant conduction electrons
with conventional masses and well-localized f-electrons. This picture
is derived from the temperature dependence of the specific heat
which exhibits pronounced Schottky anomalies corresponding to crystalline
electric field (CEF) excitations. In addition, the magnetic susceptibility
is Curie-Weiss-like reflecting the magnetic moment of the partially
filled f-shell in a CEF. The low-temperature behavior, however, observed for
$T\ll T^{*}$ is highly unusual and rather surprising: The specific
heat varies approximately linearly with temperature (that is $C = \gamma T +
\dots$), and the magnetic susceptibility, $\chi_{s}$, approaches a Pauli-like
form, becoming almost independent of temperature. Values of the coefficients
$\gamma $ are of the order of $J/molK^{2}$ and consequently two to three
orders of magnitude larger than those of ordinary metals which are
of the order of ${\frac{\pi^{2}}{2}}Nk_{B}{\frac{1}{T_{F}}}\simeq
mJ/molK^{2}$. In addition, the magnetic susceptibility $\chi_{s}$ is enhanced
by a factor of comparable magnitude. A recent survey of the experimental
properties can be found in Refs. \cite{ThalmeierP05,Thalmeier05} and references
therein.

The similarities in the behavior of Ce-based heavy-fermion systems
to that of dilute magnetic alloys have led to the assumption that
these systems are ''Kondo lattices'' where the observed anomalous
behavior can be explained in terms of periodically repeated resonant
Kondo scattering. This ansatz provides a microscopic model for the
formation of a singlet ground state and the appearance of a small
energy scale characterizing the low-energy excitations. The Kondo
picture has been confirmed by the observation of the Kondo resonance
which forms at low temperatures \cite{Reinert01}.

In contrast to the impurity case the Kondo model cannot be solved
for a periodic lattice of magnetic ions. A major difficulty is the
competition between the formation of (local) Kondo singlets and the
lifting of degeneracies by long-range magnetic order. In the high-temperature
regime the moments of the Ce 4$f$-shells are coupled by the RKKY interaction
which can favor parallel as well as antiparallel orientation of the
moments at neighboring sites. Model calculations for two Kondo impurities
\cite{Jayprakash81,Jayprakash82,Jones87} showed that antiferromagnetic
correlations between the magnetic sites weaken the Kondo singlet formation
reducing the characteristic energy scale kT$^{*}$ to rather small
values. Consequences for an extended lattice will be discussed in
the subsequent section on Quantum Phase Transitions. 

The general difficulties in understanding the low temperature properties
of Kondo lattices seem partially due to the lack of an adequate common
{}``language'' for the two regimes where local singlet formation
is dominating on one side and where the magnetic intersite interactions
dominate on the other. Such a language is usually provided by a mean-field
theory which maps the complex quantum problem onto an appropriate
classical model. In the case of the Kondo lattice separate mean-field
descriptions exist for the two regimes which cannot be reconciled
in a straightforward way to provide a unified approach.

In the present section we focus on the heavy Fermi liquid regime.
The corresponding mean-field theory was described, e.g., in Ref. \cite{Fulde88}
and references therein. The majority of recent microscopic studies
of the Kondo lattice adopted the Dynamical Mean Field Theory which
- by construction -explicitly neglects the subtle magnetic intersite
correlations. It accounts for the complex local dynamics in terms
of a local self-energy which has to be determined self-consistently.
The applications include model calculations and a study of the $\gamma
$-$\alpha$-transition in Ce \cite{McMahan03}. A major restriction on the
general validity is imposed by the fact that the 4$f$ valence has to be kept
fixed at unity, i.e., $n_{f} = 1$, throughout the calculation.

The novel feature observed in stoichiometric Ce-compounds is the formation
of narrow coherent bands of low-energy excitations. They give rise
to the temperature dependence of the electrical resistivity which
approximately follows $\rho (T) = \rho_{0} + AT^{2}$. While these findings
unambiguously show that the low-energy excitations are heavy quasiparticles
involving the $f$ degrees of freedom, they nevertheless do not provide
conclusive information on how the latter have to be incorporated into
a Fermi liquid description. A characteristic property of a Fermi liquid
is the existence of a Fermi surface whose volume is determined by
the number of itinerant fermions. It is rather obvious that at high
temperatures the $f$ electrons should be excluded from the Fermi
surface due to their apparent localized character. The latter, however,
is lost at low temperatures. The conjecture that the $f$-degrees
of freedom have to be treated as itinerant fermions and, consequently,
have to be included in the Fermi surface was met with great scepticism
\cite{Kasuya92}\emph{.} This hypothesis implies that the strong local
correlations in Kondo lattices lead to an observable many-body effect,
i. e., a change with temperature of the volume of the Fermi surface.
At high temperatures, the $f$-degrees of freedom appear as localized
magnetic moments, and the Fermi surface contains only the itinerant
conduction electrons. At low temperatures, however, the $f$ degrees
of freedom are now tied into itinerant fermionic quasiparticle excitations
and accordingly, have to be included in the Fermi volume following
Luttinger's theorem. Consequently the Fermi surface is strongly modified.
This scenario \cite{Zwicknagl93a} was confirmed experimentally by
measurements of the de Haas-van Alphen (dHvA) effect
\cite{Lonzarich88,Aoki93,Tautz95} and recent photoemission studies
\cite{Denlinger00,Denlinger01}. 

The present section is mainly devoted to the theoretical description
of the Fermi liquid state at low temperatures. We briefly introduce
the renormalized band scheme which has been devised for calculating
realistic quasiparticle bands of real materials. This is achieved
by combining ab-initio electronic structure methods and phenomenological
concepts in the spirit of Landau theory. Concerning the applications
we shall not elaborate on the results for the Fermi surface and anisotropic
effective masses in CeRu$_{2}$Si$_{2}$ for which we refer to
\cite{Zwicknagl92,Zwicknagl93a}\emph{.} We rather present recent results
concerning the instabilities of the Fermi liquid state in CeCu$_{2}$Si$_{2}$. 

We would like to emphasize the predictive power of the renormalized
band method. In both cases mentioned above (Fermi surface of CeRu$_{2}$Si$_{2}$
and SDW instability in CeCu$_{2}$Si$_{2}$) the effects were first
calculated theoretically and later confirmed experimentally - sometimes
with a delay of up to several years.

\subsection{Fermi-Liquid State and Heavy Quasiparticles: Renormalized Band
  Theory} 

The energy dispersion $E({\bf k})$ of a dilute gas of noninteracting
quasiparticles is parameterized by the Fermi wave vector ${\bf k}_{F}$
and the Fermi velocity ${\bf v}_{F}$
%3.2
\begin{equation}
E({\bf k}) = {\bf v}_{F} (\hat{\bf k}) \cdot ({\bf k} - {\bf k}_{F})
\label{eq:Qtdis}
\end{equation}
 where $\hat{\bf k}$ denotes the direction on the Fermi surface.
The key idea of the renormalized band method is to determine the quasiparticle
states by computing the band structure for a given effective potential.
Coherence effects which result from the periodicity of the lattice
are then automatically accounted for. The quantities to be parameterized
are the effective potentials which include the many-body effects.
The parameterization of the quasiparticles is supplemented by information
from conventional band structure calculations as they are performed
for {}``ordinary'' metals with weakly correlated electrons. The
periodic potential leads to multiple-scattering processes involving
scattering off the individual centers as well as to propagation between
the centers which mainly depends on the lattice structure and is therefore
determined by geometry. The characteristic properties of a given material
enter through the information about single-center scattering which
can be expressed in terms of a properly chosen set of phase shifts $\{\eta_{\nu
}^{i}(E)\}$ specifying the change in phase of a wave incident on site i with
energy E and symmetry $\nu$ with respect to the scattering center. Within
the scattering formulation of the band structure problem the values
of the phase shifts at the Fermi energy $\{\eta_{\nu }^{i}(E_{F})\}$
together with their derivatives $\left\{ \left(d\eta _{\nu
}^{i}/dE\right)_{E_{F}}\right\} $ determine the Fermi wave vectors ${\bf
  k}_{F}$ and the Fermi velocity ${\bf v}_{F}$.

A detailed description of the renormalized band method is given in
Ref. \cite{Zwicknagl92}. The first step is a standard LDA band-structure
calculation by means of which the effective single-particle potentials
are self-consistently generated. The calculation starts, like any
other ab-initio calculation, from atomic potentials and structure
information. In this step, no adjustable parameters are introduced.
The effective potentials and hence the phase shifts of the conduction
states are determined from first principles to the same level as in
the case of {}``ordinary'' metals. The f-phase shifts at the lanthanide sites,
on the other hand, are described by a resonance type expression 
%3.3
\begin{equation}
\tilde{\eta }_{f}\simeq \arctan \frac{\tilde{\Delta }_{f}}{E-\tilde{\epsilon
  }_{f}}
\label{eq:efren}
\end{equation}
 which renormalizes the effective quasiparticle mass. One of the two
remaining free parameters $\widetilde{\epsilon }_{f}$ and $\tilde{\Delta }_{f}$
is eliminated by imposing the condition that the charge distribution
is not significantly altered as compared to the LDA calculation by
introducing the renormalization. The renormalized band method devised
to calculate the quasiparticles in heavy-fermion compounds thus is
essentially a one-parameter theory. We mention that spin-orbit and
CEF splittings can be accounted for in a straightforward manner
\cite{Zwicknagl92}.

\subsection{Heavy Fermions in CeRu$_{2}$Si$_{2}$ and CeCu$_{2}$Si$_{2}$}

\begin{figure}
\includegraphics[width=0.20\textwidth]{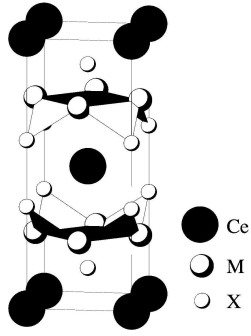}
\caption{Conventional unit cell of the ThCr$_{2}$Si$_{2}$ and CeM$_{2}$X$_{2}$
structure where M = Cu, Ni, Ru, Rh, Pd, Au, .. and X = Si, Ge. }
\label{fig:ThCr2Si2Structure}
\end{figure}

The archetype heavy fermion superconductor CeCu$_{2}$Si$_{2}$ as
well as CeRu$_{2}$Si$_{2}$ crystallize in the tetragonal ThCr$_{2}$Si$_{2}$
structure. The unit cell is shown in Fig.~\ref{fig:ThCr2Si2Structure}. 

%fig3.1%%%%%%%%%%%%%%%%%%%%%%%%%%%%%%%%%%%%%%%%%%%%%%%%%%%%%%%%%%%%%%%%%%
%%%%%%%%%%%%%%%%%%%%%%%%%%%%%%%%%%%%%%%%%%%%%%%%%%%%%%%%%%%%%%%%%%%%%%%%%
 
To study the electronic structure, we compare the results of two different
models, i.e., treating the Ce 4$f$ degrees of freedom as localized
(atomic like) states and as delocalized yet strongly renormalized
electrons. The first procedure provides a good quantitative description
of the properties at elevated temperatures, high excitation energies,
and above the metamagnetic transition. The latter ansatz yields a
model for the Fermi liquid state.

The low-temperature behavior of CeRu$_{2}$Si$_{2}$ is well described
by a paramagnetic Fermi liquid with weak residual interactions. The
relevant low-energy excitations are heavy quasiparticles as inferred
from the linear specific heat coefficient $\gamma \simeq$ 350 mJ/molK$^{2}$
\cite{Steglich85b}. In the local moment regime, the Fermi surface
is determined exclusively by the conduction states. The strongly renormalized
Fermi liquid state, on the other hand, is described by the renormalized
band method using $\tilde{\Delta }_{f}\simeq$ 10 K in Eq.~(\ref{eq:efren})
for the intrinsic width of the quasiparticle band. The value is consistent
with inelastic neutron data \cite{Regnault88} as well as thermopower
and specific heat data \cite{Steglich85b}. CEF effects are accounted
for by adopting a $\Gamma_{7}$ ground state. The details of the
calculation are described in Ref. \cite{Zwicknagl92}.

The renormalized band scheme gives the correct Fermi surface topology
for CeRu$_{2}$Si$_{2}$ and thus consistently explains the measured
dHvA data \cite{Zwicknagl90,Zwicknagl93a,Zwicknagl92}. The character
of quasiparticles in CeRu$_{2}$Si$_{2}$ varies quite strongly over
the Fermi surface. The validity of the Fermi liquid picture is concluded
from a comparison of the effective masses on Fermi surface sheets
with large $f$ contribution. From the large linear specific heat
the renormalized band scheme deduces a characteristic energy kT$^{*}\simeq $
10 K and predicts heavy masses of order m$^{*}$/m $\simeq $ 100.
This value was confirmed by experiments \cite{Albessard93} where
the $\psi $ orbit with m$^{*}$/m $\simeq $ 120 was observed. The
corresponding Fermi surface cross-section is in agreement with estimates
from the renormalized band theory. This proofs that the heavy quasiparticles
exhaust the low-energy excitations associated with the $f$-states. 

\begin{figure}
\includegraphics[width=0.40\textwidth]{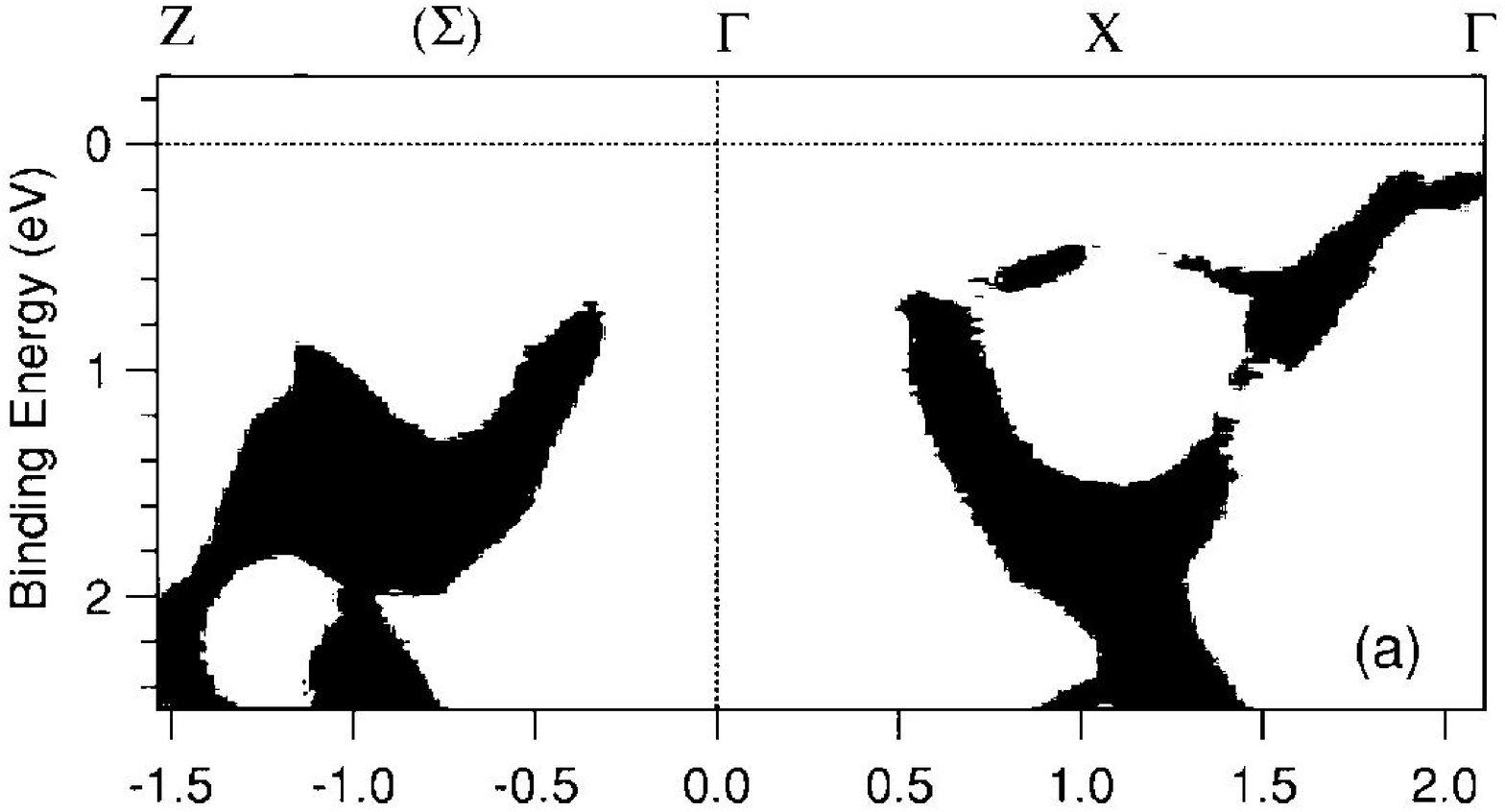}
\includegraphics[width=0.40\textwidth]{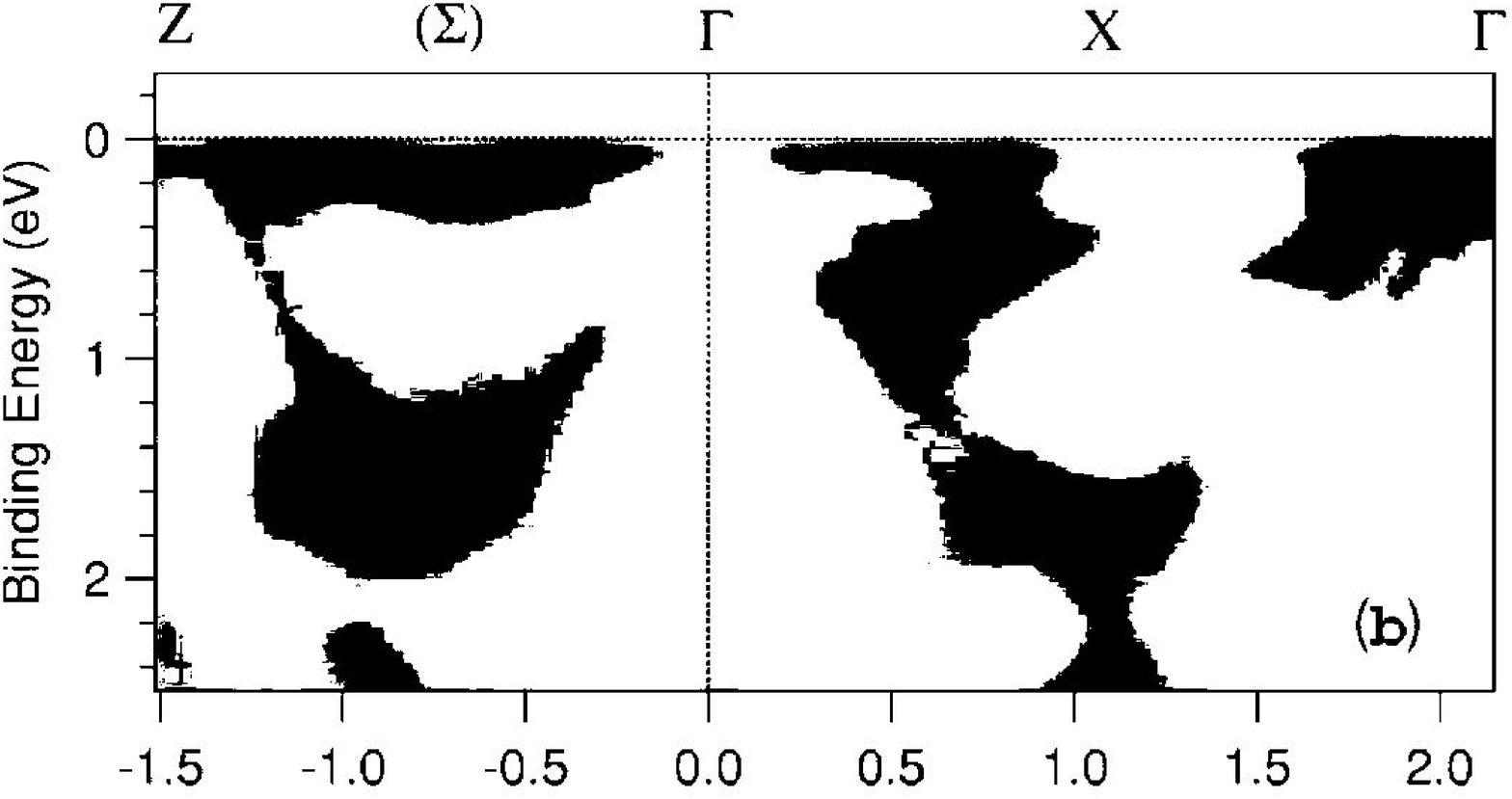}
\caption{Photoemission results for (a) LaRu$_{2}$Si$_{2}$ in comparison
to (b) CeRu$_{2}$Si$_{2}$ at T = 25 K, i.e., above the Kondo temperature
T$^{*}$ = 15 K of that system. Band structures are very similar for both
compounds. (After \cite{Denlinger01}) .} 
\label{fig:CeRu2Si2flocalizedPES}
\end{figure}

The change in volume of the Fermi surface when going from $T\ll T^{*}$
to $T\gg T^{*}$ is observed by comparing the Fermi surface of
CeRu$_{2}$Si$_{2}$ to that of its ferromagnetic isostructural counterpart
CeRu$_{2}$Ge$_{2}$ where the $f$-states are clearly localized. In a series of
beautiful experiments \cite{King91} it was demonstrated that the Fermi surfaces
of these two compounds are rather similar. However, the enclosed Fermi
volume is smaller in the case of CeRu$_{2}$Ge$_{2}$, the difference
being roughly one electron per unit cell. More direct evidence is
provided by recent photoemission experiments (see
Fig.~\ref{fig:CeRu2Si2flocalizedPES}). Denlinger et al. \cite{Denlinger01} have
shown that at temperatures around 25 K, the Fermi surface of
CeRu$_{2}$Si$_{2}$, is that of its counterpart LaRu$_{2}$Si$_{2}$ which has no
$f$ electrons. 

%fig3.2%%%%%%%%%%%%%%%%%%%%%%%%%%%%%%%%%%%%%%%%%%%%%%%%%%%%%%%%%%%%%%%
%%%%%%%%%%%%%%%%%%%%%%%%%%%%%%%%%%%%%%%%%%%%%%%%%%%%%%%%%%%%%%%%%%%%%%%

%fig3.3%%%%%%%%%%%%%%%%%%%%%%%%%%%%%%%%%%%%%%%%%%%%%%%%%%%%%%%%%%%%%%%%
\begin{figure}
\includegraphics[width=0.50\linewidth]{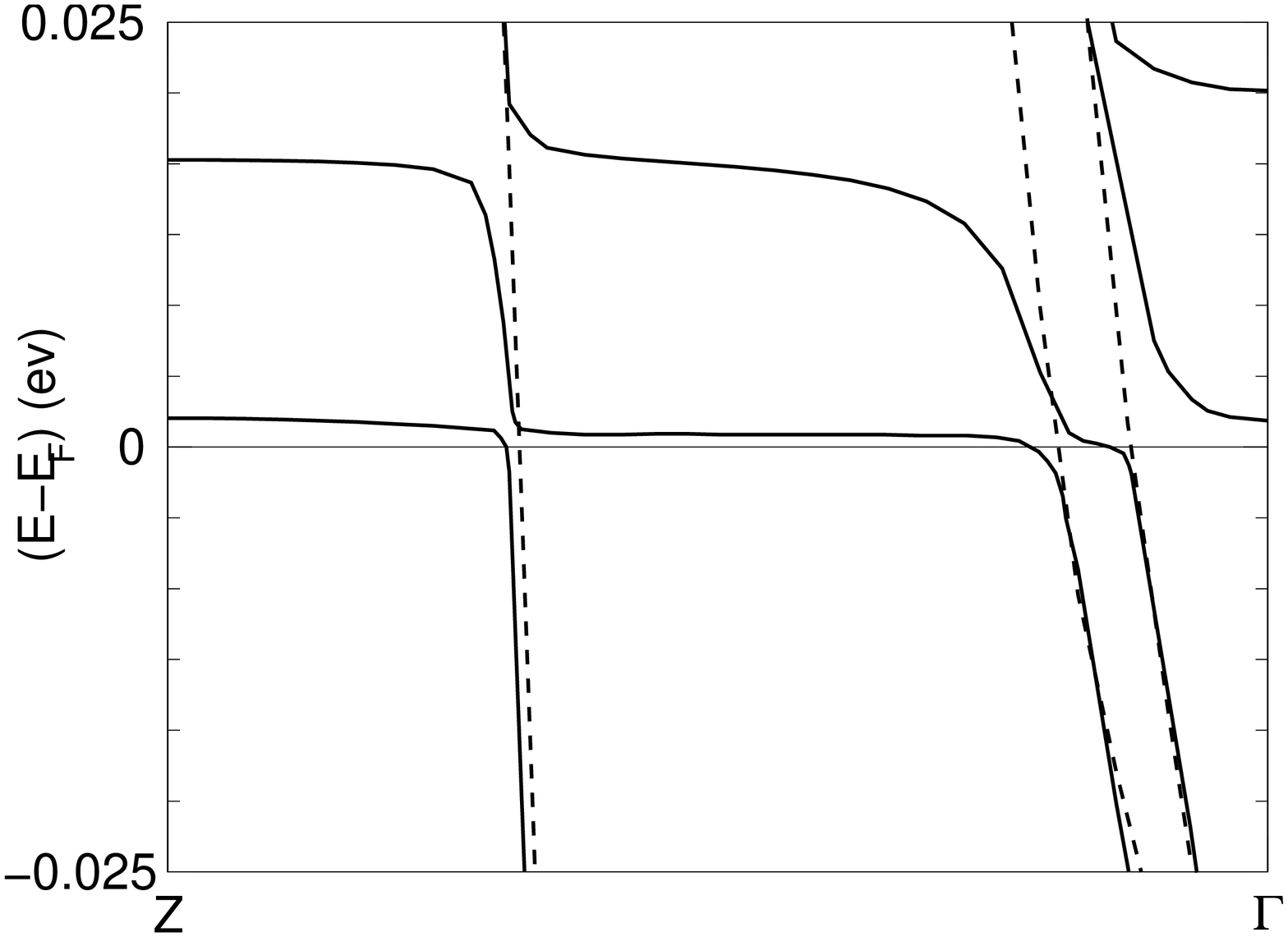}
\vspace*{-8cm}
\caption{Band dispersion for CeCu$_2$Si$_2$ along $Z-\Gamma$ for low
  temperatures $T \ll T^*$ 
  (full lines) and high temperatures (dashed lines). The formation of the heavy
  quasiparticles leads to a characteristic bending in the occupied part of the
  spectrum. 
\label{fig:CeRu2Si2HeavyBand}}
\end{figure}
%%%%%%%%%%%%%%%%%%%%%%%%%%%%%%%%%%%%%%%%%%%%%%%%%%%%%%%%%%%%%%%%%%%%%%%%
 
At this point the general question arises how the formation of heavy
quasiparticles is reflected in the angular resolved photoelectron
spectroscopy (ARPES) data. A major difficulty stems from the fact
that photoemission experiments probe the occupied part of the spectrum.
The most dramatic changes, however, are expected to occur in the empty
part. In Figure \ref{fig:CeRu2Si2HeavyBand} we compare
the dispersion of the heavy quasiparticle band at low temperatures
to its light high-temperature counterpart. In the occupied part the
main difference is a bending close to the Fermi energy which changes
the volume of the Fermi surface. The characteristic bending was recently
observed in CeCoIn$_{5}$ \cite{Koitzsch05}.

Let us now turn to the heavy fermion superconductor CeCu$_{2}$Si$_{2}$
which exhibits a highly complex phase diagram at low temperatures which is
discussed in Sec. \ref{Sect:LowTemp}.
It results from an extreme sensitivity of the physical properties
with respect to variations of the stoichiometry and external magnetic
fields.

To calculate the quasiparticle bands in CeCu$_{2}$Si$_{2}$ by means
of the renormalized band method, we adopt the doublet-quartet CEF
scheme suggested in Ref. \cite{Goremychkin93}. The ground state is separated
from the excited quartet by $\delta \simeq$ 330 K.

%fig3.4%%%%%%%%%%%%%%%%%%%%%%%%%%%%%%%%%%%%%%%%%%%%%%%%%%%%%%%%%%%%%%
\begin{figure}
\includegraphics[clip,width=10cm]{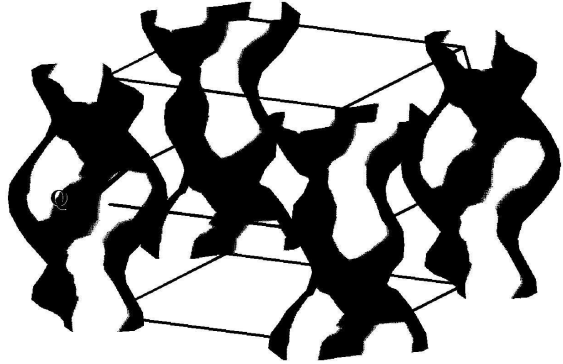}
%\vspace*{-.5cm}
\caption{CeCu$_{2}$Si$_{2}$: Main Fermi surface sheet of heavy quasiparticles
  (m$^{*}$/m $\simeq$ 500) calculated with the renormalized band method. It
  consists of modulated columns which are oriented parallel to the tetragonal
  axis. The calculations adopt the CEF scheme of Ref. \cite{Goremychkin93}
  consisting of a singlet ground state separated from an excited quartet by a
  CEF splitting $\delta \simeq $ 330 K. Therefore $\delta \gg $ T$^{*}\simeq$
  10 K (obtained from the $\gamma $-value). The nesting vector ${\bf Q} =
  (0.23, 0.23, 0.52)$ connects flat parts (''nesting'') of the Fermi
  surface. (After \cite{Zwicknagl93}) 
\label{fig:CeRu2Si2HeavyBandDispersion}}
\end{figure}
%%%%%%%%%%%%%%%%%%%%%%%%%%%%%%%%%%%%%%%%%%%%%%%%%%%%%%%%%%%%%%%%%%%%%%

The results for the Fermi surface \cite{Zwicknagl93,Pulst93} can
be summarized as follows: We find two separate sheets of the Fermi
surface for heavy and light quasiparticles. The light quasiparticles
have effective masses of the order of $m^{*}/m\simeq $ 5. They can
be considered as weakly renormalized conduction electrons. Of particular
interest are heavy quasiparticles of effective masses 
$m^{*}/m\simeq$ 500 which are found on a separate sheet. This surface
whose shape (see Figs. \ref{fig:CeRu2Si2HeavyBandDispersion} and
\ref{fig:Gamma3OrderParameter}) is rather 
different from the corresponding LDA surface mainly consists of columns
parallel to the tetragonal axis and of small pockets. The topology
of the Fermi surface suggests that the strongly correlated Fermi liquid
state should become unstable at sufficiently low temperatures. Firstly,
it exhibits pronounced nesting features which may eventually lead
to the formation of a ground state with a spin-density modulation.
This will be discussed in detail below. Secondly, the topology of
this surface depends rather sensitively on the position of the Fermi
energy. The band filling and hence the $f$-valence are critical quantities.
Reducing the $f$-occupancy from the initial value of $n_{f}\simeq 0.95$
by approximately $2 \%$ leads to changes in the topology as shown
in Refs. \cite{Pulst93,Zwicknagl93}. As a result, the quasiparticle density
of states (DOS) exhibits rather pronounced structures in the immediate
vicinity of the Fermi energy which, in turn, can induce instabilities
\cite{Kaganov79}.

\subsection{Low-Temperature Phase Diagram of CeCu$_{2}$Si$_{2}$}
\label{Sect:LowTemp}

%fig3.5%%%%%%%%%%%%%%%%%%%%%%%%%%%%%%%%%%%%%%%%%%%%%%%%%%%%%%%%%%%%%
\begin{figure}
\includegraphics[width=0.40\columnwidth,angle=270,origin=c]{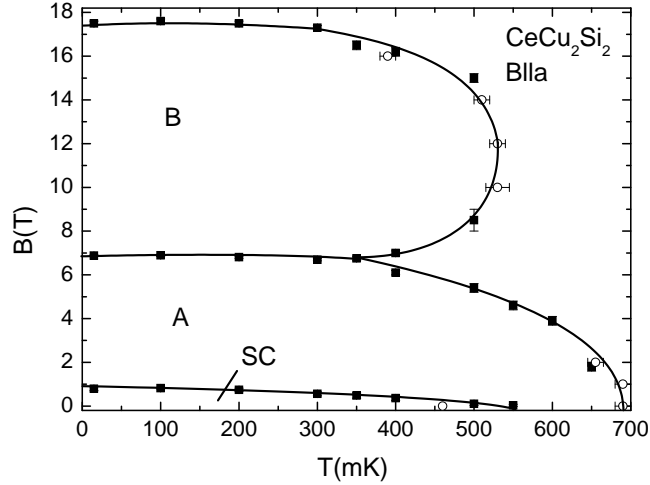}   
\caption{B-T phase diagram of CeCu$_{2}$Si$_{2}$ for B $\parallel$ a. Original
  version from Ref. \cite{Bruls90}, completed version from
  Ref. \cite{Weickert03}. In this sample the A-phase is expelled from the
  superconducting (SC) region (no coexistence).}  
\label{fig:CeCu2Si2BTPhaseDiagram}
\end{figure}
%%%%%%%%%%%%%%%%%%%%%%%%%%%%%%%%%%%%%%%%%%%%%%%%%%%%%%%%%%%%%%%%%%%%%%
 
The phase diagram of CeCu$_2$Si$_2$ contains three different phases: the A and
the B phase and a superconducting phase. In some samples the superconducting
phase expels the A phase while in other samples the two phases may
coexist. While the A phase has been identified as a spin-density wave phase as
discussed below, the character of the B phase has remained unknown. The
instability may result from a reconstruction of the Fermi surface
(\cite{Zwicknagl93}). Much effort has been devoted to the characterization of
the A phase which originally had the appearance of a 'hidden order'
phase. However, later a spin-density wave character was first inferred from
resistivity results \cite{Gegenwart98} and was supported by specific heat and
high-resolution magnetization measurements \cite{Steglich00}. The transition
temperature T$_{A}$ is suppressed by increasing the 4$f$-conduction electron
hybridization and eventually vanishes. This can be achieved by applying
hydrostatic pressure or choosing a few percent excess of Cu. The ordered
moments are expected to be rather small. 

\begin{figure}
\includegraphics[width=0.40\linewidth]{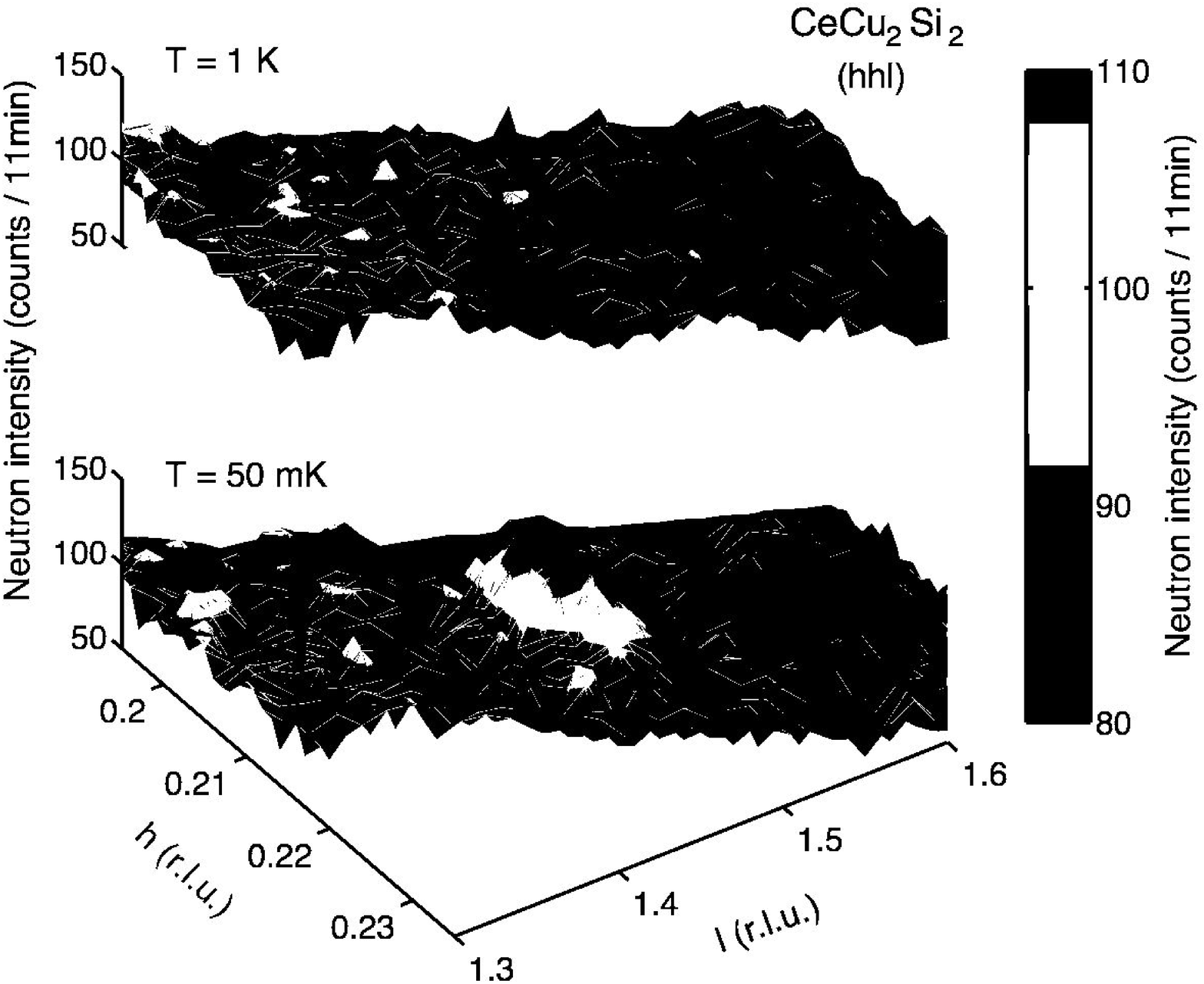}\hspace{1cm}
\includegraphics[width=0.40\linewidth]{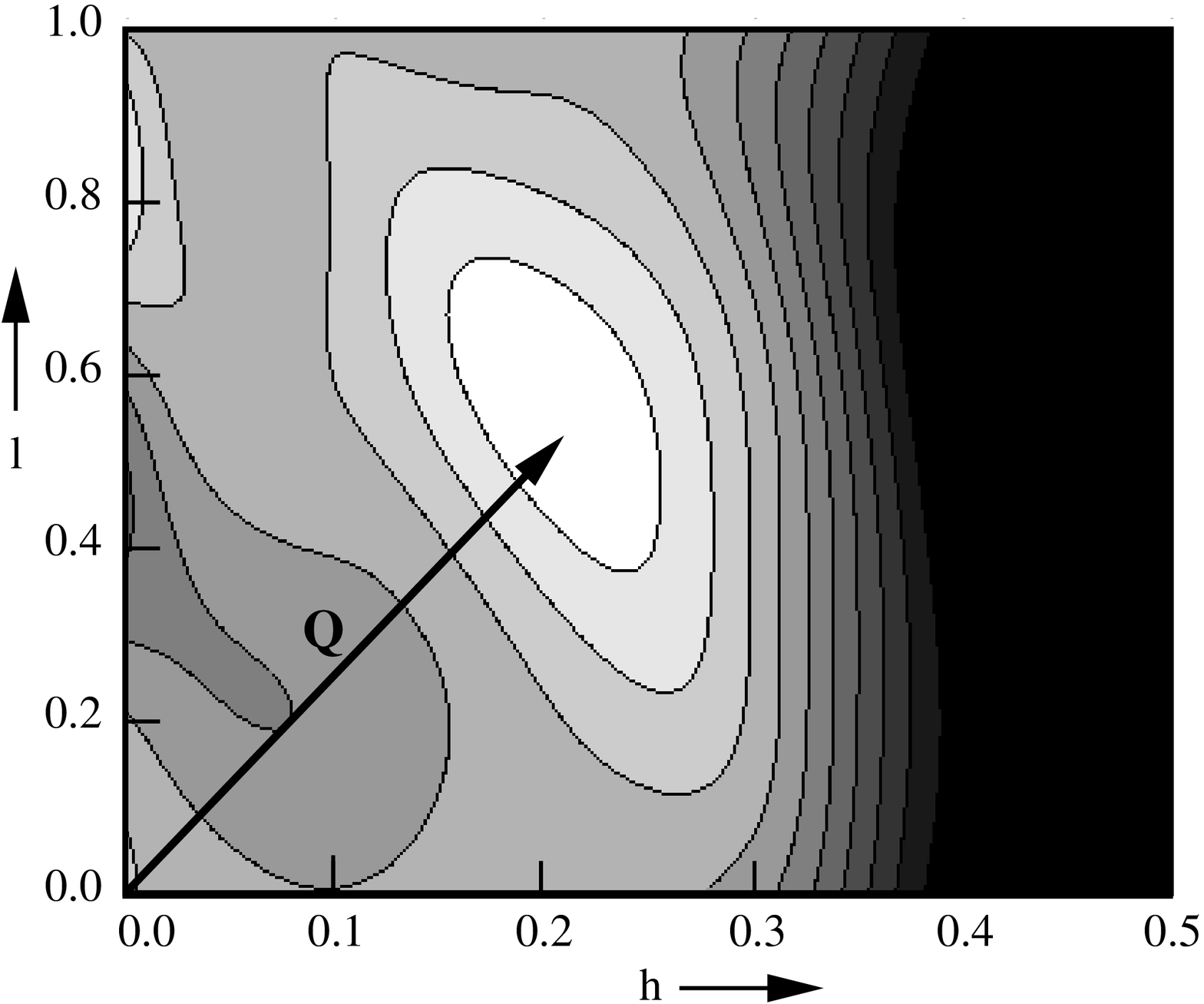}
\caption{Left panel: Neutron diffraction intensity in CeCu$_{2}$Si$_{2}$
at temperature above and below the A- phase transition temperature
T$_{A}$. Incommensurate peak is at ${\bf Q}$ = (0.22, 0.22, 0.55). 
(After \cite{Stockert04}). Right panel: Nesting of heavy FS columns
(Fig.~\ref{fig:CeRu2Si2HeavyBandDispersion}) leads to a peak in the static
susceptibility $\chi ({\bf q})$ at {\bf q} = {\bf Q}. Intensity map of $\chi
({\bf q})$ (value increasing from dark to bright) in the reciprocal (h, h,
l)-plane as calculated for the renormalized bands at T = 100 mK. The
\emph{experimental} ${\bf Q}$ at 50 mK from the left panel shows perfect
agreement with the calculated maximum position of $\chi ({\bf q})$.}
\label{fig:CeCu2Si2fBandFermiSurfaceNesting}
\end{figure}

The important question in this context is: What is the origin of the
antiferromagnetic correlations, showing up in the A phase? How do they arise in
the heavy fermion state and finally, how do they affect the heavy
quasiparticles? The key to the answers comes from the Fermi surface of
CeCu$_{2}$Si$_{2}$ and its nesting properties. As shown in
Fig.~\ref{fig:CeRu2Si2HeavyBandDispersion} there are parallel portions which
are connected by a wave vector close to $\left(1/4, 1/4, 1/2\right)$ . As a
consequence, the static susceptibility $\chi ({\bf q})$ exhibits a maximum for
momentum transfer ${\bf q}$ close to the nesting vector
(Fig~. \ref{fig:CeCu2Si2fBandFermiSurfaceNesting}). 

%fig3.6%%%%%%%%%%%%%%%%%%%%%%%%%%%%%%%%%%%%%%%%%%%%%%%%%%%%%%%%%%%%%%
%%%%%%%%%%%%%%%%%%%%%%%%%%%%%%%%%%%%%%%%%%%%%%%%%%%%%%%%%%%%%%%%%%%%
 
Recent neutron scattering experiments \cite{Stockert04}
(Fig.~\ref{fig:CeCu2Si2fBandFermiSurfaceNesting}) for the stoichiometric
compound (x = 0) show a spin-density wave (SDW) which forms below T$_{N}\simeq
$ 0.7 K. The experimental propagation vector $\bf Q$ is close to (0.22, 0.22,
0.55) and the ordered moment amounts to $\mu \simeq 0.1\mu_{B}$. These findings
show that the SDW in CeCu$_{2}$Si$_{2}$ arises out of the renormalized Fermi
liquid state. The transition is driven by the nesting properties of the heavy
quasiparticles. 

\begin{figure}
\includegraphics[width=0.40\columnwidth]{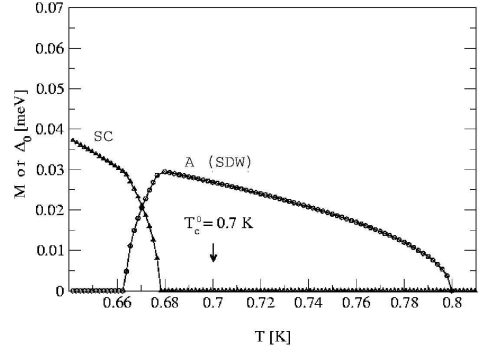}
\caption{Temperature dependence of the superconducting (SC) and magnetic A
  (SDW) order parameters in a AS crystal. The sublattice magnetization $M =
  h_0(T)$ and the amplitude $\Delta_0$ of the SC order parameter are calculated
  from Eq. (\ref{eq:FinalSelfcon}) using realistic
  quasiparticles and the experimentally determined propagation vectors for the
  A phase. A SC order parameter $\Delta({\bf k}) = \Delta_0 (\cos{k_{x}a} -
  \cos {k_{y}a})$ can form in the A phase. Below the superconducting transition
  temperature $T < T_{c} < T_{N}^{(1)}$ the two order parameters coexist and
  compete. Itinerant SDW antiferromagnetism is expelled at a temperature
  $T_{N}^{(2)} < T_{c}$. (After \cite{Neef04,Neef04a})} 
\label{fig:OrderParameterAS}
\end{figure}

Having classified the nature of the A phase we next turn to the question
how the latter competes with superconductivity. Itinerant electron
antiferromagnetism as realized in the A phase and superconductivity
both form in the system of the heavy quasiparticles. Their interplay
therefore depends sensitively on the geometric properties of the paramagnetic
Fermi surface and the symmetries of the ordered phases. This can be
seen from realistic model calculations investigating the variation
with temperature of the two order parameters in crystals where the
two ordering phenomena coexist in some temperature range (''AS-type'' crystals)
\cite{Neef04,Neef04a}. Fig. \ref{fig:OrderParameterAS} summarizes
the results for an unconventional superconducting state with $\Gamma_{3}$
(d-wave) symmetry of the gap function, i.e., $\Delta({\bf k}) \sim (\cos
{k_{x}a} - \cos {k_{y}a})$. 

%fig3.7%%%%%%%%%%%%%%%%%%%%%%%%%%%%%%%%%%%%%%%%%%%%%%%%%%%%%%%%%%%%%%%%
%%%%%%%%%%%%%%%%%%%%%%%%%%%%%%%%%%%%%%%%%%%%%%%%%%%%%%%%%%%%%%%%%%%%%%%5

The theory leading to the results displayed in Fig. \ref{fig:OrderParameterAS}
starts from the model Hamiltonian 
%3.4
\begin{equation}
H = H_{0} + H_{int}
\label{eq:HamDef}
\end{equation}
 where the free quasiparticles are described by 
%3.5
\begin{equation}
H_{0} = \sum_{{\bf k}\sigma} E\left({\bf k}\right) c_{{\bf k}\sigma }^{\dagger
} c_{{\bf k}\sigma } \quad .
\end{equation}
 The creation (annihilation) operators for quasiparticles with wavevector
${\bf k}$, (pseudo) spins $\sigma = \pm 1$ and energy $E({\bf k})$
are denoted by $c_{{\bf k}\sigma }^{\dagger}$($c_{{\bf k}\sigma }$).
The energies which are measured relative to the Fermi level are calculated
within the Renormalized Band Scheme. The residual interactions in
the strongly renormalized Fermi liquid are assumed to be repulsive
for short separations while being attractive for two quasiparticles
of opposite momenta on neighboring sites. The former favors the formation
of a SDW while the latter gives rise to a superconducting instability.
Adopting a mean-field approximation yields 
%3.6
\begin{equation}
H_{int} \rightarrow H_{SDW} + H_{SC}
\end{equation}
 where 
%3.7
\begin{equation}
H_{SDW} = -\sum _{{\bf k}\sigma} \frac{\sigma }{2} \sum_{{\bf Q}_{j}} \left( h
\left( {\bf Q}_{j} \right) c_{{\bf k}\sigma}^{\dagger } c_{{\bf k} + {\bf
	Q}_{j}\sigma} + h.c. \right)
\label{eq:HSDW}
\end{equation}
 and 
%3.8
\begin{equation}
H_{SC} = \frac{1}{2} \sum_{{\bf k} \sigma \sigma'} \left( \Delta_{\sigma
  \sigma'} \left( {\bf k} \right) c_{{\bf k} \sigma}^{\dagger } c_{-{\bf k}
  \sigma'}^{\dagger } + h.c. \right)\; .
\label{eq:HSC}
\end{equation}
 The periodically modulated magnetization associated with the SDW
with propagation vectors ${\bf Q}_{j}$ as well as the superconducting
pair potential $\Delta_{\sigma \sigma'}$ have to be determined selfconsistently %3.9 
\begin{equation}
h \left( {\bf Q}_{j} \right) = \frac{U}{L} \sum_{{\bf k} \sigma} \frac{\sigma
}{2} \left\langle c_{{\bf k} + {\bf Q}_{j} \sigma}^{\dagger} c_{{\bf k} \sigma
} \right\rangle 
\label{eq:SelfconSDW}
\end{equation}
 and 
%3.10
\begin{equation}
\Delta_{\sigma \sigma'} \left( \bf k\right) = \frac{1}{L} \sum_{{\bf k}'
  \sigma'' \sigma'''} g_{s \sigma \sigma';\sigma''\sigma'''} \left( {\bf k},
  {\bf k}' \right) \left\langle c_{-{\bf k}'\sigma''} c_{{\bf k}'\sigma'''}
  \right\rangle \; . 
\label{eq:SelfconSC}
\end{equation}
 where the strength $U$ of the local Hubbard-type repulsion is of
the order of the quasiparticle band width $k_{B}T^{*}$ and $g_{\sigma \sigma';
  \sigma'' \sigma'''} \left( {\bf k}, {\bf k}' \right)$ is the effective pair
attraction. The $\bf k$-summation runs over the entire paramagnetic Brillouin
zone and $L$ denotes the number of lattice sites. The expectation values
denoted by $\left\langle \ldots \right\rangle$ have to evaluated with the
eigenstates of the mean-field Hamiltonian $H_{MF} = H_{0} + H_{SDW} + H_{SC}$,
and consequently depend upon the order parameters. Therefore the
self-consistency equations are coupled. 

The mean-field Hamiltonian implicitly assumes that the amplitudes
of both order parameters are small. In particular, we neglect here
the pairing amplitudes of the form $\left\langle c_{-{\bf k}\sigma''} c_{{\bf
	k} + {\bf Q}_{j}\sigma'''}\right\rangle$. The latter are important when the
gaps introduced by the antiferromagnetic order into the quasiparticle spectrum
are large on the scale set by superconductivity. For a discussion of this point
we refer to Refs. \cite{Zwicknagl81,Fulde82}. 

The periodically modulated magnetization associated with the SDW acts
on the conduction electrons like a periodic spin-dependent potential
which we approximate by 
%3.11
\begin{equation}
h({\bf r}) = \sum_{{\bf Q}_{j}}h \left( {\bf Q}_{j} \right) e^{i{\bf Q}_{j}
  \cdot {\bf r}} 
\label{eq:AFMolecularField}
\end{equation}
 with the same amplitudes $h\left( {\bf Q}_{j} \right) = h_{0}$ for
the eight commensurate wave vectors ${\bf Q}_{j}\in \left\{ \left(\pm \frac{\pi
}{2a}, \pm \frac{\pi }{2a}, \pm \frac{\pi }{c} \right) \right\} $.
The magnetic superstructure breaks the translational invariance of
the underlying lattice but preserves the point group symmetry. The
mean-field Hamiltonian, however, is invariant under translations with
%3.12
\begin{equation}
{\bf a}'_{1} = \left(2a, 2a, 0 \right)\, ;\, {\bf a}'_{2} = \left( 2a, -2a, 0
\right)\,;\, {\bf a}'_{3} = \left( 2a, 0, c \right)\; .
\label{eq:AFLattice}
\end{equation}
 The volume of the magnetic supercell is 16 times the volume of the
paramagnetic unit cell. As a result the Brillouin zone is reduced
and the quasiparticle states are modified by extra Bragg reflections.
The opening of new gaps is important at sufficiently low temperatures
$T\ll T_{N}$ where $T_{N}$ is the ordering (N\'eel) temperature.

The order parameter $\Delta_{\sigma \sigma'}$ behaves as a two-fermion
wave function in many respects. This is expressed by the fact that
an off-diagonal long-range order (ODLRO) parameter is not the thermal
expectation value of a physical observable but rather a complex pseudo-
wave function describing quantum-phase correlations on the macroscopic
scale of the superconducting coherence length. Its phase is a direct
signature of the broken gauge invariance in the superconducting condensate.

Experiment (strongly) suggests that superconductivity in CeCu$_{2}$Si$_{2}$
occurs with anisotropic even-parity (pseudo-) spin singlet pairing.
We therefore restrict ourselves to this case characterized by a scalar
order parameter 
%3.13
\begin{equation}
\Delta_{\sigma \sigma'} ({\bf k}) = \phi ({\bf k}) \left( i\sigma
_{2} \right)_{\sigma \sigma '}
\label{eq:DeltaEvenParity}
\end{equation}
\noindent where $\phi ({\bf k})$ is a complex amplitude and $\sigma_2$ denotes
the Pauli matrix in spin space.

Since the two ordering temperatures, i.e., the antiferromagnetic
N\'eel temperature T$_{A}$ and the superconducting T$_{c}$ are very
close to each other we focus on pair states which are compatible with
the translational symmetry of the paramagnetic lattice. The corresponding
functions are listed in Refs. \cite{Sigrist91,Konno89,Ozaki89,Zwicknagl92}.
In the explicit calculations we restrict ourselves to one-dimensional
representations for simplicity. The generalization to multi-dimensional
representations is rather straightforward \cite{Neef04}.

Finally, the variation with momentua $\mathbf{k}$ and $\mathbf{k}'$
of the quasiparticle attraction $g(\mathbf{k},\mathbf{k}')$ is expanded
in terms of the basis functions belonging to the $\kappa $-th row,
$\kappa =1,\ldots ,d(\Gamma )$ of the $d(\Gamma )$-dimensional irreducible
representation representation of the symmetry group
%3.14
\begin{equation}
g(\mathbf{k},\mathbf{k}') = \sum _{\Gamma }g_{\Gamma }\sum_{\kappa =
  1}^{d(\Gamma )}\varphi_{\Gamma_\kappa }(\mathbf{k})\varphi_{\Gamma_\kappa
}^{*}(\mathbf{k}')\quad .
\end{equation}

We further simplify the problem by focusing on the states $\phi_{\Gamma }({\bf
  k})$ with the symmetry which yields the strongest quasiparticle attraction
  $g_{\Gamma }$. Assuming that this most stable order parameter is
  non-degenerate the self-consistency condition is
%3.15
\begin{equation}
\phi ({\textbf k})= -g_{\Gamma _{0}}\varphi_{\Gamma _{0}}({\textbf
  k})\frac{1}{L}\sum_{{\textbf k}'}\varphi_{\Gamma _{0}}^{*}({\textbf
  k}')\left\langle c_{-{\textbf k}'\downarrow }c_{{\textbf
	k}'\uparrow}\right\rangle \quad .
\end{equation}

To solve the mean-field Hamiltonian Eqs.(\ref{eq:HSDW}) and (\ref{eq:HSC})
we adopt the Nambu formalism which allows us to reduce the mean-field
Hamiltonian to single-particle form in particle-hole space, i.e., 
%3.16
\begin{equation}
H = \sum_{\bf k}^{AFBZ} \Psi_{\bf k}^{\dagger } \left\{ {\hat{\bf E}} \left(
\bf k \right) \hat{\tau }_{3} + {\hat{\bf h}} 1 + {\hat{\bm \Delta}} \left(
	{\bf k} \right) \hat{\tau}_{1} \right\} \Psi _{\bf k}\; .
\label{eq:HamiltonNambu}
\end{equation}
 The Nambu spinors $\Psi _{\bf k}$ have 32 components and are
defined as 
%3.17
\begin{equation}
\Psi_{\bf k}^{\dagger} = \left( c_{{\bf k} \uparrow}, c_{{\bf k} + {\bf Q}_{1}
	\uparrow}, \ldots, c_{-{\bf k} \downarrow}^{\dagger }, c_{-{\bf k} - {\bf
	Q}_{1} \downarrow}^{\dagger}, \ldots \right)~~~.
\label{eq:NambuSpinor}
\end{equation}
 They account for the coherent superposition of particles and holes
which is the characteristic feature of the superconducting state.
Here $\hat{1}$, $\hat{\tau }_{1}$ and $\hat{\tau }_{3}$ denote
the unit matrix and the Pauli matrices in particle-hole space. The
sixteen wave vectors ${\bf Q}_{0} = 0, {\bf Q}_{1}, \ldots, {\bf Q}_{15}$
are the reciprocal lattice vectors appearing in the antiferromagnetic
phase. The set includes the eight propagation vectors of the SDW and
their higher harmonics. The $\bf k$-summation is restricted to the reduced
Brillouin zone (AFBZ) of the antiferromagnetic state defined by the
SDW. The structure of the Hamiltonian in particle-hole space is 
%3.18
\begin{equation}
{\hat{H}} ({\bf k}) = 
\left(\begin{array}{cc}
 {\hat{\bf E}} \left( {\bf k} \right) + {\hat{\bf h}} & \hat{\bm \Delta} \\
 \hat{\bm \Delta}  & - {\hat{\bf E}} \left( {\bf k} \right) + {\hat{\bf h}}
\end{array}
\right)
\label{eq:HamiltonParticleHole}
\end{equation}
 where the $16\times 16$-diagonal matrix contains the quasiparticle
energies of the paramagnetic normal phase 
%3.19
\begin{equation}
\left({\hat{\bf E}}({\bf k})\right)_{{\bf Q}_i{\bf Q}_j} = \delta_{{\bf
	Q}_{i}{\bf Q}_{j}} E \left( {\bf k} + {\bf Q}_{i} \right)~~~.
\end{equation}
The SDW acts like an effective magnetic field. The modulated spin density leads
to Umklapp scattering which is accounted for by the matrix
%3.20
\begin{equation}
\hat{{\bf h}} = -h_{0} \left( T \right) \hat{{\bf m}}
\label{eq:MagFieldMatrix}~~~.
\end{equation}
The $16\times 16$ matrix $\hat{\bf m}$ is a purely geometric quantity
specifying the possible Umklapp processes while the temperature-dependent
amplitude $h_{0}\left(T\right)$ has to be determined self-consistently. The
diagonal matrix   
%3.21
\begin{eqnarray}
\left({\hat{\bm \Delta}}({\bf k})\right)_{{\bf Q}_i{\bf Q}_j} & = &
  \delta_{{\bf Q}_{i}{\bf Q}_{j}} \Delta_{0}(T) \varphi_{\Gamma} \left( {\bf
  k} + {\bf Q}_{i} \right)\nonumber\\ 
& \equiv & \Delta_0 (T) \left(\hat{{\bm \Phi}}({\bf k})\right)_{{\bf Q}_{i}{\bf
  Q}_{j}}      
\label{eq:DeltaMatrix}
\end{eqnarray}
contains the superconducting order parameters with the given ${\bf
  k}$-dependent function $\phi_{\Gamma }({\bf k})$ and a temperature-dependent
amplitude $\Delta_{0}(T)$. 

%fig3.8%%%%%%%%%%%%%%%%%%%%%%%%%%%%%%%%%%%%%%%%%%%%%%%%%%%%%%%%%%%%%%%
\begin{figure}
\includegraphics[width=0.50\columnwidth]{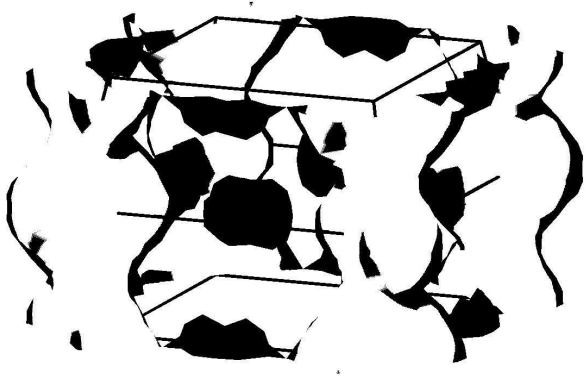}  
\caption{CeCu$_2$Si$_2$: Variation of the SC gap function amplitude $|\Delta({\bf
	k})|/\Delta_0(T) = |\phi_{\Gamma_3}({\bf k})|$ for the (pseudo-) singlet
	wave function with $\Gamma_3$-symmetry $\phi_{\Gamma_3} ({\bf k}) \sim {\rm
	cos}k_x a - {\rm cos}k_y a$ on the heavy quasiparticle sheet of the
	paramagnetic Fermi surface. The amplitude of this SC order parameter is
	maximal on the kidney-shaped surfaces centered along the $\Sigma$ direction
	which are almost unaffected by the formation of the A phase. The dominant
	contributions to the latter come from the nesting parts on the heavy
	columns where the superconducting amplitude is small. Dark and light grey
	indicate large and small amplitudes, respectively.
\label{fig:Gamma3OrderParameter}}
\end{figure}
%%%%%%%%%%%%%%%%%%%%%%%%%%%%%%%%%%%%%%%%%%%%%%%%%%%%%%%%%%%%%%%%%%%%%%

The self-consistency equations Eqs. (\ref{eq:SelfconSDW}) and
(\ref{eq:SelfconSC}) can be formulated in terms of the off-diagonal elements of
the $32\times 32$-matrix Green's function
%3.22
\begin{equation}
\hat{G} \left( i\epsilon_{n}, {\bf k} \right) = \left( i\epsilon_{n} \hat{1} -
\hat{H} ({\bf k}) \right)^{-1}
\label{eq:NambuGreen}
\end{equation}
 according to 
%3.23
\begin{eqnarray}
h_{0}(T) & = & \frac{U}{L}T \sum_{\epsilon_{n}}^{\epsilon_{c}} \sum_{\bf
  k}^{AFBZ} \frac{1}{16} \rm{Tr} \left[ \hat{{\bf m}} \hat{1} \hat{G} \left(
  i\epsilon_{n}, {\bf k} \right) \right] \nonumber\\
\Delta _{0}(T) & = & -\frac{g_{\Gamma }}{L} T
  \sum_{\epsilon_{n}}^{\epsilon_{c}} \sum_{{\bf k}'}^{AFBZ} \frac{1}{2} \mathrm{Tr} \left[ \hat{{\bm \Phi}} \left( {\bf k}' \right) \hat{\tau}_{1}
  \hat{\tau }_{3} \hat{G} \left( i\epsilon_{n}, {\bf k}' \right) \hat{\tau}_{3}
  \right] 
\label{eq:FinalSelfcon}
\end{eqnarray}
Here $\epsilon_{n} = \pi T \left( 2n+ 1 \right)$ denote the T-dependent
Matsubara frequencies and $\epsilon_{c}$ is the energy cut-off required
in weak-coupling theory. The coupling constants $U$ and $g_{\Gamma}$
as well as the cut-off $\epsilon_{c}$ are eliminated in the usual
way in favor of the observable quantities $T_{N}^{(0)}$ and $T_{c}^{(0)}$.
Solving the self-consistency equations for $\phi_{\Gamma_3} ({\bf k}) \sim {\rm
  cos}k_x a - {\rm cos}k_y a$ yields the results displayed in
Fig. \ref{fig:OrderParameterAS}. It shows that both order parameters coexist in
a finite temperature interval. This is due to the fact they have their maximum
amplitudes on different parts of the Fermi surface (see
Fig. \ref{fig:Gamma3OrderParameter}). At sufficiently low temperatures the
A-phase is finally expelled by superconductivity as shown in
Fig. \ref{fig:OrderParameterAS}. This was confirmed by neutron diffraction
which shows a suppression of magnetic Bragg penks further below $T_c$.  

\newpage
 
\section{Quantum Phase Transitions}
\resetdoublenumb 
\resetdoublenumbf

\label{Sect:QuantumPhaseTransitions}

Most phase transitions in condensed matter are governed by the
appearance of a spontaneously broken symmetry below a certain
transition temperature T$_c$. The low temperature phase is then
characterized by an order parameter belonging to a single nontrivial
representation of the high temperature symmetry group. In the simplest
case the order parameter is of the density type (diagonal long range
order), for instance charge density n(\v r) or spin density \v m(\v
r). Approaching T$_c$ from above the correlation length of spatial order
parameter fluctuations, the associated order parameter susceptibility and other
thermodynamic quantities diverge. The divergence is characterized by
critical exponents that depend only on spatial dimension d and number
of order parameter components n which define the universality class of
the model. In such finite temperature or 'classical' phase transitions
the underlying microscopic quantum fluctuations of charge and spin
densities etc. are not important ingredients for the long-range order
because their coherence is destroyed by thermal fluctuations over time
scales longer than $\hbar/kT_c$.

However, the broken symmetry state may not only be reached by lowering the
temperature. Instead at T = 0 the tuning of a physical control parameter
X, e.g., due to applied hydrostatic or chemical pressure via doping
may drive the compound from the disordered to the ordered state and
vice versa (see Fig.~\ref{fig:qcrphase}). The corresponding value
X$_c$ defines the quantum critical point (QCP) where a quantum phase
transition (QPT) takes place. In the latter the T = 0 quantum
fluctuations can be coherent over arbitrary long time scales. For this
reason the effective dimension for order parameter fluctuations close
to a QPT is given by d$_{eff}$ = d + z where z is the dynamic exponent
which characterizes the scaling of energies of quantum fluctuations
with system size. The contribution of quantum fluctuations has
therefore profound consequences for the critical exponents of
thermodynamic and transport quantities.  For example, in strongly
correlated metals close to an antiferromagnetic QCP along the
quantum critical line (X = X$_c$ or $|r| = 0$) in Fig.~\ref{fig:qcrphase} an
anomalous non-Fermi liquid (NFL) temperature dependence of physical
quantities like $\chi$(T), C(T)/T and $\rho$(T) emerges. Its origin
and theoretical description has been the subject of much recent
investigations and controversy \cite{Sachdevbook,Continentinobook}.
An additional important discovery is the observation that
superconductivity often appears in a dome-like shape around the
QCP. Viewed differently, it may be a successful strategy to look for
unconventional superconducting states in searching around magnetic
QCP's of suitable materials. 

The presence of strong electronic correlations is by no means
essential for the appearance of a quantum phase
transition. Historically they have been first studied in weakly
correlated metals without calling them explicitly with their modern
name. As an introduction we will briefly discuss some early examples
of QPT's. Subsequently well understood model theories for QPT's in local
moment systems, notably the Ising model in a transverse field 
will be discussed. Then we come to QPT's in the strongly correlated
systems where theoretical work has focused on the Kondo-lattice type
models both with and without charge degrees of freedom. Finally the
phenomenological scaling and Ginzburg-Landau theories applicable close
to the QCP will be discussed. They are important for the
interpretation of a large body of experimental work near the QCP of
heavy-fermion compounds.

\subsection{Quantum Phase Transition in Localized and Itinerant Magnets}

\label{Sect:QuantPhaseTransItinMagn}

The idea that tuning of a control parameter may drive an insulator or
metal from the paramagnetic state to a magnetically ordered state at
zero temperature is indeed a very old one. The classical
Stoner-Wolfarth theory of itinerant ferromagnetism (FM)
\cite{Stoner38,Stoner39} identifies this control parameter as X = IN(E$_F$)
where I is the exchange integral of itinerant conduction electrons and N(E$_F$)
the conduction electron DOS per spin direction. In the paramagnetic regime
with $X < X_c = 1$ the exchange interaction is too weak to cause an exchange
splitting of conduction bands and stabilize a spin polarization. However the
incipient FM order has its effect on the spin fluctuation
spectrum. The typical life time $\tau_{sf}$ of a quantum fluctuation
of magnetic moments diverges when $X\rightarrow X_c$. This may be seen from the
dynamical susceptibility, calculated from the single-band Hubbard
model for a parabolic band within random-phase approximation
(RPA). Close to the QCP (X $\apprle X_c$) the spectrum of FM spin
fluctuations with q $\rightarrow 0$ is given by \cite{Berk66,Doniach66}:
%4.1
\begin{eqnarray}
Im\chi_{-+}(\omega)=\frac{(\pi/4)N(E_F)\omega/qv_F}
{[1-IN(E_F)]^2+[(\pi/4)N(E_F)I\omega/qv_F]^2}
\label{PARSPEC}
\end{eqnarray}
where v$_F$ is the Fermi velocity. When $X\rightarrow X_c$ this 
spectrum is strongly peaked at the 'paramagnon' frequency (S $\gg$ 1)
%4.2
\begin{eqnarray}
\omega_P=\frac{4}{\pi}\frac{v_F}{S}q . 
\label{PARAMAGNON}
\end{eqnarray}
Here S = (1 - IN(E$_F$))$^{-1}$ is the 'Stoner parameter' which governs the
softening of the spin fluctuation spectrum in Eq.~(\ref{PARSPEC}) on
approaching the QCP. It also gives the enhancement of the static susceptibility
$\chi$ = S$\chi_0$ as compared to the free Pauli susceptibility $\chi_0$. In
addition the paramagnon excitations lead to a deviation from the linear
specific heat behavior of the Fermi liquid described by
\cite{Doniach66,Enzbook} 
%4.3
\begin{eqnarray}
C(T)/T=\gamma_0[m^*/m+S(T/T_{SF})^2ln(T/T_{SF})]
\end{eqnarray}
where m*/m is the mass enhancement due to paramagnons \cite{Doniach66}
and T$_{SF}$ is the spin fluctuation temperature given by T$_{SF}$ = E$_F$/S.

Equivalently one may say that $\chi_{-+}(\omega)$ has a pole at the
purely imaginary frequency $i\omega_P$. This represents a collective
overdamped spin fluctuation mode. When X is tuned through the critical
value X$_c = 1$ by pressure or alloying, the imaginary spin fluctuation
pole moves to the real axis and becomes the FM spin-wave pole with
frequency (q $\ll$ k$_F$, m$_s$ $\ll$n)
%4.4 
\begin{eqnarray}
\omega(q)= Dq^2\quad ; \quad
D \sim (I^2/E_F)k_F^{-2}nm_s.
\end{eqnarray}
Simultaneously a spontaneous FM moment caused by the spin polarization
m$_s$ = n$_\uparrow$ - n$_\downarrow$ of conduction bands appears.  Here
D is the spin wave stiffness constant and n = n$_\uparrow$ + n$_\downarrow$ is
the number of conduction electrons per site. This
collective spin excitation is undamped for small q because it is the
Goldstone mode associated with the continuous SO(3) symmetry of the FM
order parameter and therefore in the hydrodynamic limit it is protected by a
conservation law against decay. For X only marginally above the FM
QCP X$_c$ = 1 one has only weak ferromagnetism (WFM) and anomalous
thermodynamic and transport behavior due to paramagnon excitations
was observed \cite{Berk66,Doniach66}. The first theory beyond the
Hartree-Fock RPA level to address such quantum critical phenomena in
itinerant ferromagnets was Moriya's self consistent renormalization (SCR)
theory \cite{Moriya73,Moriya73a,Moriya85} for the WFM
and the theory by Hertz \cite{Hertz76} on which much of the later
developments are based.

The best known example of an enhanced paramagnetic metal which is
close to a FM quantum critical point is palladium metal. In this case one
has a large susceptibility enhancement of $S\simeq 10$
\cite{Gerhardt81}. Indeed, alloying with only 0.5\% Fe immediately leads
to FM order \cite{Mydosh68}. According to band structure
calculations a (linear) expansion of the lattice of pure Pd by $\sim$ 7 \% 
should lead to a FM ground state. Experimentally Au-Pd-Au sandwiches
have been prepared where an estimated volume expansion of 2.3\% of Pd
due to the larger lattice constant of Au leads to huge Pd Stoner factors of
S $\sim$ 10$^3$ - 10$^4$ \cite{Brodsky80}, but FM order is still not achieved. 

The classical, enhanced paramagnetism in Pd and associated QPT in Pd
alloys has been complemented by Laves phase compounds AB$_2$ like
TiBe$_2$ \cite{Acker81} and ZrZn$_2$ \cite{Matthias58} which are slightly
on the overcritical side,
i.e., IN(E$_F$)$\geq$1 and thus weak itinerant ferromagnets. By doping
with Cu the alloy series TiBe$_{2-x}$Cu$_x$ exhibits a QCP at
x$_{c}\simeq$ 0.155 where ferromagnetism disappears. The compound
ZrZn$_2$ has already been known for a long time and has recently been
investigated with renewed interest because it exhibits a FM QCP as function of
hydrostatic pressure at  p$_{c}$ = 21 GPa. As in UGe$_2$ it was found
that surprisingly superconductivity coexists within the FM phase,
albeit with a small T$_c$ \cite{Pfleiderer01}. 

Quite another quantum phase transition to a magnetically ordered state
has been known since a long time in localized moment systems. It was
studied under the name 'induced moment magnetism' without actually
stressing that it is a generic type of quantum phase transition as we
shall see. In localized (4f- or 5f-) magnetic compounds with uniaxial symmetry
the lowest CEF states of non-Kramers ions may consist of a nonmagnetic singlet
ground state $|1\rangle$ and a nonmagnetic singlet (or doublet) excited state
$|2\rangle$ at an energy $\Delta$. They are both characterized by a
vanishing moment, i.e., $\langle n| J_z|n\rangle$ = 0 (n = 1,2) with
J$_z$ denoting the total angular momentum component of the localized
4f or 5f states. Therefore there are no pre-existing localized moments
that might order as in a conventional AF phase transition. The local moments
themselves have to be induced at any given site at T$_N$. This is possible if a
nondiagonal matrix element $\alpha$ = $\langle 1| J_z|2\rangle$ exists.
The effective RKKY inter-site exchange J(\v q) then
mixes the two states, thereby creating an \emph{induced} ground state
moment. If J(\v q) is maximal at a wave vector \v Q this happens
spontaneously at a temperature
%
%%%%%%%%%%%%%%%%%%%%%%%%%%%%%%%%%%%%%%%%%%%%%%%%%%%%%%%%%%%%%%%%%%%%%%%%%%%%
\begin{figure}[tb]
\begin{center}
\includegraphics[width=7cm,clip]{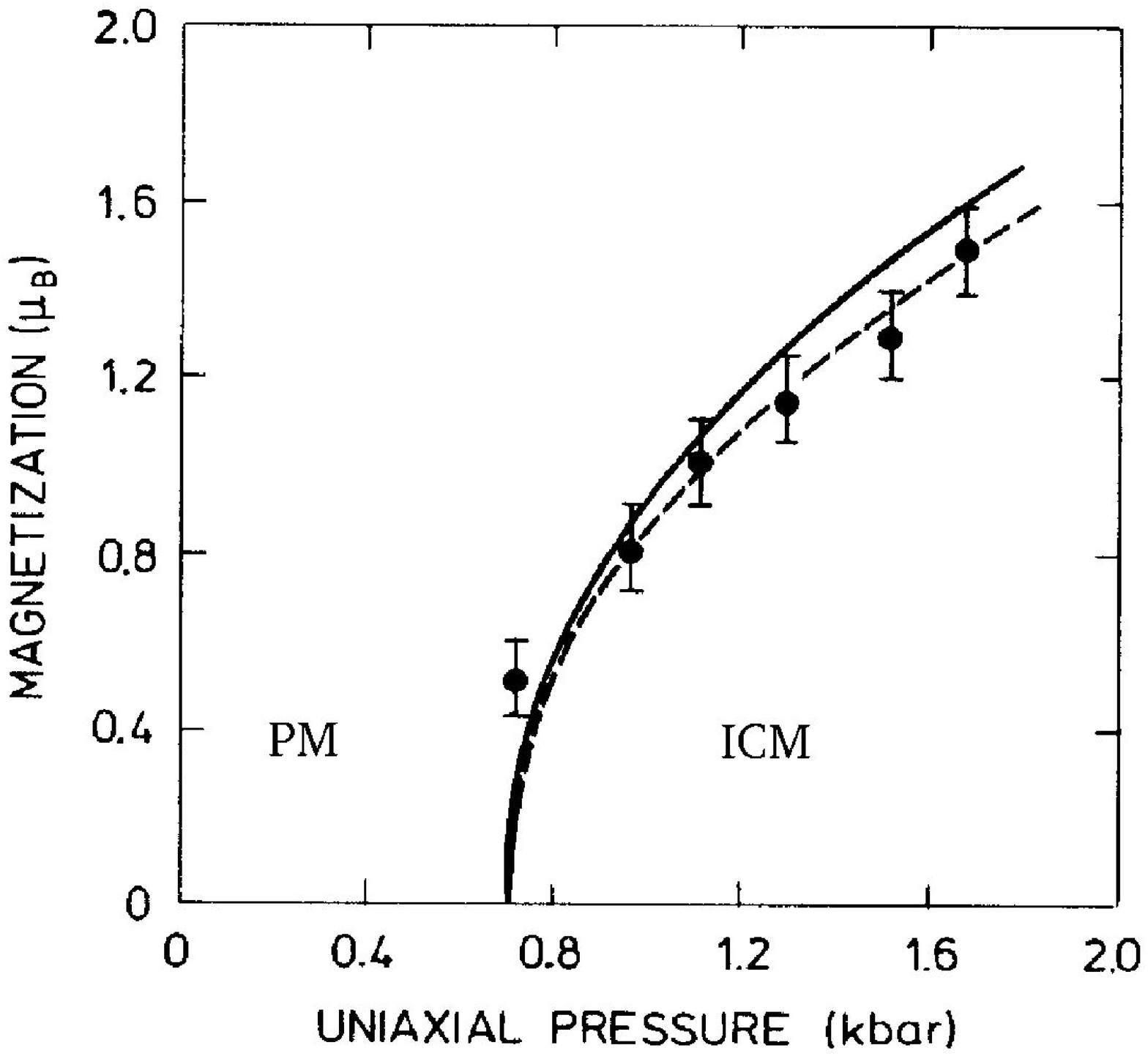}\hfill
\raisebox{-0.5cm}
{\includegraphics[width=7cm,clip]{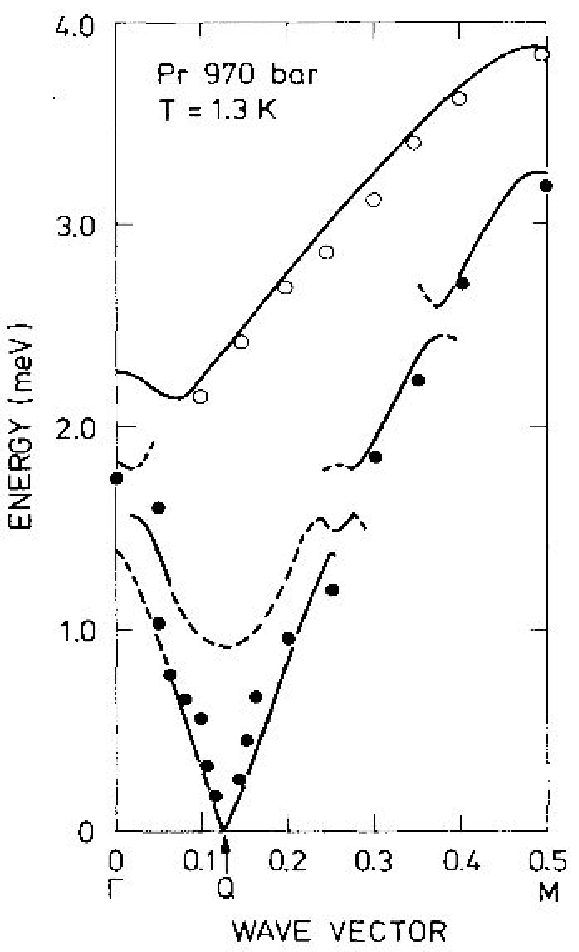}}
\end{center}
\vspace{0.5cm}
\caption{Left panel: paramagnetic (PM) to incommensurate magnetic
(ICM) quantum phase transition as function of uniaxial pressure in
dhcp Pr. The quantum critical point is at p$_c\sim$ 0.7 kbar. 
Circles are the experimental first harmonic of the ICM moment  
corresponding to modulation wave vector \v Q = 0.12 \v a$^*$. Dashed
and solid lines correspond to model calculations.
Right panel: Optic (open circles) and acoustic (closed circles)
exciton mode frequencies which give the energy  scale of quantum
fluctuations. Close to the quantum critical point at p$_c$ and for 
the ordering wave vector \v Q this energy scale vanishes. (After
\protect\cite{Jensen87})}
\label{fig:Prmode}
\end{figure}
%%%%%%%%%%%%%%%%%%%%%%%%%%%%%%%%%%%%%%%%%%%%%%%%%%%%%%%%%%%%%%%%%%%%%%%%%%%%
%
%4.5
\begin{equation}
T_N=\frac{\Delta}{2\tanh^{-1}(\frac{1}{\xi})};
\quad \mbox{here} \quad
\xi=\frac{\alpha^2J(\v Q)}{2\Delta}
\end{equation}
is the control parameter of the quantum phase transition to an induced
moment state. It takes place for $\xi > \xi_{c} = 1$. In general the
ordered state has an incommensurate modulation with wave vector \v Q. In the
paramagnetic phase the magnetic singlet-singlet excitations disperse
into a magnetic exciton band given by 
%4.6
\begin{equation}
\omega(\v q)=\Delta[1-\frac{\alpha^2J(\v q)}{2\Delta}]
\end{equation}
The onset of the QFT is signified by a softening of the exciton mode
as function of $\xi$ at the incipient ordering vector \v Q which is
given by $\omega(\v Q) = \Delta[1 - \xi]$ and vanishes at $\xi_{c} = 1$.
The control parameter contains the CEF splitting $\Delta$ which is
susceptible to pressure. The latter may therefore be used to tune the
singlet-singlet system through the QCP. Phase transitions of the
induced moment type under ambient conditions as well as under pressure
have been found in a number of rare earth systems like Pr$_3$Tl, TbSb
(for a review see \cite{Fulde79}) and dhcp Pr metal
\cite{Jensen87,Jensenbook}. It also describes the AF order in the
actinide compound \UPD with partly itinerant and partly localized
5f-electrons (see Sec. \ref{sec:PartialLocalization}). In fact dhcp
Pr, which is approximately a singlet-doublet CEF system, is of special
interest. There the critical mode softening and modulated moment
appearance under uniaxial pressure shown in Fig.~\ref{fig:Prmode}
correspond to the modern concept of a quantum phase transition due to
variation of a microscopic control parameter. However, this aspect was
not stressed or realized at that time. 

This is remarkable because it was found earlier that the simplest
Hamiltonian which describes induced moments in the
singlet-singlet system (though not precisely dhcp Pr) is of the type
%4.7
\begin{equation}
H=\Delta\sum_iT_i^x
+I\sum_{\langle ij\rangle}T_i^z T_i^z
\label{ITF}
\end{equation}
where the two orientations $|\uparrow\rangle$, $|\downarrow\rangle$ of
the pseudo-spin $T_z$ correspond to the CEF singlets n = 1,2
respectively. Indeed, this is the n.n. Ising model in a transverse field (ITF),
a genuine model for quantum phase transitions. The last term establishes AF
order of Ising spins with a twofold degenerate ground state due to Z$_2$
symmetry. The first transverse field term introduces quantum fluctuations of
the spins and destroys long-range order if the control parameter $\xi$ =
I/(2$\Delta$) exceeds $\xi_c = 1$. This model is exactly solvable in 1D
\cite{Lieb61,Pfeuty70} and therefore is a reference point for the theory of
QPT's. If one adds an infinitesimal staggered field term to Eq.~(\ref{ITF}) one
obtains a finite order parameter $\langle T_z\rangle$ given 
in Sec. \ref{subsect:NAV} provided $\xi >\xi_c$. The
critical exponent of the order parameter is 1/8. In addition the
correlation functions may be calculated exactly. In good accuracy they
are given by \cite{Lieb61,Pfeuty70} $(j=i\pm1)$:
%4.8
\begin{eqnarray} 
\langle T^z_i T^z_j\rangle = k_z; \quad \langle T^x_i T^x_j\rangle = m_x^2
\label{ITFCORR1}
\end{eqnarray}
where 
%4.9
\begin{eqnarray} 
k_z = -\frac{1}{4\pi}\int^\pi_0 dq \frac{\xi + \cos(q)}{\Lambda_q}; \qquad
m_x = \frac{1}{2\pi}\int^\pi_0 dq \frac{1 + \xi\cos(q)}{\Lambda_q} =
\langle T^x_i\rangle
\label{ITFCORR2}
\end{eqnarray}
% 
%
%%%%%%%%%%%%%%%%%%%%%%%%%%%%%%%%%%%%%%%%%%%%%%%%%%%%%%%%%%%%%%%%%%%%%%%%%%%%
\begin{figure}[tb]
\begin{center}
\includegraphics[width=7.0cm,clip]{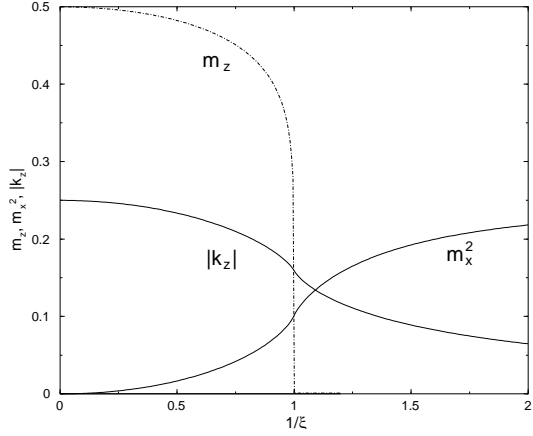}\hfill
\end{center}
\vspace{0.5cm}
\caption{Pseudo spin correlation functions for the ITF according to
Eqs.~(\ref{ITFCORR1},\ref{ITFCORR2}) as function of inverse control parameter
$1/\xi$=$2\Delta/I$. Order parameter m$_z$=$|\langle T_z\rangle|$ with
critical (pressure) exponent 1/8 is also shown. (After
\protect\cite{Yushankhai01})}
\label{fig:ITFcorr}
\end{figure}
%%%%%%%%%%%%%%%%%%%%%%%%%%%%%%%%%%%%%%%%%%%%%%%%%%%%%%%%%%%%%%%%%%%%%%%%%%%%
%
with $\Lambda_q = (1 + \xi^2 + 2\xi\cos q)^\frac{1}{2}$. These  correlation
functions together with the order parameter are plotted in
Fig.~\ref{fig:ITFcorr}. It shows nicely that long range order
m$_z$ is destroyed when the transverse quantum fluctuations characterized
by m$_x^2$ overwhelm the longitudinal correlations given by k$_z$. 
The ITF is a generic model for QPT's that may be applied to quite
different physical systems. For example, in the case of induced
magnetic moment ordering discussed above the
two pseudo-spin states correspond to the two CEF singlet states, the
Ising interaction is due to the nondiagonal exchange between them and
the transverse field is associated with the CEF splitting energy.
On the other hand the same model may be applied to the problem of
charge ordering in insulators with a 2D ladder type structure like
\NAV. In this case the pseudo-spin describes resonating, singly
occupied 3d states within the rung of a ladder, the Ising interaction
corresponds to the inter-site Coulomb interaction between d-electrons
in different rungs and the transverse field is provided by the
intra-rung kinetic (hopping) energy. The ITF then describes a quantum
phase transition where charge ordering in the rungs is destroyed by the
increase of the intra-rung kinetic energy. This will be discussed in detail in
Sec.~\ref{subsect:NAV}.

The induced moment magnetism can also appear in a different context in
compounds with localized 3d-electrons. Instead of singlet-singlet CEF
states as before one may have here singlet-triplet level systems with
a splitting $\Delta$ due to preformed dimers of S = 1/2 3d-spins. When the
inter-dimer exchange coupling is slightly subcritical the magnetic exciton mode
with a minimum at the wave vector \v Q has a finite but small energy and the
compound is paramagnetic. Application of a magnetic field splits off one
triplet component and the excitation energy at \v Q is driven to zero
at a critical field H$_c$. There a quantum phase transition to an
incommensurate magnetic phase takes place. An example is TlCuCl$_3$ where the
Cu$^{2+}$ spins form dimers \cite{Ruegg03}. Since for 3d spins the
orbital degrees are quenched, the inter-dimer coupling is of the
Heisenberg- rather than Ising type where the latter has a discrete Z$_2$
symmetry as discussed above. This makes an essential difference since the
continuous SU(2) symmetry of the former allows the exchange
Hamiltonian to be mapped to a hard-core boson Hamiltonian for the
singlet-triplet boson excitations \cite{Matsumoto04}. Then the field-driven QPT
in TlCuCl$_3$ can be interpreted as Bose-Einstein condensation (BEC) of the
singlet-triplet excitations. Another, even cleaner example of magnetic order
through BEC, though with different microscopic details was recently identified
in Cs$_2$CuCl$_4$ \cite{Radu05}. Actually the mapping to the boson model is
already possible for the xy- exchange model with U(1) symmetry in a transverse
field which applies to dhcp Pr \cite{Jensenbook}. Thus the appearance of an
incommensurate phase under pressure in Fig.~\ref{fig:Prmode} may perhaps also
be described within the BEC framework. This has not been investigated yet.

\subsection{Quantum Criticality in the Kondo Lattice}

\label{Sect:QuantCritKondoLatt}

As mentioned in the introduction a great part of the interest on QPT's
is focused on strongly correlated metallic systems which exhibit
pronounced non-Fermi liquid (NFL) behavior. This is found
frequently close to pressure- (hydrostatic or chemical) and field-
induced QPT's from the paramagnetic to the antiferromagnetic phase of
Ce- or Yb- based heavy fermion metals. The schematic phase diagram for such
compounds is shown in Fig.~\ref{fig:qcrphase}. Prominent examples are found
among the class of Ce122 , Ce115 and Ce218 intermetallic compounds and alloys
\cite{Thalmeier05}. Classical cases are CePd$_2$Si$_2$ \cite{Grosche01},
CeNi$_2$Ge$_2$ \cite{Kuechler03} and more recently YbRh$_2$Si$_2$
\cite{Custers03}. Most Ce compounds  also exhibit dome-shaped
superconductivity, sometimes with very small T$_c$. It appears in the NFL
regime around the quantum critical point of AF order. For an example see
Fig.~\ref{fig:qcrphase} (right panel). 

The generic model to describe the quantum critical Ce HF compounds is
the Kondo lattice model given by
%4.10
\begin{equation}
H_{KL}=\sum_{\v k\sigma}\epsilon_{\v k}c^\dagger_{\v k\sigma}c_{\v k\sigma}
+J_K\sum_i\v s_i\v S_i.
\label{HKL}
\end{equation}
The first term describes conduction electrons with a dispersion
$\epsilon_{\v k}$ and bandwidth W. The second term is a local AF (J$_K > 0$)
coupling of conduction electron spins $\v s_i$ to localized spins \v
S$_i$. There are two competing effects. Firstly, the solutions of the
Kondo impurity model (a single spin \v S$_{i = 0}$ coupled to the Fermi
sea) shows that  below the Kondo temperature $T_K = W\exp(-1/J_KN(0))$ the
local moment is screened. A singlet is formed which extends to a distance
$\xi_K\sim\hbar$v$_F$/T$_K$ from the impurity site \cite{Hewson93}. Secondly,
in the lattice the polarization of conduction electrons due to an on-site
exchange induces an effective RKKY-type interaction between localized
spins. This leads to a tendency to magnetic order at a temperature of the order
$T_{RKKY}\sim J_K^2\chi(2k_F)$ where $\chi(\v q)$ is the conduction electron
susceptibility. When $T_{RKKY}\ll T_K$ local but overlapping singlets form a
nonmagnetic state and below a coherence temperature $T_{\rm coh} < T_K$ they
disperse into quasiparticle bands as indicated in Fig.~\ref{fig:qcrphase} (see
also Sec. \ref{Sect:KondoLattice}). For $T_{RKKY}\gg T_K$ the singlet formation
is inhibited and magnetic order sets in at T$_N$. When the two
temperature scales are about the same size one expects a quantum phase
transition between the magnetically ordered and nonmagnetic heavy
Fermi liquid state. This criterion is only a heuristic
guide because T$_K$ is the non-perturbative energy scale (the singlet
binding energy) of the impurity problem, whereas $T_{RKKY}$ is the
perturbative energy scale on the lattice. The control parameter of the model is
X = J$_K$/W which may be assumed to vary linearly with pressure . This
behavior is illustrated in the Doniach-type phase diagram around the QCP
(Fig.~\ref{fig:qcrphase}). The wedge above the QCP is the region where
NFL behavior of thermodynamic coefficients and transport properties
is observed.

This qualitative picture is hard to quantify. In fact the Kondo
lattice model is an unsolved problem and only various approximative
and numerical methods have been applied to it. The problem may be
somewhat simplified by eliminating the charge degrees of freedom. This
was proposed in \cite{Doniach77} for the 1D Kondo chain. Using a
Jordan-Wigner transformation the 1D conduction electrons may be replaced by a
second spin system with xy-type inter-site coupling in addition to the local
Kondo spins. The model then reads
%4.11
\begin{equation}
H_{KN}=2J\sum_{\langle ij\rangle}
(\tau^x_i\tau^x_j+ \tau^y_i\tau^y_j+\delta\tau^z_i\tau^z_j)
+J_K\sum_i\v \boldtau_i\v S_i.
\label{HKN}
\end{equation}
The 1D 'Kondo necklace' model based on the Jordan-Wigner transformation of
H$_{KL}$ has $\delta =0$. But later this was generalized to $\delta >0$
in arbitrary dimension and treated as a model for the competition of
AF order (J) and local singlet formation (J$_K$) in its own right. The
control parameter is now X = $J_K/2J$. The constant 2J associated with
interacting spins corresponds to the bandwidth W of conduction electrons in the
original H$_{KL}$. The approximate quantum critical phase diagram of this model
may be obtained with the help of the bond operator method \cite{Zhang00}. It
starts from the observation that the four singlet-triplet basis states of a
local pair of spins $\boldtau_i$,\v S$_i$ may be represented by singlet-triplet
boson creation operators according to $|s\rangle = s^\dagger|0\rangle$ and
$|t_\alpha\rangle = t_\alpha^\dagger|0\rangle$ ($\alpha$ = x,y,z). The
spin operators may then be expressed in terms of singlet and triplet
boson operators:
%4.12
\begin{eqnarray}
S_{n,\alpha}&=&
\frac{1}{2}(s_n^\dagger t_{n,\alpha} + t_{n,\alpha}^\dagger s_n
-i\epsilon_{\alpha\beta\gamma}t_{n\beta}^\dagger t_{n\gamma}) \nonumber\\
\tau_{n,\alpha}&=&
\frac{1}{2}(-s_n^\dagger t_{n,\alpha} - t_{n,\alpha}^\dagger s_n
-i\epsilon_{\alpha\beta\gamma}t_{n\beta}^\dagger t_{n\gamma}). 
\label{BOSON}
\end{eqnarray}
%
%%%%%%%%%%%%%%%%%%%%%%%%%%%%%%%%%%%%%%%%%%%%%%%%%%%%%%%%%%%%%%%%%%%%%%%%%%%%
\begin{figure}[tb]
\begin{center}
\includegraphics[width=7.0cm,clip]{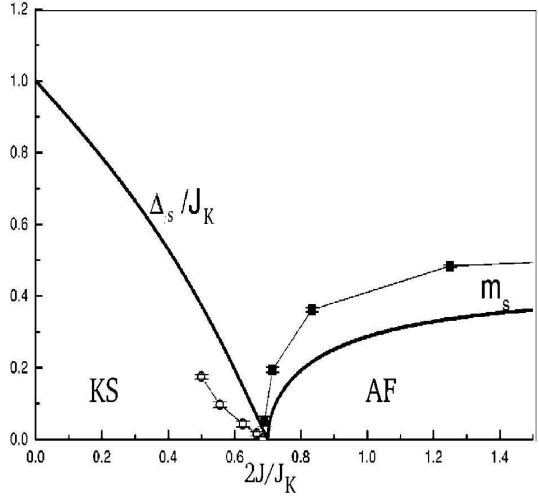}
\end{center}
\vspace{0.5cm}
\caption{Spin gap $\Delta_s$ in the Kondo singlet (KS) phase and staggered
magnetization m$_s$  in the AF phase as function of the {\emph inverse}
control parameter 1/X = 2J/J$_K$. Full line is the mean field result for
H$_{KN}$ in 2D for the xy case ($\delta = 0$). The dotted line is obtained from
Monte Carlo simulations. The QCP is at 1/X$_c\simeq$ 0.7. (After
\cite{Zhang00})} 
\label{fig:KNphaseMF}
\end{figure}
%%%%%%%%%%%%%%%%%%%%%%%%%%%%%%%%%%%%%%%%%%%%%%%%%%%%%%%%%%%%%%%%%%%%%%%%%%%%
%
These operators have to fulfill the local constraint
$s_n^\dagger s_n+\sum_\alpha t_{n\alpha}^\dagger t_{n\alpha}=1$ at every site
n. Using the above transformation in H$_{KN}$ one obtains various bosonic
interaction terms which may be decoupled by a mean field approximation both in
the Hamiltonian and in the constraint. In the strong coupling region where X =
J$_K$/2J is large the ground state is characterized by the molecular field
$\bar{s}$ = $\langle s\rangle$ corresponding to a condensation of the local
singlet bosons. The decoupling then leads to a bilinear Hamiltonian in the
triplet bosons which may be diagonalized and yields the triplet excitation
energies $\omega_{\v k}$. From the excitation spectrum the singlet amplitude
$\bar{s}(X)$ and the chemical potential $\mu(X)$ (to satisfy the mean-field
constraint) are determined self-consistently. Then the minimum triplet
excitation energy is at the AF zone boundary vector \v q = \v Q, and it is
equivalent to the spin gap in the Kondo singlet phase. It is given by (z =
coordination number) 
%4.13
\begin{eqnarray}
\Delta_{SP}&=&J_K(\frac{1}{4}+\mu/J_K)\sqrt{1-zd/2}\nonumber\\
d&=&\frac{2J}{J_K}\frac{\bar{s}^2}{(\frac{1}{4}+\mu/J_K)}
\label{SPINGAP}
\end{eqnarray}
where the dimensionless parameter d(X) is determined by the self-consistent
equation  
%4.14
\begin{eqnarray}
d&=&\frac{4J}{J_K}\bigl[1-\frac{1}{2N}\sum_{\v k}
\frac{\omega_0}{\omega_{\v k}}\bigr]\nonumber\\
\frac{\omega_{\v k}}{\omega_0}&=&\sqrt{1+d\gamma_{\v k}};
\quad \mbox{with}\quad 
\gamma_{\vk}=\sum_\alpha\cos k_\alpha.
\label{SELFCONS}
\end{eqnarray}
When the intermediate coupling regime is approached by decreasing
X$^{-1}$ = 2J/J$_K$ the spin gap eventually collapses and a magnetically
ordered phase is established. The solution of the above equations is shown in
Fig.~\ref{fig:KNphaseMF} for the 2D case. Indeed it exhibits a QPT at a
critical value X$_c\sim$ 1.43. For larger X, i.e., in the AF regime an
analogous calculation may be performed leading to the staggered magnetization
m$_s$ shown in the previous figure. In 3D results are similar but the scaling
exponents for $\Delta_{SP}(X)$ and m$_s(X)$ close to the QCP are different. We
note that the boson representation employed here for the Kondo-necklace type
model is identical to the one used in the spin-dimer problem in TlCuCl$_3$
\cite{Matsumoto04}. Therefore the transition from spin gap to AF phase
may also be interpreted as a BEC of triplet bosons.
%
%%%%%%%%%%%%%%%%%%%%%%%%%%%%%%%%%%%%%%%%%%%%%%%%%%%%%%%%%%%%%%%%%%%%%%%%%%%%
\begin{figure}[tb]
\begin{center}
\includegraphics[width=7.0cm,clip]{FQPT/KNspec.eps}\hfill
\raisebox{-0.3cm}
{\includegraphics[width=7.0cm,clip]{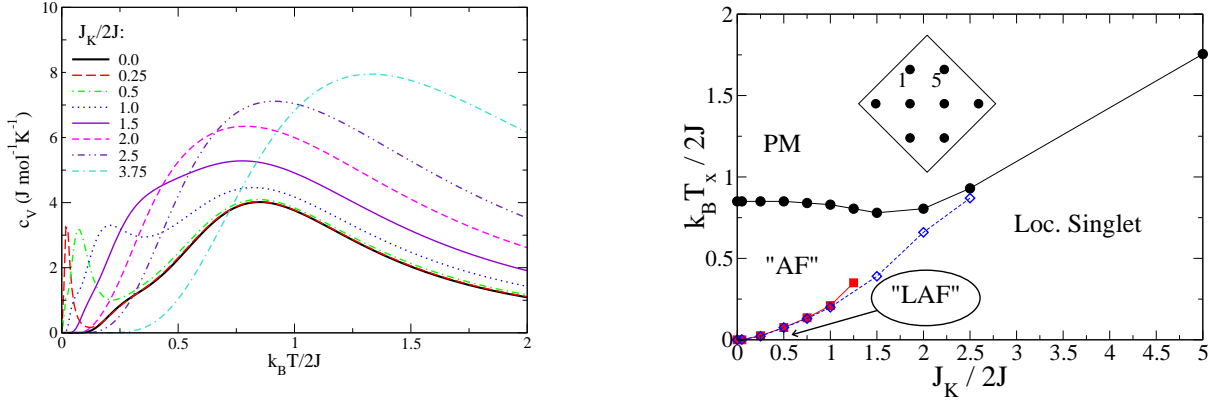}}
\end{center}
\vspace{0.5cm}
\caption{Left panel: The specific heat C$_V$ of the 8-site cluster.
Right panel: 'Phase diagram' of the 8-site cluster (inset) obtained
from tracing the characteristic maxima of C$_V$ (full lines) and
inflection points of correlation functions (from
Fig.~\ref{fig:KNcorr}) as function of control parameter X = J$_K$/2J. PM
= paramagnetic phase, 'AF'= phase with AF correlations of
$\boldtau$-spins. 'LAF'= phase with AF correlations of local spins \v
S. 'Local Singlet'- phase with Kondo spin gap. (After \cite{Zerec05})}
\label{fig:KNphaseLAN}
\end{figure}
%%%%%%%%%%%%%%%%%%%%%%%%%%%%%%%%%%%%%%%%%%%%%%%%%%%%%%%%%%%%%%%%%%%%%%%%%%%%
%
The mean-field boson treatment confirms the qualitative conjectures made above
on the quantum critical phase diagram by simply comparing the energy scales of
singlet formation and magnetic order. Because it respects the local constraint
for bosons only on the average this method is however completely inadequate to
investigate how the on-site singlet and inter-site magnetic correlations
compete as function of control parameter X and also as function of temperature.  
For this purpose one has to use advanced numerical approaches like the finite
temperature Lanczos method \cite{Jaclic00} for finite size clusters. This
method has recently been employed to Kondo lattice like models given by
H$_{KL}$ and also H$_{KN}$ with $\delta$ = 1, i.e., in the Heisenberg limit for
the interacting $\boldtau$-spins \cite{Zerec05}. Because each site has 4
states, the possible cluster sizes for exact diagonalization are limited. The
8-site cluster of the square lattice with periodic boundary conditions was
investigated and correlation functions and specific heat were calculated
\cite{Zerec05}. The Lanczos procedure is repeated 400 times for random starting
vectors in the 4$^N$ (N = 8)- dimensional Hilbert space until convergence for
thermodynamic quantities is achieved.

\begin{figure}[tb]
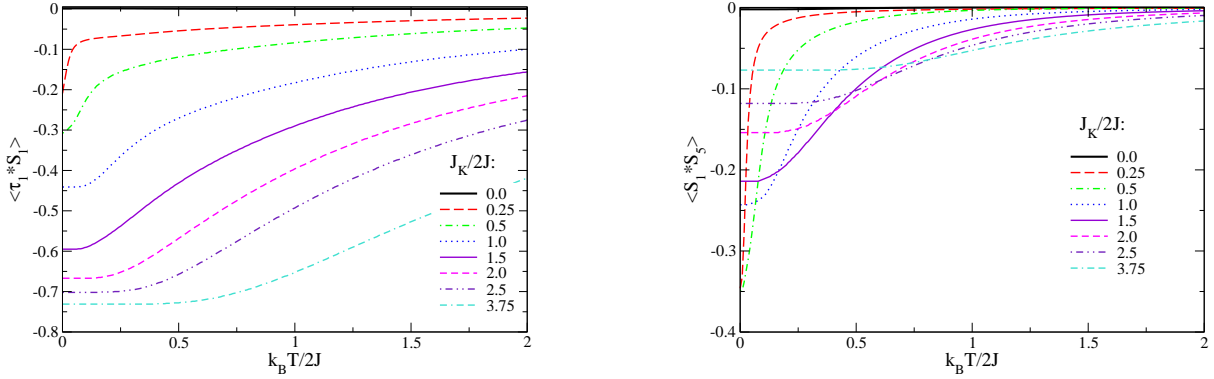

\begin{center}
\includegraphics[width=7.0cm,clip]{FQPT/KNcorr1.eps}\hfill
\raisebox{0.0cm}
{\includegraphics[width=7.0cm,clip]{FQPT/KNcorr2.eps}}
\end{center}
\vspace{0.5cm}
\caption{Left panel: On-site correlations show singlet formation at
low temperatures. Right panel: Simultaneously inter-site 'RKKY'
correlations are induced for moderate J$_K$ and suppressed again for large
J$_K$. (After \cite{Zerec05})}
\label{fig:KNcorr}
\end{figure}

The specific heat results are shown in Fig.~\ref{fig:KNphaseLAN} (left
panel) for various values of the control parameter X = J$_K$/2J. For
J$_K$ = 0 the broad upper maximum corresponds to the AF correlations of the
'itinerant' $\boldtau$-spins. When J$_K$ is turned on a lower much sharper
maximum rapidly evolves which shifts to higher temperatures and for
X$\sim$ 1.5 eventually merges with the upper maximum. The origin of the lower
sharp maximum becomes clear if one monitors the on-site singlet
correlation $\langle\boldtau_1\cdot\v S_1\rangle$ and the induced 'RKKY'
correlations $\langle\v S_1\cdot\v S_2\rangle$ between the localized
(but non-interacting) spins at n.n. sites 1 and 2. These correlations
are shown in Fig.~\ref{fig:KNcorr} as function of temperature. For
J$_K$=0 both correlations are absent. When J$_K$ is turned on
$\langle\boldtau_1\cdot\v S_1\rangle$ develops AF on-site singlet
correlations with a limiting value of -0.75 in the strong coupling
limit (left panel). Due to the inter-site coupling of $\boldtau$-spins
the local singlet formation also induces AF inter-site correlations of the
previously uncoupled \v S$_i$ spins (right panel). In the strong
coupling limit when on-site singlets are formed, however, the
inter-site correlations of \v S$_i$ spins are diminished again. The
sharp maximum in C$_V$ is well correlated with the inflection point of
the induced inter-site correlations. Therefore in the thermodynamic
limit it may be interpreted as the C$_V$-anomaly due to AF order of partially
Kondo-screened local moments. For larger J$_K$ the sharp lower C$_V$
maximum merges with the broad upper maximum but the inflection point in
$\langle\v S_1\cdot\v S_2\rangle$ may still be identified. Following these
characteristic temperatures as function of X  a 'phase diagram' of the
Kondo-lattice type model H$_{KN}$ ($\delta$ = 1) may be constructed as shown in
Fig.~\ref{fig:KNphaseLAN} (right panel). For small values of J$_K$ one
finds a correlated state of 'itinerant' $\boldtau$ spins ('AF') below
the temperature of the broad maximum in C$_V$. Below the sharp maximum
temperature one has induced antiferromagnetic correlations of the
local \v S spins ('LAF'). For larger values of J$_K$ both maxima merge
but correlations of the LAF state are still visible (dotted line). For
even larger J$_K$ the 'Local Singlet' formation dominates correlation
and specific heat behavior. Of course, due to the finite cluster size
one may not strictly speak about thermodynamic phases and phase
boundaries. However, the qualitative evolution of correlations and
thermodynamic anomalies as function of control parameter X may be expected to
survive in the thermodynamic limit. It is interesting to compare these findings
with the previous mean-field  calculation for the quantum phase transition as
function of X. In the cluster calculation one would identify the 'QCP' with the
value of X = J$_K$/2J where the correlations show a cross-over from inter-site
('LAF') to 'Local Singlet' type behavior along the T = 0 line in
Fig.~\ref{fig:KNphaseLAN} (right panel). This is found to be at X$_c\sim$ 1-1.5
and  compares reasonably well with the value X$_c$ = 1.43 of the mean-field
calculation (where however $\delta = 0$ was taken).

\subsection{Scaling Theory close to the Quantum Critical Point}

\label{Sect:ScalingTheory}

%
%%%%%%%%%%%%%%%%%%%%%%%%%%%%%%%%%%%%%%%%%%%%%%%%%%%%%%%%%%%%%%%%%%%%%%%%%%%%
%%%%%%%%%%%%%%%%%%%%%%%%%%%%%%%%%%%%%%%%%%%%%%%%%%%%%%%%%%%%%%%%%%%%%%%%%%%%

\begin{figure}%[tbh]
\begin{center}
\includegraphics[width=7cm,height=7.5cm]{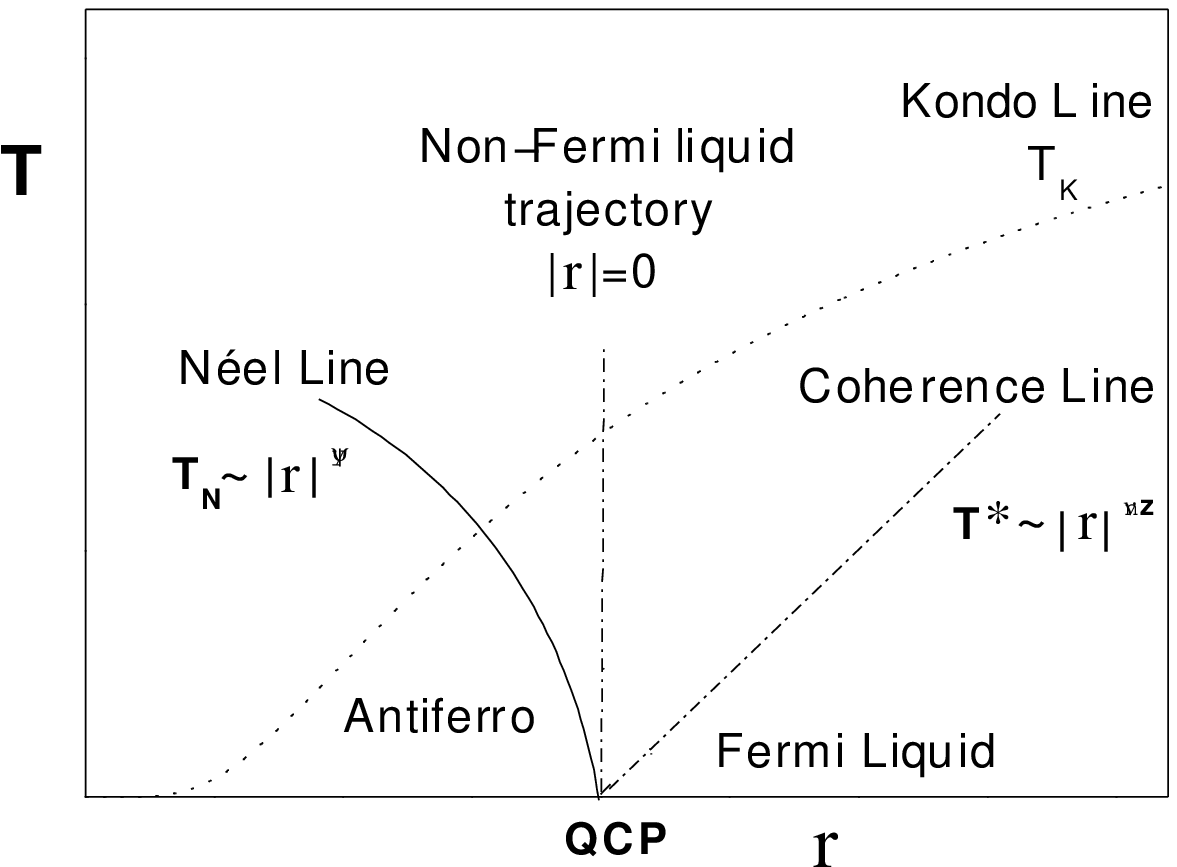}\hfill
\raisebox{0cm}
{\includegraphics[width=7.0cm,clip]{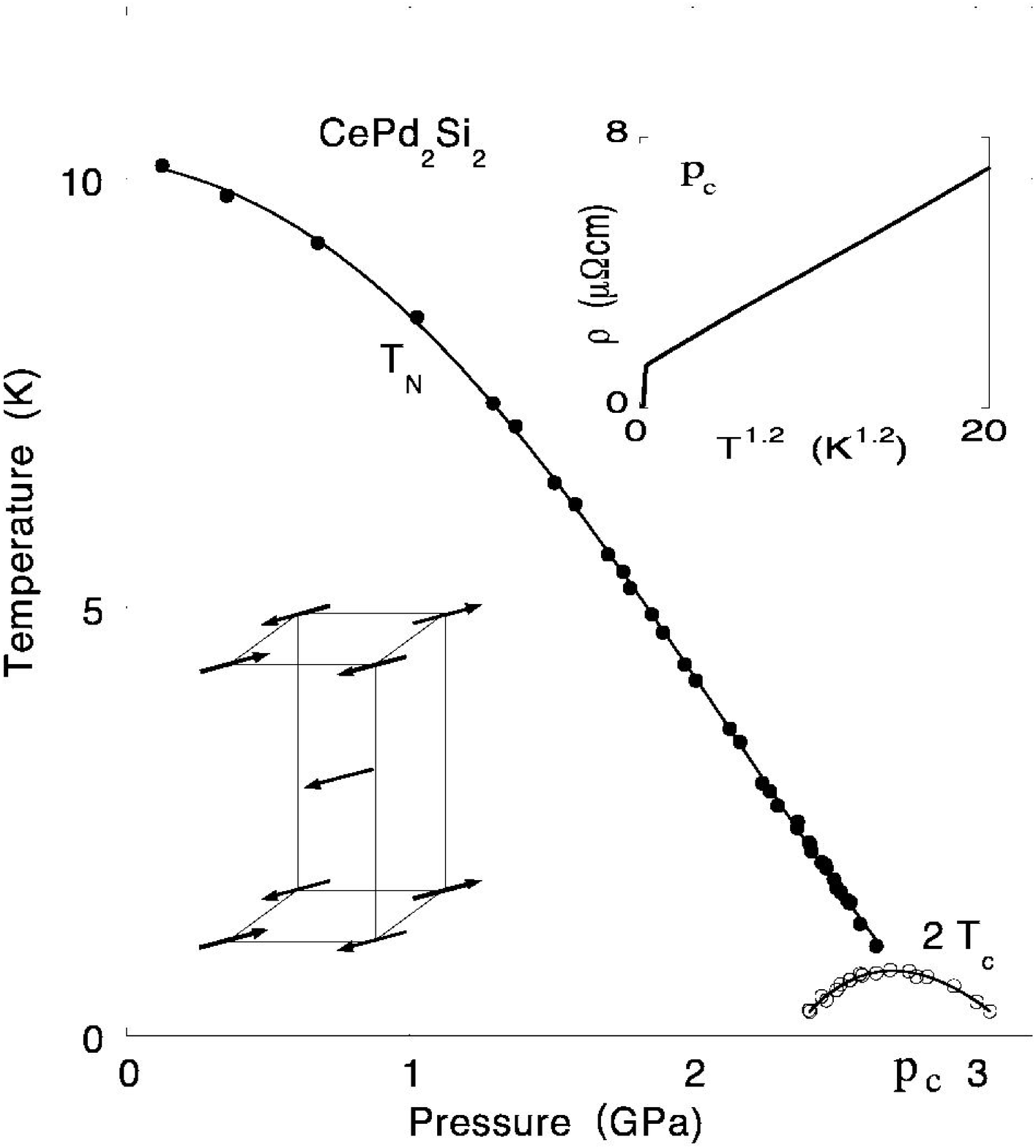}}
\end{center}
\caption{Left panel: Schematic phase diagram for Kondo compounds with
a QCP (r = 0) separating AF (left) and LFL (right) phases (full line). Scaling
of characteristic temperatures is indicated (broken lines). (After
\protect\cite{Continentino04a}). Right panel: Hydrostatic pressure induced AF
to HF liquid QPT in CePd$_2$Si$_2$. The AF structure is indicated.  At the
critical pressure p$_c$ = 2.86 GPa the resistivity shows pronounced NFL
behavior $\Delta\rho(T)\sim T^n$ with n = 1.2 down to the superconducting
transition (inset). (After \protect\cite{Grosche01})}
\label{fig:qcrphase}
\end{figure}

The previous calculations give an insight into the microscopic mechanism
of singlet formation vs magnetic order in the Kondo lattice. These
results are, however, still far removed from explaining the most common
experiments around the QCP, notably the temperature and field scaling
for specific heat, susceptibility, resistivity etc. Taking the
existence of a QCP for granted some insight into the dependence of physical
quantities on temperature and control parameters close to it (see
Fig.~\ref{fig:qcrphase}) may be obtained within a simple phenomenological
scaling theory. A Kondo impurity in a metallic host shows all the signatures of
a local Landau fermi liquid  state \cite{Hewson93} at temperatures T $\ll$
T$^*$, notably a scaling of the free energy density with T/T$^*$. This leads to
universal relations among low temperature thermodynamic quantities irrespective
of the microscopic details. This idea has been successfully extended
to the fermi liquid  phase of heavy fermion and mixed valent compounds in a
phenomenological scaling ansatz \cite{Takke81,Thalmeier86,Kaiser88}. The aim
was to explain observed relations between quantities like specific heat,
thermal expansion, magnetostriction and others. It is natural to apply these
ideas also to the vicinity of the QCP, where the characteristic energies
T$^*$(p) and T$_c$(p) themselves depend on the distance r to the QCP which
then  appears as a further scaling variable \cite{Zhu03}. The associated
correlation length ($\xi$) and time scales of quantum fluctuations ($\tau$)
diverge on approaching the phase transition. Their critical exponents are
universal, depending only on dimension and the degrees of freedom of
the order parameter. We define the quantities
%4.15
\begin{eqnarray}
r=\frac{X-X_c}{X_c}, \qquad t=\frac{T-T_c}{T_c}\qquad  
(X=p \quad \mbox{or} \quad H)
\label{REDUCED}
\end{eqnarray}
which measure the distance to the critical control parameter X$_c$ and the
transition temperature T$_c$ respectively. On approaching the QCP at
T = 0, r = 0 the correlation length, fluctuation time and free energy scale
like \cite{Continentino04a,Continentinobook}
%4.16
\begin{eqnarray}
\xi\sim|r|^{-\nu},\qquad \tau\sim|r|^{-\nu z}, \qquad
f\sim|r|^{2-\alpha}\tilde{f}(\frac{T}{T^*},\frac{H}{H^*}).
\label{SCALING1}
\end{eqnarray}
%
%
%%%%%%%%%%%%%%%%%%%%%%%%%%%%%%%%%%%%%%%%%%%%%%%%%%%%%%%%%%%%%%%%%%%%%%%%%%
%%%%%%%%%%%%%%%%%%%%%%%%%%%%%%%%%%%%%%%%%%%%%%%%%%%%%%%%%%%%%%%%%%%%%%%%%%
%
Here H$^*$ has the meaning of 'metamagnetic' field scale. For fields
H $\gg$ H$^*$ the heavy quasiparticle state is destroyed by breaking the Kondo
singlet state.  For QPT's the hyperscaling relation which relates critical
exponents to the effective dimension is given by \cite{Continentinobook}
%4.17
\begin{eqnarray}
2-\alpha=\nu d_{eff}, \qquad d_{eff}=d+z.
\label{HYPER}
\end{eqnarray}
In the case of a Gaussian fix point appropriate for d$_{eff}>4$ one has $\nu =
\frac{1}{2}$.  In the free energy of Eq.~(\ref{SCALING1}) which is a
generalization of the one used in \cite{Takke81,Thalmeier86,Kaiser88} the
characteristic temperature (T$^*$) and metamagnetic field (H$^*$) have scaling
relations 
%4.18
\begin{eqnarray}
T^*\sim|r|^{\nu z}, \qquad H^*\sim|r|^{\phi_h}.
\label{SCALING2}
\end{eqnarray}
In the magnetically ordered regime T$^*$ has to be replaced by the magnetic
transition temperature which scales as $T_c\sim|r|^\psi$ where $\psi$ is the
shift exponent. Below the upper critical dimension, i.e., for d$_{eff}<$ d$_c$
= 4 the hyperscaling relation Eq.~(\ref{HYPER}) is equivalent to the assumption
$\psi = \nu z$ \cite{Continentino04a}. In this case T$_c$(r) and T$^*$(r) scale
symmetrically around the QCP (Fig.~\ref{fig:qcrphase}), however for $d_{eff}>4$
in general one has $\psi\neq\nu z$. This is known as 'breakdown of
hyperscaling'. Within a generalized Landau-Ginzburg-Wilson approach this may be
understood as the effect of a dangerously irrelevant quartic
interaction. Although it scales to zero for d$_{eff} > 4$, it changes
nevertheless the scaling behavior at finite T leading to a modified shift
exponent $\psi = z/(d_{eff} - 2)$ for d$_{eff}>$ d$_c$.
% 
%%%%%%%%%%%%%%%%%%%%%%%%%%%%%%%%%%%%%%%%%%%%%%%%%%%%%%%%%%%%%%%%%%%%%%%%%%
\begin{figure}%[tbh]
\begin{center}
\includegraphics[width=7cm,height=7cm]{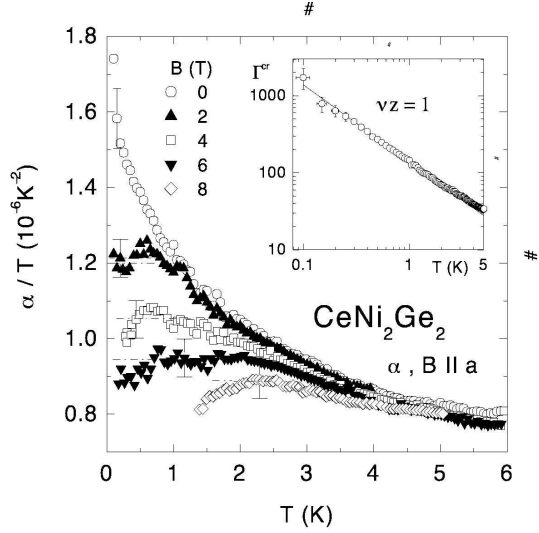}
\end{center}
\caption{Thermal expansion showing the suppression of NFL behavior as function
  of field. The inset shows that for B = 0 the critical contribution to the
  Gr\"uneisen ratio $\Gamma$ of CeNi$_2$Ge$_2$ scales like $\Gamma\sim$
  1/T$^{\nu z}$ with $\nu z$ = 1. According to Eq. (\ref{GRUEN}) this means
  (assuming z = 2 for AF SDW) a mean field correlation length exponent $\nu$ =
  1/2 which is in agreement with d$_{eff}$ = d + z = 5 for the effective
  dimension. (After \protect\cite{Kuechler03})}  
\label{fig:Grueneisen}
\end{figure}
%%%%%%%%%%%%%%%%%%%%%%%%%%%%%%%%%%%%%%%%%%%%%%%%%%%%%%%%%%%%%%%%%%%%%%%%%%
%

From an experimental viewpoint the most interesting aspect is the temperature
dependence of physical properties at the QCP(r = 0) in the non-Fermi liquid
regime. Very useful quantities are specific heat C = (T/V)($\partial
S/\partial T)_p$ and thermal expansion $\alpha$ = (1/V)($\partial
V/\partial T)_p$ \cite{Zhu03,Kuechler03}. At the QCP(r = 0) they scale
with temperature like 
%4.19
\begin{eqnarray}
C(T)\sim T^{d/z} ; \qquad \alpha(T)\sim T^{(d-\frac{1}{\nu})/z} \qquad
\mbox{and} \qquad \Gamma=\frac{\alpha}{C}\sim T^{-\frac{1}{\nu z}}
\label{GRUEN}
\end{eqnarray}
This means that the temperature dependence of the critical 'Gr\"uneisen ratio'
$\Gamma$(r = 0) is controlled by the exponent which directly determines the
time scale of quantum fluctuations in Eq.~(\ref{SCALING1}). Using this
important relation a consistent explanation of experiments in the NFL compound
CeNi$_2$Ge$_2$ can indeed be given (Fig.~\ref{fig:Grueneisen}). Tables
of the scaling behavior of the quantities in Eq.~(\ref{GRUEN}) for
various d,z have been given in \cite{Zhu03}.

The exponent $\nu z$ determines at the same time pressure scaling
(Eq.(~\ref{SCALING2})) of the characteristic temperature T$^*$ on the
nonmagnetic side of the QCP. On the other hand the pressure scaling exponent
$\phi_h$ of the characteristic field H$^*$ is an independent quantity within
the scaling ansatz. Experimentally it has been investigated in detail for
CeRu$_2$Si$_2$ which has a metamagnetic field scale H$^*$(p = 0) = 7.8T
(\v H$\parallel$ c). It was found empirically that $\phi_h = 2-\alpha = \nu z$
is fulfilled. According to the free energy in Eq.(~\ref{SCALING1}) this
implies with m = ($\partial f/\partial H$) that m(H$^*$) = $const$
independent of pressure. This was indeed found experimentally
\cite{Lacerda89}. The empirical relation $2-\alpha = \nu z$ may be
interpreted as quantum hyperscaling relation with dimension $d = 0$ according
to Eq.~(\ref{HYPER}). The empirical validity of such a relation points to a
dimensional crossover as function of pressure close to the QCP which
is caused by the different divergence of spatial and temporal
correlations \cite{Continentino04a}.

An explicit calculation of scaling exponents close to the QCP demands
the use of effective field theories based on Ginzburg-Landau type
action functionals for the spatial and temporal order parameter
fluctuations. Such theories are not specific for strongly correlated
electron systems and therefore are beyond the scope of this review. As
mentioned before they have indeed first been constructed for weakly correlated
metals in \cite{Moriya73,Moriya73a,Moriya85} and \cite{Hertz76} and
have been reinvestigated later in hindsight of QPT's in strongly
correlated metals \cite{Millis93}. Until now the results and even starting
assumptions of these field theoretical approaches are 
controversial and will not be discussed here.

\section{Partial Localization}
\resetdoublenumb 
\resetdoublenumbf

\label{sec:PartialLocalization}

The concept of orbital-selective localization applies to correlated
systems with orbital degeneracies. Important examples are transition
metal oxides \cite{Anisimov02,Koga04,Liebsch05,Laad05a} and $5f$
compounds. In these materials, the intra-atomic correlations as described
by Hund's rules play an important role. Nevertheless the physics of partial 
localization in transition metal oxides and 5$f$ systems is quite different. 
In compounds with $d$ electrons the large crystalline electric field (CEF) set 
up by the surroundings of a transition metal ion plays a major role. It is
often larger than the bandwidth as, for example in the manganites
(Sec. \ref{subsect:Polaron}). In case of a cubic lattice it splits the five
$d$ orbitals into a t$_{2g}$ triplet and an e$_g$ doublet and the corresponding
subbands are well separated. When the Hund's rule energy is larger than this 
splitting and when the orbital energy of the t$_{2g}$ is lower than that of
the e$_g$ states the t$_{2g}$ states will be occupied by the first three $d$
electrons. Those three $d$ electrons remain localized in a high-spin state with
$S = 3/2$. Additional $d$ electrons occupy e$_g$ states and remain
delocalized. The situation differs when the CEF splitting is larger than Hund's
rule coupling. In that case the t$_{2g}$ subband can accommodate six
electrons. When the $d$ electron count per ion n$_d$ is larger than six,
i.e., n$_d>6$ only (n$_d$-6) $d$ electrons will be itinerant and contribute
to metallic behavior. 

In 5$f$ compounds we are facing a different situation. Since the 5$f$ atomic 
wavefunctions are closer to the nuclei than $d$ electron wavefunctions are, 
CEF splittings are smaller and less important. But Hund's rule energies are
larger. Therefore when we deal with a situation where the 5$f$ count per
actinide ion n$_f$ exceeds two, i.e., n$_f >$ 2 only those 5$f$ electrons will
delocalize which enable the remaining ones to form a Hund's rule
state. Otherwise the Coulomb interaction is increased so much that
delocalization is disadvantageous as far as energy is concerned. Therefore
Hund's rule correlations may strongly enhance anisotropies in the kinetic
energy and eventually lead to the co-existence of band-like itinerant
5$f$ states with localized atomic-like ones.  

The central focus of the present section is the dual model for actinide-based 
heavy fermion compounds which assumes the co-existence of delocalized and 
localized 5$f$ electrons. 

Initially, the dual character has been conjectured for UPd$_{2}$Al$_{3}$ where
the variation with temperature of the magnetic susceptibility points
to the coexistence of CEF-split localized 5$f$ states in a heavy fermion
system with 5$f$-derived itinerant quasiparticles. Direct experimental evidence
for the co-existence of 5$f$-derived quasiparticles and 
local magnetic excitations is provided by recent neutron scattering
experiments \cite{Hiess04}. There is clear evidence that the presence
of localized $5f$ states is responsible for the attractive interaction
leading to superconductivity \cite{Sato01}. In addition the dual model could
allow for a rather natural description of heavy fermion superconductivity
co-existing with 5$f$-derived magnetism. For a recent review of experimental
facts see \cite{Thalmeier05,ThalmeierP05}.

Heavy quasiparticles have been observed by de Haas-van Alphen (dHvA)
experiments in a number of U compounds. The experiments unambiguously confirm
that some of the U 5$f$ electrons must have itinerant character. It has been
known for quite some time that the 5$f$-states in actinide intermetallic
compounds cannot be considered as ordinary band states. Standard bandstructure
calculations based on the Local Density Approximation (LDA) to Density
Functional Theory fail to reproduce the narrow quasiparticle bands. On the
other hand the LDA bandwidths are too small to explain photoemission data
\cite{Allen92,Fujimori99}. These shortcomings reflect the inadequate treatment
of local correlations within ordinary electron structure
calculation. Theoretical studies aiming at an explanation of the complex
low-temperature structures lay emphasis on the partitioning of the electronic
density into localized and delocalized parts \cite{Petit03,Wills04}. Concerning
the low-energy excitations it has been shown that the dual model allows for a
quantitative description of the renormalized quasiparticles - the heavy
fermions - in UPd$_{2}$Al$_{3}$. The measured dHvA frequencies for the heavy
quasiparticle portions as well as the large anisotropic effective
masses can be explained very well by treating two of the 5$f$ electrons
as localized. 

The central goal of the present section is (1) to demonstrate that
the dual model allows to determine the heavy quasiparticles in U compounds
without adjustable parameters and (2) 
to give a microscopic justification for the underlying assumptions.

Before turning to a discussion of the dual model, its results and their
implications we should like to add a few comments. In referring to the dual
model one has to keep in mind that the latter provides an effective Hamiltonian
designed exclusively for the low-energy dynamics. As such it seems appropriate
for typical excitation energies $\hbar \omega $ below $\sim$ 10 meV.
In general, effective low-energy models are derived from the underlying
microscopic Hamiltonians - to borrow the language of Wilson's renormalization
group - by integrating out processes of higher energies. In the case
of $5f$ systems the conjecture is that the hybridization between
the conduction electrons and the 5$f$ states effectively renormalizes
to zero for some channels while staying finite for others. We shall
show that the physical mechanism leading to the orbital-dependent 
renormalization of the hybridization matrix elements are the 
intra-atomic correlations which are often described by Hund's rules. 
To focus on the role of the intra-atomic correlations we consider 
model Hamiltonians for the 5$f$ subsystem where the hybridization 
with the conduction electrons is accounted for by introducing 
effective 5$f$ hopping. The orbital-dependent suppression of 
hybridization then translates into orbital-selective localization.

The concept of correlation-driven partial localization in U compounds
has been challenged by various authors (see, e.g., \cite{Opahle04}).
The conclusions are drawn from the fact that conventional band structure
calculations within the Local Density Approximation (LDA) which treat
all 5$f$-states as itinerant can reproduce ground state properties
like Fermi surface topologies and densities. The calculation of ground
state properties, however, cannot provide conclusive evidence for
the delocalized or localized character of the 5$f$-states in actinides.
First, the presence of localized states can be simulated in standard
band calculations by filled bands lying (sufficiently far) below the
Fermi level. Second, the Fermi surface is mainly determined by the
number of particles in partially filled bands and the dispersion of
the conduction bands which, in turn, depends mainly on the geometry
of the lattice. A change in the number of band electrons by an even
amount does not necessarily affect the Fermi surface since a change
by an even number may correspond to adding or removing a filled band.
As such, the Fermi surface is not a sensitive test of the microscopic
character of the states involved. Unambiguous proof of the dual character
can be provided by an analysis of the spectral function. Of particular
importance are characteristic high-energy features associated with
transitions into excited local multiplets. A detailed discussion of these
features will be given in Sec. \ref{Sect:HighEnergyExcitations}.

\subsection{Heavy Quasiparticles in UPd$_{2}$Al$_{3}$ }
\label{sec:UPd2Al3}
%%%%%%%%%%%%%%%%%%%%%%%%%%%%%%%%%%%%%%%%%%%%%%%%%%%%%%%%%%%%%%%%%%
\begin{figure}
\includegraphics[
  width=0.40\columnwidth]{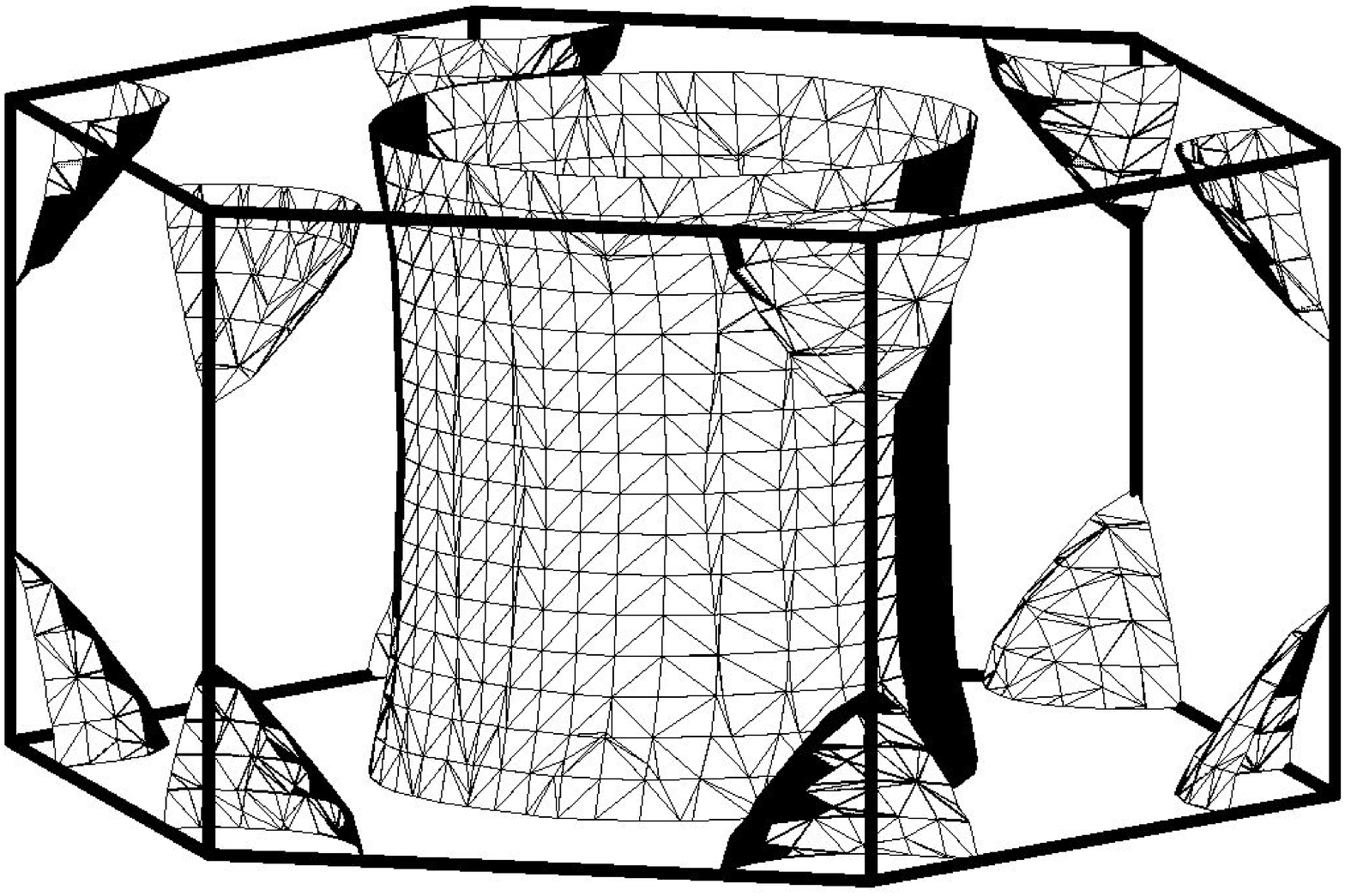} 
\includegraphics[
  width=0.40\columnwidth]{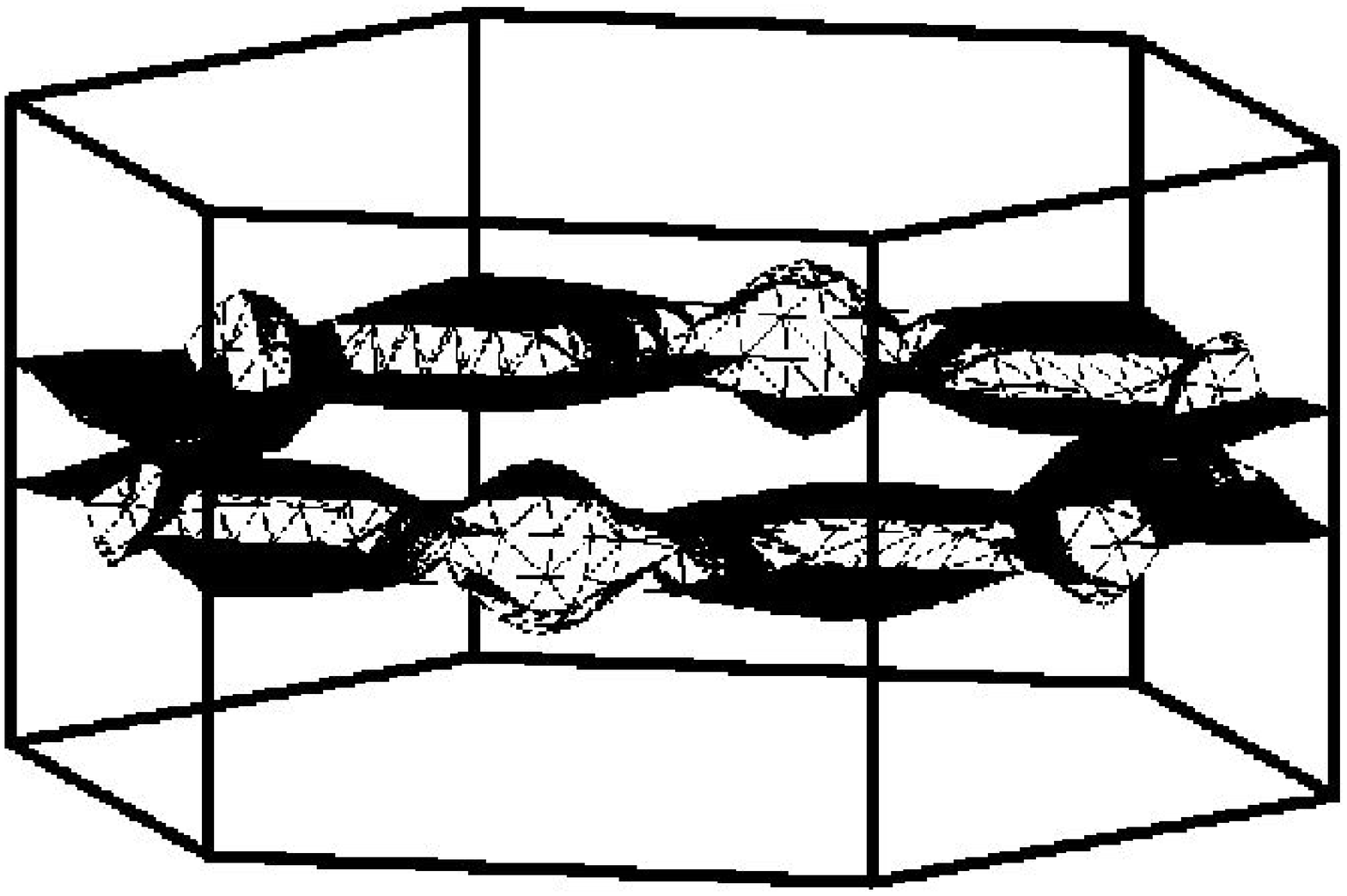} 
\includegraphics[  
  width=0.40\columnwidth]{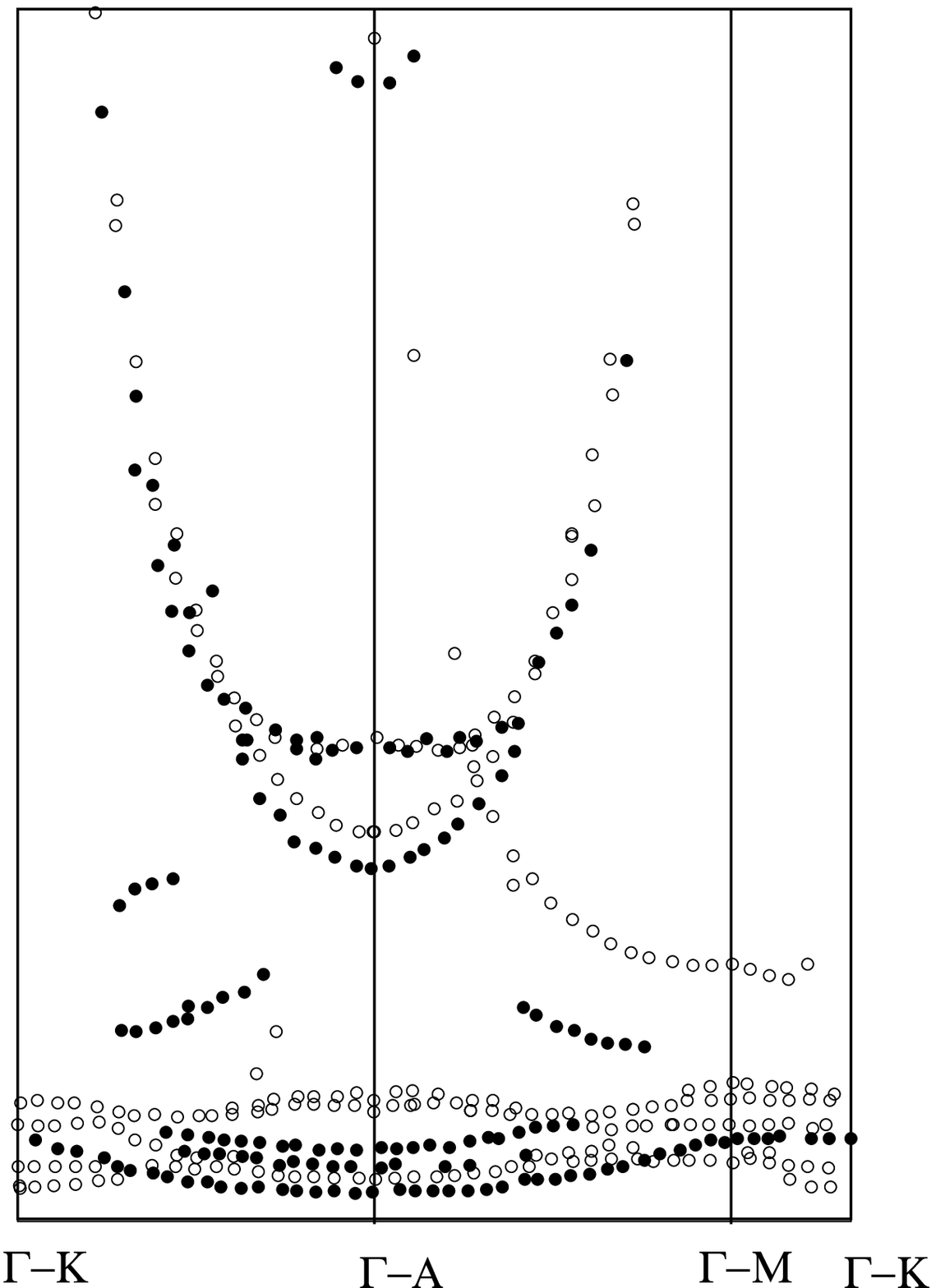} 
\caption{Upper panel Fermi surface of UPd$_{2}$Al$_{3}$ calculated within the
  dual model \cite{Zwicknagl03}. The main cylinder part has effective masses
with m$^{*}$ = 19 - 33 m, the highest masses are found on the torus.
Lower panel: Comparison of experimental dHvA frequencies (black symbols)
from \cite{Inada99} and calculated frequencies (open symbols) (After
\cite{Zwicknagl03}). The large parabola corresponds to the main FS cylinder.
\label{fig:UPd2Al3FermiSurface}}
\end{figure}
%%%%%%%%%%%%%%%%%%%%%%%%%%%%%%%%%%%%%%%%%%%%%%%%%%%%%%%%%%%%%%%%%%%%%

%
\begin{table}
\caption{Effective masses in UPd$_{2}$Al$_{3}$ for $\mathbf{H} \parallel$c.
Notation for FS sheets and experimental values from \cite{Inada99} $m_0$ is the
free electron mass. Theoretical values from \cite{Zwicknagl03}} 
\vspace{.5cm}
\begin{tabular}{ccc}
\hline 
FS sheet & m$^{*}$/m (exp.) & m$^{*}$/m (theory) \\
\hline
$\zeta$ & 65 & 59.6 \\
$\gamma$ & 33 & 31.9 \\
$\beta$ & 19 & 25.1\\
$\epsilon_{2}$ & 18 & 17.4\\
$\epsilon_{3}$ & 12 & 13.4\\
$\beta$ & 5.7 & 9.6  \\
\hline
\end{tabular}
\label{tab:UPd2Al3EffMasses}
\end{table}

Within the dual model the strongly renormalized quasiparticles in
$U$-based heavy fermion compounds are described as itinerant 5$f$
electrons whose effective masses are dressed by low-energy excitations
of localized 5$f$ states. We refer to \cite{Zwicknagl03,Thalmeier05}  
for a detailed description which proceeds in three steps. The latter include
(a) a band structure calculation to determine the dispersion of the bare
itinerant 5$f$ states (b) a quantum chemical calculation which yields the
localized 5$f$ multiplet states and, in particular, their coupling to their
itinerant counterparts and, finally, (c) a standard (self-consistent)
many-body perturbation calculation to determine the renormalized effective
mass. We should like to emphasize, however, that we treat all 5$f$ electrons as
quantum mechanical particles obeying Fermi anticommutation relations. 

The scheme has been successfully applied to UPd$_{2}$Al$_{3}$
and UPt$_{3}$ as can be seen from the comparison between the calculated
and measured dHvA frequencies and the effective masses in Figure
\ref{fig:UPd2Al3FermiSurface} and Table \ref{tab:UPd2Al3EffMasses},
respectively. It is important to note that the data are derived from
a parameter-free calculation. To show this we examine the individual
steps as described above. 

First, the bare 5$f$ band dispersion is
determined from a parameter-free ab-initio calculation by solving
the Dirac equation for the self-consistent LDA potentials but excluding
two U 5$f$ (j = $\frac{5}{2}$) states from forming bands. The apparent 
absence of Kramers' degeneracy in this compound suggests to treat an 
{\it even} number of 5$f$ electrons as localized. The calculations yield the 
dHvA frequencies which can be directly compared with experimental data.
At this point we should like to briefly comment on the strategy to
account for long-range antiferromagnetic order. The two examples,
UPd$_{2}$Al$_{3}$ and UPt$_{3}$, represent two different categories.
In UPd$_{2}$Al$_{3}$ localized 5$f$ moments order antiferromagneticly at
T$_{N}\simeq 14.5$ K with the induced moment being $\simeq 0.83\mu _{B}$ per
U. The heavy quasiparticles form in the magnetically ordered state. As a
consequence, the calculation of the bare 5$f$ bands employs the experimentally
observed antiferromagnetic structure. The superstructure strongly affects the
Fermi surface topology for the heavy quasiparticles. The corresponding
paramagnetic model cannot reproduce the heaviest orbit. 

In the second step, the localized U 5$f$ states are calculated by diagonalizing
the Coulomb matrix in the restricted subspace of the localized 5$f$
states. Assuming the j-j coupling scheme, the Coulomb matrix elements are
evaluated using the radial functions of the ab-initio band structure
potentials. The coupling between the localized and delocalized 5$f$ electrons
is directly obtained from the expectation values of the Coulomb interaction
in the 5$f^{3}$ states. 

Finally, the renormalization of the effective masses which results from the
coupling between the two $5f$ subsystems is estimated. The itinerant $5f$
states scatter off the low-energy excitations of the localized $5f^{2}$
configurations. The situation resembles that in Pr metal where a mass
enhancement of the conduction electrons by a factor of 5 results from virtual
crystal field (CEF) excitations of localized 4$f^{2}$ electrons \cite{White81}.
The  effective masses in Table \ref{tab:UPd2Al3EffMasses} are obtained from an 
isotropical renormalization of the band mass m$_{\textrm{b}}$ is given by
%5.1
\begin{equation}
\frac{m^{*}}{m_{b}}=\left.1-\frac{\partial \Sigma }{\partial \omega
}\right|_{\omega =0}
\label{eq:mmb1}
\end{equation} 

\begin{figure}
\includegraphics[width=0.50\columnwidth]{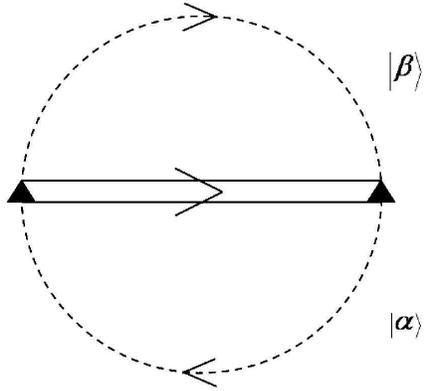}
\caption{Self-energy diagram due to local intra-atomic excitations leading to a
  mass enhancement. Solid double line: full Green's function of the conduction
  electrons. Dashed lines: effective intra-atomic two-level states $| \alpha
  \rangle$, $| \beta \rangle$ separated by an excitation energy $\bar{\delta}$
  = 7 meV. Triangles: matrix elements $a|M|$.  
\label{fig:CEFSelfEnergy}}
\end{figure}

\noindent The local self-energy of the delocalized 5$f$ states
$\Sigma (\omega )$ is displayed in Figure \ref{fig:CEFSelfEnergy}.
The explicit expressions are given in \cite{Zwicknagl03}. The mass enhancement
is calculated self-consistently inserting values for the density
of states at the Fermi level N(0) = 2.76 states /(eV cell spin) obtained
from the bandstructure, when two 5$f$ electrons are kept localized.
The vertex is given by $a | M |$ = 0.084 eV where the prefactor $a$ denotes
the 5$f$ weight per spin and U atom of the conduction electron states
near E$_{\textrm{F}}$. The matrix element M describes the transition
between the localized states $|\Gamma _{4}\rangle $ and $|\Gamma _{3}\rangle $
due to the Coulomb interaction U$_{\textrm{Coul}}$ with the delocalized
5$f$ electrons. These are the two lowest eigenstates of the localized 5$f^2$
system in the presence of the CEF. They have $J = 4$ in accordance with Hund's
rule and are combinations of $|J_z \rangle = |\pm 3 \rangle$. Finally, the
dynamical susceptibility is approximated by that of an effective two-level
system with an excitation energy $\bar{\delta }\simeq 7meV.$

%%%%%%%%%%%%%%%%%%%%%%%%%%%%%%%%%%%%%%%%%%%%%%%%%%%%%%%%%%%%%%%%%%%
%%%%%%%%%%%%%%%%%%%%%%%%%%%%%%%%%%%%%%%%%%%%%%%%%%%%%%%%%%%%%%%%%%%%%

\subsection{Microscopic Model Calculation}

To illustrate the orbital selection by intra-atomic correlations we
consider a simple molecular model consisting of two actinide atoms
at sites $a$ and $b$. The strong Coulomb interaction among
the 5$f$ electrons at the same site leads to  well-defined ionic configurations
$f^{n}$ with energies $E(f^{n})$. To model $U$ compounds we assume
that the total number of 5$f$ electrons in the cluster be five corresponding
to an averaged $f$-occupation of 2.5 per $U$ site. The ground state will be
a linear combination of states $| a; f^{3} \rangle | b; f^{2} \rangle$ and $|
a; f^{2} \rangle | b; f^{3} \rangle$, respectively. These two sets of states
are coupled by the hopping term. Since both atoms have more than one electron
in their 5$f$ shells intra-atomic correlations come into play. The two sets of
basis functions split into groups of states characterized by the total angular
momenta $J(a)$ and $J(b)$, respectively, the energy differences being of the
order of the 5$f$ exchange constant, i.e., approximately $1eV$. 
Since the spin-orbit interaction is large we use j-j coupling and restrict 
ourselves to 5$f$ states with j = 5/2. We are aiming at the low-energy subspace
which is spanned by the states $| a; f^{3}, J(a) = 9/2 \rangle | b; f^{2}, J(b)
= 4 \rangle$ and $| a; f^{2}, J(a) = 4 \rangle | b; f^{3}, J(b) = 9/2 \rangle$,
in close analogy to Hund's rules. Transferring an electron from site $a$ to
site $b$ changes the local $f$ occupation and the total angular momenta
%5.2
\begin{equation}
\left| \left. a; f^{3}, J(a) = 9/2 \right> \right. \left| \left. b; f^{2}, J(b)
= 4 \right> \right. \rightarrow \left| \left. a; f^{2}, J'(a) \right>
\right. \left| \left. b; f^{3}, J'(b) \right> \right.
\label{eq:Transfer}
\end{equation}

and the resulting final state will usually contain admixtures from excited
multiplets. The transfer of a 5$f$ electron from site $a$ to site $b$ causes
intra-atomic excitations against which the gain in kinetic energy has to be
balanced. The crucial point is that the overall weight of the low-energy
contributions to the final state depends upon (a) the orbital symmetry of the
transferred electron, i.e., on j$_z$ and (b) on the relative orientation of
$\mathbf{J}(a)$ and $\mathbf{J}(b)$. The latter effect closely parallels the
``kinetic exchange'' well-known from transition metal compounds. The
requirement that the gain in energy associated with the hopping be maximal
leads to orbital selection. The dynamics in the low-energy subspace is
described by an effective single-particle Hamiltonian where some of the
transfer integrals are renormalized to zero while others are reduced yet remain
finite. 

These qualitative considerations are the basis for microscopic model
calculations which proceed from the simple model Hamiltonian 
%5.3
\begin{equation}
H = H_{\rm band} + H_{\rm Coul} \; .
\label{eq:hamilton}
\end{equation}

The local Coulomb repulsion part
%5.4
\begin{eqnarray}
H_{\rm Coul} & = & \frac{1}{2}\sum_{a} \sum_{j_{z_{1}}, \dots, j_{z_{4}}}
U_{j_{z_{1}}, j_{z_{2}} j_{z_{3}} j_{z_{4}}} c_{j_{z_{1}}}^{\dag} (a)
c_{j_{z_{2}}}^{\dag } (a) c_{j_{z_{3}}} (a) c_{j_{z_{4}}} (a)
\label{eq:hcoul}
\end{eqnarray}

is written in terms of the usual fermionic operators $c_{j_{z}}^{\dagger }(a)$
($c_{j_{z}}(a)$) which create (annihilate) an electron at site $a$
in the 5$f$-state with total angular momentum $j = 5/2$ and $z$-projection
$j_{z}$. Considering the fact that the spin-orbit splitting is large
we neglect contributions from the excited spin-orbit multiplet $j = 7/2$
and adopt the j-j coupling scheme. The Coulomb matrix element
$U_{j_{z_{1}} j_{z_{2}} j_{z_{3}} j_{z_{4}}}$ for $j_{zi} = -5/2, \ldots, 5/2$
%5.5
\begin{equation}
U_{j_{z_{1}} j_{z_{2}} j_{z_{3}} j_{z_{4}}} = \sum _{J}U_{J}\, C_{5/2, j_{z1};
  5/2, j_{z2}}^{JJ_{z}} C_{5/2, j_{z3}; 5/2, j_{z4}}^{JJ_{z}}
\end{equation}

are given in terms of the usual Clebsch-Gordan coefficients $C_{\ldots
}^{\ldots }$ and the Coulomb parameters $U_{J}$. Here $J$ denotes the total
angular momentum of two electrons and $J_{z} = j_{z1} + j_{z2} = j_{z3} +
j_{z4}$. The sum is restricted by the antisymmetry of the Clebsch-Gordan
coefficients to even values $J =0, 2, 4$. 

The kinetic energy operator describes the hopping between all pairs at
neighboring sites $\langle ab\rangle $
%5.6
\begin{eqnarray}
H_{\rm band} & = & -\sum_{\langle ab \rangle, j_{z}} t_{j_{z}} \left(
c_{j_{z}}^{\dag } (a) c_{j_{z}} (b) + \mbox {h.c.} \right) + \sum_{a, j_{z}}
\epsilon_{f} c_{j_{z}}^{\dag } (a) c_{j_{z}} (a) \quad. 
\label{eq:hband}
\end{eqnarray}

We assume the transfer integrals $t_{j_{z}}$ to be diagonal in the
orbital index $j_{z}$. While this is certainly an idealization it
allows us to concentrate on our main interest, i.e., the interplay
of intra-atomic correlations and kinetic energy. Finally, we account
for the orbital energy $\epsilon _{f}$ which determines the $f-$valence
of the ground state.

Due to the local degeneracy, the Hilbert space increases rapidly with
the number of lattice sites and exact (numerical) solutions are possible
only for relatively small clusters. For extended systems, i.e.,
for periodic solids the ground state and the low-lying excitations
can be determined within a mean-field approximation \cite{Jedrak05}.
The slave-boson functional integral method allows for a discussion of various
ground states and co-operative phenomena starting from realistic bare
electronic band structures. The orbital-dependent separation of the low-energy
excitations into dispersive quasiparticle bands and incoherent background
is observed in the spectral functions of a linear chain calculated
by means of Cluster Perturbation Theory \cite{Pollmann05}. Itineracy is
reflected in a discontinuity of the orbital-projected momentum distribution
function
%5.7
\begin{equation}
n_{j_{z}}(\mathbf{k}) = \int d\omega \, f(\omega ) A_{j_{z}}(\mathbf{k}, \omega
) 
\label{eq:MDC}
\end{equation}

where $A_{jz} ({\bf k}, \omega)$ is the single-particle spectral function while
$f(\omega)$ denotes the Fermi distribution. Here we discuss the qualitative
features derived for two-site clusters where we can find simple approximate
forms for the ground-state wavefunction in limiting cases. 

In order to quantify the degree of localization or, alternatively,
of the reduction of hopping of a given $j_{z}$ orbital by local correlations,
the ratio of the $j_{z}$- projected kinetic energy $T_{j_{z}}$ and
the bare matrix element $t_{j_{z}}$ 
%5.8
\begin{equation}
\frac{T_{j_{z}}}{t_{j_{z}}} = \sum_{\langle ab \rangle,} \langle \Psi
_{\mathrm{g}s} | (c_{j_{z}}^{\dagger } (a) \, c_{j_{z}}(b) + h.c.) | \Psi
_{\mathrm{g}s} \rangle 
\label{eq:TjzRenormalization}
\end{equation}

%%%%%%%%%%%%%%%%%%%%%%%%%%%%%%%%%%%%%%%%%%%%%%%%%%%%%%%%%%%%%%%%%%%%%%
\begin{figure}[h]
\includegraphics[width=0.40\columnwidth]{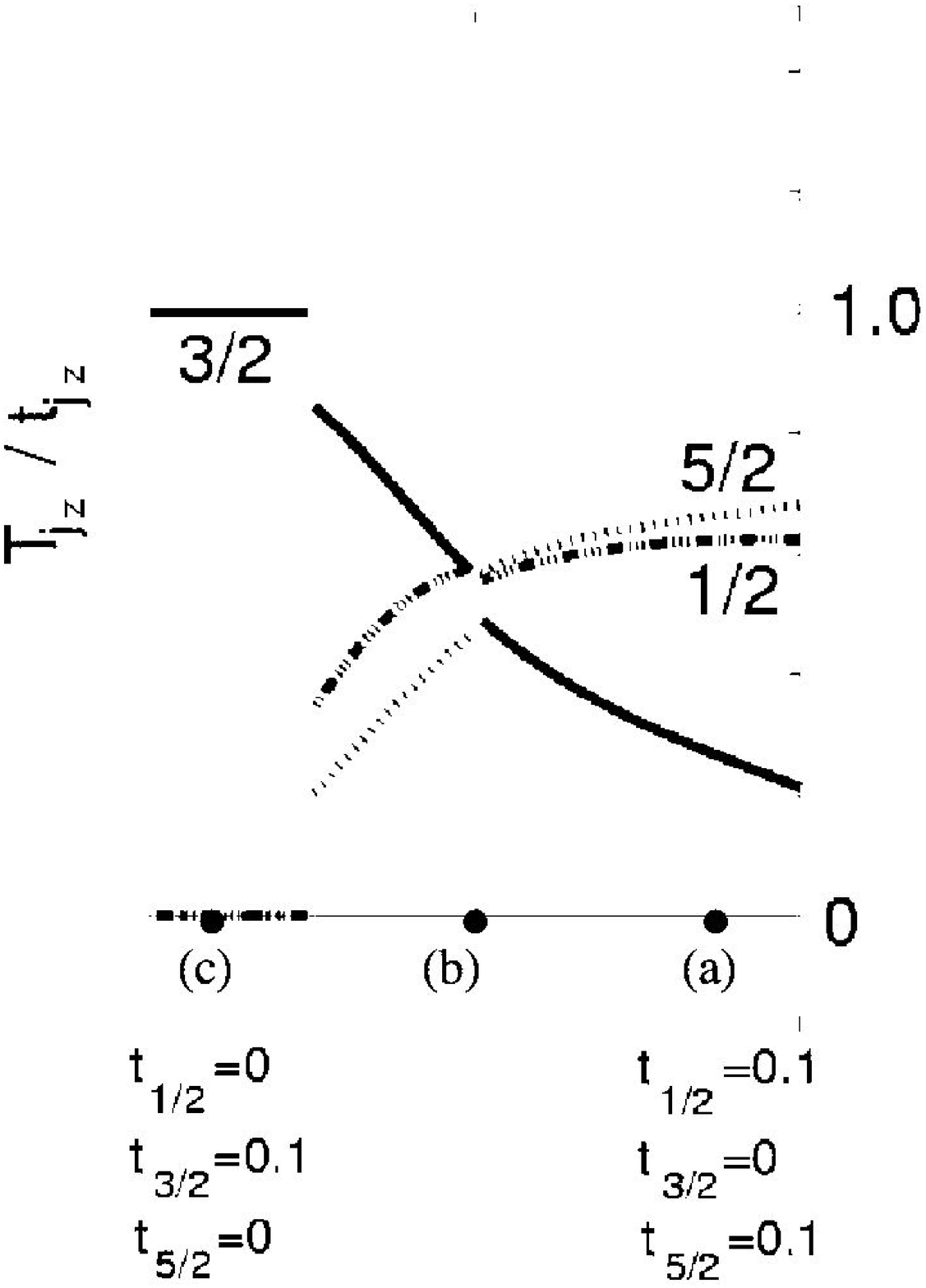}
\hspace{1cm} 
\raisebox{2.5cm}{\includegraphics[width=0.40\columnwidth]{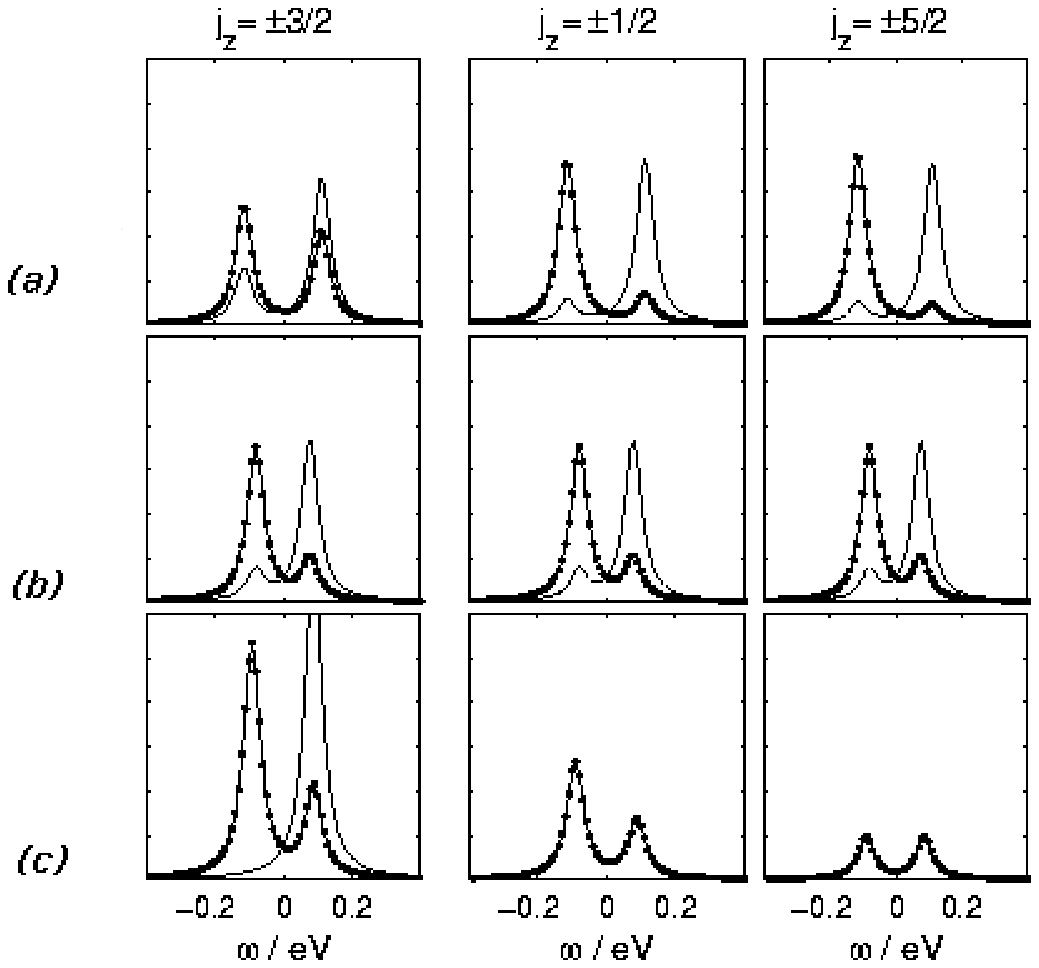}} 
\caption{Left panel: Values $T_{j_{z}}/t_{j_{z}}$ for a two-site cluster
  along a line connecting linearly the points written below the
  figure. (After \cite{Efremov04}). Right panel: Variation with wave number of 
$\left(A_{j_{z}}(k_{\ell },\omega )+A_{-j_{z}}(k_{\ell },\omega )\right)$
in the low-energy regime calculated for a two-site cluster with five
electrons. The corresponding values of the transfer integrals are those of
  the left panel. The full line and the dotted line refer to $k=0$ and $k=\pi
  $, respectively. Spectral weight is transferred to local excitations
(valence transitions and transitions into excited atomic multiplets)
which are not displayed here. The Lorentzian broadening is $\eta =0.03$ (After
  \cite{Pollmann05}) 
\label{cap:LowEnergySpectraTwoSiteCluster}.}
\end{figure}
%%%%%%%%%%%%%%%%%%%%%%%%%%%%%%%%%%%%%%%%%%%%%%%%%%%%%%%%%%%%%%%%%%%%%%%

is calculated \cite{Zwicknagl03}. The ground-state wavefunction
$|\Psi _{\mathrm{g}s}\rangle $ contains the strong on-site correlations.
A small ratio of $T_{j_{z}}/t_{j_{z}}$ indicates partial suppression
of hopping for electrons in the $\pm j_{z}$ orbitals. Two kinds of
correlations may contribute to that process. The first one is based
on the reduction of charge fluctuations due to the large values of
the isotropically averaged Coulomb repulsion which results in an isotropic
renormalization of the kinetic energy. As this is a typical high-energy
effect we defer the discussion to Sec. \ref{Sect:HighEnergyExcitations}. In
the strong-coupling limit the reduction of charge fluctuation is accounted for
by restricting the ground state to the well-defined atomic configurations. The
quantity of interest here is the orbital-dependent reduction
$T_{j_{z}}/t_{j_{z}}$ which is due to intra-atomic correlations. As the latter
are local in nature, even small clusters should adequately describe the
important qualitative features. The results for $T_{j_{z}}/t_{j_{z}}$
\cite{Zwicknagl03} - initially obtained perturbatively for a two-site cluster
- as well as their interpretation are confirmed by detailed calculations based
on exact diagonalization for small clusters \cite{Efremov04}. Figure
\ref{cap:LowEnergySpectraTwoSiteCluster} displays the reduction factors for a
two-site cluster. The model parameters  
%5.9
\begin{eqnarray}
\Delta U_{4} & = & U_{J = 4} - U_{J = 0} = -3.79\textrm{eV}\nonumber \\
\Delta U_{2} & = & U_{J = 2} - U_{J = 0} = -2.72\textrm{eV}\quad .
\label{eq:HundEnergies}
\end{eqnarray}

are chosen appropriate for UPt$_{3}$. These findings demonstrate
that in particular Hund's rule correlations strongly enhance anisotropies
in the hopping. For a certain range of parameters this may result
in a complete suppression of the effective hopping except for the
largest one, which remains almost unaffected. This provides a microscopic
justification of partial localization of 5$f$ electrons which is
observed in a number of experiments on U intermetallic compounds. 

As the relevant correlations are local, the general results qualitatively
agree with those found for a three-site cluster and four-site clusters
\cite{Pollmann03}. The magnetic character, however, is affected by
finite size effects. This can be seen by varying the cluster sizes
and the boundary conditions. Although the total angular momentum component
$\mathcal{J}_{z}$ may be different for periodic and open boundary conditions we
can identify the following different regimes in the strong-coupling limit.
In the strongly anisotropic limit with dominating transfer integral
$\left|t_{3/2}\right|\gg \left|t_{1/2}\right|=\left|t_{5/2}\right|$
the high-spin states with ferromagnetic inter-site correlations are
energetically most favorable. In the two-site cluster, the ground state has a
very simple form 
%5.10
\begin{eqnarray}
\mid \Psi \rangle & = & \frac{1}{\sqrt{2}} \left( c_{3/2}^{\dagger} (a) +
\frac{t}{|t|} \, c_{3/2}^{\dagger} (b) \right) c_{5/2}^{\dagger} (a)
c_{1/2}^{\dagger} (a) c_{5/2}^{\dagger} (b) c_{1/2}^{\dagger} (b) \mid 0
\rangle  
\label{eq:Psi15by2}
\end{eqnarray}

\noindent being simultaneously an eigenstate of the Coulomb energy and of the
kinetic energy. It can be considered as a bonding $j_{z} = 3/2$ state in a
ferromagnetically aligned background. The high-spin phases are followed
by complicated intermediate-spin phases as the isotropic limit
$t_{1/2} = t_{3/2} = t_{5/2}$ is approached. In the case with subdominant
$| t_{3/2} | \ll | t_{1/2} | = | t_{5/2} |$ low-spin phases with
antiferromagnetic intersite correlations are formed. In a two-site cluster,
they involve linear combinations of  
\begin{eqnarray*}
c_{1/2}^{\dagger} (a) c_{\pm 5/2}^{\dagger} (a) c_{\pm 3/2}^{\dagger} (a)
c_{\mp 5/2}^{\dagger}(b)c_{\mp 1/2}^{\dagger} (b) \left| 0 \right\rangle & ; &
c_{\pm 5/2}^{\dagger} (a) c_{\pm 3/2}^{\dagger} (a) c_{\mp 5/2}^{\dagger} (b)
c_{1/2}^{\dagger} (b) \left| 0 \right\rangle ~~~.
\end{eqnarray*}

The splitting of the low-energy excitations into dispersive quasiparticle
states and incoherent background is reflected in the single-particle
spectral functions $A_{j_{z}}(k_{\ell },\omega )$ where the discrete
set of quantum numbers $k_{\ell }=0,\pi $ labels the single-particle
eigenstates of the two-site cluster. The variation with the transfer
integrals is displayed in Figure \ref{cap:LowEnergySpectraTwoSiteCluster}.
The $j_{z}$-channels with dominant hopping exhibit dispersive narrow
peaks while those with subdominant hopping yield an incoherent background.
Considerable spectral weight is transferred to the high-energy excitations
not shown here.

\subsection{Superconductivity mediated by Intra-Atomic Excitations}
\label{subsect:Supercond}

Since the discovery of the isotope effect \cite{Maxwell50,Reynolds50} and the
work of Fr\"ohlich \cite{Froehlich50} the electron-phonon interaction has been
considered the main cause of Cooper-pair formation. By exchanging virtual
phonons, electrons may attract each other and form Cooper pairs. Later it was
pointed out that phonons need not be the only bosons the exchange of which
results in electron attraction. Also magnetic excitations such as paramagnons
were considered as candidates for generating superconductivity, although not
necessarily in a conventional $s$-wave pairing state \cite{Berk66,Fay77}. Also
it had been pointed out that crystalline electric field (CEF) excitations in
rare-earth ions like Pr have a pronounced effect on superconductivity
\cite{Fulde70,Keller71,Keller73} when such ions are added to a conventional
superconductor and furthermore, that those excitations can be either
pair-breaking or pair forming depending on matrix elements between different
CEF levels. The experimental observation of those effects, e.g., in doped
LaPb$_3$ and LaSn$_3$ \cite{Heiniger75,McCallum75} demonstrated not only the
reality but also the magnitude of that kind of boson exchanges between
conduction electrons. After the discovery of high-temperature superconductivity
in some of the cuprate perovskites and even before for some of the
heavy-fermion superconductors numerous suggestions of non-phononic pairing
interactions were made
\cite{Scalapino86,Miyake86,Anderson87,Spalek88,Moriya90,Monthoux91,Monthoux99,Monthoux01,Monthoux02}.
But they were mainly qualitative rather than quantitative and therefore
remained inconclusive. For reviews see
Refs. \cite{Plakida95,Thalmeier05}. Therefore it is of interest that for
UPd$_2$Al$_3$ experimental evidence exist for a non-phononic
mechanism is causing superconductivity. It has a transition temperature T$_c$
of T$_c$ = 1.8 K; which is below the onset of antiferromagnetic (AF) order with
a N\'eel temperature of T$_N$ = 14.3 K. Strong evidence for a non-phononic
pairing mechanism is provided by UPd$_2$Al$_3$ - Al$_2$O$_3$ - Pb tunneling
measurements \cite{Jourdan99}. The differential conductivity dI/dV shows
structure in the regime of 1 meV demonstrating that there are low-energy bosons
which result in a frequency dependence $\Delta(\omega)$ of the order
parameter. For phonons this structure would be an order of magnitude higher in
energy. The Debye energy of UPd$_2$Al$_3$ is 13 meV. In addition inelastic
neutron scattering (INS) experiments show that the CEF based magnetic
excitation energy $\omega_E({\bf q})$ at ${\bf q} = {\bf Q}$ is between 1 - 1.5
meV depending on temperature
\cite{Hiess04,Sato97,Sato01,Bernhoeft98}. It has been argued that these
excitations  show up in $\Delta(\omega)$ and cause superconductivity. This is
seen in
Fig. \ref{eq:hcoul} which shows the INS data as well as the tunneling density
of states. The strong coupling of those AF excitons to conduction electrons is
also demonstrated in these experiments. 

The AF structure of UPd$_2$Al$_3$ consists of ferromagnetic hexagonal planes
with a moment of $\mu = 0.83 \mu_B$ per U ion pointing in [100] direction and
stacked antiferromagnetically along the $c$-axis \cite{Krimmel92,Kita94}. This
corresponds to an AF wave vector ${\bf Q}$ = (0, 0, 1/2). The large moment
supports the dual model with two localized 5$f$ electrons. As discussed before
the Hund's rule ground-state multiplet of the 5$f^2$ localized electrons is J
= 4. In a CEF only the two lowest singlets $| \Gamma_3 \rangle$ and $| \Gamma_4
\rangle$ have to be taken into account. The Hamiltonian is then of the form

%5.11
\begin{eqnarray}
H & = & \sum_{{\bf k}\sigma} \epsilon_{{\bf k}\sigma} c^\dagger_{{\bf k}\sigma}
c_{{\bf k}\sigma} + \delta \sum_i \left| \left. \Gamma_4 \right> \left<
\Gamma_4 \right. \right|_i \nonumber \\  
&& - J_{\rm ff} \sum_{\left< ij \right>} {\bf J}_i {\bf J}_j - 2 I_0 \left(
g_{\rm eff} - 1 \right) \sum_i {\bf s}_i {\bf J}_i
\label{Hksig}
\end{eqnarray}

\noindent where

%5.12
\begin{equation}
\epsilon_{{\bf k}\sigma} = \epsilon_\bot \left( k_\bot \sigma \right) -
2t_{\|} {\rm cos}~k_z  
\label{epsbot}
\end{equation}  

\noindent serves as a model for the Fermi surface of UPd$_2$Al$_3$ in the
paramagnetic state. 

Figure \ref{cap:LowEnergySpectraTwoSiteCluster} shows that the Fermi surface in
the AF state consists of a cylindrical part and a torus. The torus has the
highest effective mass. For simplicity we will neglect it here and model the
paramagnetic Fermi surface by a cylinder. The antiferromagnetic coupling to the
conducting electrons is considerably larger than in systems like TbMo$_6$S$_8$
\cite{Ishikawa77} and 
HoNi$_2$B$_2$ \cite{Amici00}. One should also keep in mind that in accurate
determination of $T_c$ is anyway out of reach and that the aim is here to
demonstrate that the non-phononic pairing mechanism yields the right order of
magnitude for $T_c$.  

Returning to the Eqs. (\ref{epsbot}) we note that $t_{\|}$ determines the
amount of corrugation of the cylinder. The second term in Eq. (\ref{Hksig})
denotes the CEF splitting with $\delta$ = 6 - 7 meV. The coupling between
localized 5$f$ electrons on nearest-neighbor sites is given by $J_{\rm ff}$ and
the on-site exchange between localized and itinerant electrons is described by
the last term of Eq. (\ref{Hksig}). The effective Land\'e factor $g_{\rm eff}$
refers to the localized 5$f$ electrons. The total intersite exchange is
therefore 

%5.13
\begin{equation}
J \left( {\bf q} \right) = J_{\rm eff} \left( {\bf q} \right) + I^2_0 \left(
g_{\rm eff} -1 \right)^2 \chi_e \left( {\bf q} \right) 
\label{JqJeff}
\end{equation} 

\noindent where $\chi_e ({\bf q})$ is the spin susceptibility of the itinerant
electrons. Due to that intersite interaction the susceptibility of the system
becomes

%5.14
\begin{equation}
\chi \left( {\bf q}, \omega \right) = \frac{u (\omega)}{1 - J({\bf q})
  u(\omega)}  
\label{chiqom}
\end{equation}

\noindent where $u(\omega)$ is the single-ion susceptibility \cite{Cooper67}
and the CEF excitation energy goes over into an excitation band (magnetic
exciton). When $J(q)$ has its maximum value at ${\bf q} = {\bf Q}$ and $J(Q)
\equiv J_e$ exceeds a critical value, i.e., $J_e > J_{\rm crit}$ the system
becomes an induced AF. In that case the N\'eel temperature $T_N$ is given by

%5.15
\begin{equation}
T_N = \frac{\delta}{2~{\rm tan} h^{-1} \left( J_c/J_e \right)} ~~~.
\label{TNdelta}
\end{equation}

UPd$_2$Al$_3$ is an induced AF and one finds that $J_e/J_c$ = 1.015. The
critical value is given by $J_c = \delta/2M^2$ where $M = 2 \langle \Gamma_4 |
J_x | \Gamma_3 \rangle_i$. For temperatures $T < T_N$ the susceptibility is
again of a form similar to Eq. (\ref{chiqom}) but now the single-ion
susceptibility contains the effect of the AF molecular field acting on the $|
\Gamma_4 \rangle, | \Gamma_3 \rangle$ states. The magnetic excitations form a
band of AF magnons in that case. For a review see, e.g.,
Refs. \cite{Fulde79,Jensen91}. They have originally been measured
\cite{Mason97} with relatively low resolution and later with much better one,
see Refs. \cite{Hiess04,Sato01} and \cite{Bernhoeft00}. Their dispersion has
also been derived theoretically in Ref. \cite{Thalmeier02} by including the molecular field as well as the
anisotropic exchange and agreed nicely with the measured ones. An approximate
form is 

%5.16
\begin{equation}
\omega_E \left( q_z \right) = \omega_{\rm ex} \left[ 1 + \beta~ {\rm cos}
  \left( cq_z \right) \right]
\label{omeE}
\end{equation} 

\noindent with $\omega_{\rm ex}$ = 5 meV, $\beta$ = 0.8 and $c$ denoting the
lattice constant perpendicular to the hexagonal planes. The corresponding boson
propagator $K (q_z, \omega_\nu)$ in Matsubara frequency notation ($\omega_\nu =
2 \pi T, \nu$ = integer) replaces the phonon propagator when the
superconducting properties are calculated \cite{McHale04}. It is of the form

%5.17
\begin{equation}
K \left( q_z, \omega_\nu \right) = g \frac{\omega^2_{\rm ex}}{\left( \omega_E
  \left( q_z \right) \right)^2 + \omega^2_\nu}
\label{Kqzom}
\end{equation} 

\noindent where $g$ denotes the coupling constant between conduction electrons
and magnetic excitons. Their interaction Hamiltonian $H_{c-f}$ can be written
in a pseudospin notation by introducing for the two levels $| \Gamma_3
\rangle_i$ and $| \Gamma_4 \rangle_i$ of a $U$ site $i$ the pseudospin
$\bm{\tau}_i$, so that $\tau_{iz} | \Gamma_{3(4)} \rangle_i = \pm
\frac{1}{2} | \Gamma_{3(4)} \rangle_i$. Then we may write

%5.18
\begin{equation}
H_{c-f} = I \sum_i \sigma_{iz} \tau_{ix}
\label{HcfIi}
\end{equation}

\noindent where $\sigma_i$ refers to the itinerant 5$f$ electron. The two
coupling constants $g$ and $I$ are related through

%5.19
\begin{equation}
g = \frac{I^2}{4} \left( \frac{1}{c} \frac{p^2_0}{2 \pi} \right)
\frac{1}{\omega_{\rm ex}}
\label{gI241c}
\end{equation}

\noindent where $p_0$ is the radius of the circle in the $p_x, p_y$ plane which
contains the same area as the hexagon defining the Brillouin zone.

%%%%%%%%%%%%%%%%%%%%%%%%%%%%%%%%%%%%%%%%%%%%%%%%%%%%%%%%%%%%%%%%%%%%%%%%%%%%
\begin{figure}[tb]
\begin{center}
\includegraphics[width=9.5cm,clip]{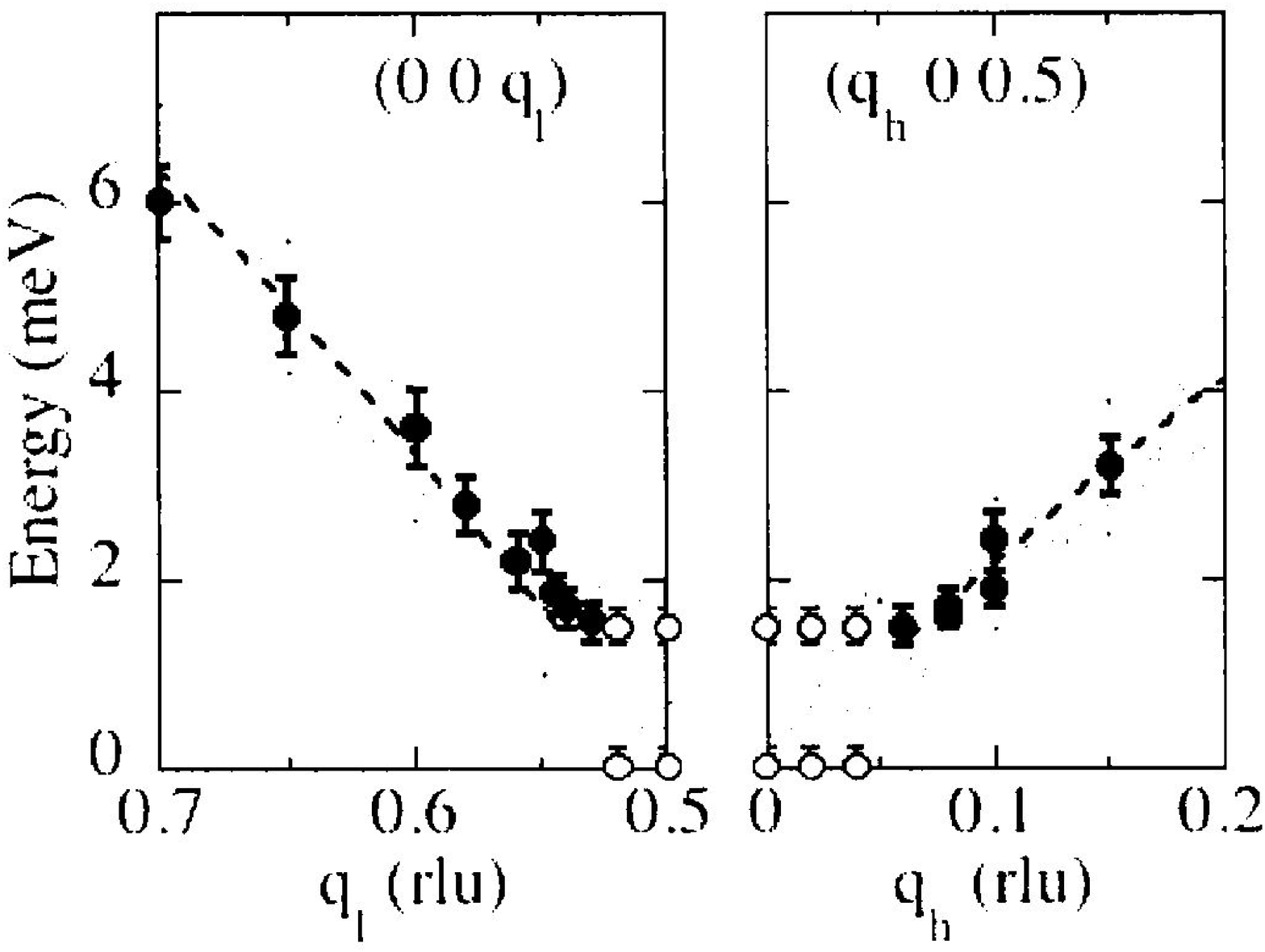}\hfill
\raisebox{1cm}
{\includegraphics[width=6.0cm,clip]{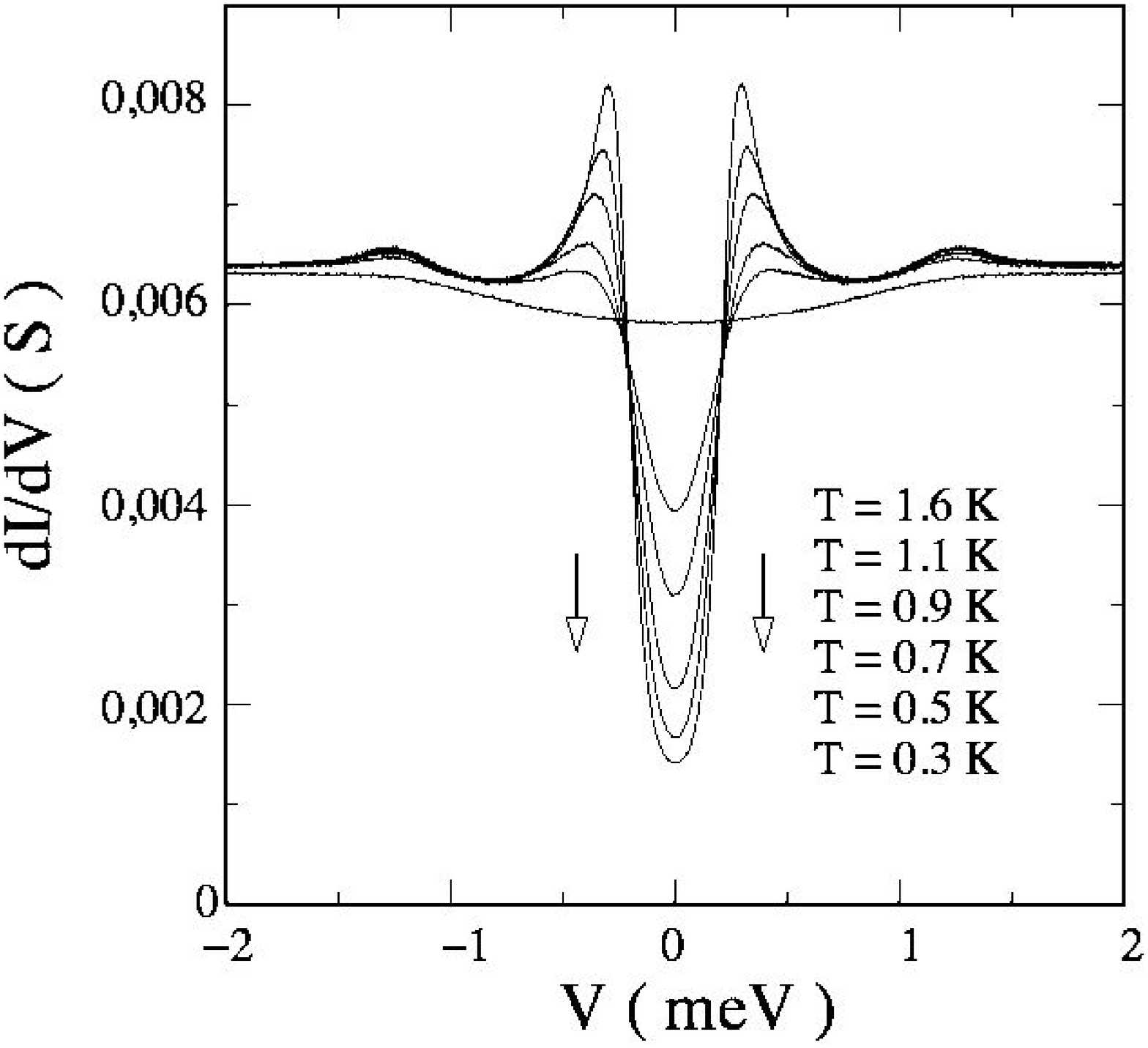}}
\end{center}
\vspace{0.5cm}
\caption{Upper panel: Magnetic exciton dispersion from INS along c$^*$ 
(0,0,q$_l$) and a$^*$ (q$_h$,0,0.5) around  the AF zone center
\v Q = (0,0, 0.5) for T = 2K just above T$_c$ = 1.8 K. A flat dispersion with a
corresponding high DOS is observed for \v q $\simeq$ \v Q.
For T $<$ T$_c$ the additional 
quasielastic response (open circles at zero energy) evolves into 
sharp resonance peaks within the the gap. (After\protect\cite{Hiess04})
Right panel: Differential conductivity dI/dV for tunneling current
along c. The additional hump at V $\simeq$ 1.5 meV has been associated with the
magnetic exciton mode at \v Q in the upper panel. This 
feature is due to a frequency dependent gap $\Delta(\omega)$ caused by a
strong coupling of quasiparticles to magnetic excitons. However the evaluation
of the tunneling data needs additional justification (J. Geerk private
communication). (After\protect\cite{Jourdan99})}
\label{fig:EXCITON}
\end{figure}
%%%%%%%%%%%%%%%%%%%%%%%%%%%%%%%%%%%%%%%%%%%%%%%%%%%%%%%%%%%%%%%%%%%%%%%%%%%%

We are now in the position to write down and solve Eliashberg's equations for
the conduction electron self-energy $\Sigma (p_z, \omega_n)$ and order
parameter $\Delta (p_z, \omega_n)$. 

%5.20
\begin{eqnarray}
\Sigma \left( p_z, \omega_n \right) & = & \frac{T}{N_z} \sum_{p'_z, m} K \left(
p_z - p'_z; \omega_n - \omega_m \right)\nonumber \\
&& \int \frac{dp'_\perp}{(2 \pi)^2} G \left( p'_\perp, p'_z, \omega_m
\right)\nonumber \\ 
\Delta \left( p_z, \omega_n \right) & = & -\frac{T}{N_z} \sum_{p'_z, m} K
\left( p_z - p'_z; \omega_n - \omega_m \right) \Delta (p'_z, \omega_m)\nonumber
\\ 
&& \int \frac{dp'_\perp}{(2 \pi)^2} \left| G \left( p'_\perp, p'_z, \omega_m
\right) \right|^2  
\label{sumpzomega}
\end{eqnarray}   

\noindent where $N_z$ is the number of lattice sites along the z-axis and
$\omega_n = 2 \pi T (n + 1/2)$. It has been assumed that the order parameter
has even parity (singlet channel). The electron Green function has the usual
form 

%5.21
\begin{equation}
G^{-1} ({\bf p}, \omega_n) = i \omega_n - \epsilon_{\bf p} - \Sigma(p_z,
\omega_n)
\label{Gqomegan}
\end{equation} 

\noindent with $\epsilon_{\bf p}$ given by Eq. (\ref{epsbot}).

After in Eqs. (\ref{sumpzomega}) the $dp'_\perp$ integration has been done the
equations reduce to a one-dimensional problem. Thereby it is essential that the
kernel $K (q_z, \omega_\nu)$ is strongly peaked at $q_z = \pi/c$ and
$\omega_\nu = 0$. Therefore, loosely speaking the gap equation is of the form

%5.22
\begin{equation}
\Delta \left( p_z, \pi T \right) = -C \left( p_z \right) \Delta \left( p_z -
\frac{\pi}{c}, \pi T \right)
\label{Deltapz}
\end{equation} 

\noindent where $C(p_z)$ is a smooth positive function. This suggests the form

%5.23
\begin{equation}
\Delta \left( {\bf p} \right) = \Delta {\rm cos} \left( cp_z \right) 
\label{coscpz}
\end{equation} 

\noindent with $A_{1g}$ symmetry. The symmetry allows also
for a multiplication of the right hand side by a fully symmetric function
$f (p_x, p_y)$. The order parameter has lines of nodes at the AF zone boundary
$p_z = \pm Q_z/2 = \pm \pi/(2c)$. 

One finds that the Eliashberg equation yield also an odd-parity solution of the
form

%5.24
\begin{equation}
\Delta \left( p_z \right) = \Delta {\rm sin} \left( cp_z \right) 
\label{sincpz}
\end{equation} 

\noindent with a spin part $| \chi \rangle = (2)^{-1/2} (| \uparrow \downarrow
\rangle + | \downarrow \uparrow \rangle)$ and the same $T_c$. Note that because
of the Ising-like interaction (\ref{HcfIi}) rotational symmetry in spin space
is broken. A more general study of possible order parameters due to pair
potentials based on magnetic excitons was undertaken in
Ref. \cite{Thalmeier02}. It turned out that in the weak coupling limit, i.e.,
without taking retardations into account, one of the odd-parity triplet states
has lower energy than the singlet one. However this could be a consequence of
the weak coupling assumption which does not apply to UPd$_2$Al$_3$. There is
strong experimental evidence discussed below that the order parameter has
indeed $A_{1g}$ symmetry (see Eq. (\ref{coscpz})) and therefore we discard the
$A_{1u}$ solution (\ref{sincpz}).

In the following we discuss the parameters which are required to explain the
anisotropic effective mass within this simplified model and the ones which are
needed to explain the observed $T_c$. For the DOS at the Fermi energy a value
of $N(E_F) \simeq$ 2 states/(eV-cell-spin) seems appropriate. This includes
also the torus. The local intra-atomic excitations responsible for the mass
enhancement are characterized by $\delta \simeq$ 7 meV \cite{Zwicknagl03} and
$\omega_{\rm ex} \simeq$ 5 meV \cite{McHale04}. With these values it turns out
that a value of

%5.25
\begin{equation}
I^2N(E_F)=0.026 eV
\label{I2NPEF}
\end{equation} 

\noindent corresponding to $g/\epsilon_\perp = 2$ is required in order to
reproduce the experimentally observed mass enhancement within the simplified
scheme applied to Eliashberg's equations. With this value a superconducting
transition temperature $T_c$ = 2.9 K is obtained while the true value is 1.8
K. The dependence of the superconducting T$_c$ on the dimensionless coupling
strength g/$\epsilon_\perp$ is shown in Fig. ~\ref{fig:EXCITON}. Taking into
account that the strong mass anisotropies were derived theoretically {\it
  without} adjustable parameters (see Tab. \ref{tab:UPd2Al3EffMasses}) it is
gratifying to find that when the parameter $I^2 N(E_F)$ (or g/$\epsilon_\perp$)
of a simplified model is adjusted so as to reproduce the mass anisotropies, a
superconducting transition temperature of the right order of magnitude is
obtained.  

%%%%%%%%%%%%%%%%%%%%%%%%%%%%%%%%%%%%%%%%%%%%%%%%%%%%%%%%%%%%%%%%%%%%%%%%%%%%
\begin{figure}[tb]
\begin{center}
\includegraphics[width=9.0cm,clip]{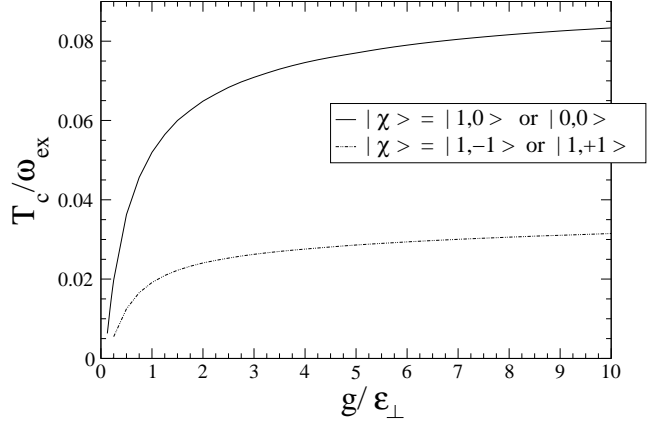}
\end{center}
\vspace{0.5cm}
\caption{The dependence of the superconducting T$_c$ on the
  electron-magnetic exciton coupling constant g (Eq.~\ref{gI241c}). Parameter
  values are $\omega_{ex}$ = 0.01$\epsilon_\perp$, $\alpha$ = -0.1, $\beta$ =
  0.8. The full curve corresponds to opposite spin pairing states
  $|\chi\rangle$ (S$_z$ = 0) A$_{1g}$ and A$_{1u}$ with gap functions given in
  Eqs.~(\ref{coscpz},\ref{sincpz}) and node lines at k$_z$ = $\pm\frac{\pi}{c}$
  and k$_z$ = 0 respectively. The dashed curve is a less favorable state with
  $|S_z| = 1$ and
  more node lines. For a value g/$\epsilon_\perp\sim$ 2 which leads to the
  observed mass enhancement m$^*$/m$_b\simeq$ 10 one obtains a calculated
  T$_c\simeq$ 2.9 K (from the full curve). This value is somewhat larger than
  the experimental T$_c$ = 1.8 K. (After \protect\cite{McHale04})} 
\label{fig:TCcalc}
\end{figure}
%%%%%%%%%%%%%%%%%%%%%%%%%%%%%%%%%%%%%%%%%%%%%%%%%%%%%%%%%%%%%%%%%%%%%%%%%%%%

As regards the experimental situation the superconducting state of
UPd$_2$Al$_3$ has been studied in great details for which we refer to the
review \cite{Thalmeier05}. Here we point out merely those experiments which
allow for a determination of the symmetry of the order parameter. This applies
in particular to the studies of the anisotropic thermal conductivity in an
applied magnetic field. As pointed out in \cite{Watanabe04} measurements of
the thermal conductivity under rotating magnetic field with heat current
perpendicular to the rotation plane yield information on the \v k - space
position of gap nodes. Corresponding measurements on single crystals of
UPd$_2$Al$_3$ \cite{Watanabe04} and their analysis \cite{ThalmeierPe05} have
indeed shown that $\Delta ({\bf p})$ has node lines perpendicular to the
c-axis. However these experiments cannot distinguish between a node line at
p$_z$=$\pm\frac{\pi}{2c}$ at the AF zone boundary or at p$_z$ = 0 at the zone
center, i.e., which one of the theoretically favored gap functions of
Eq.~(\ref{coscpz}) or Eq.~(\ref{sincpz}) is realized. Other evidence for node
lines was pointed out by Bernhoeft \cite{Bernhoeft00}. He argued that the
symmetry property $\Delta ({\bf p} \pm {\bf Q}) = -\Delta ({\bf p})$ is
required in order to explain the large intensity of a low energy
quasiparticle-like peak in the inelastic neutron scattering spectrum below
$T_c$. This again implies a line of nodes perpendicular to the $c$ axis but
cannot distinguish between the two candidates. However already before these
results became available a pronounced Knight shift reduction below T$_c$ was
found in 
UPd$_2$Al$_3$ \cite{Kitaoka00}. A naive interpretation of this result advocates
for the even parity A$_{1g}$ gap function in  Eq.~(\ref{coscpz}) whose
antisymmetric spin function corresponds to the spin singlet state in models
with isotropic pairing interaction. 

In concluding this section we want to summarize the above findings. It has been
shown that in UPd$_2$Al$_3$ superconductivity is due to non-phononic
bosons. Intra-atomic excitations of localized 5$f$ electrons in U ions, a
consequence of strong correlations and described within the dual model provide
the glue for the Cooper pairs. We want to stress that their interaction with
the itinerant 5$f$ electrons (see Eq. (\ref{HcfIi})) is {\it not} time-reversal
invariant. Nevertheless Cooper pairs may form but a sign change of the order
parameter along the c-direction is mandatory. Otherwise the interaction would
not be pair forming but rather act as a pair breaker. It is likely that a
similar pairing mechanism, mediated by the exchange of quadrupolar excitons,
is operative in the Pr-skutterudite cage compound.

%%%%%%%%%%%% letzte Bearbeitung: 15.11.2005 %%%%%%%%%%%%%%%%%%%
\section{Charge Ordering}
\resetdoublenumb 
\resetdoublenumbf

\label{sect:ChargeOrdering}

%\subsection{subsect:WC}
%\subsection{subsect:CDW}
%\subsection{subsect:MH}
%\subsection{subsect:Polaron}
%\subsection{subsect:YBA}
%\subsection{subsect:NAV}
%\newpage

The concept of charge order in electronic systems was introduced by Wigner in
the early 1930´s \cite{Wigner34}. He considered a homogeneous electron
gas, i.e., a system in which the positively charged background is distributed
uniformly over the sample. Wigner showed that in the limit of low densities the
Coulombic repulsion energy between the electrons will always dominate their
gain in kinetic energy due to delocalization. Therefore at sufficiently low
temperatures electrons will form a lattice. This way they minimize their
mutual repulsions. 

Following the original proposal of Wigner a number of systems have been
discovered which lead to electronic charge order. The cleanest realization of
Wigner crystallization is observed in a classical 2D sheet of electrons
generated on a liquid-He surface where the electron density can be varied by an
applied electrical field \cite{Grimes79}. In constrained geometries like
quantum dots clear signatures of Wigner crystallization may be found already
for small electron numbers \cite{Filinov01}.
The most common systems which exhibit this type of electronic charge order 
have an underlying atomic lattice as an important ingredient in
distinction to Wigner's homogeneous positive background. When valence
electrons are situated close to an atomic nucleus the overlap with orbitals
from neighboring sites is expected to be small. This implies that the
associated kinetic energy gain is small when electrons delocalize. Therefore
the mutual repulsions between electrons on neighboring sites will dominate the
kinetic energy gain at a higher density than it is the case for a uniformly
distributed positive background. But a resulting electronic charge order must
here be commensurate with the underlying atomic lattice. 

The 3d-valence electrons of transition metal compounds are most amenable to
this kind of charge order. The latter may occur within a metallic (or
at least conducting) state as in the prominent example of magnetite
Fe$_3$O$_4$ or within an already insulating state as in \NAV. Since
the number of d-electrons in 3d-compounds may commonly be changed by
doping, a large variety of commensurate 3d-charge ordered states can be
achieved, for example in the cuprate (parent compound LaCu$_2$O$_4$)
and single layer and bilayer manganite families (parent compounds LaMnO$_3$
and half-doped compound LaSr$_2$Mn$_2$O$_7$, respectively). The amplitude of
the total charge order 
parameter or charge disproportionation on the inequivalent 3d-site is 
typically small of the order $0.1 el$ per site or less. This is due to
the screening of the large 3d-orbital occupation changes by the
valence electrons of ligands, as can be clearly seen from LSDA+U
calculations. Experimentally the amount of charge disproportionation is
extremely difficult to determine. This is mostly done via the empirical
valence-bond analysis of X-ray results in the ordered phase, where the change
in bond distances is linked to the valence charge disproportionation.
In some 3d-oxide compounds like, e.g., the bilayer-manganites one must
be aware that 3d-charges are dressed by strong distortions of the
surrounding lattice leading to (small) polaron formation. It is then
the latter which exhibit the ordering transition.

Charge ordering is less common in 4$f$-compounds, because
the intersite Coulomb interaction necessary for charge order is well
screened and in genuine metallic compounds the hybridization with
conduction electrons tends to favor a site-independent 4f-occupation.
This leads to a metallic mixed valent or heavy fermion state. There is
however an important case where 4f-charge order may occur. In
insulating or semimetallic 4f compounds which are homogeneous
valence fluctuators at high temperatures, the intersite Coulomb
interaction may be strong enough to lead to a 4f-charge
disproportionation, i.e., an inhomogeneous mixed valent state at low
temperatures with different 4f-orbital occupations on inequivalent
sites. Again the amplitude of the total charge order including the
effect of ligand screening charges is much smaller than the bare
4f-charge disproportionation. An important class of compounds where
this charge-order transition occurs are members of the R$_4$X$_3$
series (R = rare earth Yb, Sm, Eu and X = As, Bi, P, Sb), notably the
semimetal \YBA~ 
%\ref{tab:R4X3} 
which is discussed in Sec. \ref{subsect:YBA}. 

An attractive feature of CO transitions in these 3d or 4f compounds is
the possibility of lowering the effective dimension of the arrangement
of magnetic ions, for example to a family of 1D chains, planes of
zig-zag chains or ladders and stripes. The effect of low
dimensionality has then important consequences for the spin excitation
spectrum, e.g., the appearance of a two-spinon continuum in the case of
spin chains.

When this type of electronic charge order in 3d and 4f systems is
compared with the original suggestion of Wigner one notices two differences. As
pointed out before, the lattice structure of the positive background is very
important but also the repulsions between electrons on neighboring sites are
not purely Coulombic. Instead they may be modified due to the electrons in
inner closed shells which leads to strongly screened Coulomb interactions that
usually extend only to nearest and next nearest neighbors.

Another type of charge ordering which does not break spatial symmetries is
obtained in systems which can be described by a Hubbard Hamiltonian at half
filling. Here it is the on-site Coulomb interaction expressed by an energy $U$
which is competing with electron hopping processes and the associated energy
gain. While the repulsion energy $U$ favors single occupancy of sites, and
suppresses double occupancies the kinetic energy gain favors a sizeable
fraction of sites with double occupations. Without electron correlations 25 \%
of them would be doubly occupied in order to optimize the kinetic energy. For
sufficiently large values of $U$ charge order in the form of strongly
suppressed on-site charge fluctuations will take place leading to a
Mott-Hubbard metal to insulator (M-I) transition. Again, similarities to the
previously considered cases are obvious but so are the differences. Repulsions
between electrons on different sites are completely neglected in the Hubbard
model and it is crucial to have precisely one electron per site in order to
obtain charge order. A similar requirement does not exist in the previously
considered cases. 

Finally, charge order can also occur via formation of a charge density wave
(CDW) in metals. In this case the instability is driven by minimizing the
kinetic (band) energy of conduction electrons leading to a
reconstruction of the Fermi surface. A prerequisite is the presence of
nesting properties in the Fermi surface. The generally incommensurate 
vector \v Q which connects the nesting parts determines
wavelength and direction of a corresponding CDW. Note that strong electron
correlations are not required for a CDW to form. Summarizing we
distinguish between the following electronic charge ordering processes:

\begin{itemize}
\item[(a)] Wigner crystallization in the homogeneous electron gas or
liquid %(sect.~\ref{subsect:WC})
\item[(b)] Charge order due to weak hybridizations and strong intersite
interactions %(sects.~\ref{subsect:YBA},\ref{subsect:NVA})
\item[(c)] Mott-Hubbard charge order due to strong on-site interaction
%(sect.~\ref{subsect:MH})
\item[(d)] Charge density waves due to nesting properties of the Fermi
surface %(sect.~\ref{subsect:CDW})
\item[(e)] Charge ordering in polaronic systems 
%(sect.~\ref{subsect:Polaron})
\end{itemize}

In this overview we will not address the genuine Mott-Hubbard M-I
transition which is reviewed in existing articles, e.g.,
Ref. \cite{Georges96}. It demands a considerable technical effort based on the
recently developed dynamical mean field theory (DMFT). Low dimensional metallic
CDW systems will also not be included, since this topic is well represented in
the literature \cite{Gruenerbook} and should be discussed together with
spin-density waves and superconductivity which is beyond the scope of this
article. 

\subsection{Wigner Crystallization in Homogeneous 2D Electron Systems}
\label{subsect:WC}

The electron gas may be subject to many different instabilities
depending both on the background (lattice) potential and the resulting
shape of the Fermi surface as well as the strength and range of the
screened Coulomb interactions. A convenient way to characterize the relevant
regime is the Brueckner parameter $r_s =
\frac{1}{a_0}(\frac{4\pi}{3}n)^{-\frac{1}{3}}$ which is the ratio of average
electron distance to the Bohr radius a$_0$ (n = electron density). In the two
limits of small and large r$_s$ the electron system exhibits radically
different behavior.  

At large density (small r$_s$) the system is in a metallic state dominated by
the kinetic (Fermi) energy and the Coulomb interactions between
electrons are well screened. If the Fermi surface has parallel
(quasi-1D) sections, 'nested' by a wave vector \v Q, the residual
Coulomb interactions may lead to a condensation of electron-hole
(Peierls) pairs into a charge- or spin-density wave state with
translational and possibly other symmetries broken. Part or all of the
nested Fermi surface sheets are then removed by the selfconsistent potential in
the condensed state.  For a conventional density wave spontaneous
modulation of the charge- or spin density with period 2$\pi/Q$ takes
place which may be identified by common methods like
X-ray or neutron diffraction. Since real metallic materials have high
densities and small or moderate r$_s$ this type of instability is
frequently encountered. It often competes or coexists with an
alternative condensation mechanism, namely electron-electron (Cooper)
pair formation leading to a superconducting state.

In the opposite limit of small density and large r$_s$ the long range
Coulomb interaction is badly screened and it dominates the small
kinetic energy gain due to delocalizations. At low enough temperatures
the electron liquid condenses in real space, forming a Wigner solid
\cite{Wigner34} which also breaks translational symmetry. This is complementary
to condensation in the \v k-space as found in the previous density wave
case. For a two dimensional (2D) homogeneous electron liquid in a
uniform positive background the appropriate Hamiltonian H=T+V is
%6.1
\begin{equation}
H = \sum_{\v k\sigma} \epsilon(\v k)c^\dagger_{\v k\sigma} c_{\v k\sigma}
+ \frac{1}{2\Omega} \sum_{\v p\v k\v q;\sigma\sigma'}
v_{\v q}c^\dagger_{\v p+\v q\sigma} c^\dagger_{\v k-\v q\sigma'}
c_{\v k\sigma'}c_{\v p\sigma}
\label{WLHAM}
\end{equation}
where $\epsilon(\v k) = \v k^2/2m$ is the kinetic energy and $v_{\v
q} = \frac{4\pi e^2}{\v q^2}(1-\delta_{\vq 0})$ is the Coulomb
interaction. The Kronecker delta $\delta_{\v q 0}$ ensures that due to
the positive background v$_{\v q}$ = 0. Furthermore $\Omega$ is the
volume. For practical calculations on a lattice and for finite systems
the Ewald summation technique has to be used to obtain the real space
Coulomb potential V(\v r). 

\begin{figure}[tb]
\begin{center}
\includegraphics[clip,width=7.0cm]{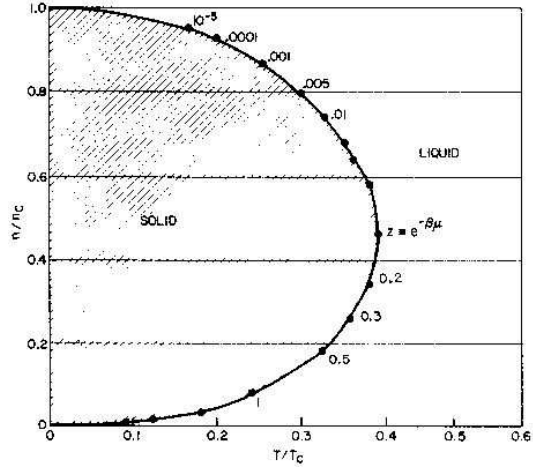}
\end{center}
\vspace{0.5cm}
\caption{Schematic Fermi liquid - Wigner solid phase diagram of the 2D
electron system. Estimates of critical density and temperature
are given by n$_c$ = (4/$\pi$a$_0^2$)(1/$\Gamma_0^2$) and
T$_c$ = (2e$^4$m/$\Gamma_0^2$). (After \protect\cite{Platzmann74})}
\label{fig:WLphaseTn}
\end{figure}

The qualitative shape of the n-T phase boundary for the liquid-solid transition
has been given in \cite{Platzmann74}. It is derived from the intuitive notion
that at the phase transition the average potential and kinetic energies of
electrons should be comparable, i.e., $\langle V \rangle/\langle
T\rangle\equiv\Gamma_0$. This leads to a parametrically ($z=\exp(-\beta\mu)$)
determined n-T phase boundary ($\mu$ = chemical potential) shown in
Fig.~\ref{fig:WLphaseTn}. In the classical limit (kT $\gg \mu$) one has
$\langle T\rangle$ = kT, furthermore $\langle V\rangle = e^2\sqrt{\pi n}$. Then
the melting curve is simply given by n(T) = (kT$\Gamma_0/\sqrt{\pi}e^2)^2$ with
an unknown parameter $\Gamma_0$.
%fig6.1
%%%%%%%%%%%%%%%%%%%%%%%%%%%%%%%%%%%%%%%%%%%%%%%%%%%%%%%%%%%%%%%%%%%%%%%%%%%%
%%%%%%%%%%%%%%%%%%%%%%%%%%%%%%%%%%%%%%%%%%%%%%%%%%%%%%%%%%%%%%%%%%%%%%%%%%%%
%
In a homogeneous background the 2D electron liquid solidifies in
a trigonal (hexagonal) lattice structure \cite{Bonsall77}. In the
static approximation without kinetic energy the ground-
state energy is E$_{GS}^\bigtriangleup$ = -3.921034 e$^2$a$_c^{-1/2}$
which is lower than for any other of the five 2D Bravais lattices.
(Here a$_c$ = ($\sqrt{3}$/2)a$_0^2$ is the hexagonal cell area with a$_0$
denoting the lattice constant, the density is n = 1/a$_c$). 
This result is simple to understand because of all 2D lattices the
trigonal one has the largest lattice spacing for a given density and
therefore minimizes the total energy if the latter is dominated by the
Coulomb repulsion as in the small density regime.
Stability analysis shows that the trigonal lattice is stable under
longitudinal and transverse (shear) distortions and the corresponding two
phonon branches have real frequencies. In the long wavelength limit they are
isotropic, i.e., they depend only on q = (q$_x^2$+q$_y^2$)$^{-1/2}$ and are
given by $\omega_1(q)$ = $\omega_p$(a$_0$q)$^\frac{1}{2}$ and $\omega_2(q)$ =
0.19 $\omega_p$a$_0$q for longitudinal and transverse modes respectively. Here
$\omega_p$ 
%=$\frac{4\pi e^2}{m^*}$($2a_ca_0$)$^{-1}$
is the 3D plasma frequency of a slab of thickness 2a. 
%fig6.2
%%%%%%%%%%%%%%%%%%%%%%%%%%%%%%%%%%%%%%%%%%%%%%%%%%%%%%%%%%%%%%%%%%%%%%%%%%%%
\begin{figure}[tb]
\begin{center}
\includegraphics[clip,width=8.0cm]{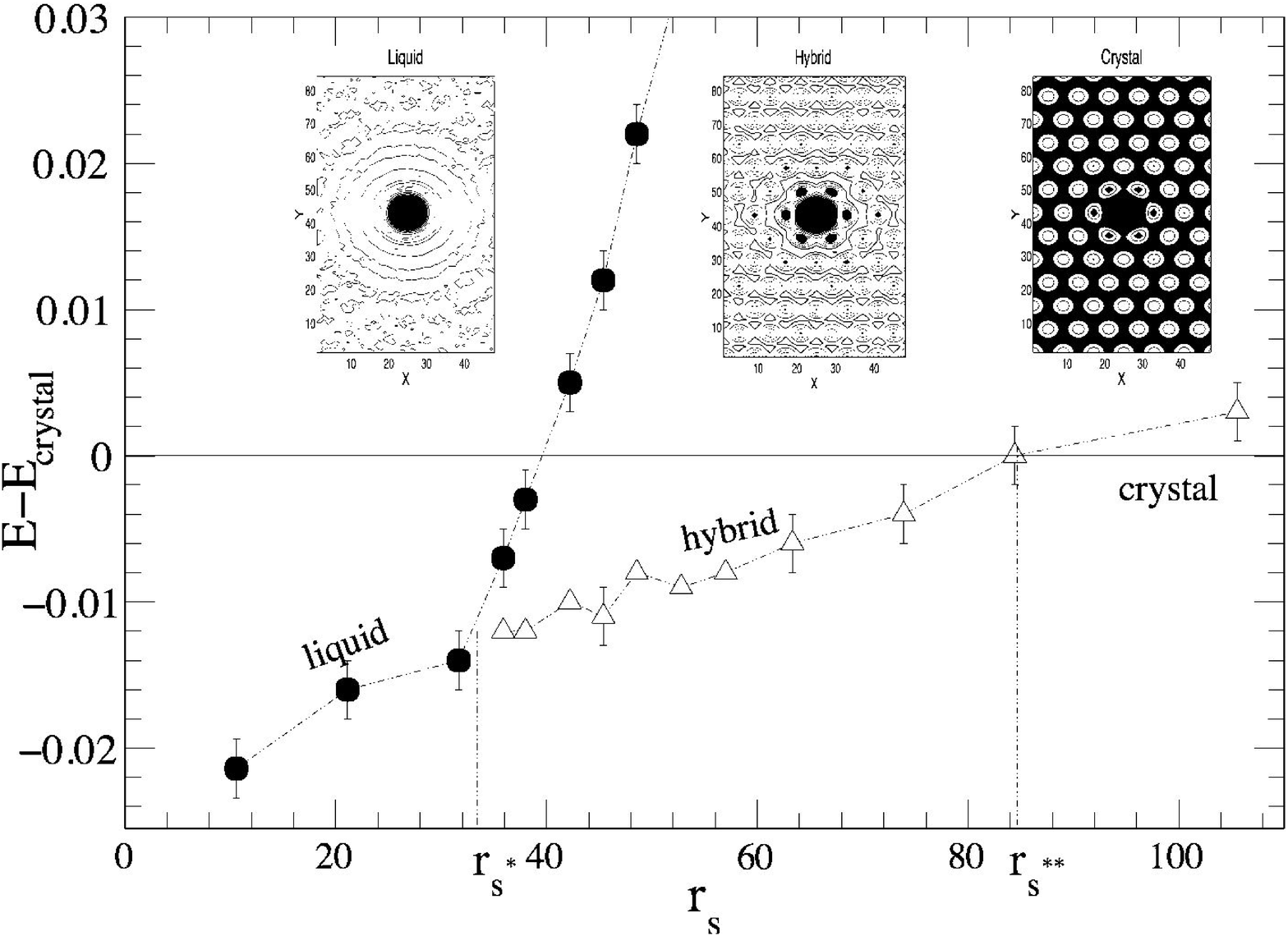}
\end{center}
\vspace{0.5cm}
\caption{Energy differences in units of 2$\pi$Nnt (t = hopping energy,
n = N/L$_x$L$_y$ = average density) of liquid and intermediate (or
hybrid) phases with respect to the Wigner crystal phase.  Here
E$_{liquid}$-E$_{crystal}$ (circles) and E$_{hybrid}$ - E$_{crystal}$
(triangles) are plotted as function of r$_s$ for N = 72 electrons on a
L$_x$ = 48 $\times$ L$_y$ = 84 grid. Critical values are r$^*_s\sim$ 30
and r$^{**}_s\sim$ 80. Inset: density-density correlation function
with reference particle in the center. (After \protect\cite{Falakshahi05})}
\label{fig:WLphase_rs}
\end{figure}
%%%%%%%%%%%%%%%%%%%%%%%%%%%%%%%%%%%%%%%%%%%%%%%%%%%%%%%%%%%%%%%%%%%%%%%%%%%%
%
While the Wigner lattice structure and stability is well understood,
the melting into the liquid state, both classical melting (as function
of T) and quantum melting at T = 0 (as function of the control parameter
r$_s$) is much less clear. The situation sketched in Fig.~\ref{fig:WLphaseTn}
is certainly oversimplified in one aspect: In both cases presumably an
intermediate phase between Wigner solid and Fermi liquid appears which is
characterized by the loss of long range translational order but preserves
quasi-long range orientational order. In the case of classical melting of 2D
trigonal lattices this intermediate phase is known as the 'hexatic phase'
\cite{Halperin78}. In the case of quantum melting (at T = 0) as function of
density (r$_s$) a similar precursor phase to the liquid seems to
appear. This can only be investigated by numerical techniques like
Quantum Monte Carlo (QMC) simulation. It was known from earlier QMC
results \cite{Tanatar89} that for $r_s>37\pm 5$ the Fermi liquid state becomes
instable. A more recent investigation of phases as function of r$_s$ was
undertaken in \cite{Falakshahi05} with a fixed-node QMC approach. A variational
trial or guiding wave function for the ground state of H (Eq.~(\ref{WLHAM})) on
a grid (L$_x$,L$_y$) was used. It has the following form
%6.2
\begin{equation}
\Psi(\v r_1,\v r_2,\dots\v r_N) = \mbox{Det} [\phi_i(\v r_j)] \prod_{i<j}J (|\v
r_i - \v r_j|)~~~ .
\label{JASTROW}
\end{equation}
The single particle wave functions $\phi_i(\v r_j)$ in the Slater
determinant Det$[\phi_i(\v r_j)]$ are taken as plane waves in
the liquid state and localized Gaussian orbitals on trigonal sites in
the Wigner lattice case. The Jastrow function $J(|\v r_i-\v r_j|)$
describes correlations and consists of modified Yukawa functions. It
takes the Coulomb correlations into account by keeping the electrons
apart. The range of correlations scales with the average electron
distance d = 1/$\sqrt{\pi n}$ where n = N/L$_x$L$_y$ denotes the
density.  With this trial wave function one would obtain indeed the
instability of the liquid state at $r^*_s\simeq$ 40 since for
$r_s>r_s^*$ its energy exceeds that of the Wigner solid (see
Fig.~\ref{fig:WLphase_rs}). However, the latter with its localized wave
functions is still not the most stable state in that range of r$_s$. This may
be seen if one uses instead trigonal lattice Bloch states for the single
particle functions $\phi_i(\v r_j)$ but with \v k constrained to the first
BZ. This implies an orientational symmetry breaking in \v k space, i.e., a
trigonally shaped Fermi surface. But the static density correlation function
has still no fully developed Bragg peaks like in the crystal in analogy to
the classical melting scenario. This intermediate or hybrid state has
indeed a lower energy than the Wigner solid in the range
$r^*_s<r<r^{**}_s$ with r$^*_s\simeq$ 30 and $r^{**}_s\simeq 80$. For
still higher values of $r_s$ the Wigner solid with fully localized
functions $\phi_i(\v r_j)$ finally becomes stable. There is no symmetry
change involved in going from the intermediate to the Wigner solid
state. Therefore it is not clear whether there is a real quantum phase
transition at $r^{**}_s$, as there is at $r^{*}_s$, or simply a
crossover to more pronounced density correlations (see inset of
Fig.~\ref{fig:WLphase_rs}). The picture of the intermediate phase is also
supported by QMC calculations for mesoscopic 2D electron systems in a harmonic
trap. There a similar  two-step transition is found from liquid to
orientational and finally to fully developed lattice correlations
\cite{Filinov01}. Finally we note that at even larger values of r$_s$ the
Wigner lattice becomes spin-polarized due to a ferromagnetic ring exchange in
the trigonal lattice \cite{Ceperley04}.

The experimental realization of the genuine Wigner lattice formation in
the 2D electron liquid has been attempted along two alternative
approaches. The more successful route is the accumulation of electrons
in a monolayer on the surface of liquids such as helium. The
accumulation is achieved by applying an electrical pressing field
perpendicular to the surface. Its strength allows to vary the electron density
in the monolayer over several orders of magnitude. The lower part of the
classical melting curve in Fig.~\ref{fig:WLphaseTn} has been determined in
\cite{Grimes79} by a rf-resonance method.  For the ratio of potential to
kinetic energy at the phase boundary a surprisingly large value of
$\Gamma_0\sim$ 137 was found in agreement with early MC simulations. Due to a
finite surface tension the pressing field deforms the liquid surface around
the electrons thereby leading to single electron 'dimples'. This
induces an effective attraction between electrons which has to be
added to the Coulomb repulsion. It has been proposed that this
attraction may lead to a structural phase transition of the Wigner
lattice from trigonal to square lattice for sufficiently low density
and high pressing fields \cite{Haque03}. The second route to generate 2D
electron systems are semiconductor heterojunctions. It has been much less
conclusive because the evidence for Wigner lattice formation in
transport properties is obscured by the 2D localization effects caused
by impurities \cite{Abrahams01}.

Computational results presented before show that the genuine Wigner
lattice formation takes place at r$_s$ values which are more than an
order of magnitude larger than those found in real solids.
But in real crystals with inhomogeneous electron densities the
condition for a Wigner type of lattice formation may be much easier to
fulfill. In case that the overlap of atomic wavefunctions  of
neighboring atoms is small the gain in kinetic energy due to electron
delocalization is also small. The mutual Coulomb repulsion of
electrons on neighboring sites can become more easily dominant in that
case than in a homogeneous electron gas. Neighboring rare earth ions
have a particularly small overlap of their 4f wavefunctions. Therefore they are
particularly good candidates for the formation of charge ordered Wigner like
lattices \cite{Fulde97} as will be discussed in the next section.

Furthermore in 3d/4f compounds the Coulomb interactions are strongly
screened whereas in the two-step classical and quantum melting of a 2D
Wigner lattice the long range part of the Coulomb interactions is most
important. Indeed, charge ordering in compounds usually takes place at
a well defined temperature and is mostly of first order. It is driven
by a competition between short range next and next nearest neighbor Coulomb
interactions and the kinetic energy. Also some compounds are already in a
Mott-Hubbard insulating state due to large on-site Coulomb repulsion before
charge order caused by the inter-site Coulomb interactions appears. Therefore,
for real compounds like vanadates and manganites the extended Hubbard type
models are a better starting point than the Hamiltonian in Eq.~(\ref{WLHAM}) to
describe the multitude of charge ordering phenomena in solids. Nevertheless,
loosely speaking one may refer to them as a kind of generalized Wigner
lattice formation \cite{Fulde97}.

\subsection{Generalized Wigner Lattice: Yb$_{4}$As$_{3}$}
\label{subsect:YBA}

The intermetallic compound Yb$_4$As$_3$ is a perfect example of a system in
which charge order takes place, here of 4$f$ holes. Yb$_4$As$_3$ has a cubic
anti-Th$_3$P$_4$ structure with a $I\bar{4}3d$ space group. Due to the special
lattice structure charge order in this three-dimensional system results in the
formation of well separated chains of Yb$^{3+}$ ions. They act like
one-dimensional spin chains. The net result is 
that Yb$_4$As$_3$ shows all the signs of a low-carrier-density heavy
fermion system which here is due to the properties of the spin
chains. This is a good example for the formation of heavy quasiparticles caused
by electronic charge order. The Kondo effect plays no role in this compound.

\begin{figure}[t b]
%\begin{center}
\includegraphics[clip,width=7.0cm]{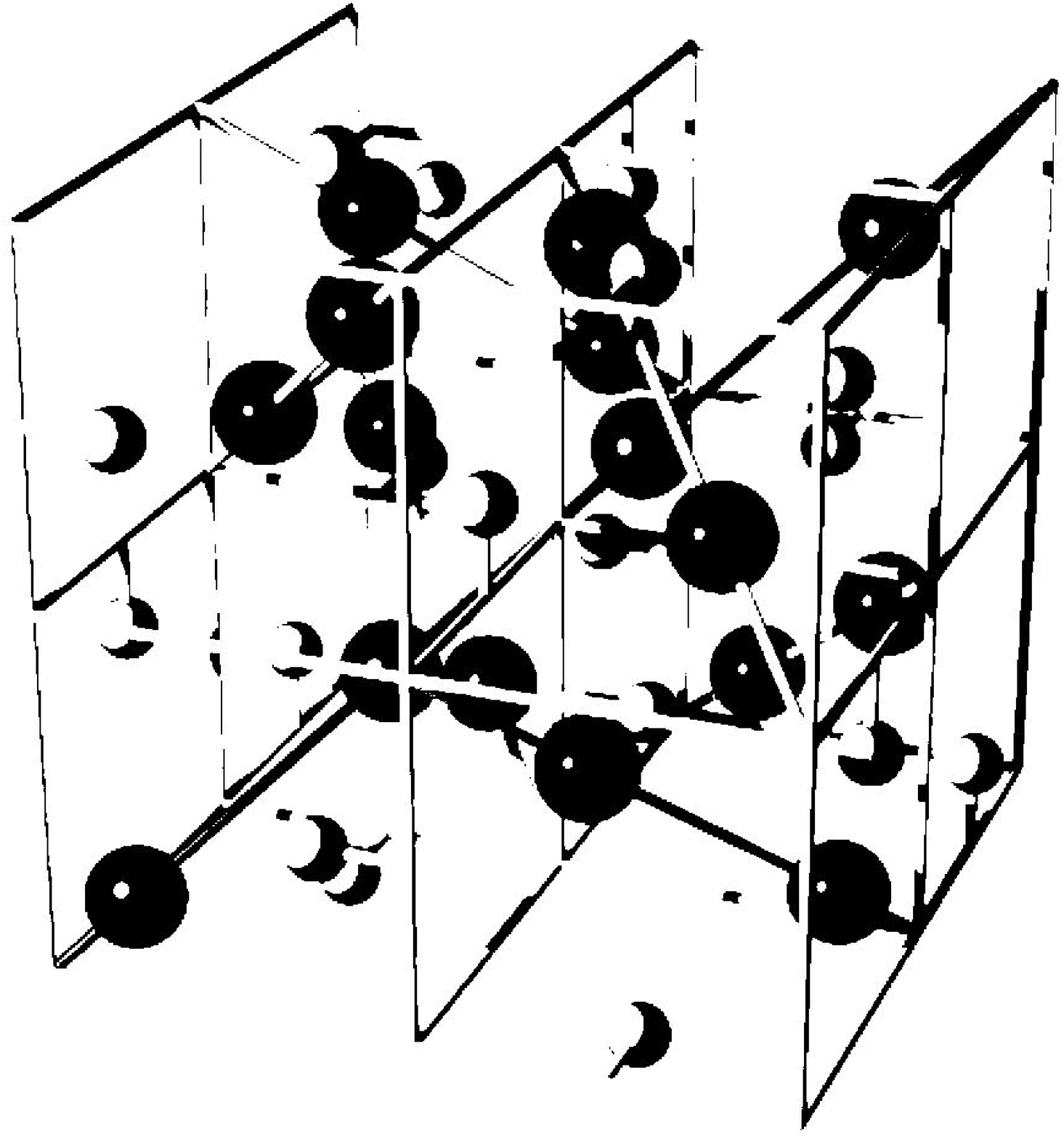}
\hfill
\includegraphics[clip,width=7.0cm]{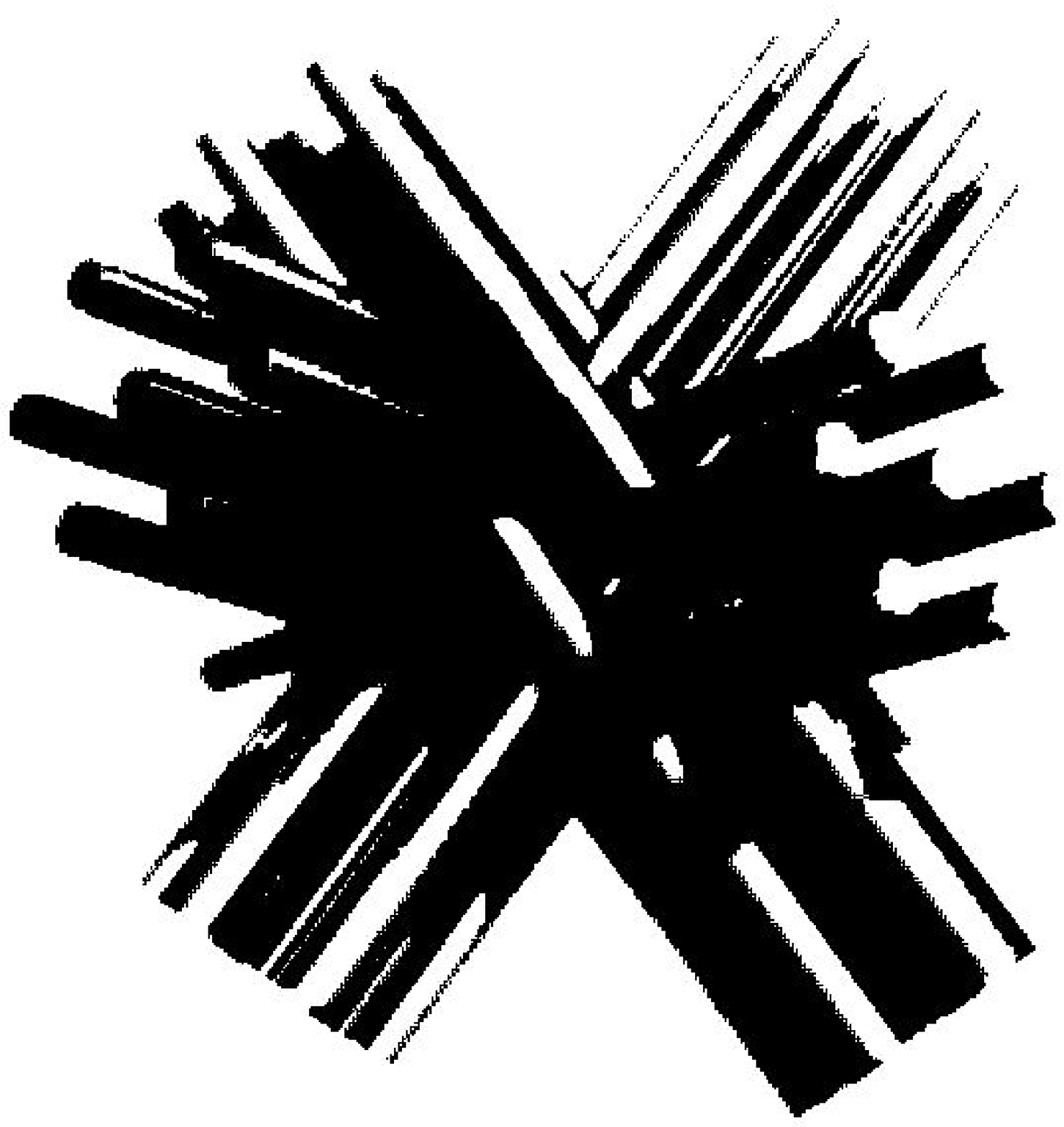}
%\end{center}
\vspace{0.5cm}
\caption{Left panel: Anti-Th$_3$P$_4$ structure of Yb$_4$As$_3$. Large and
small spheres symbolize Yb and As ions, respectively. The Yb ions are
residing on four interpenetrating families of chains oriented along
$\la 111\ra$ space diagonals. Right panel: Dense rod packing representation of
the Yb-chains. In the CO structure only \emph{one} family of chains
carries Yb$^{3+}$ ions with pseudo-spin S = 1/2 whereas the other three
families are occupied with S = 0  Yb$^{2+}$ ions. (After
\protect\cite{Schmidt01})} 
\label{fig1}
\end{figure}

\begin{figure}[tb]
\begin{center}
\includegraphics[clip,width=7.0cm]{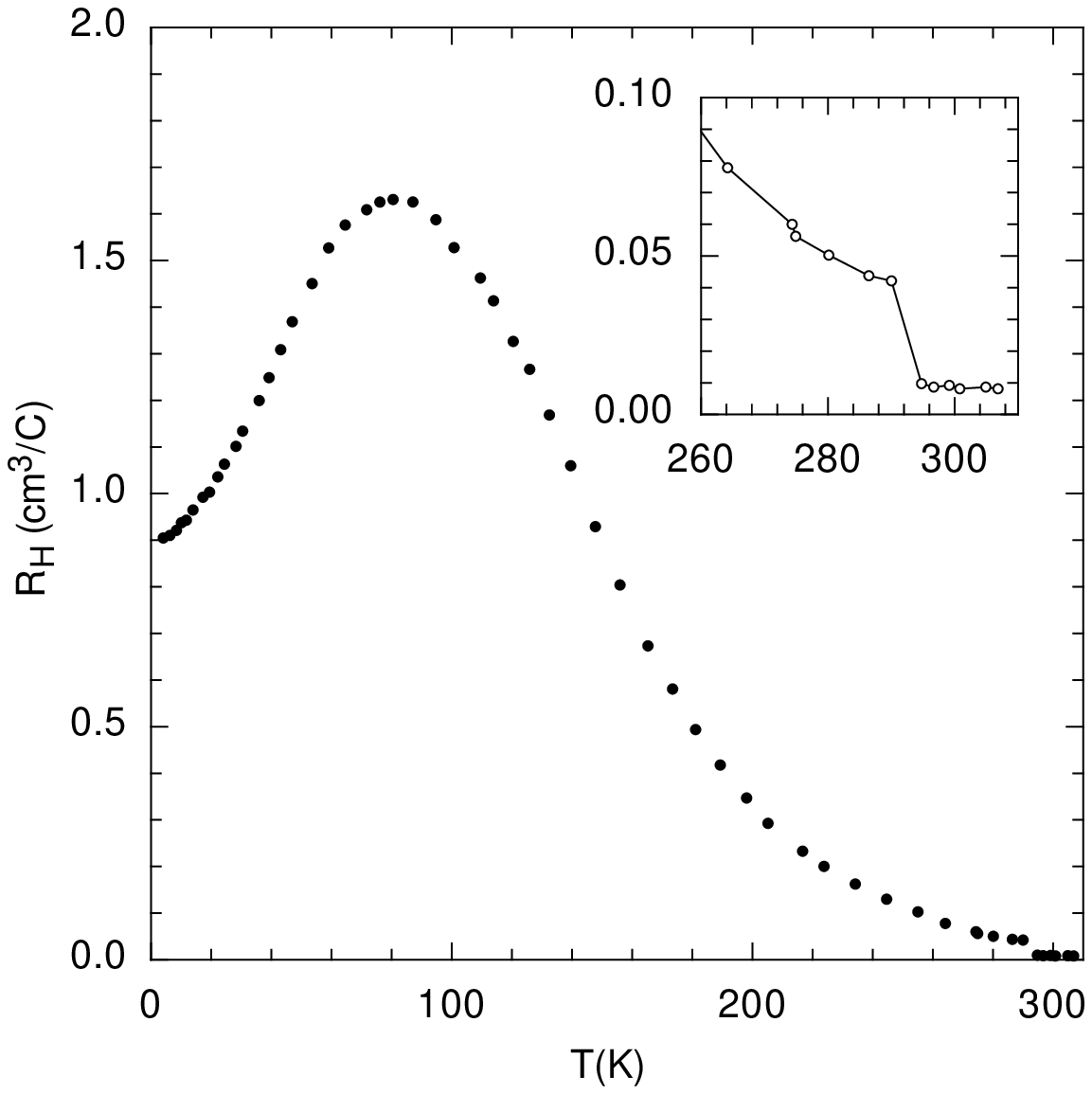}\hfill
\includegraphics[clip,width=7.0cm]{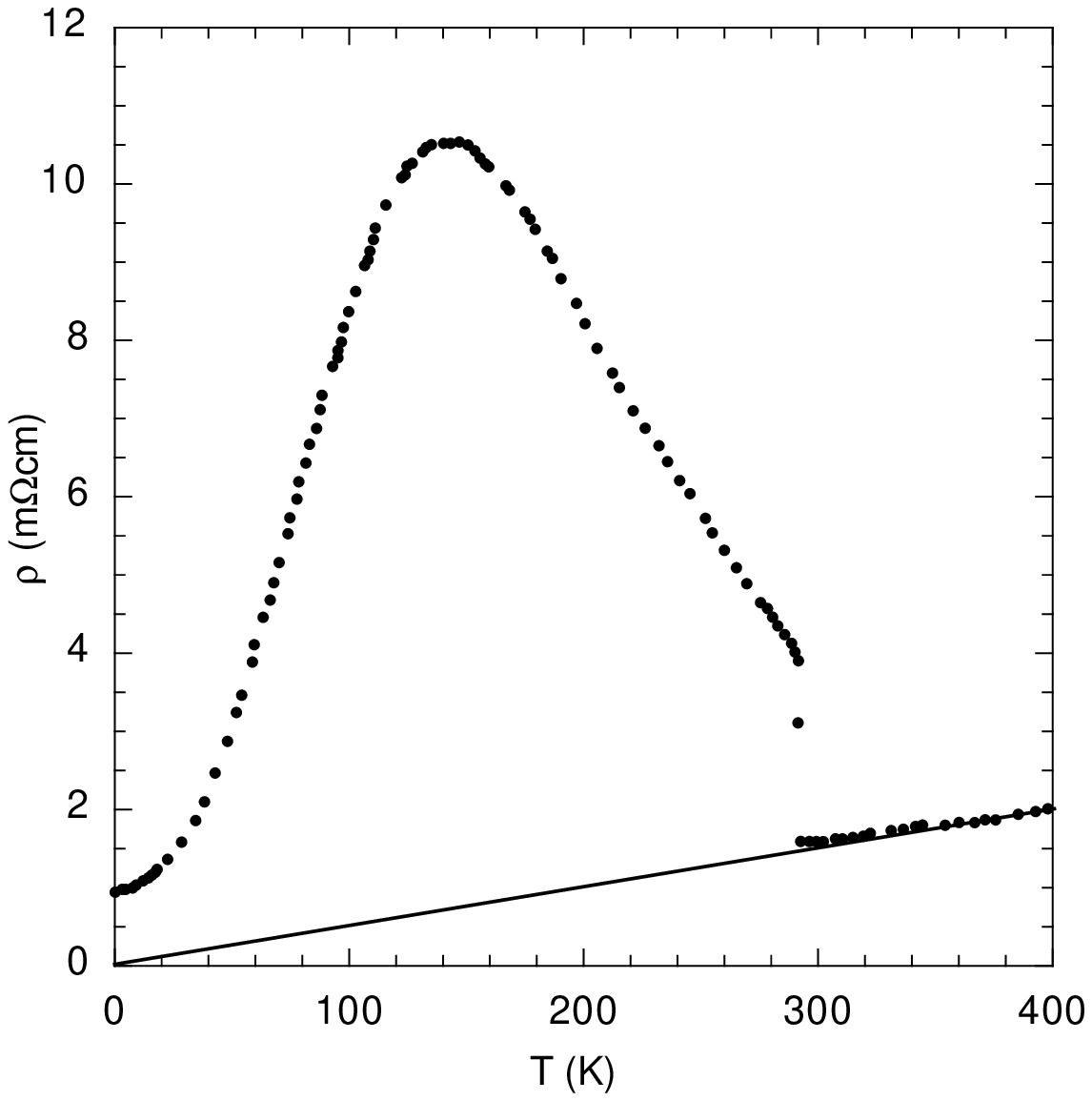}
\end{center}
\vspace{0.5cm}
\caption{Left panel: Hall coefficient R$_H$(T). The insert shows the
change at the phase transition temperature T$_c$. Right panel: resistivity
$\rho$(T) for Yb$_4$As$_3$. At T$_c$ = 295 K a phase transition due to charge
ordering is taking place. Solid line: extrapolation of $\rho$(T)$\sim$T. (After
\protect\cite{Ochiai90})} 
\label{fig2}
\end{figure}

\begin{figure}[tb]
\begin{center}
\includegraphics[clip,width=6.5cm]{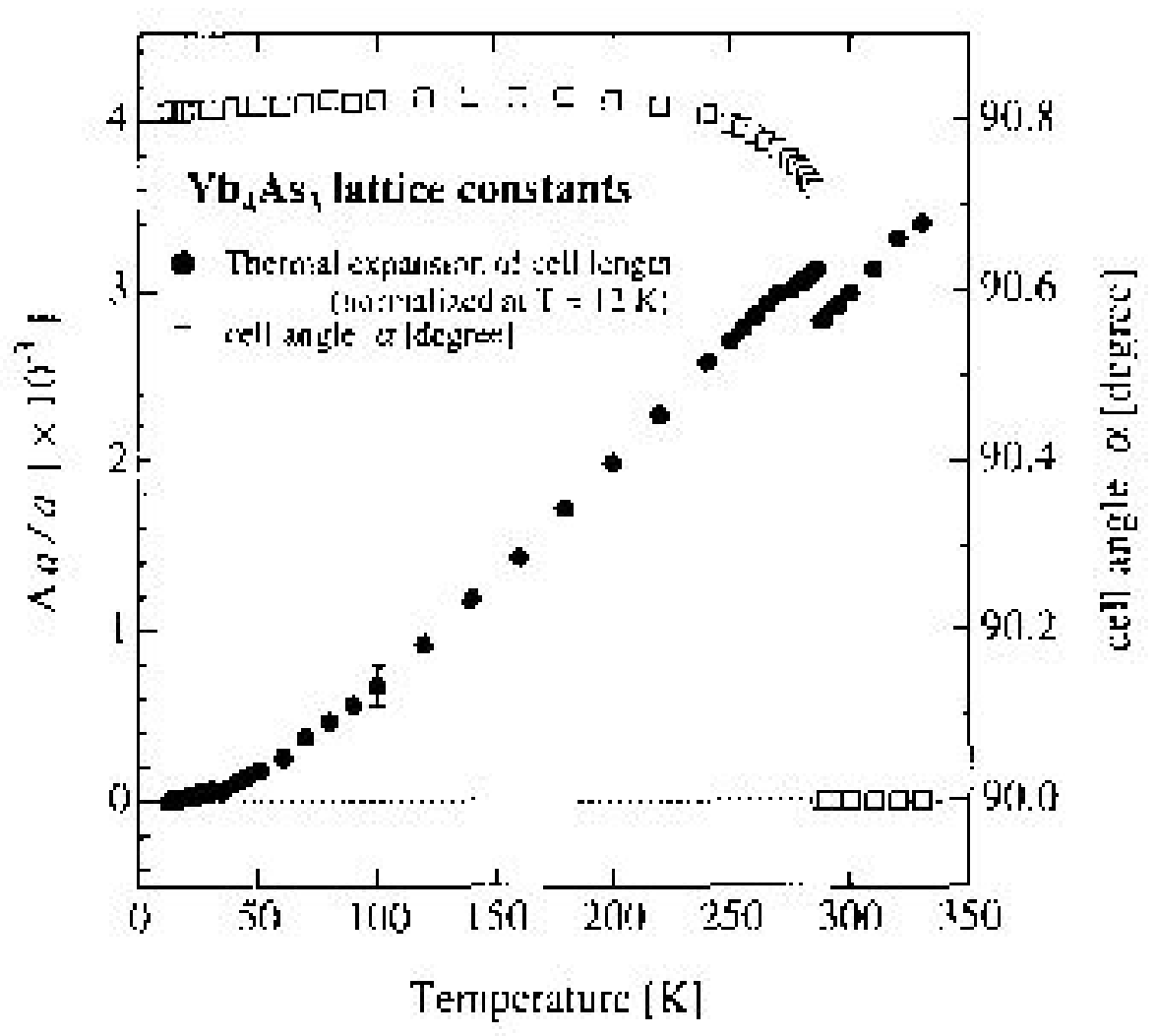}\hfill
\includegraphics[clip,width=7.5cm]{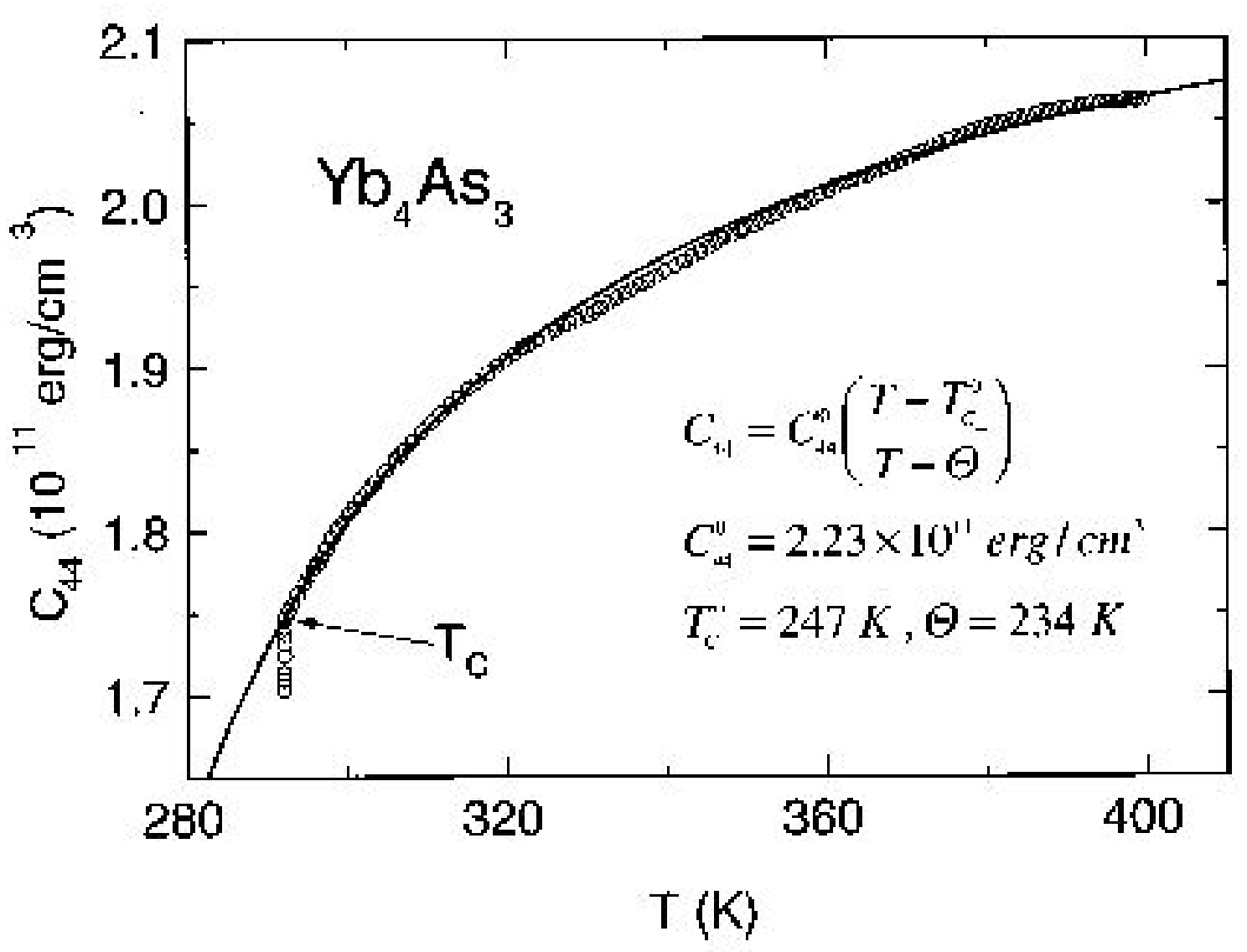}
\end{center}
\vspace{0.5cm}
\caption{Left panel: Unit cell angle $\alpha = 90^\circ +\delta$ (open
symbols) as function of T showing first order trigonal distortion at
the CO transition temperature T$_c\simeq$ 288 K together with linear thermal
expansion $\Delta a/a$ (full symbols). After \protect\cite{Iwasa98}. 
Right panel: temperature dependence of the elastic constant c$_{44}$(T). Above
the structural phase transition temperature T$_c$ a strong softening is
observed described by Eq.~(\ref{c044}) caused by coupling to the $\Gamma_5$
type charge order parameter. Due to the first-order nature of the transition
the theoretical mean field transition temperature T$_{c0}$ = 234K is smaller
than the actual T$_c$. (After \protect\cite{Goto99})}
\label{fig11}
\end{figure}

We start out by summarizing some experimental facts and results. Yb$_4$As$_3$
and other family members of R$_4$X$_3$ were first systematically investigated
by Ochiai et al. \cite{Ochiai90}. A compilation of more recent results may be
found in \cite{Schmidt01}. By counting valence electrons one notices that since
As has a valency of -3, three of the Yb ions must have a valency of +2
while one ion has a valency of +3, i.e., one expects ${\rm Yb}_4{\rm
As}_3 \rightarrow ({\rm Yb}^{2+})_3({\rm Yb}^{3+})({\rm As}^{3-})_3$. But
Yb$^{2+}$ has a filled 4$f$ shell. Thus there is one 4$f$ hole per formula
unit. The Yb ions occupy four families of interpenetrating chains 
which are pointing along the diagonals of a cube (see Fig. \ref{fig1}). This is
often referred to as body-centered cubic rod packing \cite{Li97}. It is
important to notice that the distance between two neighboring Yb ions along a
chain is larger than the distance between ions belonging to different
chains. This implies that nearest neighbor Yb ions belong to different
families of chains. At sufficiently high temperatures, i.e., above 300
K the 4$f$ holes move freely between sites and the system is
metallic. Measurements of the Hall coefficient R$_{\rm H}$ confirm that the
carrier concentration is approximately one hole per unit cell in that
temperature range. The situation is different at low temperatures where the
measured Hall coefficient has a value of ${\rm R_H} = 7 \cdot 10^{18}cm^{-3}$
implying approximately one hole per 10$^3$ Yb ions (see Fig. \ref{fig2}). 
%fig6.3
%%%%%%%%%%%%%%%%%%%%%%%%%%%%%%%%%%%%%%%%%%%%%%%%%%%%%%%%%%%%%%%%%%%%%%%%%%%
%%%%%%%%%%%%%%%%%%%%%%%%%%%%%%%%%%%%%%%%%%%%%%%%%%%%%%%%%%%%%%%%%%%%%%%%%%%
%
Thus the system changes from a metal to a semimetal as the temperature
decreases. This is particularly seen in measurements of the resistivity
$\rho$(T) (see Fig. \ref{fig2}). While for T $>$ 300 K a linear temperature
dependence is observed, one notices that near T$_{\rm c} \simeq$ 292 K
a first-order phase transition is taking place with a corresponding
increase in resistivity. At low temperatures $\rho$(T) = $\rho_0$ +
AT$^2$ is found, i.e., the semimetal is a Fermi liquid. Despite the
low carrier concentration of order n $\simeq 10^{-4}$ per cell
obtained from the Hall constant Yb$_4$As$_3$ shows all the signs of a
heavy-quasiparticle system at low temperatures. The
$\gamma$-coefficient of the specific heat is $\gamma \simeq$ 200
mJ/(mol $\cdot$ K$^2$), the Sommerfeld-Wilson ratio is R$_W$ =
4$\pi^2$k$_B^2$/3(g$\mu_B)^2$($\chi/\gamma$) $\simeq$ 1 implying an
equally enhanced spin susceptibility and the Kadowaki-Woods ratio
A/$\gamma^2$ is similar to that of other heavy quasiparticle
systems. If one were to postulate an origin of the heavy mass within the
Kondo lattice mechanism, this may seem very strange in view of the
fact that this semimetal has a very low density of charge carries.
%fig6.4
%%%%%%%%%%%%%%%%%%%%%%%%%%%%%%%%%%%%%%%%%%%%%%%%%%%%%%%%%%%%%%%%%%%%%%%%%%%%
%%%%%%%%%%%%%%%%%%%%%%%%%%%%%%%%%%%%%%%%%%%%%%%%%%%%%%%%%%%%%%%%%%%%%%%%%%%%
%
The phase transition at T$_c$ $\simeq$ 300 K is accompanied by a trigonal
distortion and a change of the space group to R3c. This structural transition
is volume conserving and is triggered by charge order of the 4$f$ holes. The
angle between orthogonal axes in the cubic phase changes to $\alpha$ =
90.8$^\circ$ in the trigonal phase for T $\ll$ T$_c$
(Fig.~\ref{fig11}). Associated with this change is a spontaneous elastic strain
below T$_c$ which is proportional to the charge order parameter. The structural
instability is accompanied by a softening of the c$_{44}$ elastic mode above
T$_c$ \cite{Goto99}. This suggests that the trigonal elastic strain
$\epsilon_{yz}, \epsilon_{zx}, \epsilon_{xy}$ with $\Gamma_5$ symmetry plays a
crucial role.  

The temperature dependence of the c$_{44}$ mode may be obtained from a
Ginzburg-Landau expansion of the free energy in terms of the strains and the
charge ordering parameter components (Q$_{yz}$,Q$_{zx}$,Q$_{xy}$)
\cite{Goto99}. The latter are defined by expanding the charge $\rho =
\rho_0 + \Delta\rho$ in the form
%6.3
\begin{equation}
\Delta\rho = Q_{yz} \rho_{yz} (\Gamma_5) + Q_{zx} \rho_{zx} (\Gamma_5) +
Q_{xy} \rho_{xy} (\Gamma_5)~~~.
\label{DeltaRho}
\end{equation} 
Here $\rho_0$ is the part of the charge distribution which remains unchanged by
the phase transition while $\rho_{ij}(\Gamma_5)$ are the charge fluctuation
modes of $\Gamma_5$ symmetry. It is Q$_{ij}$= 0 for T $>$ T$_c$ and Q$_{ij}
\neq$ 0 for T $<$ T$_c$. Up to a constant the free energy contains three
different contributions. One (F$_Q$) is due to the order parameter, a
second one (F$_{el}$) is due to the elastic energy of the lattice and the third
one (F$_{Q-el}$) describes the interactions of the order parameter with the
lattice. For the $c_{44}$ elastic constant we obtain \cite{Goto03} 
%6.4
\begin{eqnarray}
F_Q & = & F_0 + \frac{\alpha}{2} \left( Q^2_{xy} + Q^2_{xz} + Q^2_{yz}
\right)\nonumber \\ 
&& + \frac{\beta}{4} \left( Q^4_{xy} + Q^4_{xz} + Q^4_{yz} - \frac{3}{5} \left(
Q^2_{xy} + Q^2_{xz} + Q^2_{yz} \right)^2 \right)\nonumber \\
F_{el} & = & \frac{c^0_{44}}{4} \left( \epsilon^2_{xy} + \epsilon^2_{xz} +
\epsilon^2_{yz} \right)\nonumber \\
F_{Q-el} & = & -g \left( Q_{xy} \epsilon_{xy} + Q_{xz} \epsilon_{xz} + Q_{yz}
\epsilon_{yz} \right) ~~~.
\label{FQFQ}
\end{eqnarray}
Near a phase transition $\alpha = \alpha_0$(T-$\Theta$) changes sign
at a characteristic temperature $\Theta$. The fourth-order terms in
Q$_{ij}$ stabilize the ordered state of the system.  For T $>\Theta$ the
softening of the elastic constant is obtained by neglecting the terms $\sim$
Q$^4_{ij}$ in the free energy and minimizing F = F$_Q$ + F$_{el}$ +
F$_{Q-el}$ by setting $\partial$F/$\partial$Q$_{ij}$ = 0. For $\beta >$ 0 this
leads to a trigonal charge order parameter Q$_t$ =
$\frac{1}{\sqrt{3}}$(Q$_{xy}$,Q$_{xz}$,Q$_{yz}$). It also leads to a
proportionality between Q$_{ij}$ and the strain $\epsilon_{ij}$ which may be
used to rewrite the free energy in the form
%6.5
\begin{equation}
F = F_0 + \frac{1}{2} \left( c ^0 _{44} - \frac{g^2}{\alpha_0 (T - \Theta)}
\right) \left( \epsilon^2_{xy} + \epsilon^2_{xz} + \epsilon^2_{yz} \right)~~~. 
\label{FF0}
\end{equation}  
Therefore the renormalized elastic constant is
%6.6
\begin{equation}
c_{44} = c^0_{44} \left( \frac{T - T_{c0}}{T - \Theta} \right) \qquad
\mbox{where} \qquad T_{c0} = \Theta + \frac{g^2}{\alpha_0 c^0_{44}}
\label{c044}
\end{equation}
denotes the theoretical mean-field transition temperature in the presence of
the strain interaction. The explanation of a first-order phase transition at
T$_c >$ T$_{c0}$ requires the inclusion of higher order terms in Q$_{ij}$ in
(\ref{FQFQ}). A detailed group-theoretical analysis of different measured
elastic constants is found in Ref. \cite{Goto99}.
%fig6.5
%%%%%%%%%%%%%%%%%%%%%%%%%%%%%%%%%%%%%%%%%%%%%%%%%%%%%%%%%%%%%%%%%%%%%%%%%%%%
%%%%%%%%%%%%%%%%%%%%%%%%%%%%%%%%%%%%%%%%%%%%%%%%%%%%%%%%%%%%%%%%%%%%%%%%%%%%
%
At temperatures T $>$ T$_{\rm c}$ the 4$f$ holes are equally distributed over
the chains. But at low temperatures the system avoids nearest-neighbor
Yb$^{3+}$ - Yb$^{3+}$ sites in order to minimize the mutual short-range
inter-site repulsion of 4$f$ holes. This is accomplished by an accumulation of
4$f$ holes in one family of chains, i.e., by charge order. This way
nearest-neighbor repulsions of 4$f$ holes are reduced.  In the idealized case
one expects that for T $\rightarrow$ 0 one family of chains consists of
Yb$^{3+}$ sites while in the three remaining families of chains all sites are
in a Yb$^{2+}$ configuration. This explains the trigonal distortion
(Fig.~\ref{fig11}) which is accompanying charge ordering of the 4$f$
holes. Since Yb$^{3+}$ ions have a smaller ionic radius than Yb$^{2+}$ ions the
sample is shrinking in the direction of the chains containing the Yb$^{3+}$
ion. In order to keep the volume of the unit cell constant (otherwise a too
large amount of elastic energy would be necessary) the remaining three families
of chains must expand correspondingly.\\ 
%fig6.6
%%%%%%%%%%%%%%%%%%%%%%%%%%%%%%%%%%%%%%%%%%%%%%%%%%%%%%%%%%%%%%%%%%%%%%%%%%%%
\begin{figure}[tb]
\includegraphics[width=8cm]{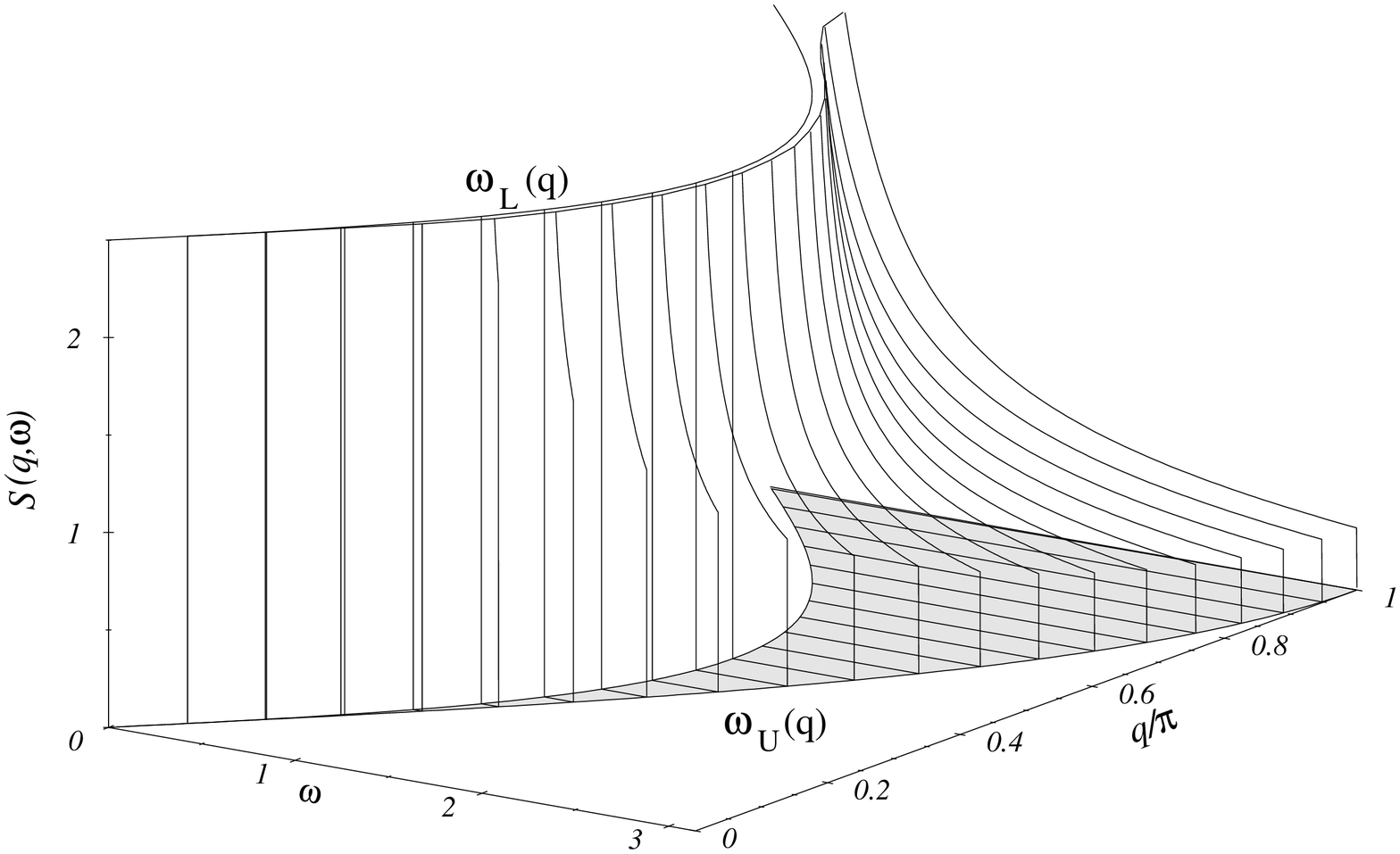}
\vspace{0.5cm}
\caption{Spectral function S(q,$\omega$) of the two-spinon excitation spectrum
  of a Heisenberg chain in the q,$\omega$-plane. It is nonzero in the shaded
  region above the lower bound $\omega_L(q)$ calculated by
  \protect\cite{Cloizeaux62} and the below the upper bound $\omega_U(q)$ given
  by Eq.(\ref{DYNSTRUC}). S(q,$\omega$) diverges at $\omega_L(q)$ approximately
  with a root singularity. The step like cutoff at $\omega_U(q)$ is an artefact
  of the approximate form in Eq.(\ref{DYNSTRUC}). The lattice constant is set
  equal to unity. (After \protect\cite{Karbach97})} 
\label{fig:spinon2a}
\end{figure}
%%%%%%%%%%%%%%%%%%%%%%%%%%%%%%%%%%%%%%%%%%%%%%%%%%%%%%%%%%%%%%%%%%%%%%%%%%%%
%
It was first pointed out in Ref. \cite{Fulde95} that the origin of the heavy
quasiparticles are spin excitations (spinons) in the chains of Yb$^{3+}$
ions and that it is not due to the Kondo effect as previously thought. This
physical model was beautifully confirmed by inelastic neutron scattering
experiments (INS) \cite{Kohgi97,Kohgi99}. They demonstrated that the magnetic
excitations are those of an isotropic antiferromagnetic Heisenberg chain. They
agree with the spectrum of two-spinon excitations lying within the lower and
upper boundaries $\omega_L(q)$ and $\omega_U(q)$ respectively, given by 
%6.7
\begin{eqnarray}
\omega_L(q) = \frac{\pi}{2}J\sin(q) \qquad \mbox{and} \qquad \omega_U(q) = \pi
J\sin \left( \frac{1}{2}q \right) 
\end{eqnarray}
with q measured in units of inverse lattice constants d$^{-1}$. The lower
boundary $\omega_L(q)$ was calculated long before by des Cloizeaux and Pearson
\cite{Cloizeaux62} (see Fig.~\ref{fig:spinon2a}). Note that unlike in the
classical spin chain there is no sharp spin wave excitation for a given
momentum q but a two-spinon continuum. Its corresponding dynamic spin-structure
factor (Fig.~\ref{fig:spinon2a}) is proportional to the INS cross section and
may be approximated by \cite{Karbach97} 
%6.8
\begin{equation}
S(q,\omega) = \frac{\Theta \left( \omega-\omega_L(q) \right) \Theta \left(
  \omega_U(q)-\omega \right)}{\sqrt{\omega^2-\omega_L^2(q)}}~~~.
\label{DYNSTRUC}
\end{equation}
%fig6.7
%%%%%%%%%%%%%%%%%%%%%%%%%%%%%%%%%%%%%%%%%%%%%%%%%%%%%%%%%%%%%%%%%%%%%%%%%%%%
\begin{figure}[tb]
\begin{center}
\raisebox{0.3cm}
{\includegraphics[width=5cm]{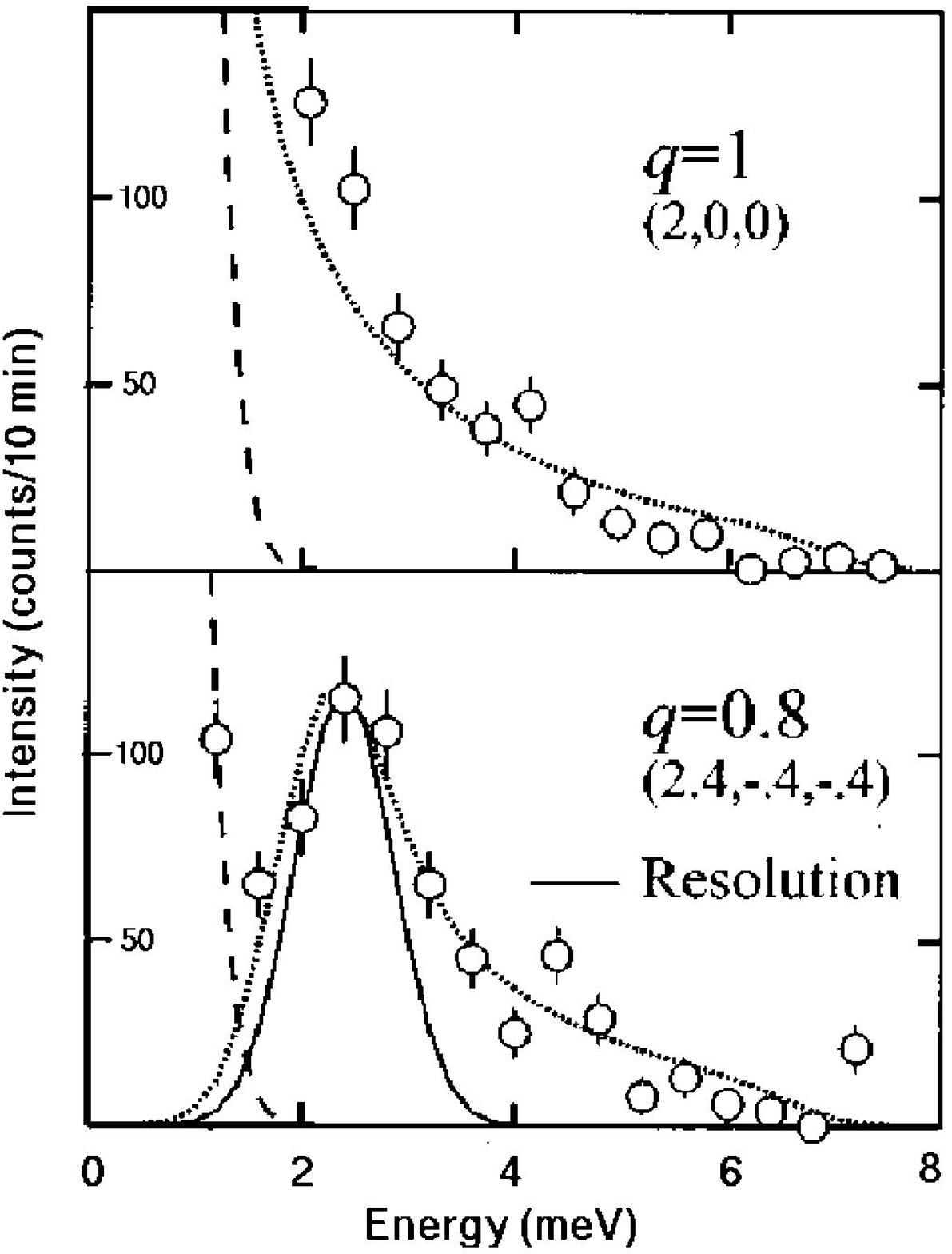}}\hfill
\includegraphics[width=8.5cm]{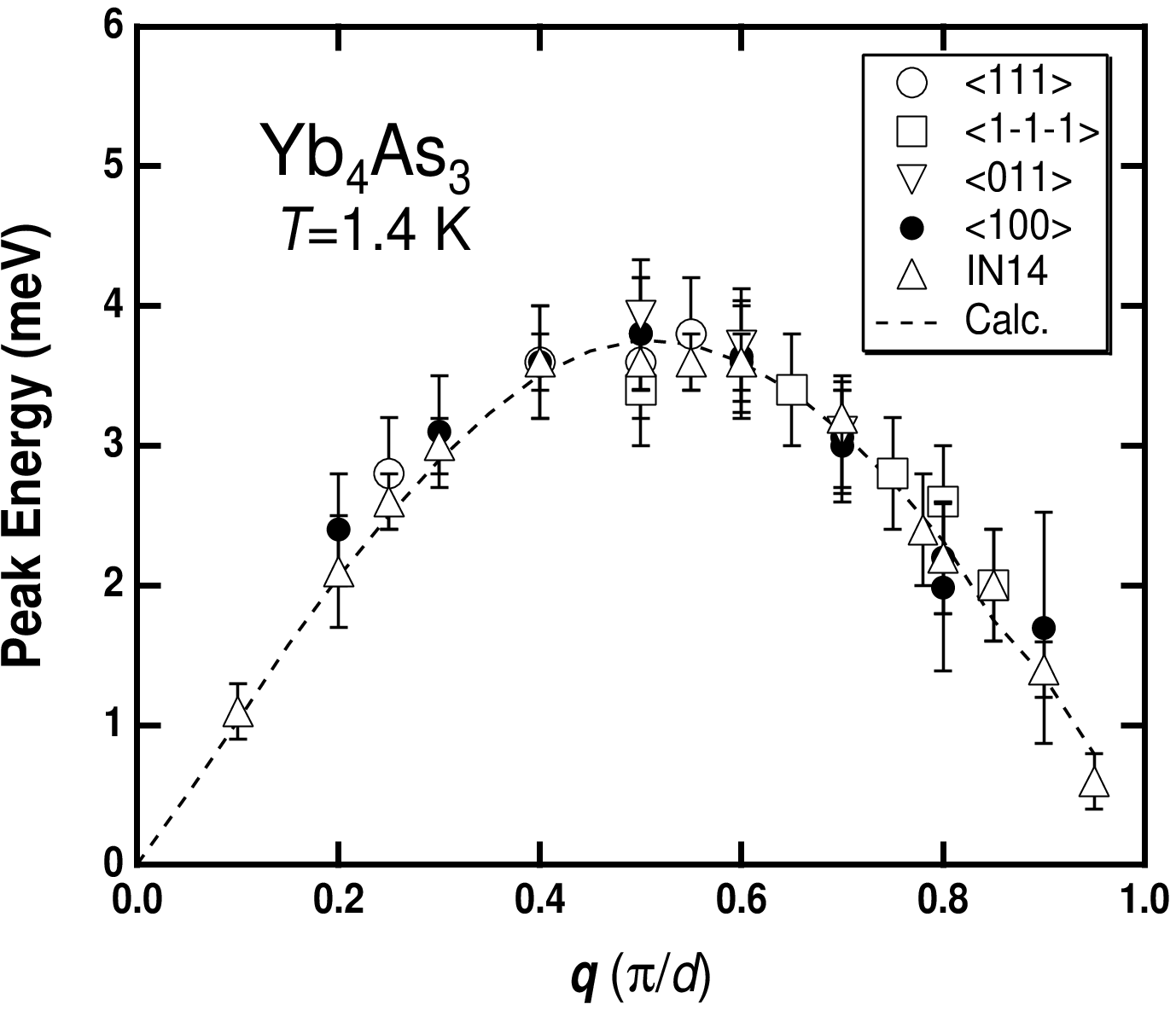}
\end{center}
\vspace{0.5cm}
\caption{Left panel: Experimental S(q,$\omega$) from INS \protect\cite{Kohgi99}
  for momenta q = 1 and 0.8 (here q is given in units of $\pi$/d where d is the
  Yb distance in the chain). The asymmetric shape of the two-spinon spectrum in
  Fig.~\ref{fig:spinon2a} is clearly observed. For q $<$ 1 the resolution
  limited peak position corresponds to the lower boundary $\omega_L(q)$ of the
  spinon spectrum. Right panel: Dispersion of $\omega_L(q)$ Yb$_4$As$_3$ from
  INS experiments for different directions of the momentum transfer. Here q is
  the projection of momenta on the $\la 111\ra$ chain direction. All data
  points fall on the dashed curve which is the theoretical $\omega_L$(q) =
  $\frac{\pi}{2}$J $\sin dq$ with $J/k_B$ = 25 K. In addition to the asymmetric
  shape of S(q,$\omega$) this proves the 1D character of magnetic excitations
  in Yb$_4$As$_3$. (After \protect\cite{Kohgi97,Kohgi99})}  
\label{fig:INSH_0}
\end{figure}
%%%%%%%%%%%%%%%%%%%%%%%%%%%%%%%%%%%%%%%%%%%%%%%%%%%%%%%%%%%%%%%%%%%%%%%%%%%%
%
This spectrum diverges at the lower boundary with a square root singularity and
therefore has a very asymmetric appearance as function of energy transfer
$\omega$ (see Fig.~\ref{fig:spinon2a}). The total integrated intensity, i.e.,
the frequency integral of S(q,$\omega$) is linear in q for q $\ll\pi$ and
diverges like $[-ln(1-q/\pi)]^\alpha$ for $q\rightarrow\pi$, where $\alpha$ = 1
for the approximation in Eq.~(\ref{DYNSTRUC}) and $\alpha$ = 3/2 is the exact
result. Therefore the two-spinon continuum should appear in INS as a spectrum
strongly peaked at  $\omega_L(q)$ with an asymmetric tail reaching up to
$\omega_U(q)$ and a strongly increasing total intensity for q
$\rightarrow\pi$. This is precisely what has been found in the INS experiments
of Kohgi et al. (see Fig.~\ref{fig:INSH_0}) and constitutes a proof for the 1D
character of spin excitations in \YBA. 

That the interacting crystal field ground-state doublets of Yb$^{3+}$ behave
like an isotropic Heisenberg system is not immediately obvious. It was shown
independently in \cite{Uimin00} and \cite{Shiba00}. Thereby the local symmetry
of the crystalline electric field was properly accounted for. Spin fluctuations
of Heisenberg chains can explain the observed heavy quasiparticles excitations
as they appear in the specific heat C = $\gamma$T and spin susceptibility
$\chi_S$. They are given by
%6.9
\begin{equation}
\gamma = \frac{2}{3}~ \frac{k_BR}{J}
\qquad\mbox{and}\qquad
\chi_S = \frac{4 \mu^2_{\rm eff} R}{(\pi^2 J)}
\label{gamma23}
\end{equation} 
(see Ref. \cite{MattisBook81}). Here J is the AF coupling constant of
nearest-neighbor sites in the effective S = $\frac{1}{2}$ spin chain described
by the 1D Heisenberg Hamiltonian  
%6.10
\begin{equation}
H = J\sum_{\langle ij \rangle} \v S_i \v S_j~~~.
\label{HJ4}
\end{equation} 
Furthermore, R is the gas constant and $\mu_{\rm eff}$ is the effective
magnetic moment of a Yb$^{3+}$ site. Therefore one finds R$_W$ = 2 for
the Sommerfeld-Wilson ratio. When the experimental value of J/k$_B$ = 25
K (from Fig. \ref{fig:INSH_0}) is used in (\ref{gamma23}) the observed size
of the $\gamma$ coefficient is well reproduced.

Having described the underlying physics we discuss a model description for the
compound. For that purpose we neglect the electrons hopping terms between
different chains as it was done, e.g., in the Labb\'e-Friedel model for A-15
compounds like V$_3$Si or Nb$_3$Sn \cite{Labbe66}. There one is dealing with
three families of intersecting chains while here there are four types of
chains. The effective 4f-model Hamiltonian \cite{Fulde95} is then written as
%6.11
\begin{eqnarray}
H = & - & t \sum_\mu \sum_{\langle ij \rangle \sigma} \left( f^+_{i \mu \sigma}
f_{j \mu \sigma} + h.c. \right) + U  \sum_\mu \sum_i n_{i \mu \uparrow} n_{i
  \mu \downarrow}\nonumber \\
& + & \epsilon_\Gamma \sum_\mu \sum_{i \sigma} \Delta_\mu n_{i \mu \sigma} +
\frac{N}{4} c_\Gamma \epsilon^2_\Gamma~~~.
\label{Htsigma}
\end{eqnarray}
The first term describes effective 4$f$-hole hopping due to hybridization with
As 4p ligand states within a chain of a family $\mu$ = 1 - 4 from site $i$ to a
nearest neighbor site $j$. From LDA calculations one can deduce that 4t
$\simeq$ 0.2 eV. The second term is due to the on-site Coulomb repulsion of
4$f$ holes with n$_{i \mu \sigma}$ = f$^+_{i \mu \sigma}$f$_{i \mu \sigma}$ and
ensures that Yb$^{4+}$ states with 4$f^{12}$ configurations, i.e., two holes
per site are excluded. The third term describes the volume conserving coupling
of the $f$ bands to the trigonal strain $\epsilon_\Gamma >$ 0 with $\Gamma =
\Gamma_5$. It leads to a deformation potential of the form 
%6.12
\begin{equation}
\Delta_\mu = \frac{\Delta}{3} \left( 4 \delta_{\mu 1} - 1 \right)
\label{DeltaMu3}
\end{equation}
for 4$f$ holes situated in chains, e.g., in [111] direction denoted by $\mu$ =
1. Since the Yb$^{3+}$ ions are smaller than the Yb$^{2+}$ ions, the distance
between Yb ions shrinks for $\mu$ = 1 while it expands in the other chain
directions denoted by $\mu$ = 2, 3, 4. As previously pointed out the origin of
the deformation potential is the Coulomb repulsion between holes on neighboring
sites. It is treated here as an effective attraction V$_{\rm eff}$ between
holes on nearest-neighbor sites of a chain. The fourth term in (\ref{Htsigma})
is the elastic energy in the presence of a trigonal distortion, where $N$ is
the number of sites and $c_\Gamma$ is the background elastic constant. A
reasonable value is $c_\Gamma/\Omega = 4 \cdot 10^{11}$ erg/cm$^3$ where
$\Omega$ denotes the volume of a unit cell with a lattice constant of a$_0$ =
8.789 \AA. 

The Hamiltonians for chains $\mu$ with an interaction V$_{\rm eff}$ can be
written in the form
%6.13
\begin{equation}
H_\mu = -t_\mu \sum_{\langle ij \rangle \sigma} f^+_{i \mu \sigma} f_{j \mu
\sigma} - V_{\rm eff} \sum_{\langle ij \rangle} \left( n_{i \mu} - \Bar{n}
\right) \left( n_{j \mu} - \Bar{n} \right)
\label{Hmut}
\end{equation}
where $\Bar{n}$ is the average occupancy of all the sites in the system. Within
a molecular-field approximation the Hamiltonian reduces to 
%6.14
\begin{equation}
H^{\rm MF}_\mu = -t_\mu \sum_{\langle ij \rangle \sigma} f^+_{i \mu \sigma}
f_{j \mu \sigma} - 2V_{\rm eff} \left( \Bar{n}_\mu - n \right) \sum_{\langle ij
\rangle} \left( n_{i \mu} - \Bar{n} \right) + \frac{N}{4} V_{\rm eff} \left(
\Bar{n}_\mu - \Bar{n} \right)^2~~~. 
\label{HMFmut}
\end{equation}
We denote the distinct chains with $\mu$ = 1 and note that with the
correspondence 
%6.15
\begin{equation}
\frac{4}{3} \epsilon_\Gamma \Delta = -2V_{\rm eff} \left( \Bar{n}_1 - \Bar{n}
\right)~~;~~ \frac{\Delta^2}{c_\Gamma} = \frac{9}{4} V_{\rm eff} 
\label{43epsilon}
\end{equation}
the Hamiltonians (\ref{HMFmut}) and (\ref{Htsigma}) become the same. This
serves as a justification for the band Jahn-Teller type of description chosen
above. When a distortion is taking place the hopping matrix elements also
depend on $\mu$, according to 
%6.16
\begin{eqnarray}
t_\mu & = & t_+ \delta_{\mu 1} + t_- \left( 1 - \delta_{\mu 1} \right)
\nonumber\\ 
t_+ & = & te^{\lambda \epsilon_\Gamma}, \quad t_- =
te^{\lambda\epsilon_\Gamma/3}~~~.  
\label{tmut}
\end{eqnarray} 
But this $\mu$ dependence of $t$ is not essential. 

The Hamiltonian (\ref{Htsigma}) is well suited for describing the lattice
distortion caused by charge ordering. The distortion is described here like a
collective band Jahn-Teller effect. The four-fold degeneracy of the $f$ bands
is lifted by the spontaneous appearance of a trigonal strain which lowers the
symmetry. For the purpose of demonstration we consider first the case of U =
0. This neglects the effects of strong on-site correlations on the band
Jahn-Teller effect. Near the charge ordering transition where each of the
chains contains nearly 25 \% of Yb$^{3+}$ sites this approximation is justified
. However, it is no longer acceptable for low temperatures when nearly all of
the sites in the $\mu$ = 1 chains are Yb$^{3+}$. The condition for a collective
band Jahn-Teller effect taking place is $\Delta^2/(t \epsilon_\Gamma) > 3$
\cite{Fulde95}. Choosing $\Delta$ = 5 eV results in a transition temperature of
T$_c \simeq$ 250 K which is close to the experimental value of 300 K. This
value of $\Delta$ corresponds to a Gr\"uneisen parameter of $\Omega_G \equiv
\Delta$/(4t) = 25 which is of a comparable size found in other 4f-mixed valence
systems. 
%fig6.8
%%%%%%%%%%%%%%%%%%%%%%%%%%%%%%%%%%%%%%%%%%%%%%%%%%%%%%%%%%%%%%%%%%%%%%%%%%%%
\begin{figure}[tb]
\begin{center}
\includegraphics[width=7.0cm]{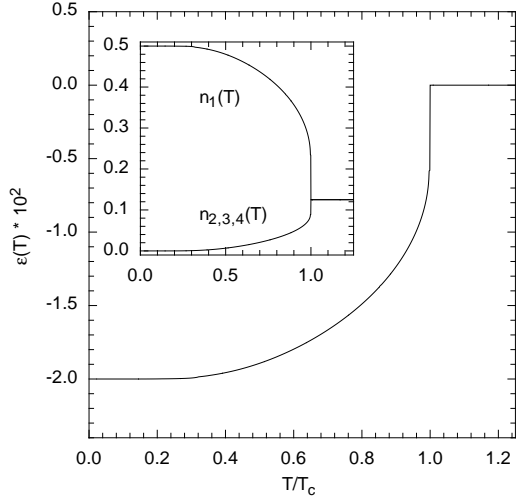}
\end{center}
\vspace{0.5cm}
\caption{Temperature dependence $\epsilon$(T) of the (secondary) strain order
parameter. Shown in the inset are the corresponding changes in the population
of the different families of chains. Q = n$_1$ - n$_2$ is the primary
charge order parameter. (After \protect\cite{Fulde95})}
\label{fig5}
\end{figure}
%%%%%%%%%%%%%%%%%%%%%%%%%%%%%%%%%%%%%%%%%%%%%%%%%%%%%%%%%%%%%%%%%%%%%%%%%%%%

The symmetry strain $\epsilon_\Gamma$ as function of reduced temperature
T/T$_c$ is shown in Fig.~\ref{fig5}. As a consequence of a finite strain the
four degenerate one-dimensional $f$ bands split into a lower and three upper
bands. Their centers of gravity differ by (4/3)$| \epsilon_\Gamma \Delta |$
with the equilibrium strain given by $\epsilon_\Gamma = -\Delta/(2 c_\Gamma)
\simeq$ 0.02. The changes in the population of the chains with $\mu$ = 1
and $\mu \neq$ 1 with decreasing temperature are shown in the inset of
Fig. \ref{fig5}. The transition is of first order due to the singular
DOS of the effective 1D f-bands. Let us reemphasize that those changes are
caused by the Coulomb repulsion of holes and drive $\epsilon$(T) and not vice
versa. Band refillings by a band Jahn-Teller transition and crystallization of
holes are alternative points of view of the respective descriptions here.

\begin{figure}[t b]
\begin{center}
\includegraphics[clip,width=7.0cm]{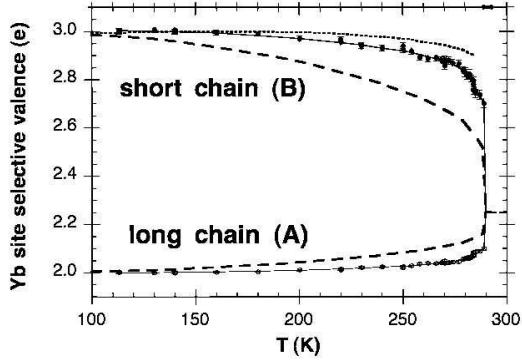}
\end{center}
\vspace{0.5cm}
\caption{Yb$_4$As$_3$: Site selective values of Yb valence/hole concentration
  (data points) as obtained from the $30\bar{3}$ reflection in the vicinity of
  the Yb L$_3$ X-ray absorption edge. Full line: guide to the eye. Dashed line:
  model calculation from \protect\cite{Li97}. Dotted line: from the angular
  deviation $\delta(T)$ \cite{Iwasa98} in Fig.~\ref{fig11}. (After
  \protect\cite{Staub01})} 
\label{fig6}
\end{figure}

There have been experiments which directly observed one-dimensional charge
order \cite{Staub01}. This was achieved, e.g., by resonant X-ray diffraction on
the Yb L$_3$ absorption edge. Below T$_c$ forbidden reflections appear in the
slightly rhombohedrally distorted cube. From their intensity one can deduce an
effective valency for the short and long chains (see Fig.~\ref{fig6}). The
discontinuities at the first-order phase transition are even larger than the
above model calculation predicts.  

We return to the Hamiltonian in Eq.~(\ref{Htsigma}) and consider the effects of
a large on-site interaction U in the charge ordered state for T $\ll$
T$_c$. For this purpose we transform to a $t-J$ Hamiltonian for the chains
$\mu$ = 1. The exclusion of Yb$^{4+}$ configurations (2 holes) due to U is
taken into account by a projector P which projects onto 0 and 1 hole occupation
of sites. To lowest order in t/U we obtain 
%6.17
\begin{equation}
H_{\mu = 1} = -t \sum_{\langle ij \rangle \sigma} P \left( f^+_{i 1 \sigma}
f_{j 1 \sigma} + h.c. \right) P + J \sum_{\langle ij \rangle} \left( {\bf S}_i
{\bf S}_j - \frac{1}{4} n_{i 1} n_{j 1} \right)
\label{Hmu1t}
\end{equation} 
where J = 4t$^2$/U and furthermore ${\bf S}_i = \sum\limits_{\alpha \beta}
f^+_{i 1 \alpha} {\boldsigma}_{\alpha \beta} f_{i 1 \beta},~n_{i 1} =
\sum\limits_\sigma n_{i 1 \sigma}$. For U = 10 eV we find J = 1$\cdot$10$^{-3}$
eV. A slight band-narrowing effect due to the lattice distortion (see
Eq. (\ref{tmut})) has been neglected here. The Hamiltonian H$_{\mu = 1}$ can be
treated within the slave-boson approximation. Thereby the projector P is
replaced by an auxiliary boson $b^+_j$ which generates a configuration without
a hole, i.e., a Yb$^{2+}$ site. By treating the boson field in mean-field
approximation one obtains an effective mass of the quasiparticles 
%6.18
\begin{equation}
\frac{m^*}{m_b} = \frac{t}{t \delta + (3/4)\chi J}
\label{mmbtt}
\end{equation} 
where $m_b$ is the bare band mass and $\chi = (\frac{2}{\pi})
\sin(\frac{\pi}{2}(1 - \delta))$. Furthermore, $\delta$ is the deviation of the
chains $\mu$ = 1 from half filling. When $\delta$ = 0, i.e., for the
half-filled case we find (J = 4t$^2$/U) 
%6.19
\begin{equation}
\frac{m^*}{m_b} = \frac{\pi}{6}~\frac{U}{t} =  \frac{2\pi}{3}~\frac{t}{J}
\label{mmbpi}
\end{equation} 
which is a large mass enhancement of order 10$^2$. In this case the heavy
quasiparticles are spinons as we are dealing with an antiferromagnetic
Heisenberg chain. The low temperature thermodynamics is completely determined
by these spin degrees of freedom and we have an example of spin-charge
separation here. A de Haas-van Alphen experiment should not yield heavy
quasiparticle masses here and the Fermi-liquid relation between the $\gamma$
coefficient and a renormalized mass of (nearly) free electrons is
violated. This becomes obvious when Yb$_4$(As$_{1-x}P_x$)$_3$ (x = 0.3 - 0.4)
\cite{Aoki05,Ochiai97} is considered. In distinction to Yb$_4$As$_3$ this is a
charge ordered insulator but it has nevertheless a similar large linear
specific heat coefficient $\gamma$. The latter is caused by spin
excitations in Heisenberg chains as in Yb$_4$As$_3$.  
%fig6.9
%%%%%%%%%%%%%%%%%%%%%%%%%%%%%%%%%%%%%%%%%%%%%%%%%%%%%%%%%%%%%%%%%%%%%%%%%%%%
%%%%%%%%%%%%%%%%%%%%%%%%%%%%%%%%%%%%%%%%%%%%%%%%%%%%%%%%%%%%%%%%%%%%%%%%%%%%
%
It is interesting to note that the Hamiltonian (\ref{Htsigma}) can be solved
exactly by adaptation of Lieb and Wu's solution of the one-band Hubbard
model. The solution is based on the Bethe ansatz. This is possible since the
different bands  are not coupled directly with each other. One interesting
finding is that the increase in the strain order parameter just below T$_c$ is
much steeper than in Fig. \ref{fig5} and therefore in better agreement with
experiments.  For more details we refer to the original literature \cite{Li98}.

\begin{figure}[tb]
\begin{center}
\includegraphics[clip,width=10.0cm]{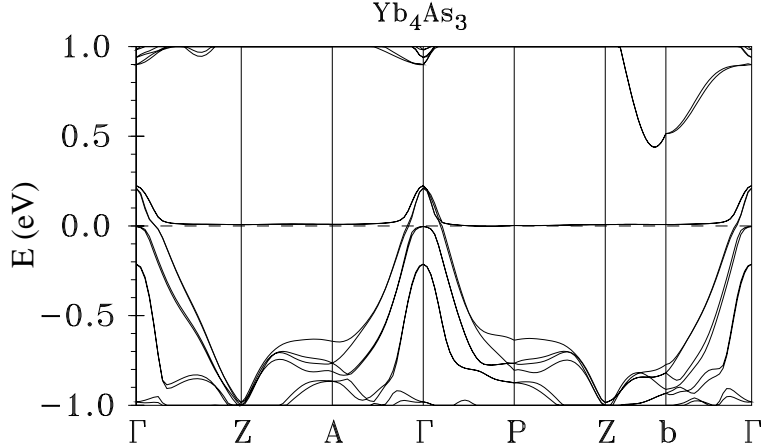}
\end{center}
\vspace{0.5cm}
\caption{LDA+U energy bands for Yb$_4$As$_3$. A small number of As 4$p$ holes
  appears at the $\Gamma$-point. The flat band has Yb 4$f$-character. (After
  \protect\cite{Antonov98})} 
\label{fig7}
\end{figure}

This brings us to the question: why is Yb$_4$As$_3$ at low temperatures still a
semimetal and not an insulator? From the present model we would naively expect
that at zero temperature Yb$_4$As$_3$ is an insulator. Charge order should be
complete and a half-filled Hubbard chain is an insulator at sufficiently low
temperatures for any positive value of U. In reality this is not the case
due to incomplete charge order, i.e. the Hubbard chains are not half-filled but
doped. Within the model one possible explanation for incomplete charge order
would be self-doping. Indeed, as function of U the exact solution of the
Hamiltonian (\ref{Htsigma}) shows a regime in the c$_\Gamma~ vs.$ total
electron concentration n$_0$ plane where charge ordering is
incomplete. However, in \YBA~ the true reason for the observed semimetallic
behavior is found when bandstructure calculations within the local spin-density
approximation plus U approach (LSDA+U) are performed. They show a rigid pinning
of the Fermi energy to states close to the top of the As 4$p$ band (see
Fig. \ref{fig7}). The filled 4$f$ shell of Yb$^{2+}$ is treated as a core shell
and the interaction U = 9.6 eV between two 4$f$ holes was adjusted to obtain
the proper number of charge carriers. Because of a large gap between the As
4$p$ and and Yb 5$d$ bands, the 4$f$ hole band is pinned to the top of the As
4$p$ band. This allows for charge  transfer between Yb and As. As a result a
small amount of As 4$p$ holes which act as mobile charge carriers is created
with a corresponding reduction of Yb 4$f$ holes. 
%fig6.10
%%%%%%%%%%%%%%%%%%%%%%%%%%%%%%%%%%%%%%%%%%%%%%%%%%%%%%%%%%%%%%%%%%%%%%%%%%%%
%%%%%%%%%%%%%%%%%%%%%%%%%%%%%%%%%%%%%%%%%%%%%%%%%%%%%%%%%%%%%%%%%%%%%%%%%%%%
%
The calculated cyclotron masses of the almost spherical hole sheet of the As
$p$ states are in the range of 0.6 to 0.8 times the free electron mass m$_0$,
in good agreement with the value of m$_{\rm exp}$ = 0.72m$_0$ obtained from
cyclotron-resonance experiments while a value of 0.275m$_0$ was obtained from
Shubnikov-de Haas oscillations \cite{Gegenwart02}. However, the calculated mass
should still be renormalized due to electron-phonon interactions. We want to
draw attention to the fact that the relation between the large $\gamma$
coefficient in the specific heat and the mass of the charged quasiparticles
responsible for charge transport is lost here. This calls in question one
fundamental requirement of Landau's Fermi liquid theory. Its basis is a one to
one correspondence between the excitations of an interacting electron system
and those of a free electron system with renormalized parameters such as the
quasiparticle mass. In the case of Yb$_4$As$_3$ this correspondence holds only
when the charge of the free electrons is renormalized to zero, i.e., when we
deal with neutral fermions. This is due to the fact that the low-lying
excitations are those of a Heisenberg chain which are of magnetic origin and
can be described either by bosons or alternatively by neutral fermions. In
addition to the neutral fermionic excitations in the chains which are
spinon-like there are charged fermions (the As 4p-holes), i.e., electrons with
a low density of states which provide for charge transport. The large A
coefficient in $\rho$(T) results from scattering of the charged fermions by the
neutral ones, while the thermal conductivity is dominated by the neutral ones
(spinons). This special feature is amplified when instead of Yb$_4$As$_3$
semiconducting mixed crystals Yb$_4$(As$_{1-x}$P$_x$)$_3$ with x = 0.3-0.4
are considered. This crystals also exhibit charge order like Yb$_4$As$_3$
forming spin chains below the ordering temperature. The Sommerfeld constant is
$\gamma$ = 250 mJ/(mol K$^2$) despite the fact that there are no charge
carriers in the system \cite{Ochiai97}. 
 
The LSDA+U calculations also show that one must be careful in finding the
correct ordered structure by comparing Madelung energies. In competition with
the trigonal phase considered here is a cubic P2$_1$3 phase with chains of
sequence - Yb$^{3+}$ - Yb$^{2+}$ - Yb$^{2+}$ - Yb$^{2+}$ - Yb$^{3+}$ -
Yb$^{2+}$ -. The Madelung energy of this structure is slightly lower than the
one of the trigonal structure. It does not couple to the $\Gamma_5$ strain and
therefore would not explain the observed softening of the elastic constant
c$_{44}$. But when the self-consistent charges are calculated one
finds a slight charge disproportionation of 0.05 electrons between the center
Yb$^{2+}$ ion and the two adjacent Yb$^{2+}$ ions. This changes the Madelung
energy difference in favor of the trigonal structure.

\begin{figure}[t b]
\begin{center}
\includegraphics[clip,width=7.0cm]{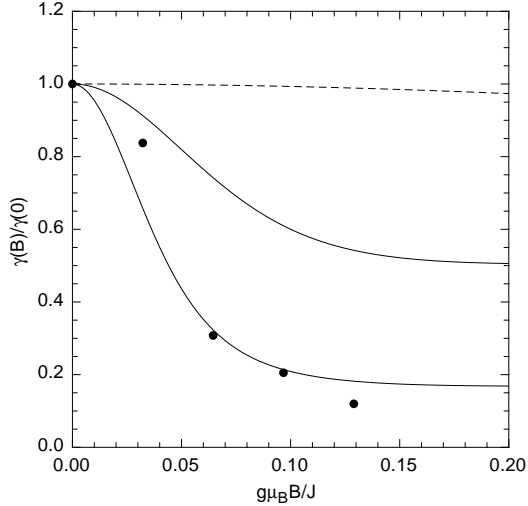}
\end{center}
\vspace{0.5cm}
\caption{Yb$_4$As$_3$: Experimental field dependence of the specific heat
  $\gamma$ coefficient (full circles) at T$\simeq$0.5 K and theoretical results
  based on interchain interactions: lower and upper solid lines: H $\bot$ [111]
  and H $\parallel$ [111]. The dashed curve is for T = 5 K. A ratio J'/J =
  10$^{-4}$ was used. (After \protect\cite{Schmidt96})}
\label{fig:gamma_h}
\end{figure}

When a magnetic field $\rm{\bf H}$ is applied to an isotropic antiferromagnetic
spin chain (S = $\frac{1}{2}$, g = 2), the excitations remain gapless as
long as the field remains smaller than g$\mu_B$ HS = J beyond which the system
becomes a fully polarized ferromagnet with a Zeeman gap \cite{Johnson81}. It
was therefore a surprise when it was found that the specific heat depends
strongly on $\rm{\bf H}$ \cite{Koeppen99}. Experimental results for $\gamma$(H)
are shown in Fig.~\ref{fig:gamma_h}. Likewise strong anomalies in the thermal
expansion which is related to the specific heat via the Ehrenfest relation were
found \cite{Koeppen99}. A number of possible  explanations of these
observations have been proposed. One is based on magnetic interchain
interactions \cite{Schmidt96}. Another model calculation advocates a staggered
field mechanism due to the Dzyaloshinsky-Moriya interaction \cite{Shiba00} and
a third one suggests that the dipolar interactions within a chain are
responsible for the decrease of $\gamma$ with increasing external field
\cite{Uimin00}.  
%fig6.11
%%%%%%%%%%%%%%%%%%%%%%%%%%%%%%%%%%%%%%%%%%%%%%%%%%%%%%%%%%%%%%%%%%%%%%%%%%%%
%%%%%%%%%%%%%%%%%%%%%%%%%%%%%%%%%%%%%%%%%%%%%%%%%%%%%%%%%%%%%%%%%%%%%%%%%%%%
%
In all three cases a magnetic field applied perpendicular to the Yb$^{3+}$
chains is opening a gap in the excitation spectrum while the excitations remain
gapless when the field is pointing along the chains. But they differ in
details, e.g., in the way the gap opens.

When a small interchain coupling J' is assumed between adjacent Yb$^{3+}$
chains a field perpendicular to the chains induces a gap
$\Delta\sim(|JJ'|)^{1/2}$. For J'/J = 10$^{-4}$ one obtains a decrease of the
$\gamma$ coefficient with increasing field of the observed size (see
Fig.~\ref{fig:gamma_h}). 

\begin{figure}[t b]
\begin{center}
\includegraphics[width=7.0cm]{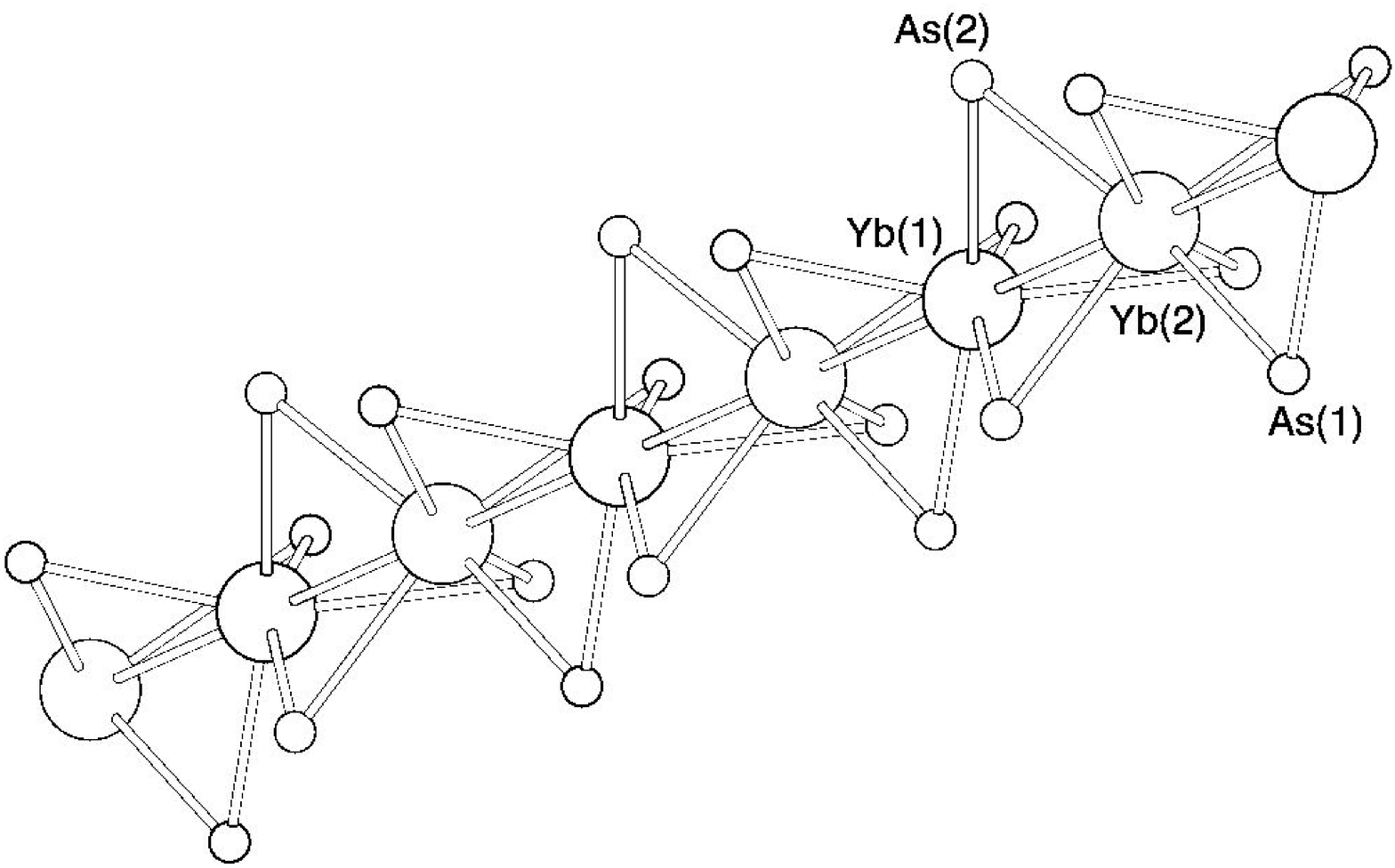}
\end{center}
\vspace{0.5cm}
\caption{Yb$_4$As$_3$: Absence of a center of inversion in Yb$^{3+}$
  chains. The distance Yb(1) - As(1) (or Yb(2) - As(2)) is smaller than the
  distance Yb(1) - As(2) (or Yb(2) - As(1)). Atoms with equal gray scale are
  equivalent. (After \protect\cite{Shiba00})}  
\label{fig:chain}
\end{figure}

The two other models are based on independent Heisenberg chains. The model
based on the Dzyaloshinsky-Moriya interaction exploits the fact that the
Yb$^{3+}$ sites are not centers of inversion (see
Fig.~\ref{fig:chain}). The local C$_3$ symmetry together with a glide reflexion
and a glide vector parallel to the chains allows for the following features: a
uniaxial anisotropy for the symmetric part of the spin-spin interaction and an
antisymmetric Dzyaloshinsky-Moriya (DM) interaction. However due to the hidden
symmetry they  are not independent \cite{Shiba00}. A transformation consisting
of staggered rotations with angle $\theta$ around the chain direction (z) makes
this more apparent: The pseudo-spin Hamiltonian for the lowest Yb$^{3+}$
Kramers doublet, expressed in rotated spin operators, is then given by 
%6.20
\begin{eqnarray}
& H_{eff} = J \sum_{\langle ij\rangle}\v S_i\v S_j -\sum_i & \left[ g_\parallel
  S_{iz}H_z +\cos\theta g_\perp \left( S_{ix}H_x + S_{iy}H_y \right) \right. \nonumber \\
&& \left. + (-1)^i\sin\theta g_\perp \left( S_{iy}H_x-S_{ix}H_y \right) \right] ~~~. 
\label{HEFF}
\end{eqnarray}
Here the C$_3$ - doublet wave function is associated with anisotropic
g-factors g$_{\parallel,\perp}$ whose ratio is 2.5 \cite{Shiba00}. The
angle $\theta$ is adjustable and in principle determined by the C$_3$ CEF
parameters. 

Thus for zero field one has indeed an {\it isotropic} spin chain (S = 1/2)
Hamiltonian despite involving strongly anisotropic wave functions of
the lowest Kramers doublet state. This leads to the gapless spinon excitation
spectrum discussed previously. 
%fig6.12
%%%%%%%%%%%%%%%%%%%%%%%%%%%%%%%%%%%%%%%%%%%%%%%%%%%%%%%%%%%%%%%%%%%%%%%%%%%%
%%%%%%%%%%%%%%%%%%%%%%%%%%%%%%%%%%%%%%%%%%%%%%%%%%%%%%%%%%%%%%%%%%%%%%%%%%%%
%
When a magnetic field is applied perpendicular to the chain a staggered
field perpendicular to both chain and applied field is induced by the
DM interaction described by the last term in Eq.(\ref{HEFF}):
%6.21
\begin{equation}
{\bf H}_{s} = g_\bot\sin\theta \left[ {\bf n}\times{\bf H} \right]~~~.
\label{HStg}
\end{equation} 
%
%fig6.13
%%%%%%%%%%%%%%%%%%%%%%%%%%%%%%%%%%%%%%%%%%%%%%%%%%%%%%%%%%%%%%%%%%%%%%%%%%%%
\begin{figure}[t b]
%\begin{center}
\raisebox{-0.3cm}
{\includegraphics[clip,width=6.5cm]{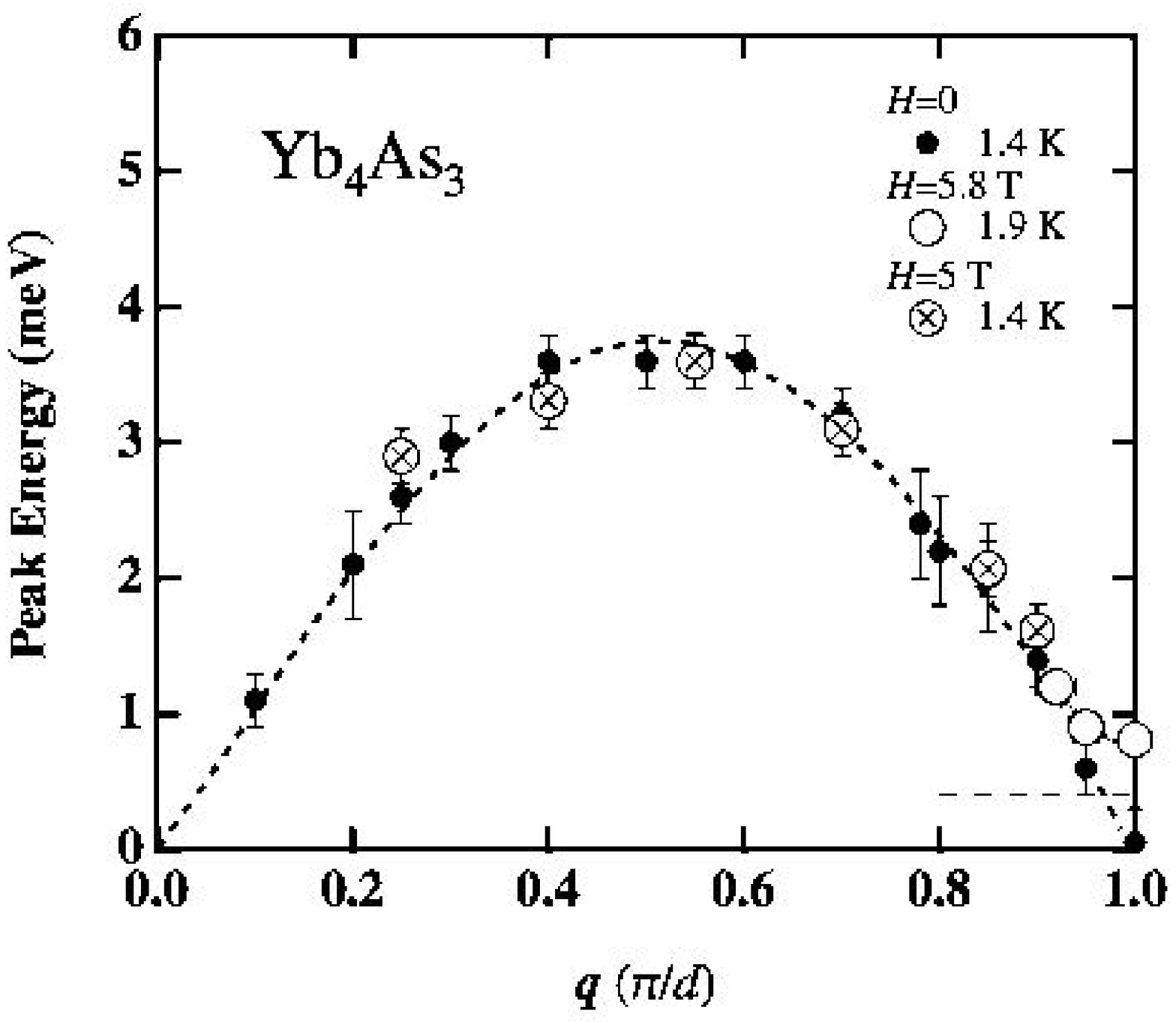}}
\hspace{1cm}
\includegraphics[clip,width=8.0cm,height=5.3cm]{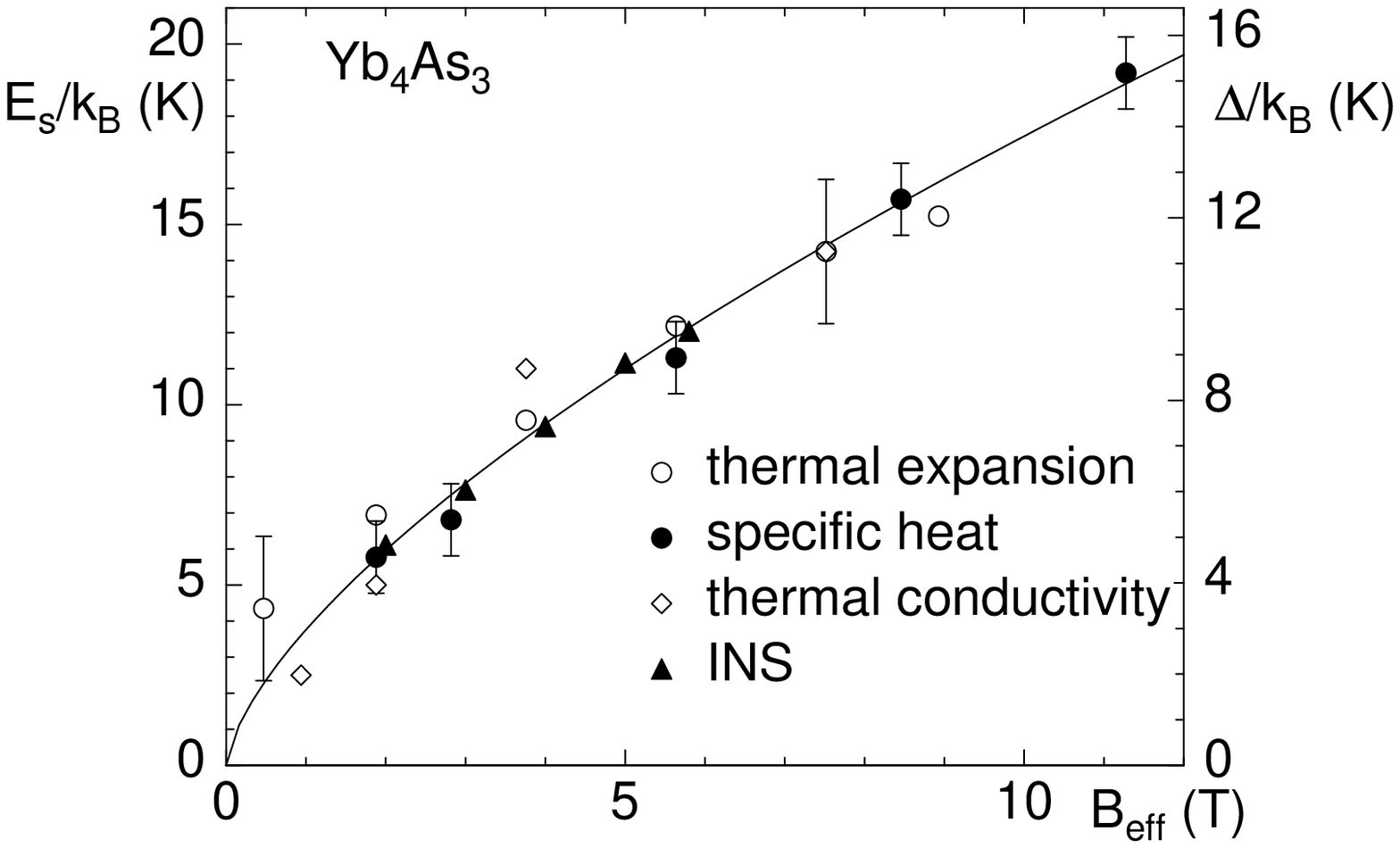}
%\end{center}
\vspace{0.5cm}
\caption{Upper panel: Spin excitations ($\omega_L(q)$) from INS in an Yb$^{3+}$
chain in a magnetic field perpendicular to the chains and without it. Note the
gap opening at q = $\pi$/d. After \protect\cite{Kohgi01}. Right panel: Gap as
function of applied field. A scaling exponent 2/3 corresponding to quantum
sine-Gordon equation
is found (full line: $\Delta$(H) $\sim$ E$_s$(H)$\sim$ H$^\frac{2}{3}$). Here
E$_s$(H) is the gap from thermodynamic and transport quantities and
$\Delta$(H) is the gap obtained in INS. (After \protect\cite{Schmidt01})} 
\label{fig:disp}
\end{figure}
%%%%%%%%%%%%%%%%%%%%%%%%%%%%%%%%%%%%%%%%%%%%%%%%%%%%%%%%%%%%%%%%%%%%%%%%%%%%
%
Here ${\bf n}$ is a unit vector in chain direction and $g_\bot$ = 1.3 is the
corresponding $g$ factor. Due to the staggered field the uniform physical
susceptibility $\chi_\perp(0)$ is a mixture of q = 0 (uniform) and q = $\pi$
(staggered) susceptibilities $\chi_{1D}$ of the (S = $\frac{1}{2}$, g = 2)
isotropic Heisenberg chain according to 
%6.22
\begin{equation}
\chi_\perp(0) = g_\perp^2 \left[ \cos^2(\theta)\chi_{1D}(0) +
  \sin^2(\theta)\chi_{1D}(\pi) \right]~~~. 
\label{CHIPHYS}
\end{equation}
Since the latter diverges $\sim$ 1/T at low temperatures this should lead to a
Curie like upturn also in the homogeneous $\chi_\perp(0,T)$. This behavior was
indeed found and in principle the value of tan$^2(\theta)$ = 0.04 may be
extracted from the analysis of the upturn. The most important effect of the
staggered field is the appearance of an induced gap $\Delta$ in the excitation
spectrum with a field dependence 
%6.23
\begin{equation}
\Delta(H) \simeq 1.8J^{1/3} H_s^{2/3} \sim H^{2/3}~~~.
\label{Deltasim}
\end{equation} 
This gap was seen in thermodynamic and transport properties like specific heat,
thermal expansion etc. \cite{Koeppen99,Gegenwart02} and the field scaling
exponent 2/3 was indeed identified (Fig.~\ref{fig:disp}). A field dependence
of this form has also been directly observed by inelastic neutron scattering
(INS) on a single crystal at T = 1.9 K. A gap was found to open up at q =
$\frac{\pi}{d}$ when the field was perpendicular to the short, i.e., Yb$^{3+}$
chains (see Fig. \ref{fig:disp}). This supports the above model. In passing we
note that the above excitations can also be described by a sine-Gordon equation
associated with moving domain walls as applied to Cu-benzoate
\cite{Dender97}. But when the magnetic field is parallel to the chain, gapless
modes at finite q = $\frac{\pi}{d}$(1$\pm 2\sigma$) ($\sigma$ = magnetization)
are expected \cite{Mueller81,Shiba02} which sofar have not been seen and this
casts some doubt on the theory.

This brings us to the third model which is based on the weak dipolar
interaction of the Yb$^{3+}$ sites. For that purpose the DM interaction is
neglected. The dipolar interaction is 
%6.24
\begin{equation}
H_{\rm dip} = g^2 \mu^2_B \sum_{i<j} \frac{{\bf J}_i {\bf J}_j - 3  \left( {\bf
	J}_i {\bf e} \right) \left( {\bf J}_j {\bf e} \right)}{\mid {\bf R}_i -
	{\bf R}_j \mid^3} 
\label{Hdipg}
\end{equation} 
where ${\bf e} = ({\bf R}_i - {\bf R}_j)/|{\bf R}_i - {\bf R}_j|$ and only the
ground state doublet of the Yb$^{3+}$ is taken into account. This leads to an
effective, slightly anisotropic Heisenberg antiferromagnet in an applied
field. In a transverse field a gap opens up \cite{Dimitriev02}. A special
feature is that at high fields the gap should close again. Measurements up to H
= 30 T reveal this effect. Indeed a strong reduction of the gap below the
$\Delta(H)\sim H^\frac{2}{3}$ scaling curve (Fig.~\ref{fig:disp}) was observed
already above H = 10T \cite{Gegenwart02}. When ${\bf H}$ is parallel to the
chains a Zeeman splitting is expected and there should be no incommensurate
peaks appearing, in agreement with experiments. In summary it seems that the
magnetic field effects are not yet fully understood.

\subsection{Charge Ordering and 1D Spin Excitations in \NAV}
\label{subsect:NAV}

In quasi-1D metals such as organic charge transfer salts or the famous KCP
chain compound \cite{Gruenerbook} it is known that the ground state exhibits a
spontaneous dimerization due to the instability of the 1D Fermi surface. It is
signified by a diverging electronic susceptibility at wave number q =
2k$_F$. In real compounds this instability, the 'Peierls-transition' takes
place at a finite transition temperature controlled by the 
interchain-coupling. Amazingly a similar 'spin-Peierls' transition may occur in
insulating quasi-1D spin chains with antiferromagnetic nearest neighbor (n.n.)
coupling. This is most conveniently understood for an xy-type exchange
interaction model which can be exactly mapped by a Jordan-Wigner transformation
(JW) to a model of free spinless 1D fermions at half filling
\cite{Beni72}. Naturally this leads to a chain dimerization via the same
'electron'-phonon coupling as in the case of real conduction electrons. Adding
the z-term of the Heisenberg exchange interaction one obtains an interacting
fermion model after the JW transformation but the basic mechanism of
dimerization is unchanged. The spin-Peierls transition has originally been
found in a number of organic spin chain compounds and has been theoretically
investigated first in Ref. \cite{Pytte74} within a mean-field treatment and
later in Ref. \cite{Cross79} using the Heisenberg exchange and its coupling to
the lattice in the fermionic (JW) representation. Surprisingly the spin-Peierls
transition rarely occurs in anorganic 1D-spin chain compounds, presumably
because magnetic order due to interchain-exchange is prefered in most
cases. The only such spin-Peierls compound known until recently was CuGeO$_3$
\cite{Hase93} where the S = $\frac{1}{2}$ spins of Cu-chains undergo
dimerization at T$_{SP}$ = 14 K. The corresponding  dimerization of the
exchange integral along the chain then creates an isotropic spin excitation
gap. The ensuing isotropic drop in the spin susceptibility below the transition
temperature is therefore the most direct method to identify the the
spin-Peierls mechanism. Actually this is not unambiguous because the
presence of n.n.n exchange interactions J' with a ratio J'/J $>$ 0.24
where J is the n.n. exchange also leads to a spin gap and this issue is
still controversial in CuGeO$_3$.\\ 
%fig6.14
%%%%%%%%%%%%%%%%%%%%%%%%%%%%%%%%%%%%%%%%%%%%%%%%%%%%%%%%%%%%%%%%%%%%%%%%%%%%
\begin{figure}[tb]
\begin{center}
\includegraphics[clip,width=8.0cm]{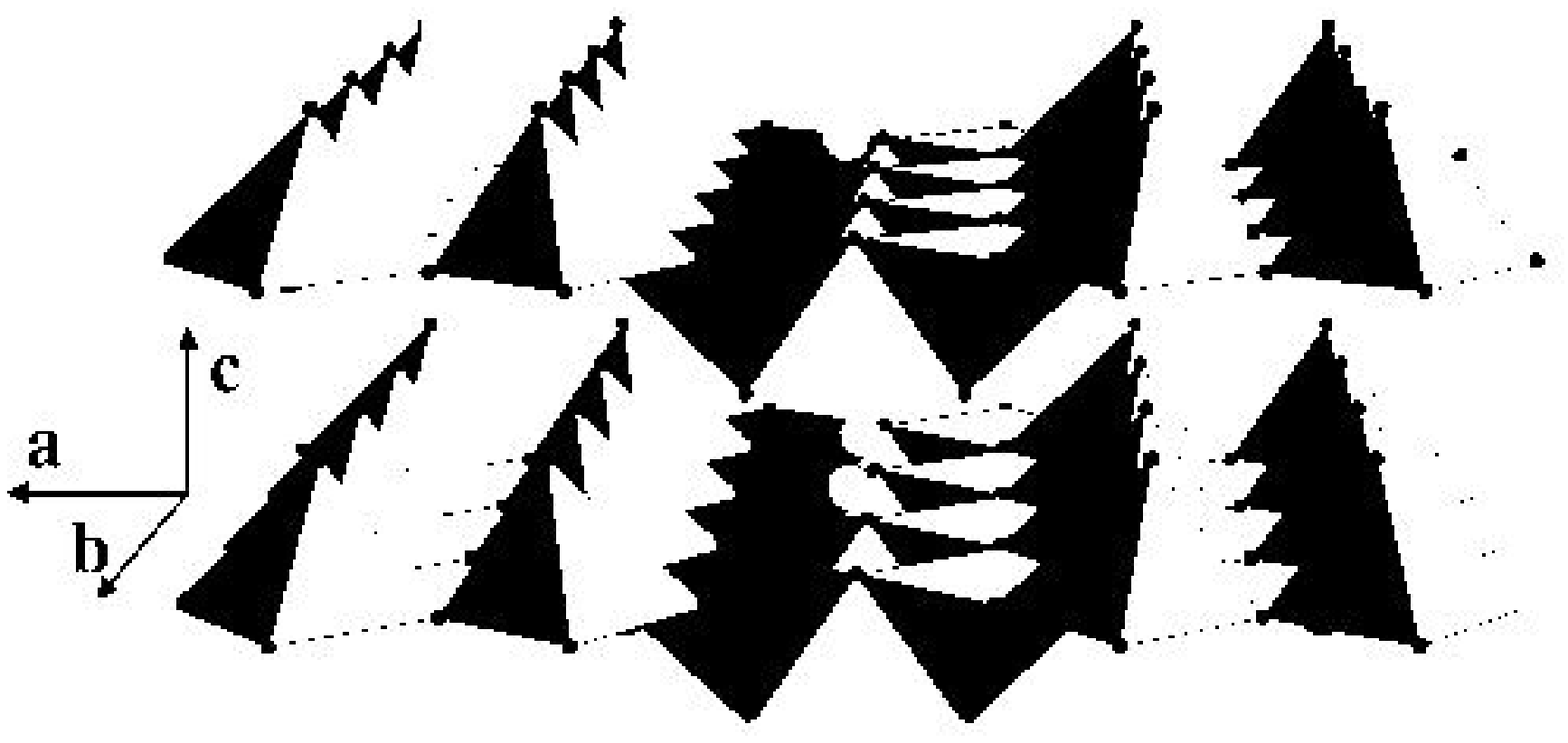}
\includegraphics[clip,width=6.0cm]{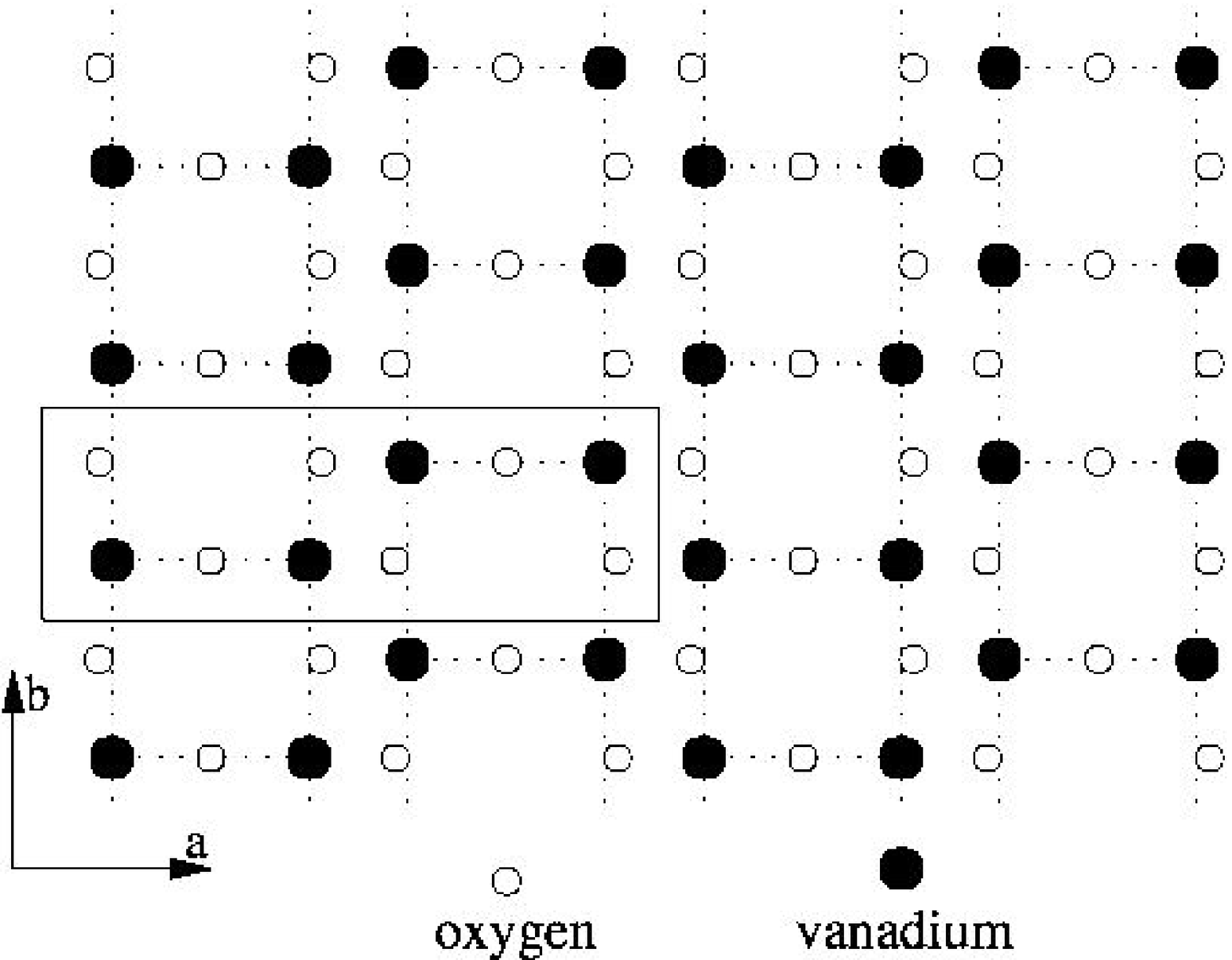}
\end{center}
\vspace{0.5cm}
\caption{Left panel: Layered perovskite structure of \NAV~consisting of chains
  of oxygen pyramids that contain the V atoms aligned along the crystal b
  axis. The layers are stacked along c axis. Na atoms (grey spheres) are
  centered above the ladder plaquettes. Right panel: ab-plane Trellis lattice
  structure of V-V ladders alternatingly shifted along b by half a lattice
  constant. This leads to a quasi-'triangular' structure for V-V rung
  units. Orthorhombic high temperature unit cell is indicated.}
\label{fig:Nastruc2}
\end{figure}
%%%%%%%%%%%%%%%%%%%%%%%%%%%%%%%%%%%%%%%%%%%%%%%%%%%%%%%%%%%%%%%%%%%%%%%%%%%%
%
Therefore the discovery of a structural phase transition in the layered
perovskite insulator \NAV~at T$_c\simeq 34K$ with a corresponding isotropic
spin-gap formation \cite{Isobe96} has created enormous interest and activity,
both experimental and theoretical. This compound was seen as a second candidate
for an anorganic spin-Peierls system which seemed to be consistent with the
original crystal structure determination \cite{Carpy75} that suggested the
existence of S = 1/2 V$^{4+}$ spin chains along the crystal b-axis (every
second V chain in Fig.~\ref{fig:Nastruc2}), isolated by intervening nonmagnetic
V$^{5+}$ (S = 0) chains. Both belong to V-V ladders formed by the oxygen
pyramids, where neighboring ladders are shifted by b/2. In the ab-plane this
leads to a Trellis lattice structure shown in
Fig.~\ref{fig:Nastruc2}. Exchange-dimerization of the S = 1/2 V chains due to
the spin-Peierls mechanism below T$_c$ would then lead to the isotropic
spin gap. This interpretation was however much too naive as became clear
subsequently. It turned out that the structural phase transition is not of the
simple chain dimerization type but leads to a very low symmetry (monoclinic)
structure that can only be understood if in addition a charge ordering
transition of V$^{4+}$/V$^{5+}$ is present, starting from a homogeneous high
temperature insulating mixed valence (V$^{4.5+}$) state, in contrast to the
first assumption in Ref. \cite{Carpy75}. Nevertheless above T$_c$ the
susceptibility exhibits the typical Bonner-Fisher maximum of 1D spin chains
\cite{Isobe96,Horsch98}. This compound therefore is another example of an
intricate relation between charge ordering and quasi-1D spin excitations which
in the present case are gapped. 

In this section we first discuss the result of the increasingly more detailed
X-ray analysis of the high and low temperature structure of \NAV, nevertheless
no final agreement on the low-T structure has been reached. Electronic
structure calculations are essential to construct an effective Hamiltonian for
the charge ordering. We will argue that this transition and the ensuing spin
gap formation requires the inclusion of lattice degrees of freedom which lead
to the appearance of two inequivalent V-V ladder types in the perovskite
layers. The gapped spin excitation spectrum in the distorted CO phase will be
discussed within a simple dimer-RPA approach and a comparison with INS results
is given which also advocates the inequivalent ladder model. Finally we briefly
mention the stacking of CO layers along c and the destruction of CO by Na
deficiency doping or substitution of Na by Li or Ca and in connection with the
nature of the insulating state in \NAV. In this section we will not review the
various alternative theoretical models that have been proposed for \NAV. We
also will exclude the discussion of most optical experiments, since this is not
the ideal method to investigate spin excitations.

The determination of the crystal structure of \NAV~, both above and below
T$_c$, is fundamental for constructing a microscopic model for the phase
transition and spin gap mechanism. This was surprisingly controversial and has
led to much confusion, mostly for the low- but even for the high-temperature
structure. The latter was originally thought to have non-centrosymmetric
symmetry \cite{Carpy75} already above T$_c$ leading to the assumption of
V$^{5+}$/V$^{4+}$ in-line charge disproportionation on each V-V rung of the
ladders from the outset. However later it was discovered \cite{Schnering98}
that the proper high temperature structure belongs to 
the centrosymmetric Pmmn space group. Due to the reflection plane containing y
both V-sites in the rungs must be equivalent, i.e., they are in a mixed valence
state V$^{4.5+}$. The structure of the low temperature phase (T$\ll$T$_c$)
proved even more controversial. In
Refs. \cite{Luedecke99,vanSmaalen00,deBoer00} an orthorhombic space group Fmm2
with a $2a\times 2b\times 4c$ supercell was proposed. In the low temperature
phase the modulation (atomic shifts of V, O and Na atoms with respect to high
temperature structure) strongly differs for neighboring ladders A and
B. Loosely speaking the atomic positions on A are modulated while on B they are
not (see Fig.~\ref{fig:Naexco}). Subsequently it was shown \cite{Sawa02} that
the symmetry is even lower, characterized by the monoclinic space group A112
which does not contain an inversion symmetry. This implies the existence of a
spontaneous dipole moment (induced ferroelectric order parameter) below the
phase transition which has indeed been observed by measuring the dielectric
function. The two space groups differ in the number of inequivalent atoms per
unit cell, e.g., 6 V sites in  two inequivalent layers a,b for Fmm2 and 8 V
sites again in two layers for A112. Likewise there are 6 vs. 8 inequivalent Na
atoms and 16 vs. 20 inequivalent O atoms in both cases
respectively. Nevertheless, the full refinement of the modulated crystal
structure with A112 and Fmm2 space group leads to almost identical atomic
positions \cite{Sawa02}.  

This raises the question how the valences, most importantly of V-atoms
should be assigned in the low temperature structure. Unfortunately the
modulated structure has not yet been investigated within LDA+U
calculations. Therefore one has to resort to the empirical valence-bond method
\cite{Brown92}. In this approach every atom at an inequivalent site is assigned
a formal valence V$_i$ (or oxidation number) which may be thought of as the
number of electrons it contributes to all the bonds connected to ligand atoms
according to the prescription  
%6.25
\begin{equation}
|V_i| = \sum_j v_{ij} \quad\mbox{with}\quad v_{ij} = \exp \left(
 \frac{R_0-R_{ij}}{B} \right) 
\label{VAL}
\end{equation}
where $v_{ij}$ is the valence of a given central atom-ligand bond
which depends on the bond length. Here R$_0$ and B are empirical
parameters characteristic of a given type of chemical bond, e.g., V-O
bond, independent of the material. The optimal parameters R$_0$ and B
may however be weakly temperature dependent. Using the sum rule in
Eq.~(\ref{VAL}) and identifying valence with the number of electrons
contributing to bonds the valence of constituents may be
determined. In \NAV~we are primarily interested in the number of
electrons in the V-d$_{xy}$ orbital because the
3d$_{xy}$-bands are well separated from all other bands as seen
below. Its orbital occupation is then given by n$_{d_{xy}}$ = 5 - V$_i$,
assuming that bonding takes place mainly with the lower lying O3p
orbitals. This quantity is the most interesting to study in view of possible
charge ordering.  
%fig6.15
%%%%%%%%%%%%%%%%%%%%%%%%%%%%%%%%%%%%%%%%%%%%%%%%%%%%%%%%%%%%%%%%%%%%%%%%%%%%
\begin{figure}[tb]
\begin{center}
\includegraphics[width=7.5cm,clip]{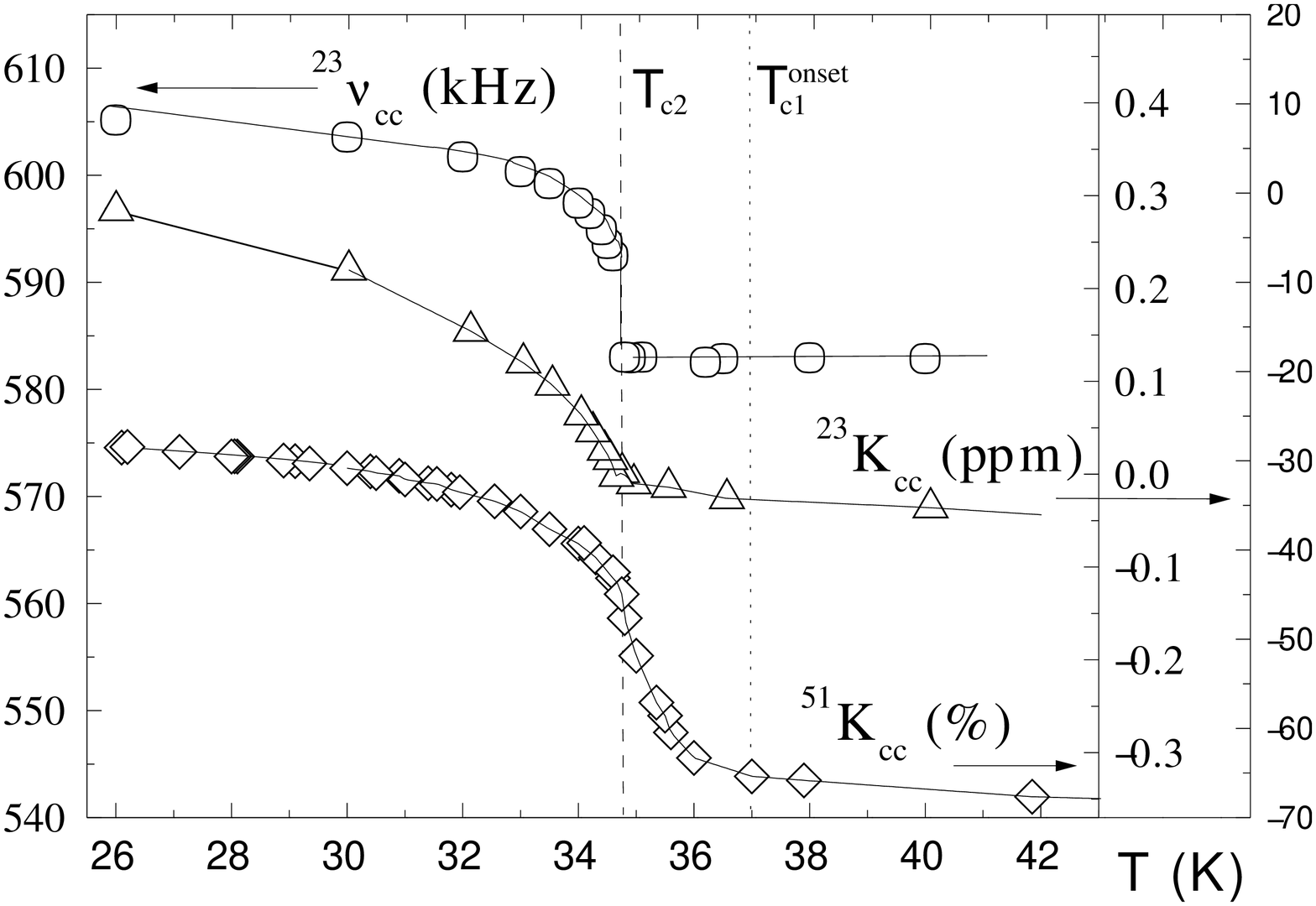}\hfill
\raisebox{5cm}
{\includegraphics[angle=-90,width=7.3cm,clip]{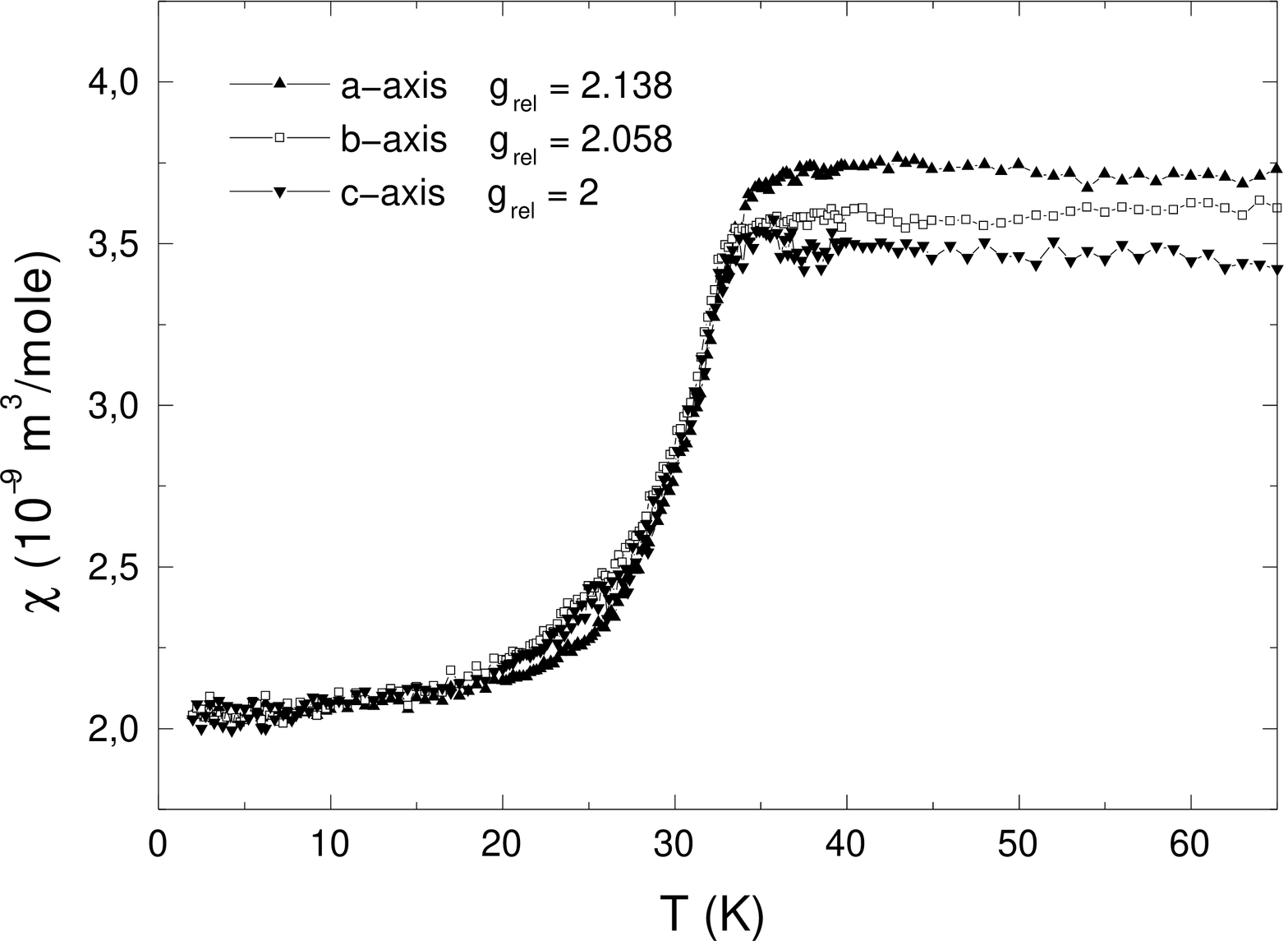}}
\end{center}
\vspace{0.5cm}
\caption{Upper panel; V-Knight shift $^{51}K_{cc}$ and Na-Knight shift
  $^{23}K_{cc}$ as well as quadrupolar frequency $^{23}\nu_{cc}$ of \NAV. The
  first reflects charge order, the second the opening of the spin gap and the
  third the lattice distortion around Na-sites. Since $^{23}K_{cc}$ and
  $^{23}\nu_{cc}$ behave synchronously below T$_{c2}$ this suggests that spin
  gap formation is connected with exchange dimerization due to Na-shifts. After
  \protect\cite{Fagot00}. Right panel: Magnetic susceptibility for single
  crystalline \NAV~showing the isotropic spin gap opening below
  T$_{c2}$. (After \protect\cite{Weiden97})} 
\label{fig:NaKnight}
\end{figure}
%%%%%%%%%%%%%%%%%%%%%%%%%%%%%%%%%%%%%%%%%%%%%%%%%%%%%%%%%%%%%%%%%%%%%%%%%%%%
%

We now discuss a few basic observations that prove the connection
between charge ordering and spin gap formation at the structural phase
transition. It has been proposed early from thermal expansion
measurements \cite{Koeppen98} that actually there are two separate but
closeby phase transitions at T$_{c1}\simeq$ 33 K  and T$_{c2}\simeq$ 32.7 K
which are of first and second order respectively. These values depend
considerably on sample quality and may be higher (see Fig.~\ref{fig:NaKnight})
than those given above. Detailed investigation of NMR frequencies and Knight
shifts have shown \cite{Fagot00} that T$_{c1}$ and T$_{c2}$ may be
associated with charge ordering on V-sites and a spin-gap formation
respectively (Fig.~\ref{fig:NaKnight}). In the following we will
denote the whole region of these close transitions simply by
'T$_c$'. At the lower transition the critical exponent of the spin-gap opening
$\Delta_{SP}\sim (1-T/T_c)^{\beta_\Delta}$ is $\beta_\Delta\sim 0.34$ close to
the theoretical value 0.33 obtained from $\Delta_{SP}\sim\delta^{\frac{2}{3}}$
\cite{Cross79} and a mean-field chain and exchange dimerization behavior
$\delta\sim (1-T/T_c)^\frac{1}{2}$. This suggests indeed a spin-Peierls
dimerization at the lower transition. The first transition on the
other hand has a 2D-Ising character, as indicated by a logarithmic
peak in the specific heat superposed by the jump of the second
transition. Furthermore, according to temperature dependence of X-ray
satellites the total lattice distortion has a temperature exponent
$\beta\sim$ 0.2 closer to the Ising value 1/8 than to the mean-field
exponent. In summary, these findings suggest that
the T$_{c1}$-transition leads to charge ordering on the V-sites and
has partly 2D Ising character. Its field dependence is smaller than for an
Ising transition but larger than for a pure structural transition. At
T$_{c1}$ the main lattice distortion takes place. At the slightly
lower T$_{c2}$ the spin gap opens with a spin-Peierls like exponent
and connected with an additional exchange dimerization along the
charge ordered chains.

\begin{figure}[tb]
\begin{center}
\includegraphics[clip,width=5.0cm]{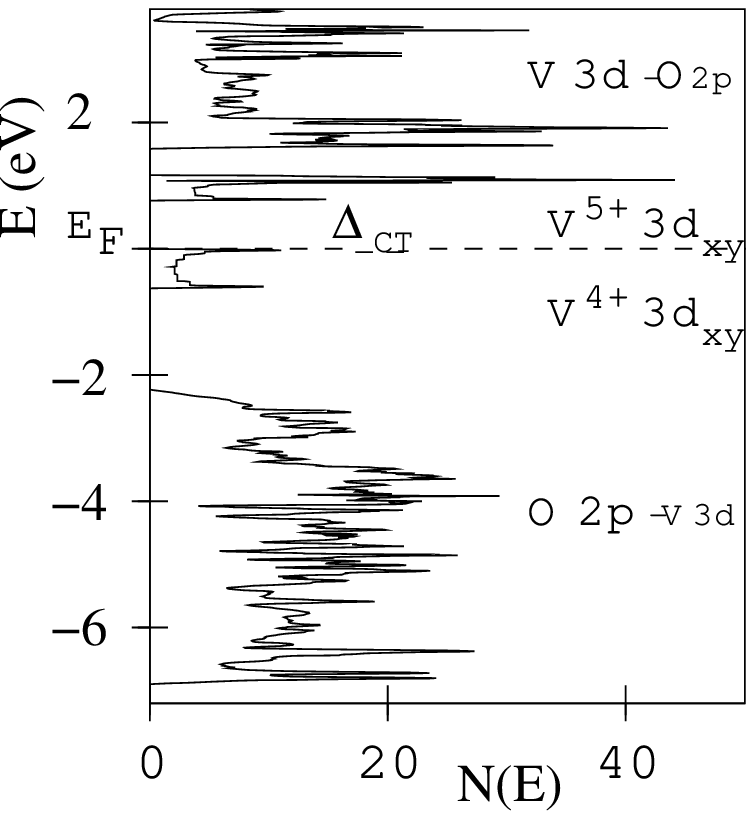}\hfill
\raisebox{0.3cm}
{\includegraphics[clip,width=8.0cm]{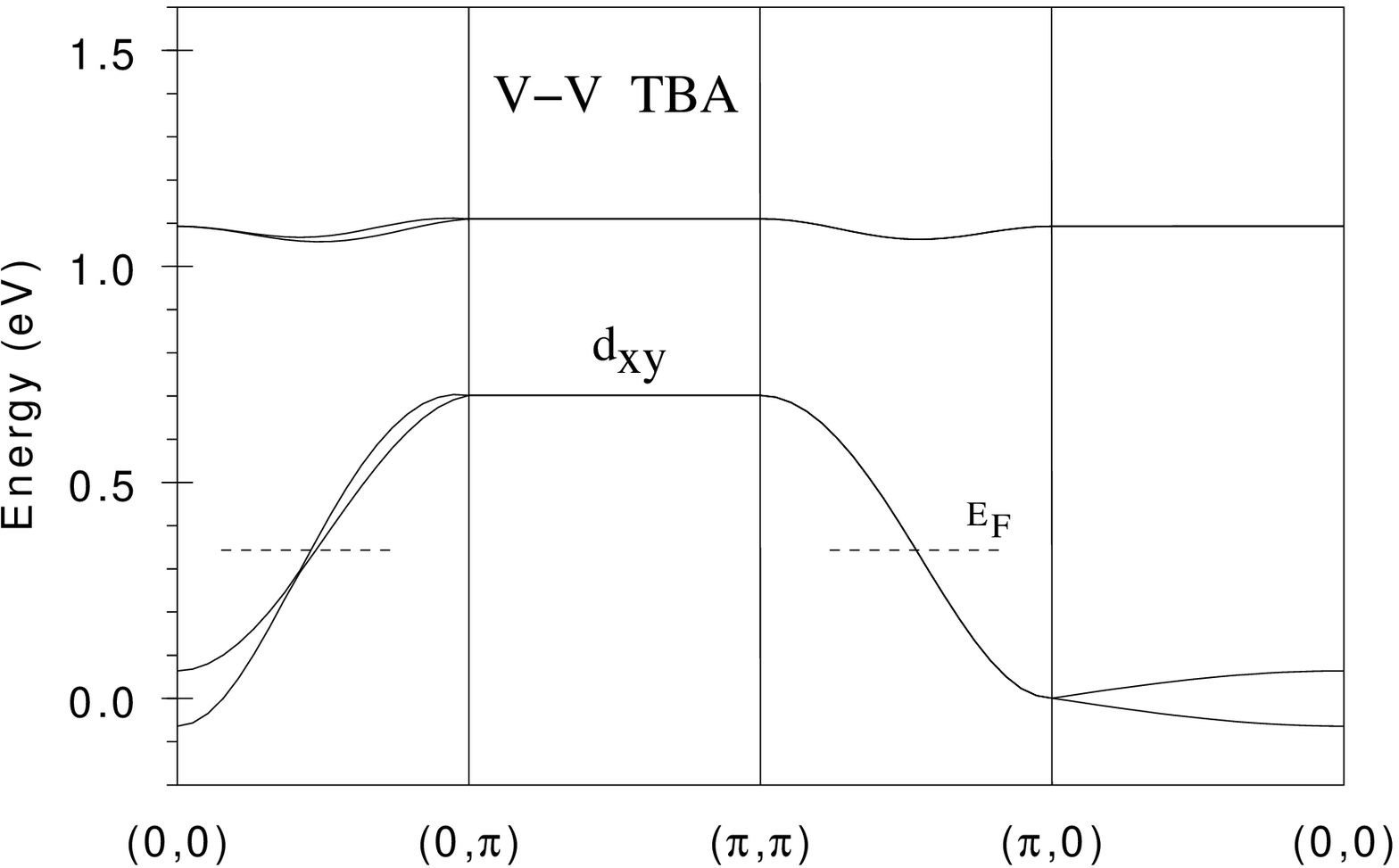}}
\end{center}
\vspace{0.5cm}
\caption{Left panel LDA+U DOS of \NAV~ with U=4.1eV and J=1.1 eV exhibits {\it
  two} 
  isolated 3d$_{xy}$ subbands due to enforced charge order with a charge
  transfer gap $\Delta_{CT}\sim$ 0.5-1 eV between them. Right panel: Tight
  binding (TB) fit to LDA bands ($\Delta_{CT}$ = 0) with full V-O basis
  corresponding  to effective hopping elements t$_R$ = 0.380 eV,  t$_L\simeq$
  t$_D$ = 0.085 eV for the effective V-V TB model with only d$_{xy}$-states
  included. Note that within LDA a metallic state is predicted with E$_F$ lying
  in the lower d$_{xy}$ band. (After \protect\cite{Yaresko00})}
\label{fig:NaLDAU}
\end{figure}

\begin{figure}[tb]
\begin{center}
\includegraphics[clip,width=7cm]{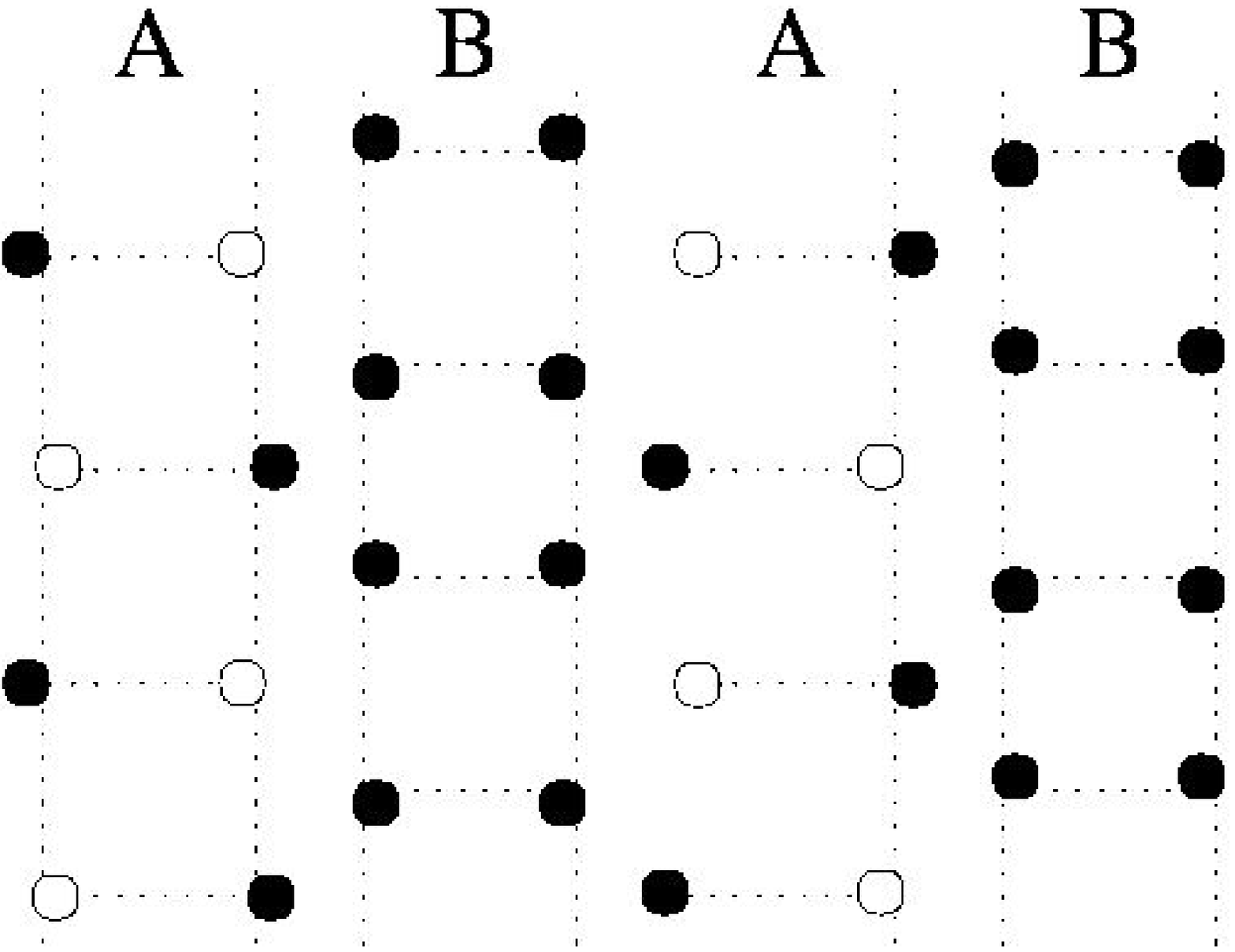}\hfill
\raisebox{-0.5cm}
{\includegraphics[clip,width=7cm]{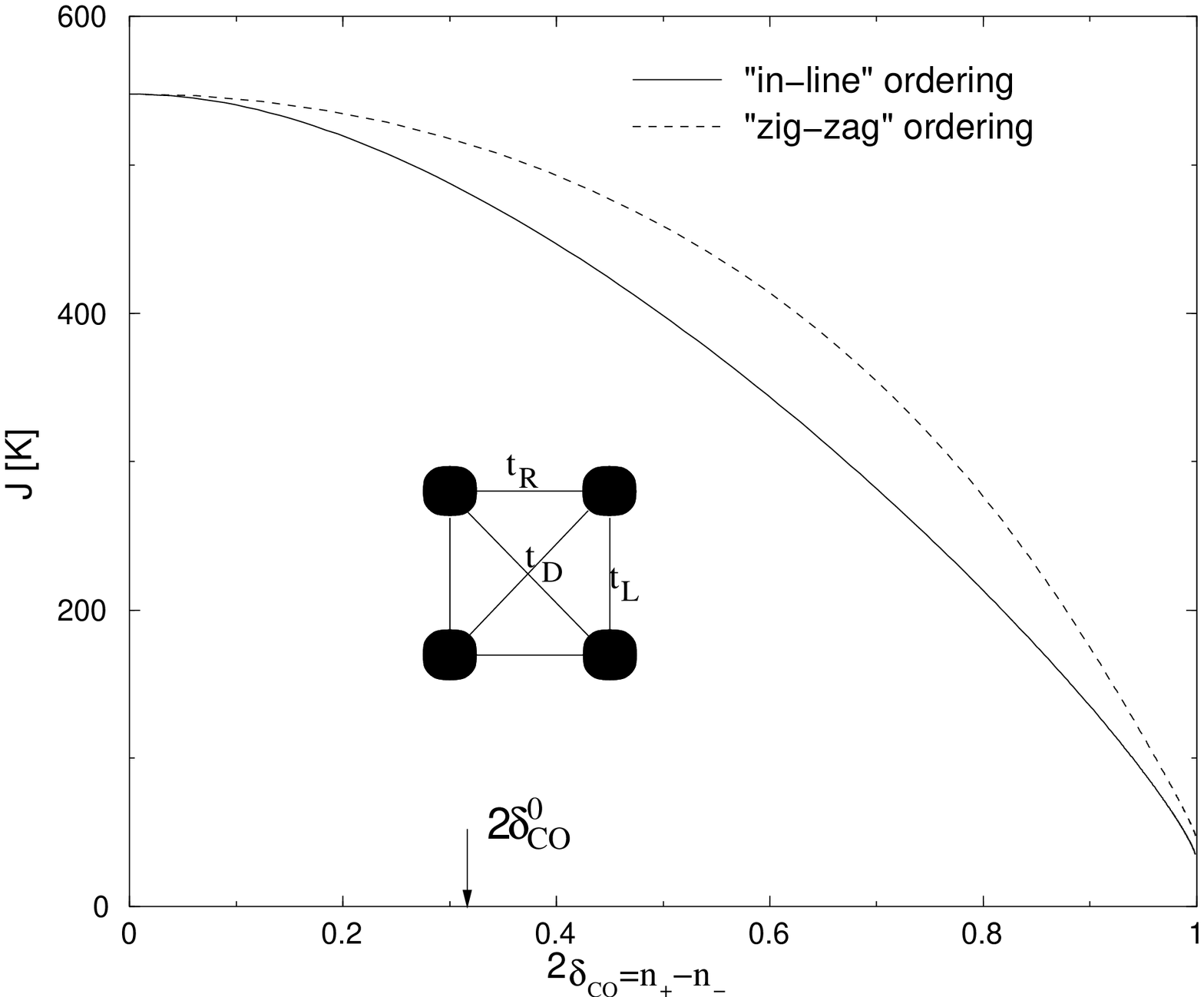}}
\end{center}
\vspace{0.5cm}
\caption{Charge order and dimerization in \NAV: Left panel: Shift of
  V-positions in distorted a-layer below 
  T$_{c1}$. Stacking along c is of aaa'a' type. In a'-layers A shifts have the
  same phase and B shifts are moved by one lattice constant along b (ladder
  direction). Each layer has three inequivalent V (full, open and gray
  circles), leading to 6 inequivalent V. Charge order (zig-zag) happens mainly
  on A, ladder dimerization on B.  Below T$_{c2}$ the A ladders also dimerize
  and V sites on B become inequivalent leading to 8 inequivalent V sites
  altogether. Right panel: Variation of effective spin exchange along the
  ladder  (b-direction) with V charge order parameter 2$\delta_{CO}$ for
  zig-zag CO (A). $\delta^0_{CO}$ is the actual charge order parameter for
  \NAV~on A-ladders corresponding to J$_A$ = 440 K (37.9 meV). Inset shows TBA
  hopping elements t$_R$ (along rung $\parallel$ a), t$_L$ (along leg
  $\parallel$ b) and t$_D$ (along diagonal) on a ladder plaquette. The same
  convention for inter-site Coulomb interactions V$_R$, V$_L$ and V$_D$ is
  used. (After \protect\cite{Bernert02})} 
\label{fig:Naexco}
\end{figure}

Before discussing microscopic models for these peculiar transitions
results of the electronic structure calculations for \NAV~ have to be
summarized. Sofar they were only done for the undistorted
structure. Both LDA calculations \cite{Smolinski98} and  spin
polarized LDA+U calculations \cite{Yaresko00} for the charge ordered
(but undistorted) case have been performed. In both cases it is found that the
planar V d$_{xy}$ bands are well separated from hybridized Vd-Op-bands
which are fully occupied or empty as shown in Fig.~\ref{fig:NaLDAU}. In the LDA
calculation the Fermi level is centered in the d$_{xy}$ band predicting a
metallic state, contrary to the fact that even without CO above T$_c$
\NAV~is an insulator. As in the undoped cuprates this is due to a
neglect of on-site Coulomb correlations in LDA. In the LDA+U treatment
they are simulated by breaking the orbital symmetry, i.e., by
using an orbital-occupation dependent one-electron potential 
%6.26
\begin{eqnarray}
V_\sigma^{LDA+U} = V_\sigma^{LSDA} + \sum_\alpha(U-J) \left( \frac{1}{2} -
n_{\alpha\sigma} \right) |\alpha\sigma\rangle \langle\alpha\sigma|
\end{eqnarray}
where U,J are the on-site Hubbard and exchange energy and $\alpha$ denotes the
3d-orbitals. In a charge ordered state the different orbital occupations of
inequivalent V-sites cause relative shifts of their orbital energies. As shown
in Fig.~\ref{fig:NaLDAU} this leads to a d$_{xy}$ subband splitting with a
charge transfer gap $\Delta_{CT}$. It moves the Fermi level to the top of the
lower subband, thus creating a charge transfer insulator in the CO
phase. The size of the gap is $\Delta_{CT}\simeq$ 0.5-1eV. A gap of this
size was indeed found in optical conductivity measurements \cite{Presura00}.
Calculation of the total LDA+U energy for various CO structure favors the
zig-zag structure in one ladder \cite{Yaresko00}, however sofar it cannot say
anything on the arrangement of adjacent CO ladders or whether only
every second ladder exhibits CO as proposed in the structure model of Ref.
\cite{Luedecke99} and others. The nature of the insulating phase above
T$_c$ is an unsolved problem since every rung is only singly occupied,
i.e., one has quarter filled V-ladders. Even without CO when
$\Delta_{CT}$ vanishes there is an excitation gap for double occupancy
of V-V rungs. An interpretation as a simple Mott-Hubbard
insulator in an effective d$_{xy}$-one band model based on the
molecular orbital (bonding) state of a rung seems inadequate. An
interpretation of the insulating state in terms of an extended
Hubbard model has been proposed in Ref. \cite{BernertDiss}.
%fig6.16
%%%%%%%%%%%%%%%%%%%%%%%%%%%%%%%%%%%%%%%%%%%%%%%%%%%%%%%%%%%%%%%%%%%%%%%%%%%%
%%%%%%%%%%%%%%%%%%%%%%%%%%%%%%%%%%%%%%%%%%%%%%%%%%%%%%%%%%%%%%%%%%%%%%%%%%%%
%
Such microscopic models require as an input the effective hopping
parameters obtained from a mapping of LDA band structure to a tight binding
model. It was shown in \cite{Yaresko00} that an adequate mapping
requires a basis of both V d$_{xy}$- and O p$_x$, O p$_y$ orbitals. The
result is shown in Fig.~\ref{fig:NaLDAU} for the full basis. Nevertheless
in a crude approximation this may be further simplified by mapping to
an effective d$_{xy}$ model containing only three hopping parameters
t$_R$, t$_L$ and t$_D$ along the rung, leg and diagonal of a single
ladder (inset of Fig.~\ref{fig:Naexco}). Their values are given in
Fig.~\ref{fig:NaLDAU}. It is found that t$_L\simeq$ t$_D$. This leads to the
essentially flat upper band in Fig.~\ref{fig:NaLDAU} (right) and therefore
t$_D$ cannot be neglected. It is also essential to obtain the proper exchange
Hamiltonian for the zig-zag CO structure \cite{Yushankhai01} which will be
discussed now. 

The description of the coupled CO and exchange dimerization in \NAV~
starts from an extended Hubbard model containing hopping terms
discussed before, the on-site effective Coulomb energy U$_{eff}\simeq
3eV$  and unknown intersite-Coulomb energies V$_R$, V$_L$ and V$_D$ for
the same bonds as the hopping integrals (inset of Fig.~\ref{fig:Naexco}). For
an investigation of CO, this model may be reduced to a much simpler one taking
for granted an insulating state with double occupancies of rungs
prohibited. Then the charge degrees of freedom are described by a pseudo spin 
T=$\frac{1}{2}$ where T$_z$ = $\pm\frac{1}{2}$ describes a d$_{xy}$
electron that occupies the left ($-\frac{1}{2}$) or right
($+\frac{1}{2}$) V-atom of a rung \cite{Thalmeier98}. The projected
Hamiltonian in the charge sector then has the form \cite{Bernert01}
%6.27 
\begin{equation} 
H_{ITF} = \sum_{\ll i,j\gg_L} K^{ij}_{Lz}T_{i}^z T_{j}^z + \sum_{\langle
i,j\rangle_{IL}} K^{ij}_{IL} T_{i}^z T_{j}^z + \sum_i 2\tilde{t}^i_R T_{i}^x
\label{HITF}
\end{equation}
which is the Ising model in a transverse field (ITF) whose role is plaid by
the renormalized intra-rung hopping $\tilde{t}^i_R$. In the spin
sector one has a Heisenberg Hamiltonian with exchange between the
spins on neighboring rungs and legs (Eq.~(\ref{HSEX})). Like the Ising
interaction 'constants' K$^{ij}_{Lz}$ (intra-ladder) and K$^{ij}_{IL}$
(inter-ladder)  $\tilde{t}^i_R$ is related to the original Hubbard parameters
which are renormalized by terms that depend on the spin configuration
\cite{Bernert02}: 
%6.28
\begin{eqnarray}
K^{ij}_{Lz} & = & 2V_L + \delta K_{Lz}(\v S_i,\v S_j)\nonumber\\
K^{ij}_{IL} & = & -V_{IL} + \delta K_{IL}(\v S_i,\v S_j)\nonumber\\
\tilde{t}_R & = & t_R + \sum_{\langle ij\rangle_L} \delta\tilde{t}_R(\v S_i,\v
S_j) ~~~. 
\label{ITFCOUPL}
\end{eqnarray}    
In this model the charge (\v T) degrees of freedom and spin (\v S)
degrees of freedom are coupled through the renormalized interaction
constants of Eq.~(\ref{ITFCOUPL}) that depend on the spin
configuration. As a consequence, even above T$_c$ the optical
conductivity which probes the (rung) charge excitations also shows a
signature of coupled spin excitations \cite{Mostovoy02}. For the moment,
considering only the possibility of charge ordering we may freeze the spins in
an AF or FM configuration along or between the ladders according to the sign of
the spin exchange obtained in LDA+U calculations \cite{Yaresko00}. This leads
to values $\tilde{t_R}$ = -0.19 eV and K$_{Lz}\simeq$ -K$_{IL}$ = 0.68
eV \cite{Bernert02}.  One might expect that the Hamiltonian in
Eq.~(\ref{HITF}) on the rigid Trellis lattice describes charge order
in a natural way, however this is not so obvious. It is true that the
ITF on a {\emph single} ladder (first and last term in
Eq.~(\ref{HITF})) has a (doubly degenerate) ground state with
staggered 'AF' (K$_{Lz}>0$) pseudo spins corresponding to zig-zag
charge order along the ladder if the magnitude of the 'transverse
field' $\tilde{t}_R$ is smaller than a critical value $4\tilde{t}^c_R$
= K$_{Lz}$. This value defines the quantum critical point $\lambda_c$
= 1 of the ITF model \cite{Sachdevbook} with the dimensionless control
parameter $\lambda$ = K$_{Lz}$/$4\tilde{t}_R$. For $\lambda
>\lambda_c$, and assuming an infinitesimal staggered field to lift the
twofold ground-state degeneracy, the order parameter of zig-zag CO is
given by the exact solution of the 1D ITF:
%6.29
\begin{equation}
\langle T^z_i \rangle = (-1)^i~ \frac{1}{2} \left[ 1- \left(
  \frac{\lambda_c}{\lambda} \right)^2 \right]^{\frac{1}{8}} =
  (-1)^i\delta_{CO}(\lambda) ~~~.
\label{ITFOP}
\end{equation}
However due to the Trellis lattice structure (Fig.~\ref{fig:Nastruc2}) the
rungs of a ladder form a trigonal covering lattice and therefore the {\it
inter}-ladder coupling K$_{IL}$ frustrates the zig-zag charge
ordering on a given ladder. Therefore the critical $\lambda_c(K_{IL})$
increases monotonously with the inter-ladder coupling which defines a
quantum critical line for zig-zag CO at T = 0 in the K$_{Lz}$-K$_{IL}$
plane. At finite temperature the presence of frustration prevents
zig-zag CO to occur. If K$_{IL}$ becomes equal to the intra-ladder K$_{Lz}$, CO
melts and for some range of K$_{IL}$ even at T = 0 a disordered state
appears. Finally if K$_{IL}$ increases even further one obtains again
a CO state, but now with in-line order with parallel alignment of
pseudo spins along a ladder, i.e., V$^{4+}$ configurations are on one
side of the ladder and  V$^{5+}$ on the other side. These conclusions
have been drawn from an exact diagonalization study of the 2D model with
finite K$_{IL}$ \cite{Langari01}. 

The LDA+U results for \NAV \cite{Yaresko00} imply a relation
2V$_L$-V$_{IL}\simeq$ 0.027 eV meaning K$_{Lz}\simeq$ K$_{IL}$. Therefore \NAV~
is indeed close to the quantum critical line for zig-zag charge order and
purely Coulombic interactions cannot lead to a phase transition at finite
temperature in this compound due to geometric frustration. This is rather
similar to the charge ordering in the pyrochlore- or spinel lattices (or their
2D analogon, the checkerboard lattice) where charge ordering is also prohibited
by the inherent geometric frustration due to corner sharing tetrahedrons
of V- or other 3d ions of different valencies (Sec.
\ref{Sect:GeometricFrustration}). In such structures CO requires the lifting of
macroscopic degeneracy of the charge configurations by a lattice distortion as
is the case in AlV$_2$O$_4$ (see Sec.
\ref{Sect:StrucTransChargDispAlV2O4}). Something similar happens in \NAV. The
driving mechanism for the lattice distortion is here the spin superexchange
energy between singly occupied rungs along a ladder. Due to the quarter-filled
ladders the superexchange does not only contain terms coming 
from intermediate states with doubly occupied sites but also contributions from
rungs with two singly occupied sites. The latter depend on the pseudospin
configurations, which is the complementary effect as compared to
Eq.~(\ref{ITFCOUPL}). Therefore the effective spin exchange constants will
depend on the degree of charge order. One obtains
%6.30
\begin {eqnarray}
\label{HSEX}
H_{ex} & = & \sum_{i,j} J_{ij} {\bf S}_{ij} {\bf S}_{i+1j}\nonumber\\ 
J_{ij} & = & \left(1 + \sum_\alpha {\bf u}_{i + \frac{1}{2},j}^{\alpha}
{\bm {\nabla}}_{i + \frac{1}{2},j}^{\alpha} \right) J \left( T_{ij}^z, T_{i +
  1j}^z \right)\nonumber\\
J \left( T_{ij}^z, T_{i + 1j}^z \right) & = & J_0^{ij} \left( 1 + f \left(
T_{ij}^z, T_{i + 1j}^z \right) \right)\nonumber \\
{\bf u}_{ij} & = & \sum_{\lambda q}\frac{1} {\left(mN \right)^{1/2}} \exp
\left( i {\bf q} {\bf R}_{ij} \right) \left( b^{\lambda\dag}_{qj} +
b^\lambda_{qj} \right)~~~. 
\end{eqnarray}
The summation runs over the atomic shifts ${\bf u}_{ij}$ of neighboring atoms
($\alpha$ = O,Na) with respect to V which modulate the superexchange.  In
addition the latter depends on the charge configuration. Therefore spin ($\v
S_i$), charge ($\v T_i$) and lattice (b$^\lambda_{qj}$) degrees of freedom are
now coupled via Eq.~(\ref{HSEX}).  The spin exchange in Eq.~(\ref{HSEX}) has to
be added to Eq.~(\ref{HITF}) to get the total Hamiltonian H$_{ISSP}$ of
the Ising-spin Peierls model for \NAV. To simplify the model the
$J\left(T_{ij}^z,T_{i+1j}^z\right)$ dependence is approximated by
J$_{ij}$ = J$_0^{ij}(1-4\la T_j^z\ra^2)$. This reduction of
superexchange (Fig.~\ref{fig:Naexco}) with increasing charge order was
explained before. The remaining spin-lattice part is treated within
the Cross-Fisher theory \cite{Cross79}. The combined spin-Peierls
transition and charge ordering takes place at T$_{c1}$. At that
temperature the renormalized frequency of the VO-bond dimerization
mode with wave vector $\rm q_0$ = $\pi/b$ along the ladder given by
%6.31
\begin {equation} 
\label{PHONON}
\tilde{\omega}^2_{q_0}(T) = \omega^2_{q_0} - 0.26 \left| g_{VO} \left( q_0
\right) \right|^2 T^{-1} - 4g_{Is}^2 \chi_{q_0} (T)
\end {equation}
becomes soft, i.e. when $\tilde{\omega}^2_{q_0}(T_{c1})=0$. Here the last
term is determined by the pseudo-spin susceptibility $\chi_{q_0}(T)$ of
H$_{ITF}$ where g$_{Is}$ is a coupling constant that describes the change of
the transverse pseudo-spin field with lattice distortion. It may be shown that
inclusion of a spin-Zeeman term leads to the proper field dependence of
T$_{c1}(H)$ observed in experiment. It is found to be much weaker than for a
pure spin Peierls transition.
%6.17
%%%%%%%%%%%%%%%%%%%%%%%%%%%%%%%%%%%%%%%%%%%%%%%%%%%%%%%%%%%%%%%%%%%%%%%%%%%%
%%%%%%%%%%%%%%%%%%%%%%%%%%%%%%%%%%%%%%%%%%%%%%%%%%%%%%%%%%%%%%%%%%%%%%%%%%%%
%
The choice of the precise order parameter and distortion pattern between the
ladders below T$_{c1}$ is determined by the minimization of the total free
energy. On a given ladder there is a competition between the exchange
dimerization energy gain of the spin Peierls distortion and the
zig-zag charge order because the former is proportional to J and the
latter reduces J as just discussed (Fig.~\ref{fig:Naexco}). This can
be avoided by generating two {\it inequivalent} types of ladders A
and B below T$_{c1}$ where one (A) is mainly charge ordered and the
other (B) mainly dimerized \cite{Bernert02}. At the same time this reduces the
symmetry of the Trellis lattice in such a way that geometric
frustration is reduced and long range charge order is stabilized at a
finite temperature. Indeed, X-ray results \cite{Luedecke99,Sawa02}
show that the distortion of B-type ladders is much stronger as of
A-type ladders, independent of whether orthorhombic Fmm2 or monoclinic
A112 space groups are used for the structure refinement (Fig.~\ref{fig:Naexco},
left panel). At this stage, slightly below T$_{c1}$ one has a large spin gap on
B ladders but none on the zig-zag spin chains on the A ladders. However the
rapidly growing charge order $\delta_{CO}$(T) on A-type ladders below T$_{C1}$
modulates the hopping integrals and orbital energies via Na- and rung O-
shifts. This occurs in such a way that an additional but smaller exchange
dimerization is also induced on the A-type ladders. It grows with
$\delta_{CO}$(T). Once it is big enough it leads to a further dimerization
below T$_{c2}$ on the A-type ladders which is now of the pure spin-Peierls type
and opens the spin gap seen in experiment (Fig.~\ref{fig:NaKnight}). This
interpretation is supported by the appearance of additional Na-NMR splittings
caused by the shifts of Na atoms on top of the A-ladders for T $<$ T$_{c2}$. 
They correspond to 8 instead of 6 inequivalent Na positions below T$_{c2}$ 
which is indeed compatible with the monoclinic A112 structure from
X-ray analysis \cite{Sawa02}. 

Although the mechanism for the phase transitions is intricate,
involving charge, spin and lattice degrees of freedom and their
competing coupling effects, the low temperature ($\rm T\ll T_c$) spin
dynamics is quite simple again. With charge order and lattice
distortion saturated only the pure spin part in Eq.~(\ref{HSEX}) of the
original ISSP Hamiltonian remains. Thereby the existence of two
inequivalent A,B-type ladders and their low temperature (T $\ll$
T$_{c1}$) distortions in the parameterization of exchange constants has
to be taken into account.
%6.18
%%%%%%%%%%%%%%%%%%%%%%%%%%%%%%%%%%%%%%%%%%%%%%%%%%%%%%%%%%%%%%%%%%%%%%%%%%%%
\begin{figure}[tb]
\begin{center}
\includegraphics[clip,width=7.0cm]{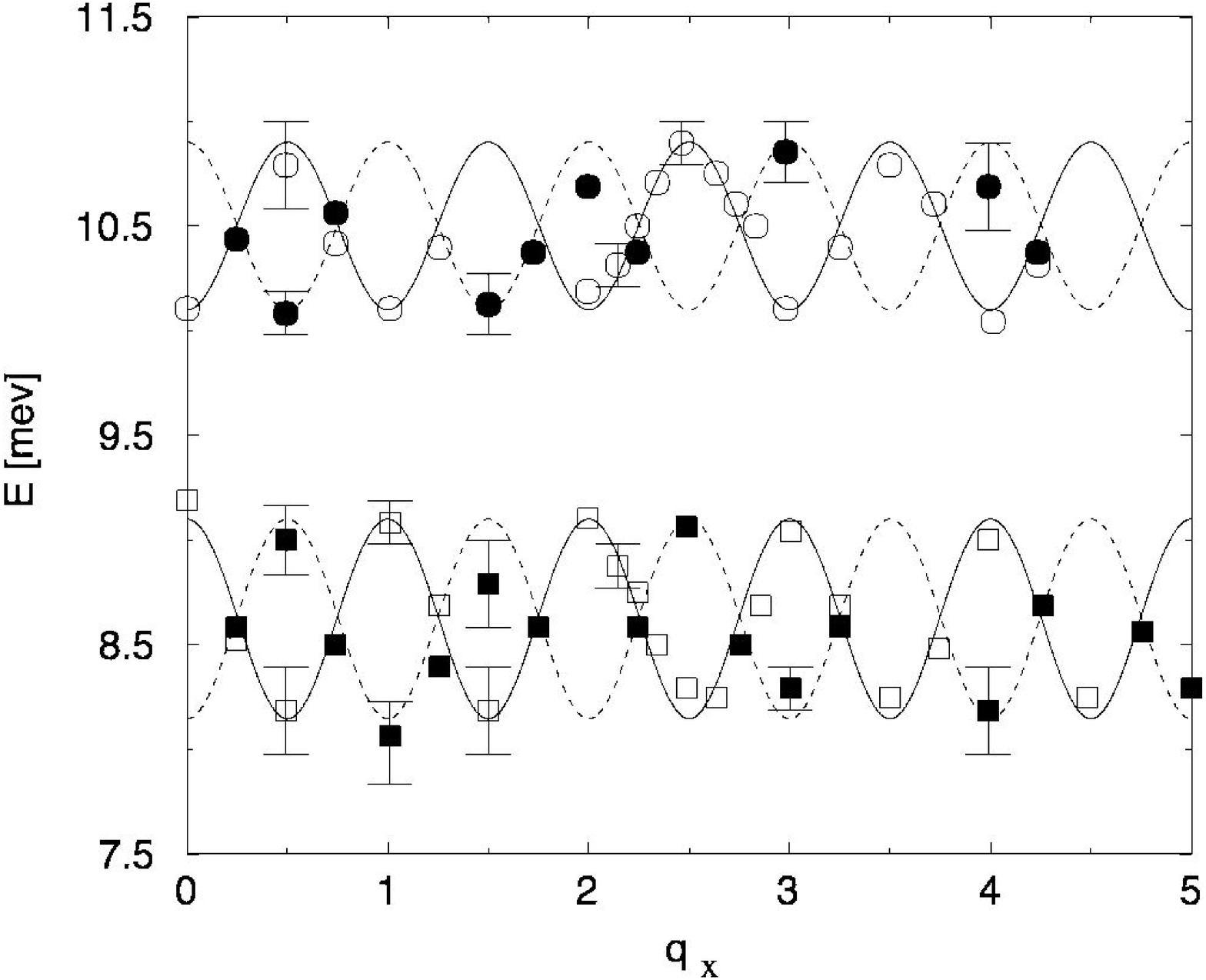}\hfill
\raisebox{0.0cm}
{\includegraphics[clip,width=7.0cm]{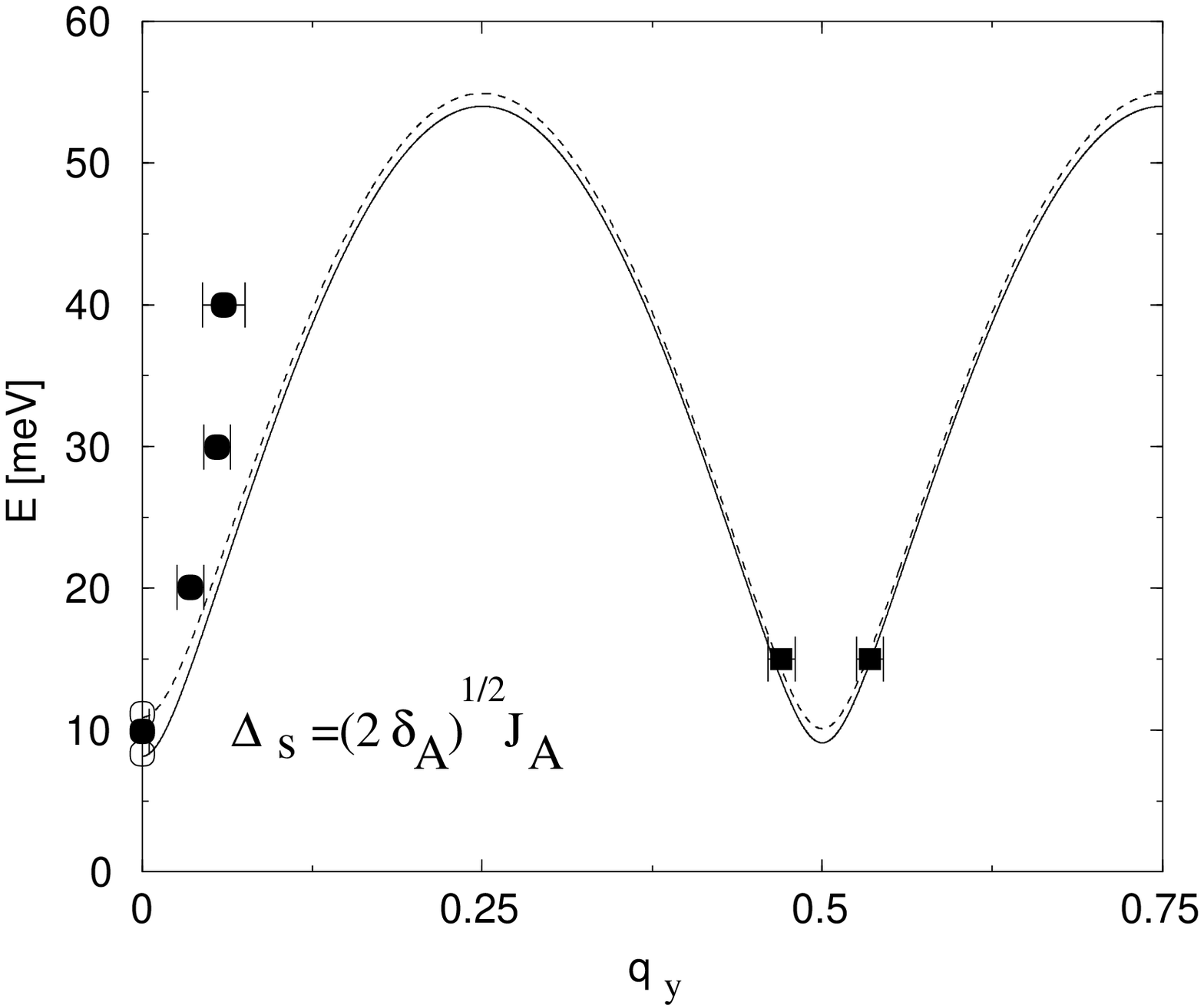}}
\end{center}
%\vspace{0.5cm}
\caption{ Dispersion of spin excitations  in \NAV~ (q$_x$ $\parallel a$ and
  q$_y$ $\parallel b$ given in r.l.u. 2$\pi/a$ and 2$\pi/b$
  respectively). Comparison of theoretical fit after Eq.~(\ref{MODES}) with
  experimental results from \cite{Grenier01}. Full lines:
  q$_y^0$=$\frac{1}{2}$, broken lines q$_y^0$ = 1. Left panel: $\omega^+(q_x)$
  - dispersion (top) and $\omega^-(q_x)$ - dispersion (bottom). From extremal
  values at q$_x$ = 0 model parameters $\delta_A$ = 0.03, J$'_a$ = 0.21 meV and
  J$_c$ = 0.43 meV are obtained. Right panel: Dispersion along b with exchange
  J$_A$ = 37.9 meV (440 K) along zig-zag chain on A. (After
  \protect\cite{Bernert02})} 
\label{fig:Nadispb}
\end{figure}
%%%%%%%%%%%%%%%%%%%%%%%%%%%%%%%%%%%%%%%%%%%%%%%%%%%%%%%%%%%%%%%%%%%%%%%%%%%%
%
Since the B ladders are strongly exchange dimerized with a
$\delta_B\sim$ 0.25 and J = J$_B$(1$\pm\delta_B$) they have a large
spin-excitation gap $\Delta_B$ = 38 meV \cite{Bernert02}. It is not
visible in the inelastic neutron scattering (INS) results which show a
minimum excitation energy around 10 meV. It must therefore result from
excitations on the more weakly dimerized ($\delta_A\ll\delta_B$) A-type CO
ladders. Consequently the low temperature spin Hamiltonian comprises only
slightly dimerized 1D zig-zag S=1/2 spin chains (on A) with an
intra-chain exchange J$_A$(1$\pm\delta_A$) along the b-direction. Since they
are separated by intervening B-ladders the A-ladders are only weakly coupled
with J$'_a\ll$ J$_A$ in the transverse a-direction. One therefore
would expect gapped quasi-1D spin excitations in \NAV. Indeed it was
found that their dispersion is much stronger along b than along a
\cite{Yosihama98,Grenier01}. As mentioned earlier the stacking of
layers along c is of the aaa'a'-type \cite{Ohwada05}. Therefore the
zig-zag chains on A are in-phase on aa (and a'a') bilayers and out-of-
phase between aa'. In the c-direction one may therefore assume an
exchange J$_c$ within the bilayers and neglect coupling between them.

Using this effective exchange model with parameter set
(J$_A$,$\delta_A$, J$'_a$,J$_c$), where J$'_a$ = J$_a$-4J$_D$ is an
effective intra-chain exchange along a, the magnetic excitations have
been calculated within a local dimer approach. It is applicable here since
J$_A\gg$ J$'_a$, J$_c$. The susceptibility of an isolated dimer pair in the
bilayer is given by 
%6.32
\begin{equation} 
\label{LOCSUSC}
u^\pm (\omega) = \frac{2 \left( J_A \left( 1 + \delta_A \right) \mp J_c
  \right)}{\left( J_A \left( 1 + \delta_A \right) \mp J_c \right)^2 -
  \omega^2}~~~~. 
\end{equation}
The collective susceptibility of the bilayers in RPA is then expressed as 
%6.33
\begin{equation} 
\label{RPASUSC}
{\chi^\pm}(\vec{q},\omega) = \left[ 1 -{J} \left( \vec{q} \right) {u^\pm}
  (\omega) \right]^{-1}{u^\pm} (\omega)~~~.
\end{equation}
The dispersion of spin excitations in the ab plane may be obtained
from the poles of ${\chi}^{\pm}\left(\vec{q},\omega\right)$ as 
%6.34
\begin{eqnarray} 
\label{MODES}
\omega^2_\pm(q_x,q_y) & = & \left[ J_A \left( 1 + \delta_A \right) \mp J_c
  \right]^2 - \left[ J_A \left( 1 + \delta_A \right) \mp J_c \right]\nonumber\\
  &&\times \left[ J_A \left( 1 - \delta_A \right) \cos 2q_y \pm \left( J_a \cos
  \left( q_x-q_y \right) -4J_D\cos q_x \cos q_y \right) \right]~~~. 
\end{eqnarray}
The comparison of mode dispersions with experimental results is shown
in Fig.~\ref{fig:Nadispb}. Using the low temperature J$_A$ = 440 K
\cite{Yosihama98} which corresponds to the charge order parameter
2$\delta_{CO}$ = 0.32 in Fig.~\ref{fig:Naexco} the remaining
parameters may be determined from three of the four observed gaps
$\omega^\pm(0,\frac{1}{2})$ and $\omega^\pm(0,1)$ in
Fig.~\ref{fig:Nadispb}. The fourth (lowest) gap is then correctly
calculated as $\omega^-(0,1)$ = 8.14 meV. The exchange dimerization
obtained is $\delta_A = 0.03 \ll \delta_B$.  The model calculation
explains a number of observations from INS: i) The dispersion along b
is much larger than along a because J$'_a\ll$ J$_A$ due to weak
coupling of A-ladders through intervening B-ladders. ii) The
a-dispersion of $\omega^+(q_x,q_y^0)$ is considerably smaller than for
$\omega^-(q_x,q_y^0)$ with q$_y^0$ = $\frac{1}{2}$ or 1. iii) Contrary
to earlier results \cite{Yosihama98} the high resolution experiments
\cite{Grenier01} presented in Fig.~\ref{fig:Nadispb} show a finite gap between
lower and upper mode $\Delta_{+-}\simeq$ 1 meV which is determined by a
combination of J$_A$ and J$_c$. It provides strong evidence for the
inequivalent A-B ladder model. If only equivalent A-ladders were
present, the gap should vanish, i.e., $\omega^+(q_x,q_y^0)$ and
$\omega^-(q_x,q_y^0)$ would touch with their minima and maxima and the
dispersion would have have twice the observed period along a \cite{Gros99}.
We conclude that \NAV~ presents another example where charge ordering may lead
to pronounced 1D character of spin excitations below T$_c$, here they are
gapped due to exchange dimerization caused by the complicated nature of the
associated lattice distortions.

It has become clear that CO in \NAV~ is severely inhibited by effects
of geometric frustration. In fact its comparatively low T$_c\simeq$ 33 K
indicates that it is close to the quantum critical point of the ITF
model, where the CO is of essentially 1D Ising type. The true 2D
ordered state at finite T is then established by a staggered longitudinal 
pseudo spin field set up by the distortions of the neighboring
ladders. In this scenario it is suggestive that even small
perturbations of the 1D Ising spin correlations along the ladder might
suppress the CO state in \NAV~.  This can be achieved by reducing the filling
factor n of the 1D ladders below $\frac{1}{4}$ by doping with holes. This
introduces "empty" rungs into the ladder which cut the Ising bonds. The ensuing
destruction of long- range 1D correlations should then strongly reduce
T$_c$ as function of the hole concentration $\delta_h$
(n = $\frac{1}{4}$-$\delta_h$). This has indeed been found by introducing
holes into the ladders through Na- deficiency doping \cite{Isobe97} where
a few per cent holes are sufficient to destroy the CO state and the
associated spin Peierls transition. Rapid suppression of charge order has
also been found in various other doping series, i.e., replacing Na by
Li and K (isoelectronic) or Ca (electron-doping) \cite{Dischner01}.
This may be observed directly by specific heat measurements \cite{Dischner01}
which show a progressive suppression of $\Delta$C(T$_c$) with increasing
doping. It is also seen in the susceptibility \cite{Isobe97}
which exhibits a closing of the spin gap associated with CO. 
Most importantly, in this doping range \NAXV~remains an insulator. This is
not easy to understand within a Hubbard like model for the quarter
filled ladder \cite{Bernert01,Bernert02}. Possibly 1D localization
and polaronic effects play a role, indeed the conductivity was found to exhibit
variable-range hopping behavior \cite{Isobe97} for hole doping. 
%6.19
%%%%%%%%%%%%%%%%%%%%%%%%%%%%%%%%%%%%%%%%%%%%%%%%%%%%%%%%%%%%%%%%%%%%%%%%%%%%
\begin{figure}[tb]
\begin{center}
\includegraphics[clip,width=6.5cm]{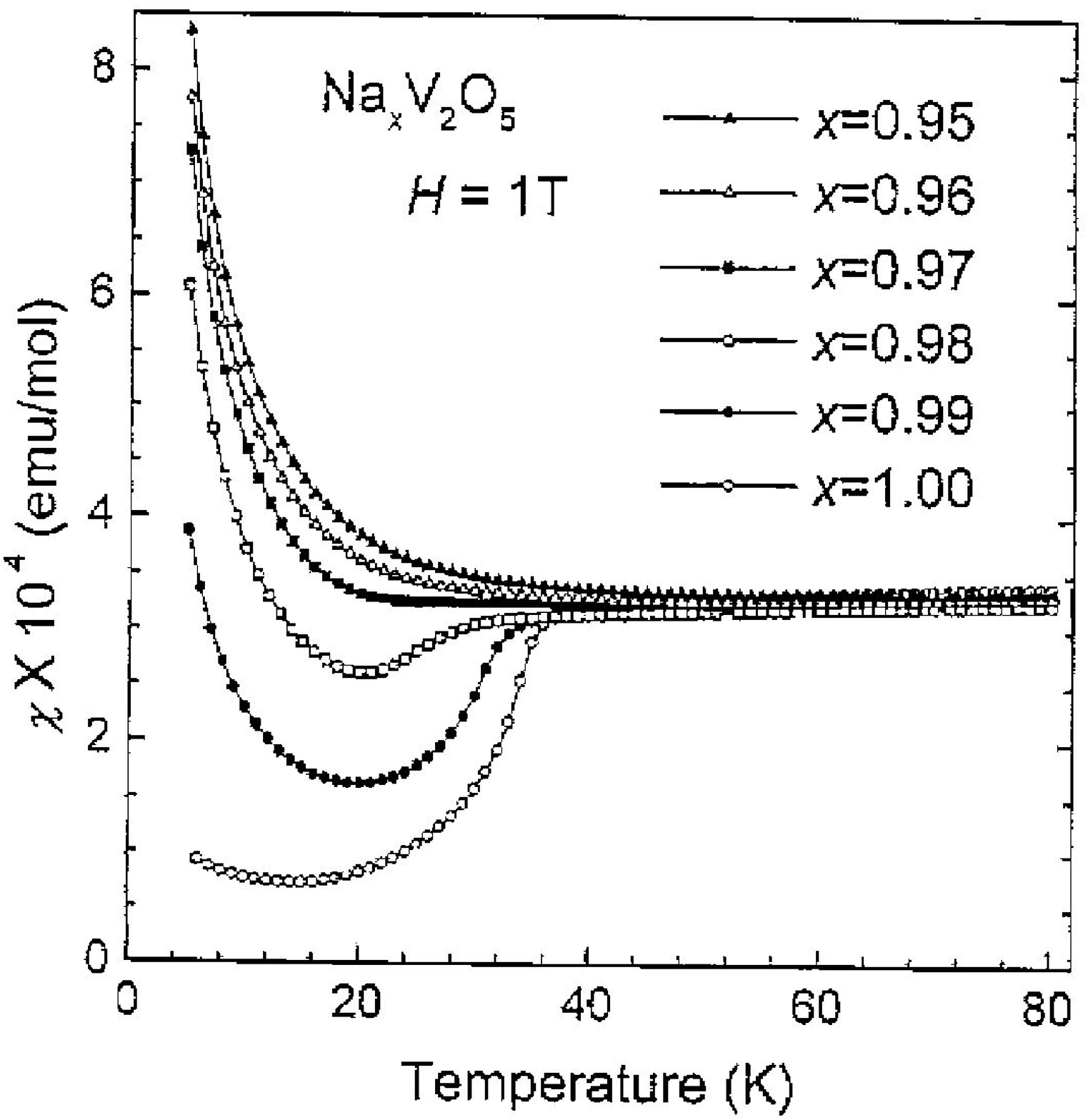}\hfill
\raisebox{0.0cm}
{\includegraphics[clip,width=7.5cm]{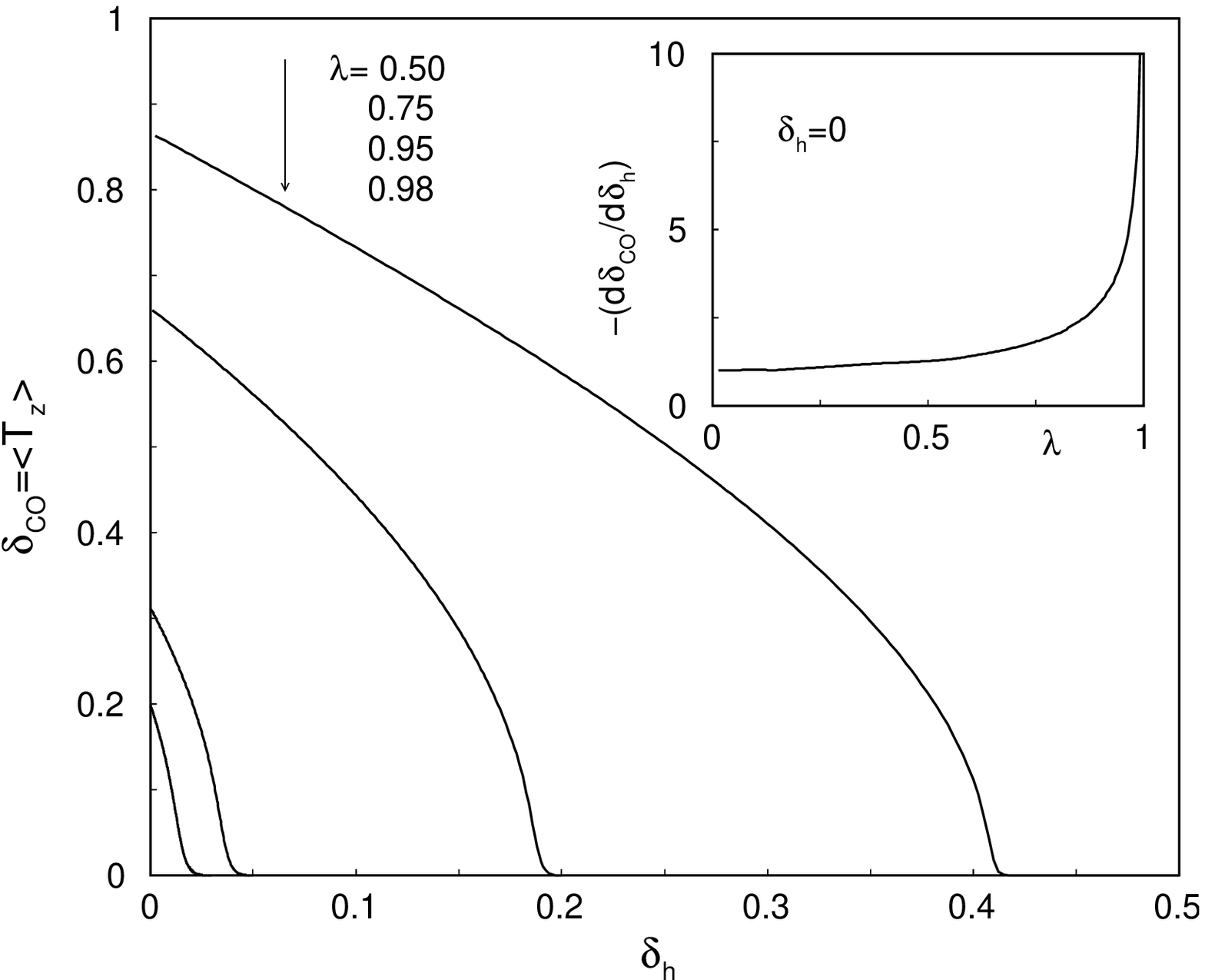}}
\end{center}
%\vspace{0.5cm}
\caption{Left panel: Melting of charge order by Na-deficiency (1-x) (hole)
  doping seen from the rapid closing of the spin gap in $\chi(T)$ with
  increasing 1-x (corresponding nominally to $\delta_h$). (After
  \protect\cite{Isobe97}). Right panel: Calculated melting of CO with hole
  doping $\delta_h$ for various $\lambda$ = 2t$_R$/V$_L$ given in decreasing
  order. The absolute slope value -(d$\delta_{CO}$/d$\delta_h$) increases
  strongly when approaching the quantum critical point $\lambda_c$ = 1 of CO;
  this is shown in the inset. (After \protect\cite{Thalmeier03}).}
\label{fig:Nadope}
\end{figure}
%%%%%%%%%%%%%%%%%%%%%%%%%%%%%%%%%%%%%%%%%%%%%%%%%%%%%%%%%%%%%%%%%%%%%%%%%%%%
%
It was shown in \cite{Thalmeier03} that even for the (hole-) doped case the CO
problem may be treated with an extended pseudo-spin model. In order to
incorporate the possibility of empty rungs on a ladder a pseudo spin
T = 1 is introduced where the $|T_z = \pm 1\ra$ states describe occupation
of the right or left Vd$_{xy}$ orbital of a rung respectively, and in
addition $|T_z = 0\ra$ the empty (hole-doped) rung. The total
effective T = 1 Hamiltonian of a single ladder is then given by
%6.35
\begin{eqnarray}
\label{HAMDOP}
H & = & \sum_i \left[ (\epsilon-\mu) T_{zi}^2 + t_R \left( T_{xi}^2 - T_{yi}^2
  \right) - h^0_{si} T_{zi} \right] + 2t_L \sum_{\la ij\ra} \left[ O^i_{zy}
  O^{j\dagger}_{zy} + O^{i\dagger}_{zy} O^j_{zy} \right]\nonumber\\ 
&& + V_-\sum_{\la ij\ra} T_{zi} T_{zj} + V_+\sum_{\la ij\ra}
  T^2_{zi}T^2_{zj}~~~. 
\end{eqnarray}
Here $\epsilon$ is the on-site orbital energy and h$^0_s$(i) = h$^0_s(-1)^i$ is
a longitudinal staggered pseudo-spin field that simulates the effect of
distortion connected with CO and the coupling to neighboring ladders, $\la
ij\ra$ denotes n.n. rungs along the ladder. Here O$_{zy}$ = iT$_z$T$_y$  may be
interpreted as quadrupolar operators in the T = 1 
pseudo-spin space. Furthermore $\epsilon$ is the d$_{xy}$ orbital energy and
$\mu$, a 'chemical potential' to fix the number of holes $\delta$ = 1-n (n =
number of d-electrons per rung) at a value determined by the
doping. Interaction parameters are defined as V$_\pm = \frac{1}{2}(V_L\pm
V_D)$. If the ladder diagonal term V$_D$ is neglected then V$_\pm$ =
$\frac{1}{2}$V$_L$ leads to a control parameter $\lambda$ = 2t$_R$/V$_L$ (to
stay in accordance with Ref. \cite{Thalmeier03} the inverse value of the
previous definition for $\lambda$ is used here). In isospin language the
(staggered) CO parameter $\delta_{CO}$ and the hole doping $\delta_h$ are given
by 
%6.36
\begin{eqnarray}
\delta_{CO}= \la T_{z1}\ra = - \la T_{z2}\ra \qquad  \delta_h=1-\la
T^2_z\ra~~~. 
\end{eqnarray}
Here i=1,2 denote the two 1D sublattices along the ladder direction
b. The selfconsistent mean-field solution of the model is shown in
Fig.~\ref{fig:Nadope} (right panel). Its usefulness relies on the
Ising type nature of the ordered state. It shows that close to the QCP
$\lambda_c$=1 the charge order parameter is rapidly suppressed with
increasing doping which is also evident from the $\delta_h$-
dependence of the slope shown in the inset. This behavior corresponds
qualitatively to the rapid reduction of CO in Na-deficiency doped
\NAXV~ where the spin-gap and hence the associated charge order is
suppressed by a few per cent Na-deficiency 1-x. For small doping one
may assume that the latter is equal to the average hole concentration
$\delta_h$ in the V-rungs.

Finally we briefly discuss the $\beta$-vanadium bronzes
$\beta$-Na$_{0.33}$V$_2$O$_5$ with large but stoichiometric (1-x = 2/3) Na
deficiency doping. Their crystal structure is different but it still contains
the Trellis lattice layers. For high temperatures these
compounds are 1D metals (only in the stoichiometric case) along the
b-axis and exhibit a CDW-instability at T$_{CDW}$ = 136 K into an
insulating state \cite{Okazaki04}. While the Wigner-lattice type CO transition
discussed for \NAV~ is due mostly to intersite Coulomb interaction
energies of localized 3d electrons, the CDW Fermi-surface instability
in $\beta$-Na$_{0.33}$V$_2$O$_5$ is driven by the kinetic energy of 1D
conduction electrons. Under pressure the CDW transition may be suppressed again
and around p$_c\simeq$ 8 GPa the T$_{CDW}$(p)-line ends in a quantum critical
point with an associated superconducting dome around a maximum superconducting
T$_c$ of about 10 K.

\subsection{Reentrant Charge Ordering and Polaron Formation in Double
Exchange Bilayer Manganites \LSXM}
\label{subsect:Polaron}

The layered perovskite manganites have been at the center of 3d-oxide
research since the discovery of the colossal magnetoresistance effect
(CMR) \cite{Moritomo96}. Its signature is a change in the resistivity
over several orders of magnitude under comparatively small magnetic
field changes for doped metallic manganites close to the ferromagnetic
phase transition. The investigation of CMR has led to a global survey
of doped manganite compounds. Their structures consist of MnO$_6$
octahedra corner-linked to layers that may be stacked in different
fashions.  As in the cuprates, the parent compounds are AF
Mott-Hubbard insulators and hole doping destroys the AF order.
However, the metallic state is not superconducting but
ferromagnetic. In the manganites the doping with holes only reduces
the Mn-moments, rather than creating non-magnetic Zhang-Rice singlets
as in the cuprates. The Mn moments then order ferromagnetically via
the double exchange (DE) mechanism \cite{Zener51,Anderson55} that
lowers kinetic energy of e$_g$ conduction electrons when their spins are
aligned with spins of localized Mn-t$_{2g}$ spins (see
Fig.~\ref{fig:Bimagstruc}). 

The physics of manganites, especially those derived from the infinite
layer parent compound LaMnO$_3$, has been reviewed in many articles,
e.g., Ref. \cite{Imada98}. In this chapter we shall focus exclusively on
aspects of the bilayer manganites with half-doped insulating compound \LSM~that
are related to charge ordering and possible polaronic effects. In addition
we discuss magnetic excitations in the doped metallic ferromagnetic bilayer
compounds which give evidence for the double exchange mechanism that
is central to the physics of magnetic phases and magnetotransport. This topic
has also been more extensively reviewed in Ref. \cite{Chatterji04}.
%6.20
%%%%%%%%%%%%%%%%%%%%%%%%%%%%%%%%%%%%%%%%%%%%%%%%%%%%%%%%%%%%%%%%%%%%%%%%%%%%
\begin{figure}[tb]
\begin{center}
\includegraphics[clip,width=3.0cm]{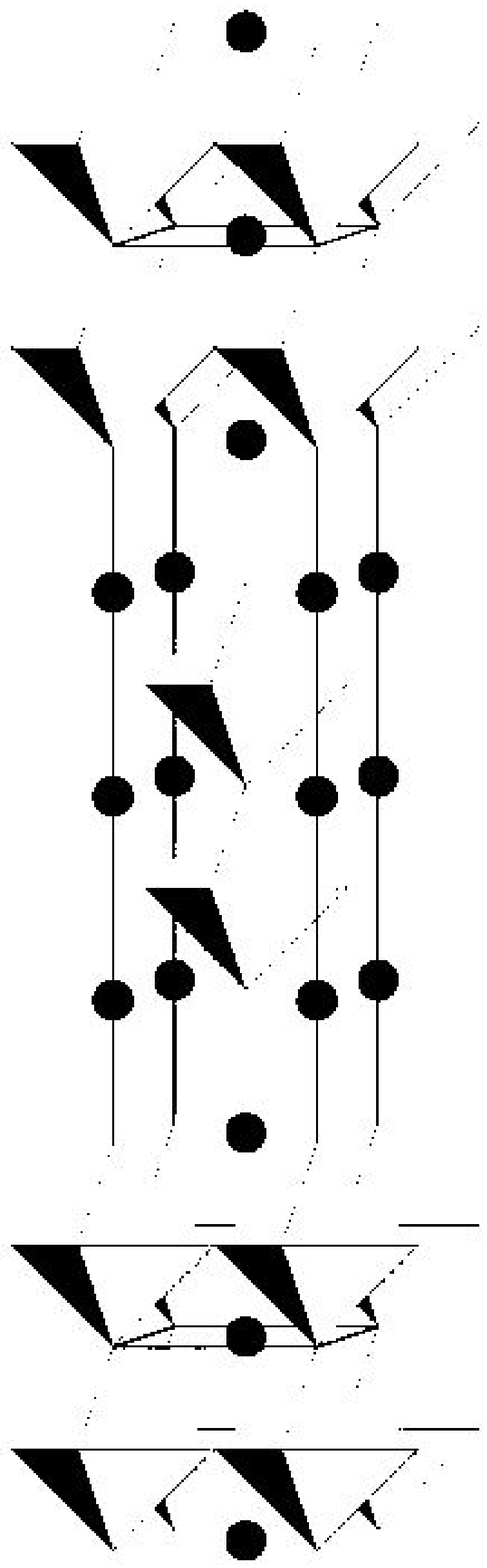}
\hfill
\raisebox{7.3cm}
{\includegraphics[clip,width=6.5cm,angle=-90]{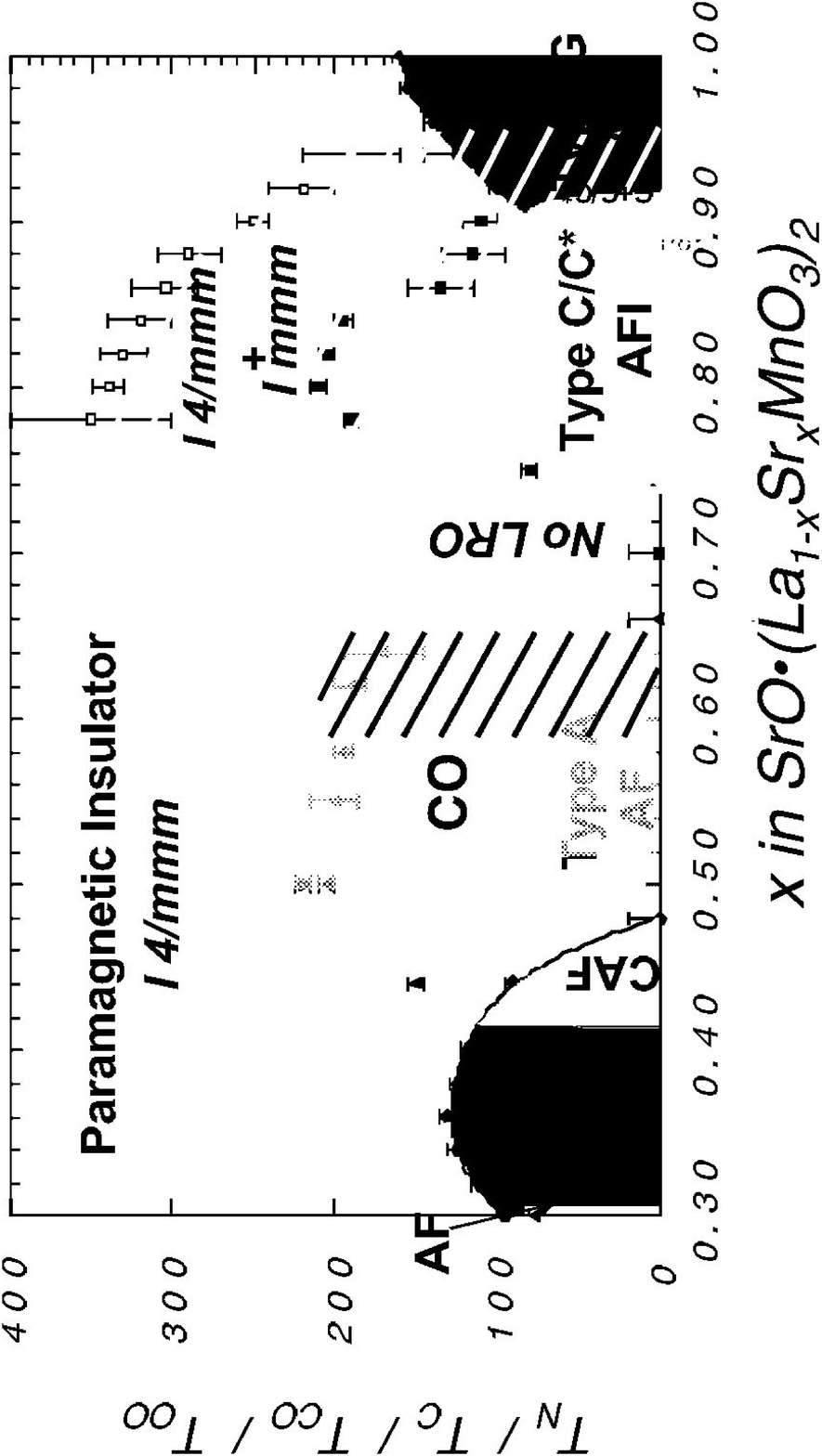}}
\end{center}
\vspace{0.5cm}
\caption{Left panel: Crystal structure of \LSM~and \LSXM~consisting of MnO$_6$
  octahedra and (La,Sr)-cations (circles). Lattice constants are (x = 0.5) a =
  3.874\AA, c = 19,972\AA. The space group is {\it I4/mmm}. Right panel:
  Structural and magnetic phase diagram of bilayer \LSXM~in the doping range
  $0.3<x<1.0$. Charge order (CO) appears for $0.5\leq x \leq 0.65$. Around x =
  0.7 magnetic order is absent. (After \protect\cite{Qiu03}.)}
\label{fig:Bistruc2}
\end{figure}
%%%%%%%%%%%%%%%%%%%%%%%%%%%%%%%%%%%%%%%%%%%%%%%%%%%%%%%%%%%%%%%%%%%%%%%%%%%%
%
We first give a brief summary of structural properties. The
manganites belong to the Ruddlesden-Popper phases which may be
described by intergrowth of rock-salt like MnO slabs and n slabs of
the perovskite LaMnO$_3$. For n = $\infty$ one has the infinite layer LaMnO$_3$
perovskite, while n = 2 corresponds to the bilayer manganite \LSXM~considered
here. Its structure is shown in Fig.~\ref{fig:Bistruc2}. In the (hypothetical)
compound with x = 0 La and Sr are tri- and divalent cations which implies a
Mn$^{3+}$ (S = 2) state for the magnetic cations. Replacing La by Sr according
to the chemical formula \LSXM~is equivalent to hole doping and creates
Mn$^{4+}$ (S = 3/2) with a nominal concentration of x holes/Mn site. For x =
0.5 one has a stoichiometric mixed valent compound \LSM~with a 1:1 ratio of
Mn$^{3+}$/ Mn$^{4+}$ ions. This suggests the possibility of charge (and
orbital) ordering leading to an insulating state. In the possible concentration
range $0.2<x<1$ the crystal structure remains the same although the Mn-O bond
lengths depend on x due to the Jahn Teller distortion of the octahedrons
containing Mn$^{3+}$. The AF (x = 0.5) and FM (x$<$0.4) structures are shown
later in Fig.~\ref{fig:Bimagstruc}. 
 
Electronic structure calculations for \LSM~within LSDA+U \cite{deBoer99} lead
to a quasi-2D band structure that is close to that of a half metal. The gap
between majority and minority spin bands is $\Delta_{\uparrow\downarrow}$ = 2.7
eV. The influence of possible charge ordering has been neglected in this
calculation.  Due to the CEF splitting of 3d states into t$_{2g}$ and e$_g$
states, the lower lying t$_{2g}$ bands, which are $\sim$ 1eV below E$_F$ are
almost dispersionless while the bands crossing E$_F$ have mainly e$_g$
character. This allows one to use a simple model for the electronic
structure: While the e$_g$ electrons are described in a nearest neighbor (n.n.)
tight-binding (TB) approximation, the t$_{2g}$ electrons are treated as
localized with the intra-atomic exchange aligning their spins to a total spin
\v S (S = 3/2) (see Fig~\ref{fig:Bimagstruc}). Furthermore, there is a Hund's
rule coupling of strength J$_H$ which tries to align localized t$_{2g}$ and
itinerant e$_g$ spins. Finally on-site (U) and inter-site (V) Coulomb
interaction terms for the e$_g$ electrons have to be added. Then one obtains
the total Hamiltonian 
%6.37
\begin{eqnarray}
H & = & \sum_{ij\sigma} t_{ij} \left( c^\dagger_{i\sigma} c_{j\sigma} +
h.c. \right) + U\sum_in_{i\ua} n_{i\da} + V\sum_{\la ij\ra} n_in_j \nonumber\\
&& - J_H \sum_i\v S_i\v s_i + J \sum_{ij}\v S_i\v S_j~~~. 
\label{MANHAM}
\end{eqnarray}
Here the first three terms describe e$_g$ conduction electrons
(c$_{i\sigma}$) where n$_{i\sigma}$ = c$^\dagger_{i\sigma}$c$_{i\sigma}$
and n$_i$ = n$_{i\ua}$ + n$_{i\da}$, the fourth term describes FM Hund's rule
coupling ($J_H>0$) between e$_g$-spins (\v s) and t$_{2g}$ spins (\v S) and the
last one a superexchange between the localized  t$_{2g}$ spins. Note that i =
(l,$\lambda)$ where $\lambda$ = 1,2 is the bilayer index and l the site within
a layer. The model is able to describe both the charge ordering and magnetism
in the manganites. The orbital degree $\alpha$ = 1,2 of e$_g$ electrons is
still missing. It is too complex to be solved in full generality, we therefore
treat CO and magnetic aspects separately and disregard the possibility of
simultaneous orbital order.  

To investigate  charge order as function of the e$_g$ band filling n
we take into account only the first three terms which constitute an extended
Hubbard model (EHM) for the e$_g$ electrons. Assuming identical CO in both
layers the problem reduces to the EHM on a 2D square lattice. Magnetic order
will be suppressed and two-sublattice (A,B) charge order is assumed. In the
limit U$\gg$t$>$V this model may be treated \cite{Hoang02} by a combination of
Hartree-Fock approximation for the inter-site term (V) and a CPA approximation
for the on-site Hubbard term (U). The latter rests on the alloy-analogy, i.e.,
the EHM is replaced by a single-particle Hamiltonian with diagonal (sites 1,2
within each layer) disorder of e$_g$ orbital energies E$^{(1,2)}_{A/B\sigma}$
and corresponding probabilities p$^{(1,2)}_{A/B\sigma}$ on each
sublattice. Accordingly, 
%6.38
\begin{eqnarray}
E^{(1)}_{A/B\sigma} & = & zVn_{A/B} \quad \mbox{with} \quad p^{(1)}_{A/B\sigma}
= 1-n_{A/B-\sigma}\nonumber\\
E^{(2)}_{A/B\sigma} & = & zVn_{A/B} +U \quad \mbox{with} \quad
p^{(2)}_{A/B\sigma} = n_{A/B-\sigma}~~~.
\end{eqnarray}
The Green's functions of the equivalent single-particle Hamiltonian is
then configuration averaged within CPA, i.e., requiring the average
T-matrix of the system to vanish. This leads to averaged Green's functions 
%6.39
\begin{eqnarray}
\bar{G}_{A/B}(\v k,\omega) = \left[ \omega - \Sigma_{A/B}(\omega) - \frac{t_{\v
	  k}^2}{\omega - \Sigma_{B/A}(\omega)}\right]  
\end{eqnarray}
where t$_{\v k}$ = (t/2)($\cos k_x$ + $\cos k_y$). The corresponding
self-energies are given by 
%6.40
\begin{eqnarray}
\Sigma_{A/B}(\omega) = \bar{E}_{A/B} - \left[ zVn_{B/A} - \Sigma_{A/B}(\omega)
  \right] \bar{G}_{A/B}(\omega) \left[ zVN_{B/A} + U-\Sigma_{A/B}(\omega)
  \right] ~~~.
\end{eqnarray}
Here $\bar{E}_{A/B}$ = zVn$_{B/A}$ + $\frac{1}{2}$ Un$_{A/B}$ are the
effective orbital energies on A/B sublattices. On the bipartite lattice CO has
to be symmetric which leads to the restriction n$_{A/B}$ = n
$\pm$n$_{CO}$. This requires only a single averaged Green's function defined by
G($\pm$n$_{CO}$, $\omega$) = $\bar{G}_{A/B}(\omega)$. From the above equations
the self energies may be eliminated. One obtains a cubic equation for G from
which the order parameter n$_{CO}$ may be calculated as function of filling n
and Coulomb interaction parameters U, V by requiring charge
conservation. Setting n$_{CO}$ = 0 the phase boundary between CO and
homogeneous phase is obtained as a surface in (n, U , V) space. It is obtained
as an implicit solution of the equations 
%6.41
\begin{eqnarray}
n = -\frac{2}{\pi}\int d\omega f(\omega)ImG(0,\omega) \quad \mbox{and}\quad
1 = -\frac{2}{\pi}\int d\omega f(\omega)ImG'(0,\omega)
\end{eqnarray}
where the derivative is defined as $G'(\omega)$ = $\partial G(n_{CO},
\omega)/\partial n_{CO}|_{n_{CO} = 0}$ and f($\omega$) is the Fermi function. 
%6.21
%%%%%%%%%%%%%%%%%%%%%%%%%%%%%%%%%%%%%%%%%%%%%%%%%%%%%%%%%%%%%%%%%%%%%%%%%%%%
\begin{figure}[tb]
\begin{center}
\includegraphics[clip,width=6.5cm]{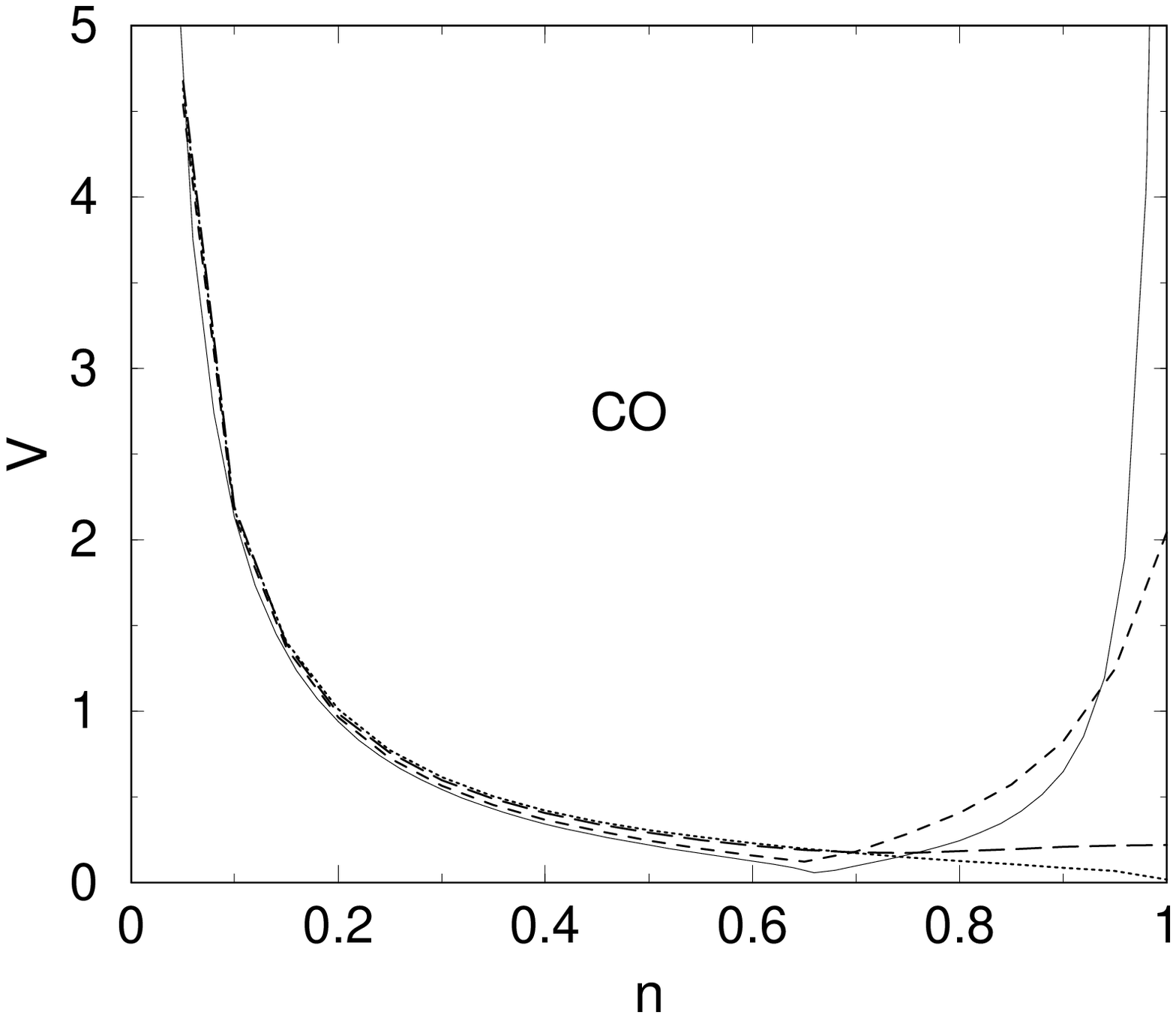}
\hfill
\raisebox{0.3cm}
{\includegraphics[clip,width=6.8cm]{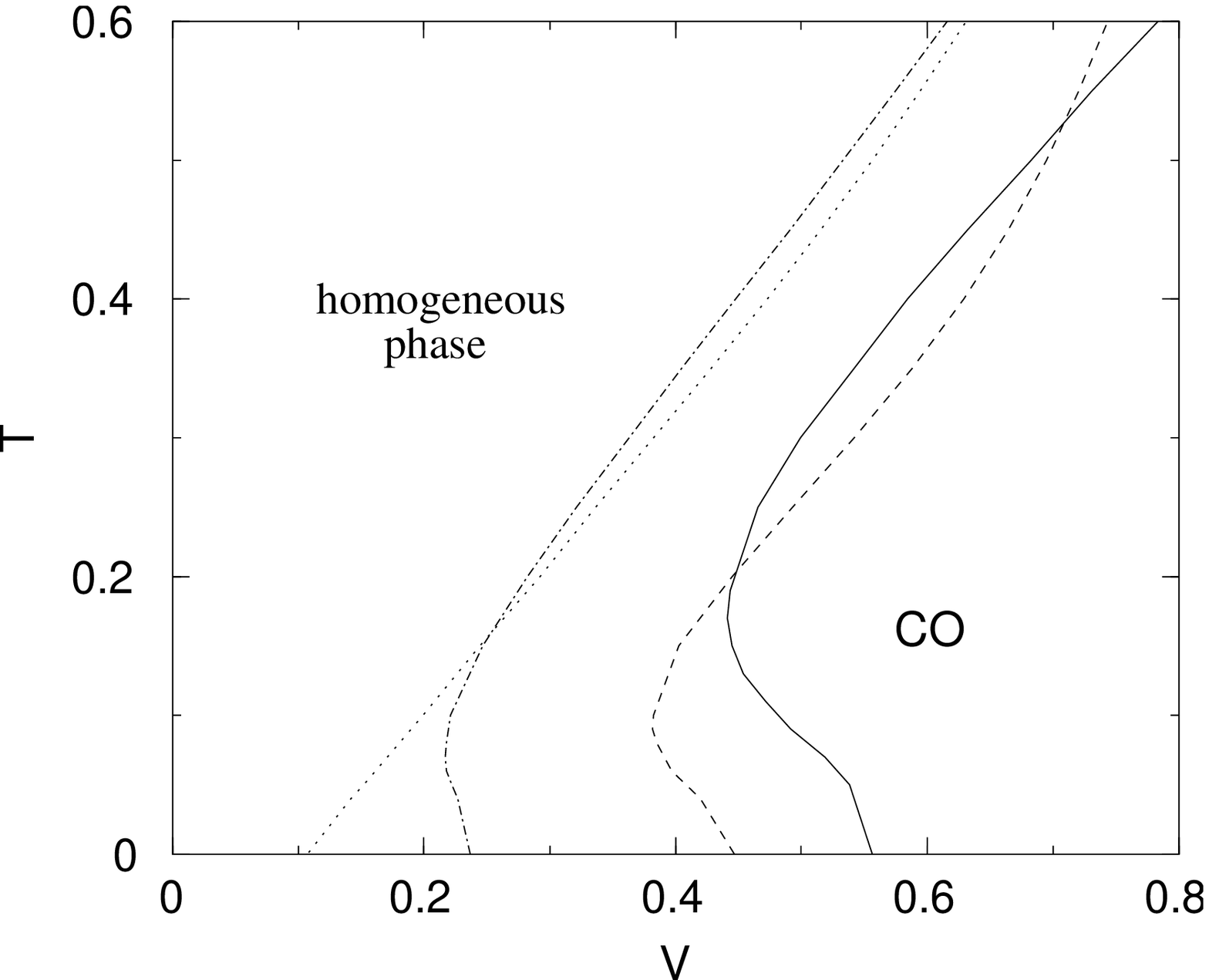}}
\end{center}
\vspace{0.5cm}
\caption{Left: n-V phase diagram of charge order (CO) for the 2D EHM (T = 0, W
  = 4t$\equiv$1) 
  and U = 0, 0.5, 1.5, $\infty$ corresponding to dotted, long-dashed, dashed
  and solid lines respectively. Right: V-T phase diagram of CO for U = 2 and
  various band filling n = 0.3, 0.5, 0.65 and 0.8 corresponding to solid,
  dash-dotted, dotted and dashed curves respectively. The CO regime is to the
  right of the boundary for each value of n. Energies and temperature are given
  in units of t. (After \protect\cite{Hoang02}.)}
\label{fig:BinPD}
\end{figure}
%%%%%%%%%%%%%%%%%%%%%%%%%%%%%%%%%%%%%%%%%%%%%%%%%%%%%%%%%%%%%%%%%%%%%%%%%%%%
%
The self-consistent solution for charge order has been determined for
the 2D square lattice (z = 4 and W = 4t = half bandwidth). The resulting n-V
and V-T phase diagrams for CO are shown in the left and right part of
Fig.~\ref{fig:BinPD} respectively. In the n-V phase diagram we notice
two regions. For $n<n^*\sim 0.67$ the CO boundary V$_c$(n) is almost
independent of U while for $n>n^*$ when the half filled case n=1 is
approached CO is strongly suppressed with increasing U. This is due to
the fact that in a Mott-Hubbard insulator charge fluctuations are
already strongly suppressed and additional spatial symmetry breaking
CO then needs a larger threshold value V$_c$. On the other hand for $n<n^*$,
e.g., quarter filling n = 1/2 the V$_c\sim$ 0.25 W = t is determined
essentially by the hopping t. The minimum V$_c$ where CO is most
easily achieved is obtained around $n = n^*$.

The V-T phase diagram shows an interesting aspect: For each value of U
a range of V values exists for which the CO transition is reentrant as
function of temperature. For such a V value the ground state is
homogeneous while CO appears in an intermediate temperature
range. This behavior cannot be obtained by treating the EHM in
Hartree Fock approximation and thus it is a genuine correlation
effect. However, as we shall see, another mechanism based on polaron
formation may also lead to reentrant CO. In the present model the
absolute value of transition temperatures is unrealistically large
because only the n.n. interaction V is assumed. This may be improved
by including also a competing n.n.n Coulomb repulsion which reduces the
CO temperature as shown later.

\begin{figure}[tb]
\begin{center}
\includegraphics[clip,width=5.0cm]{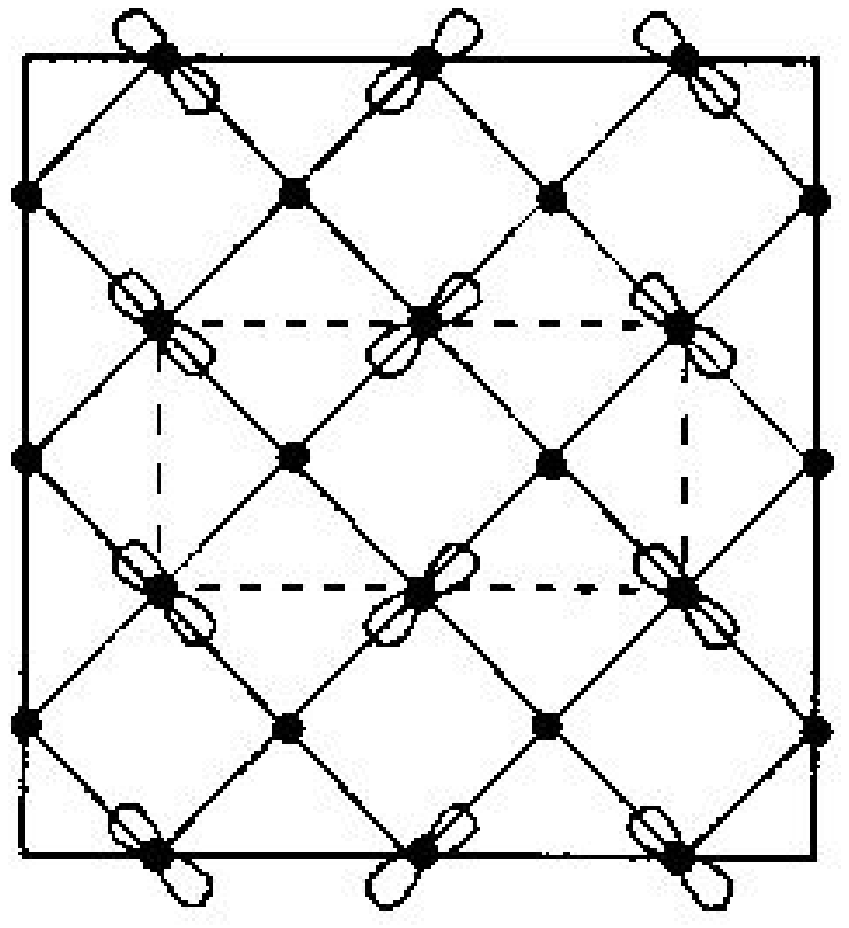}
\raisebox{2cm}
{\includegraphics[clip,width=5.0cm]{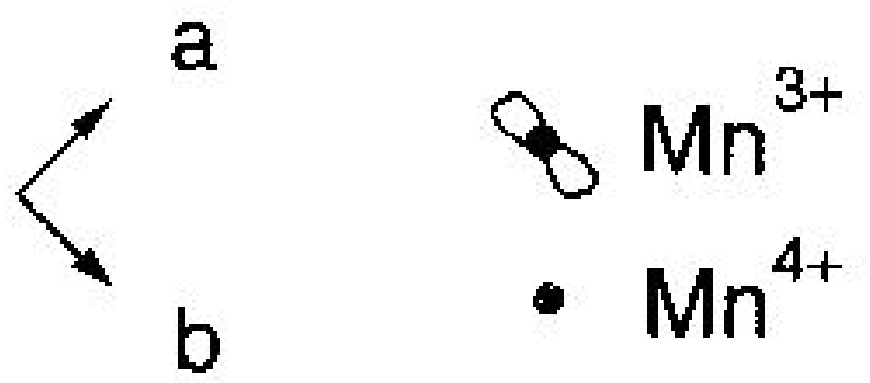}}
\end{center}
\vspace{0.5cm}
\caption{Structure of charge and orbital order in stoichiometric (x = 0.5)
  \LSM~corresponding to wave vector \v Q =
  $(\frac{1}{4},\frac{1}{4},0)$. (After
  \protect\cite{Kimura98,Wakabayashi03}.)} 
\label{fig:Biorbor}
\end{figure}

\begin{figure}[tb]
\begin{center}
\includegraphics[clip,width=6.5cm]{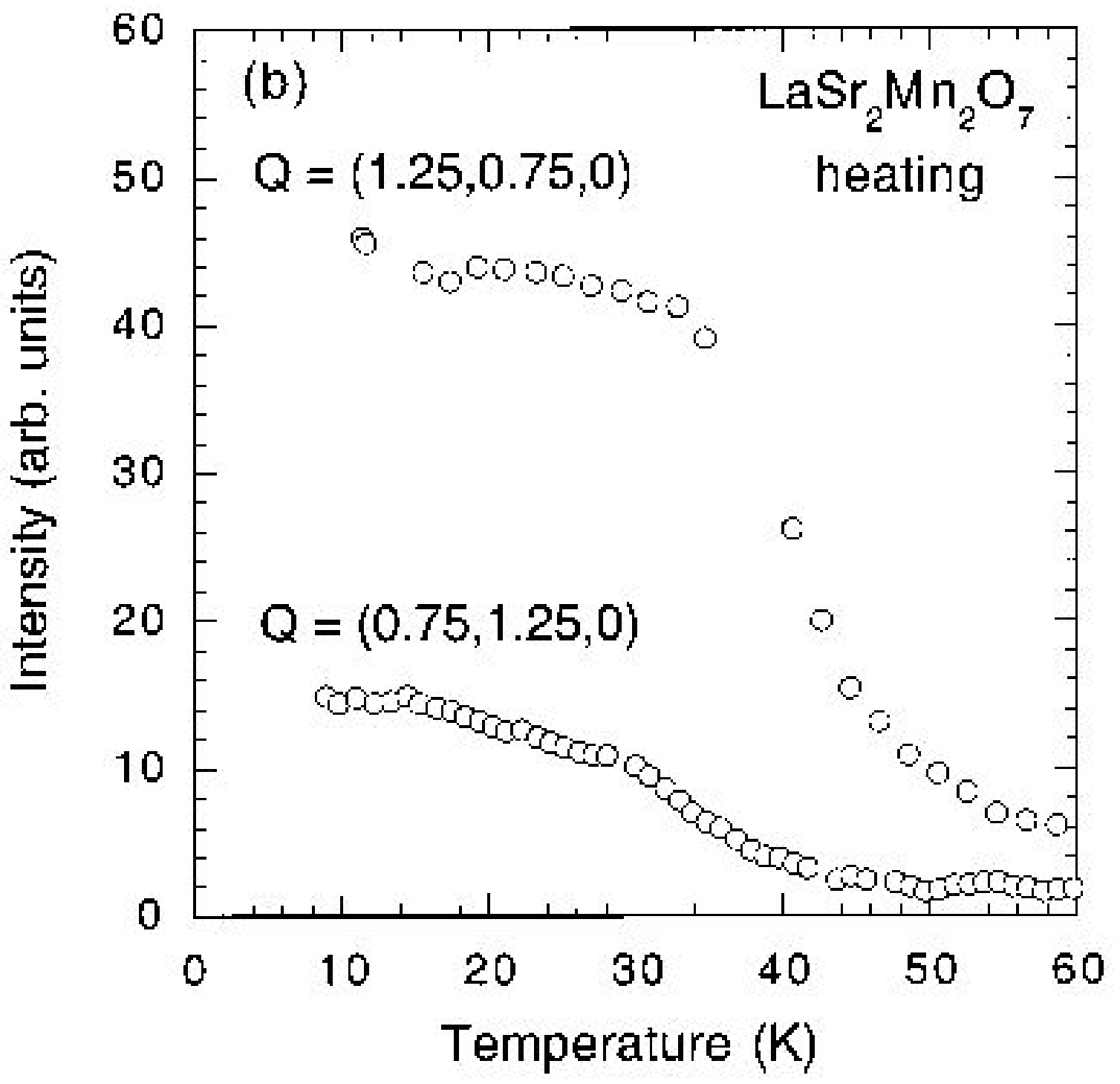}
\hfill
\raisebox{0cm}
{\includegraphics[clip,width=6.5cm]{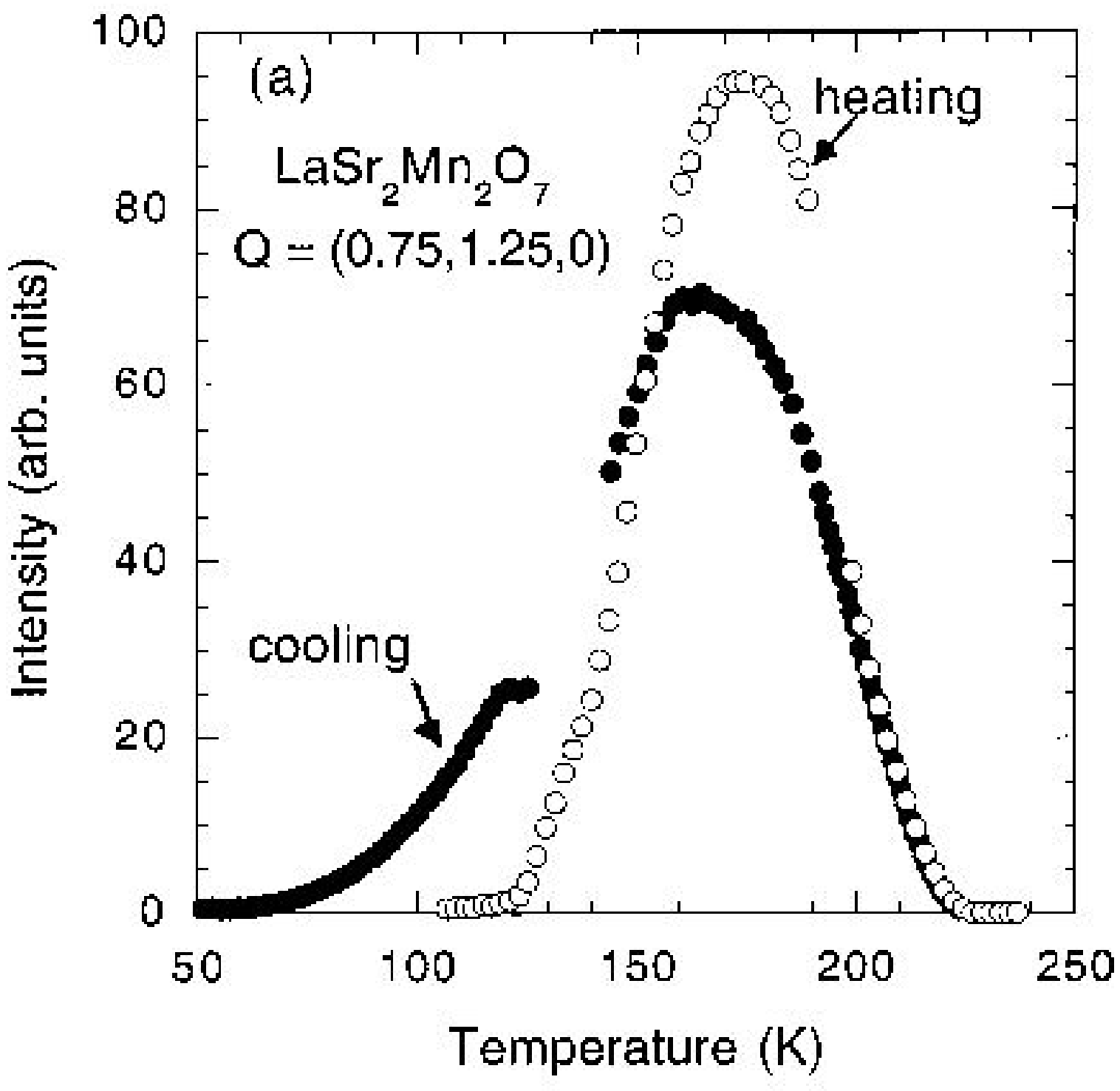}}
\end{center}
\vspace{0.5cm}
\caption{Temperature variation of superlattice reflexion intensities in
  \LSM~corresponding to wave vector \v Q = $(\frac{1}{4},\frac{1}{4},0)$
  (cf. inset of Fig.~\ref{fig:Bireentpol}). Right: CO and first reentrance of
  homogeneous state. Left: second possible reentrance of CO below 50 K. (After
  \protect\cite{Chatterji00}.)} 
\label{fig:Bireent}
\end{figure}

The predicted phase diagram of charge order qualitatively agrees with
experimental observations, keeping in mind that due to electron-hole
symmetry the calculated phase diagram can be used for the hole doped
case of \LSXM~(x = 1-n). One must note however that there is no
unanimous agreement on doping range and temperature behavior of
CO in this compound. Both depend considerably on the experimental method used
to detect CO ,e.g., X-ray, electron diffraction or transport measurements. In
the latter case one finds CO in a broad range of doping ($0.44\leq x\leq
0.8$) at T$_{CO}\sim$ 200 K \cite{Dho01}. Another observation is the
collapse of CO or reentrance of the homogeneous phase at lower
temperatures. Surprisingly a second reentrance of CO appears below 50 K as
concluded from an upturn in the resistivity. From diffraction experiments the
doping range of CO is somewhat smaller ($0.5\leq x\leq 0.65$) \cite{Qiu03}. For
the stoichiometric case x = 0.5 (Mn$^{3+}$/Mn$^{4+}$ = 1) the CO has first
been observed in \cite{Kimura98} by X-ray diffraction. If one had only
CO one would expect a simple two-sublattice structure with
Mn$^{3+}$/Mn$^{4+}$ ordering corresponding to a commensurate
superstructure with \v Q = $(\frac{1}{2},\frac{1}{2},0)$. However one
rather observes \v Q = $(\frac{1}{4},\frac{1}{4},0)$. This is due to
additional orbital ordering of Mn$^{3+}$ e$_g$ orbitals in a staggered
fashion (d$_{3x^2-r^2}$/d$_{3y^2-r^2}$) on top of CO
(see Fig.~\ref{fig:Biorbor}). The \v Q = $(\frac{1}{4},\frac{1}{4},0)$
superlattice reflexions are then due to the associated JT distortion of
the crystal structure. The additional orbital order is not contained in the
above single-orbital model. Below 100 K the superstructure reflections
vanish, indicating melting of both CO and orbital order as 
conjectured from transport \cite{Dho01}. Later X-ray experiments on \LSM~(x =
0.5) have shown that there is also evidence for a CO reentrance below 50 K
\cite{Chatterji00} in agreement with Ref. \cite{Dho01}. This evidence for the
second (CO) reentrance is seen in Fig.~\ref{fig:Bireent}. The intensity of
reflexions in the reentrant CO region below 50 K is however much smaller,
indicating much weaker CO. In fact, it seems sample dependent since it was not
observed in other experiments.
%6.22
%%%%%%%%%%%%%%%%%%%%%%%%%%%%%%%%%%%%%%%%%%%%%%%%%%%%%%%%%%%%%%%%%%%%%%%%%%%%
%%%%%%%%%%%%%%%%%%%%%%%%%%%%%%%%%%%%%%%%%%%%%%%%%%%%%%%%%%%%%%%%%%%%%%%%%%%%
%
The possible existence of the second CO reentrance cannot be explained by the
purely electronic EHM model. In fact around x = 1-n$^*$ there is not
even the first reentrance found to the homogeneous phase. Instead the charge
order parameter increases monotonously with decreasing temperature as
may be infered from the V-T phase diagram (n = 0.65) in Fig.~\ref{fig:BinPD}.
  
We note that Mn$^{3+}$/Mn$^{4+}$ charge order has also been observed in
the infinite-layer manganites, typically around the half-doped
(x = 0.5) compounds, e.g. in La$_{0.5}$Ca$_{0.5}$MnO$_3$ and
R$_{0.5}$Sr$_{0.5}$MnO$_3$ (R = Pr,Nd). For the Pr-compound CO was
observed in a large doping range ($0.3\leq x\leq 0.7$) similar as
in the bilayer \LSXM. The reentrance into the homogeneous
phase was also observed in Pr$_{0.5}$Sr$_{0.5}$MnO$_3$, again only
away from half doping (x = 0.5). It becomes especially pronounced in the
presence of magnetic fields up to 10 T \cite{Tomioka96}. There is no evidence
for a second CO reentrance in the infinite layer compounds.
%6.23
%%%%%%%%%%%%%%%%%%%%%%%%%%%%%%%%%%%%%%%%%%%%%%%%%%%%%%%%%%%%%%%%%%%%%%%%%%%%
%%%%%%%%%%%%%%%%%%%%%%%%%%%%%%%%%%%%%%%%%%%%%%%%%%%%%%%%%%%%%%%%%%%%%%%%%%%%
%

To explain these phenomena in the stoichiometric bilayer \LSM~a more
extended but still single-orbital based theory is apparently needed. 
It was proposed in \cite{Yuan99} that in this case CO is profoundly
affected by polaron formation caused by a strong coupling to the
lattice. This is suggestive since polaron formation changes the ratio
of kinetic vs. intersite Coulomb energy of the holes and thus affects the
conditions of CO. The electron-lattice coupling is due to the modulation of
pd-hybridization along the in-plane bond directions caused by vibrations of
oxygen atoms that form corner-sharing octahedra around the Mn ions. Effectively
this leads to a coupling of the Mn$^{3+}$-e$_g$ level shift to a
bond-stretching vibration of the oxygen. The latter  may be assumed
dispersionless with a frequency $\omega_0$. We note that this refers to
conventional Holstein-type polarons. They differ from the Jahn-Teller (JT)
type polarons where the e$_g$ level is split by coupling to the JT symmetry
distortions of the whole oxygen octahedron. The latter have been proposed for
the slightly doped infinite layer La$_{1-x}$Sr$_{x}$MnO$_3$ in the region
$0.1\leq x\leq 0.2$ \cite{Yamada96}. The present Holstein-type model for the
x $\simeq$ 0.5 bilayer \LSM~is described by 
%6.42
\begin{eqnarray}
H_{e-ph} & = & g\sum_{i,\delta} \left( b_{i,\delta} + b_{i,\delta}^{\dagger}
\right) \left( n_{i + \delta} - n_i \right) + \omega_0 \sum_{i,\delta}
b_{i,\delta}^{\dagger} b_{i,\delta}~~~.
\end{eqnarray}
Here b$^\dagger_{i,\delta}$ is the local vibration at oxygen site
$\delta$ associated with Mn site i and g is the coupling
constant. This has to be added to the EHM part, i.e., to the first
three terms of Eq.~(\ref{MANHAM}). The latter is also generalized by
including both n.n. interactions (V$_1$) and n.n.n. interactions
(V$_2$) to achieve a realistic T$_{CO}$ which is controlled by the
ratio (V$_1$-V$_2$)/t. In the limit of strong coupling ($\alpha\equiv
g^2/\omega_0^2\sim 1$) the phonon coordinates may be eliminated by a
combined Lang-Firsov (LF) transformation $U_1 = \exp
[-\sum_{i,\delta}g/\omega_0 (b_{i,\delta} -
  b_{i,\delta}^{\dagger})(n_{i+\delta}-n_i)]$ \cite{Lang63}, and squeezing
transformation $U_2 = \exp [\gamma \sum_{i,\delta}(b_{i,\delta}b_{i,\delta}-
b_{i,\delta}^{\dagger}b_{i,\delta}^{\dagger})]$. Here $\gamma > 0$
is a variational parameter \cite{Zheng88} determined by minimization
of the ground-state energy. This leads again to an effective
electronic model, but with renormalized hopping and interaction
parameters. Furthermore, the on-site correlation problem is simplified
by using the limit U $\rightarrow\infty$ where the layers are fully
spin polarized and we may assume spinless fermions. For x = 0.5 this
implies the spinless half-filled band is realized in \LSM. Then,
after applying U$_2$U$_1$ the effective Hamiltonian reads, up to a
constant,
%6.43  
\begin{eqnarray}
H_{eff} =  
%\omega (\tau +1/\tau) N/2 
-\tilde{t}\sum_{i,\delta} \left( c_{i}^{\dagger} c_{i + \delta} + h.c. \right)
+ \left( V_1 + 2\alpha \omega_0 \right) \sum_{i,\delta} n_in_{i + \delta} +
V_2\sum_{i,\eta}n_i n_{i + \eta}~~~.  
\end{eqnarray}
The first term describes hopping of small polarons. The essential
point is that the effective hopping element $\tilde{t} = t\exp
[-5\alpha\tau \coth (\omega_0/2T)]$ with $\tau = \exp(-4\gamma)$ is
strongly temperature dependent. It is small compared to the bare
hopping t when T $\gg\omega_0$ and increases for T $\ll\omega_0$ because
the occupation of the n-phonon modes decreases with T eventually until
only the effect of zero-point fluctuations in $\tilde{t}$ is
left. This T-dependent renormalization of $\tilde{t}$ is very
important for charge ordering to take place, since the latter is
determined by the balance of kinetic and inter-site Coulomb
energies. Treating the intersite Coulomb terms for two sublattices A,B
in Hartree Fock approximation as before and choosing $\langle
n_i\rangle = 1/2\pm n_{CO}, i\in {\rm A,B}$, the condition for the CO
instability line is
%6.24
%%%%%%%%%%%%%%%%%%%%%%%%%%%%%%%%%%%%%%%%%%%%%%%%%%%%%%%%%%%%%%%%%%%%%%%%%%%%
\begin{figure}[tb]
\begin{center}
\includegraphics[clip,width=7cm]{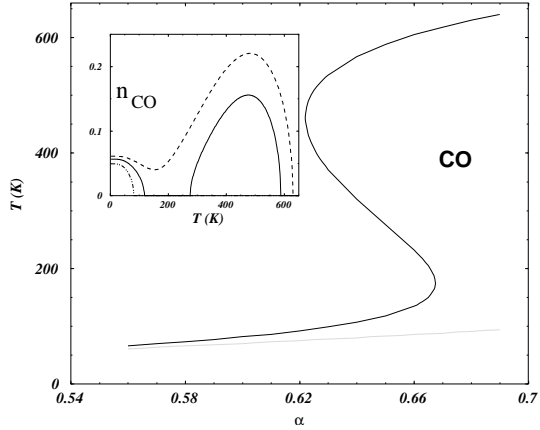}
\end{center}
\vspace{0.5cm}
\caption{$\alpha$-T phase diagram for the spinless (half filled) case of the
  Holstein-EHM model with parameters V$_{12}$ = 0.1, $\omega_0$ = 0.05 in units
  of t = 4$\cdot$10$^3$ K. For comparison, the thin line corresponds to a
  $\tilde{t}$ which is kept constant (cf. Fig.~\ref{fig:Bireent}). The inset
  shows n$_{CO}$(T). The dotted, solid and dashed lines correspond to the three
  possible types of diagrams with $\alpha$ = 0.60, 0.65 and 0.68
  respectively. (After \protect\cite{Yuan99}.)} 
\label{fig:Bireentpol}
\end{figure}
%%%%%%%%%%%%%%%%%%%%%%%%%%%%%%%%%%%%%%%%%%%%%%%%%%%%%%%%%%%%%%%%%%%%%%%%%%%%
%
%6.44
\begin{eqnarray}
1 & = & \frac{2V}{\pi^2} \int_{0}^{1} \frac{\tanh \left[ 2\sqrt{(\tilde{t}z)^2
	  + (Vn_{CO})^2}/T \right]}{\sqrt{(\tilde{t}z)^2 + (Vn_{CO})^2}}K \left(
	  \sqrt{1-z^2} \right)\ {\rm d}z   
\end{eqnarray}
where $K$ is the complete elliptic function of the first kind and 
$z = \sqrt{\varepsilon^2-(4Vn_{CO})^2}/4\tilde{t}$ with $V = V_{12} + 2\alpha
\omega$, V$_{12}$ = V$_1$ - V$_2$. For T$\rightarrow$0 this equation has always
a solution with n$_{CO}>$ 0. For a reasonable parameter set
(t,$\omega_0$,V$_{12}$) its solution leads to the $\alpha$-T phase
diagram shown in Fig.~\ref{fig:Bireentpol}. Depending on the size of $\alpha$,
both the non-reentrant and double-reentrant scenarios are possible. There is
an intermediate region for $\alpha\sim$ 0.65 where on lowering the
temperature one obtains a CO phase, then a first reentrance into the
homogeneous phase and subsequently a second reentrance into the CO phase. The
corresponding variation of the order parameter n$_{CO}$(T) is shown
in the inset of Fig.~\ref{fig:Bireentpol} (full line). The other possible
cases are also illustrated. Physically the behavior around
$\alpha\sim$ 0.65 may be explained as follows. First the effective
hopping $\tilde{t}$ is reduced from the bare value and polaronic CO
appears at relatively high temperatures. With decreasing  temperature
the kinetic energy ($\sim\tilde{t}$) increases until CO can no
longer be maintained and the homogeneous state is stable again. For
even lower temperatures CO always reappears because the
n.n. tight binding Fermi surface has a nesting instability to CO for the half
filled (spinless) case for arbitrary small values of V. The reentrance
behavior discussed here is entirely due to the temperature dependence of the
effective polaron hopping $\tilde{t} = t\exp [-5\alpha\tau \coth (\omega
/2T)]$. Without it only the lowest CO transition does occur. This agrees
with the previous study where no reentrance was observed for quarter
filling including spin or spinless half filling. 
In the discussion of the reentrance behavior in \LSXM~we have so far
invoked electronic correlation effects and polaron formation as
origin but neglected the influence of coexisting magnetic order. A
treatment of the full problem of CO would require the inclusion of
spin degrees of freedom for finite values of U together with the last
two exchange terms in Eq.~(\ref{MANHAM}) and the electron-lattice
coupling. This is still an open problem. 
%6.25
%%%%%%%%%%%%%%%%%%%%%%%%%%%%%%%%%%%%%%%%%%%%%%%%%%%%%%%%%%%%%%%%%%%%%%%%%%%%
\begin{figure}[tb]
\begin{center}
\includegraphics[clip,width=2.cm]{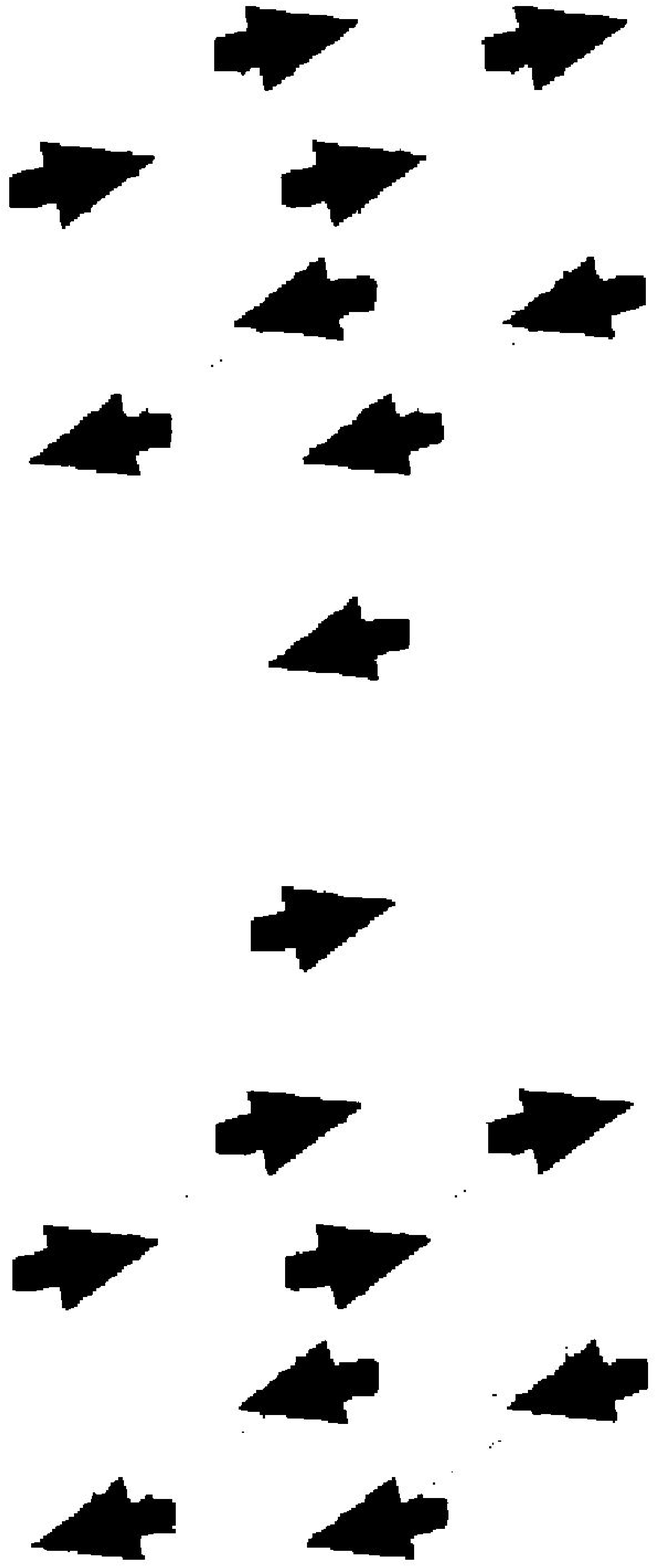}
\raisebox{1cm}{\includegraphics[clip,width=6cm]{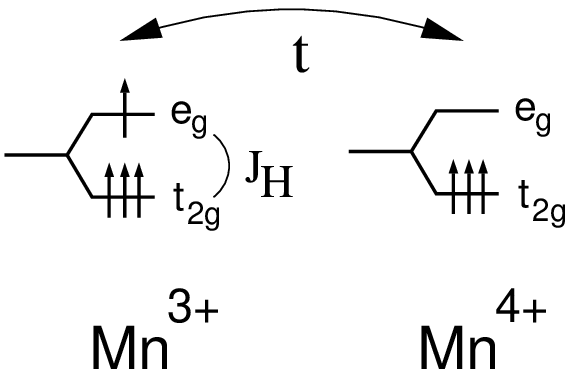}}
\raisebox{0.2cm}
{\includegraphics[clip,width=2cm]{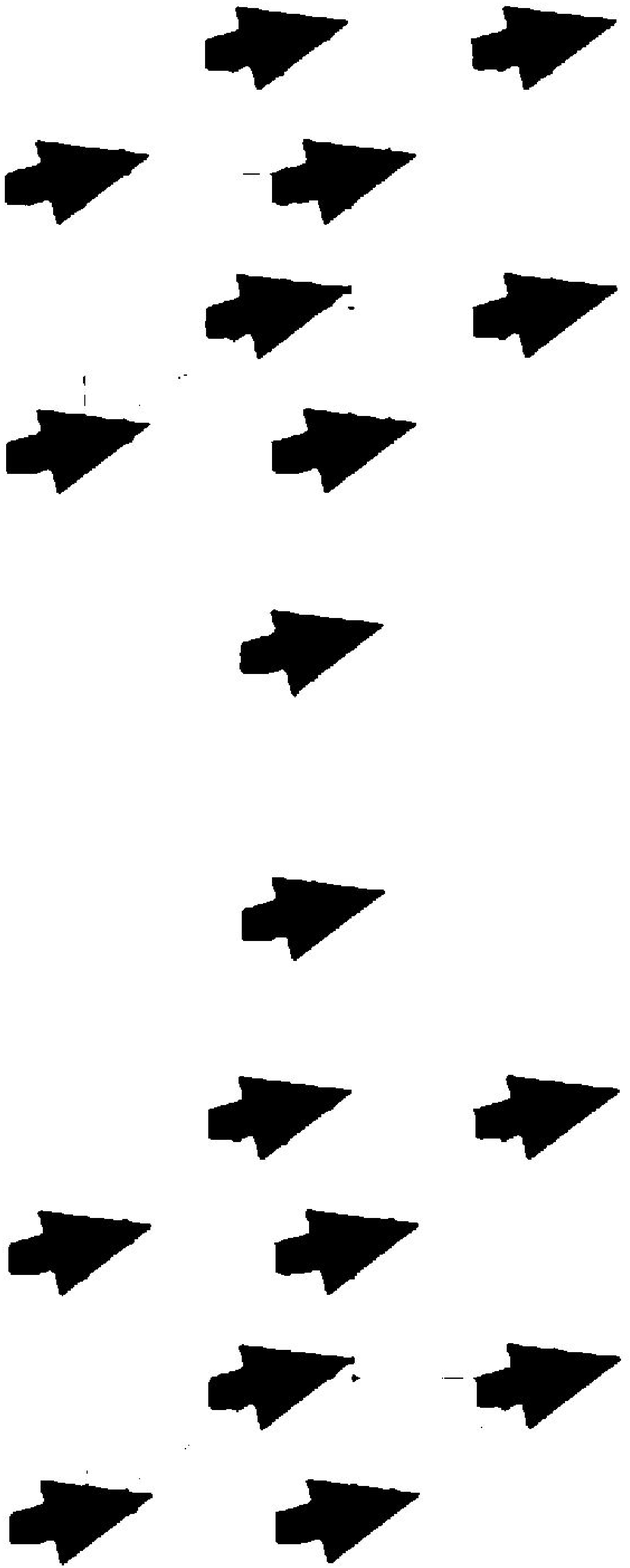}}
\end{center}
\vspace{0.5cm}
\caption{Left: AFM structure of stoichiometric (x = 0.5) CO insulator
  \LSM~consisting of AF stacked FM layers along c. Right: FM structure of
  metallic (x = 0.4) \LSXM. Moments are parallel to [110] in both
  cases. Center: Illustration of FM double exchange mechanism in doped
  manganites. FM polarization of itinerant e$_g$ spins (s = 1/2) leads to
  kinetic energy gain ($\sim$ t) due to FM Hund's rule coupling ($\sim$ J$_H$)
  to localized t$_{2g}$ spins (S = 3/2).}
\label{fig:Bimagstruc}
\end{figure}
%%%%%%%%%%%%%%%%%%%%%%%%%%%%%%%%%%%%%%%%%%%%%%%%%%%%%%%%%%%%%%%%%%%%%%%%%%%%
%

In the last part of this chapter we therefore focus on a
rather complementary case: When the doping is large enough Coulomb
correlation effects are less prominent and CO is absent for all
temperatures.  Then only the exchange terms in Eq.~(\ref{MANHAM}) are
important. Double exchange (DE) illustrated in
Fig.~\ref{fig:Bimagstruc} favors FM order within and between layers
by optimizing kinetic energy gain for parallel e$_g$ (s = 1/2)
conduction and t$_{2g}$ (S = 3/2) localized spins. In contrast superexchange
(last term in Eq.~(\ref{MANHAM})) favors AF orientation of moments
between layers leading to the AF order in Fig.~\ref{fig:Bimagstruc} for x =
0.5. For $0.4<x<0.5$ its competition with FM double exchange along the c-axis
leads to a canting of moments out of the plane \cite{Hirota98} (CAF
region in Fig.~\ref{fig:Bistruc2}) which may be interpreted as a superposition
of an AF and FM structure. For $x\leq 0.4$ the e$_g$ conduction bands become
increasingly two-dimensional with d$_{3z^2-r^2}$ character. Therefore double
exchange along c increases significantly as discussed below. Consequently only
the FM structure with FM bilayers remains and the last term in
Eq.~(\ref{MANHAM}) may be neglected. The less than half-doped ($x\leq 0.4$)
metallic \LSXM~ compounds are then described by the ferromagnetic (J$_H>$ 0)
Kondo-lattice Hamiltonian 
%6.45
\begin{eqnarray}
\label{DEX}
H_K & = & - t \sum_{\langle ij\rangle \lambda \alpha}
c_{i\lambda\alpha}^{\dagger}c_{j\lambda\alpha} - t_{\perp} \sum_{i\alpha}
\left\{ c_{i 1\alpha}^{\dagger}c_{i 2\alpha} + h.c. \right\}
- J_{H} \sum_{i \lambda \alpha\beta} \v S_{i\lambda}\cdot
c^\dagger_{i\lambda\alpha} \v s_{\alpha\beta} c_{i\lambda\beta}~~~.
\end{eqnarray}
Here t, t$_\perp > 0$ are the nearest neighbor hopping parameters within a
layer and between the two partners ($\lambda$ =1,2) of a bilayer along c
respectively. Furthermore \v S and \v s are localized and conduction electron
spins respectively with $\alpha,\beta = \uparrow,\downarrow$. In \LSXM~the
lattice constants (x = 0.4) are a = 3.87 \AA~and c = 20.14 \AA. The
intra-bilayer splitting d$\simeq$a is much smaller than the distance D = 6.2
\AA~between adjacent bilayers (Fig.~\ref{fig:Bistruc2}). Therefore
inter-bilayer hopping or exchange has been neglected above. Diagonalization of
the hopping term leads to bonding and antibonding tight binding bands split by
t$_\perp$ with identical in-plane dispersion and a DOS given by
%6.46
\begin{eqnarray}
\label{EGBAND}
\epsilon_{\sigma\pm}(\v k) & = & - \left[ \pm t_\perp + t \left( \cos k_x +
  \cos k_y \right) \right]\nonumber\\ 
N_\pm(\epsilon) & = & N \left( \epsilon\pm t_\perp \right); \quad\ N(\epsilon)
  = \frac{2}{\pi^2} \frac{1}{W} K \left( \left[ 1- \left( \frac{\epsilon}{W}
  \right)^2 \right]^\frac{1}{2} \right)~~~. 
\end{eqnarray}
The justification for this model and the size of its parameters can be obtained
by considering the spin wave excitations below the Curie temperature
T$_C$. They have been investigated by inelastic neutron scattering by various
groups \cite{Chatterji99a,Chatterji99b,Perring01,Hirota98} and analyzed in
\cite{Chatterji01,Shannon02} both in the classical limit and with quantum
corrections. In the manganites the condition J$_H\gg t,t_\perp$ is
fulfilled because the Hund's rule coupling J$_H\sim$ 2 eV is quite
large. This greatly simplifies spin wave calculations because firstly the
e$_g$ bands $\epsilon_{\sigma\pm}(\v k)$ will be spin-split such that only
majority bands are occupied, i.e. n$_{\uparrow\pm}$ = 1 and
n$_{\downarrow\pm}$ = 0 and secondly the large S = 3/2 local t$_{2g}$
spins allow for a 1/S expansion. This approach has been first applied
to the cubic manganites \cite{Furukawa96} and later used for the
bilayer manganites \cite{Chatterji01,Shannon02}. It turns out that in
the limit $J_H\rightarrow\infty$, and to order 1/S one obtains
classical spin waves of an effective Heisenberg model for t$_{2g}$
spins. The effective exchange constants are determined by the
e$_g$ conduction band dispersion and filling according to
%6.47
\begin{eqnarray}
\label{BIDISP}
\omega_A(\v q) & = & zJ^{DE}S \left[ 1 - \gamma_q \right] \nonumber\\
\omega_O(\v q) & = & zJ^{DE}S \left[ 1 - \gamma_q \right] + 2J^{DE}_{\perp}S
~~~. 
\end{eqnarray}
In two dimensions $\gamma_q = \frac{1}{2}(\cos q_x+\cos q_y)$ and
$\omega_A(\v q)$ and $\omega_O(\v q)$ are the dispersion of acoustic (A)
and optical (O) spin wave branches respectively. They have equal
dispersion in the ab-plane and are split by the A-O gap $\Delta_{AO} =
2SJ^{DE}_{\perp}$. In this approximation there is no spin-space
anisotropy and therefore the A branch has no gap at \v k = 0. The
effective Heisenberg exchange constants J$^{DE}$, J$^{DE}_\perp$ for
t$_{2g}$ spins are given by
%6.26
%%%%%%%%%%%%%%%%%%%%%%%%%%%%%%%%%%%%%%%%%%%%%%%%%%%%%%%%%%%%%%%%%%%%%%%%%%%%
\begin{figure}[tb]
\begin{center}
\includegraphics[clip,width=6.5cm]{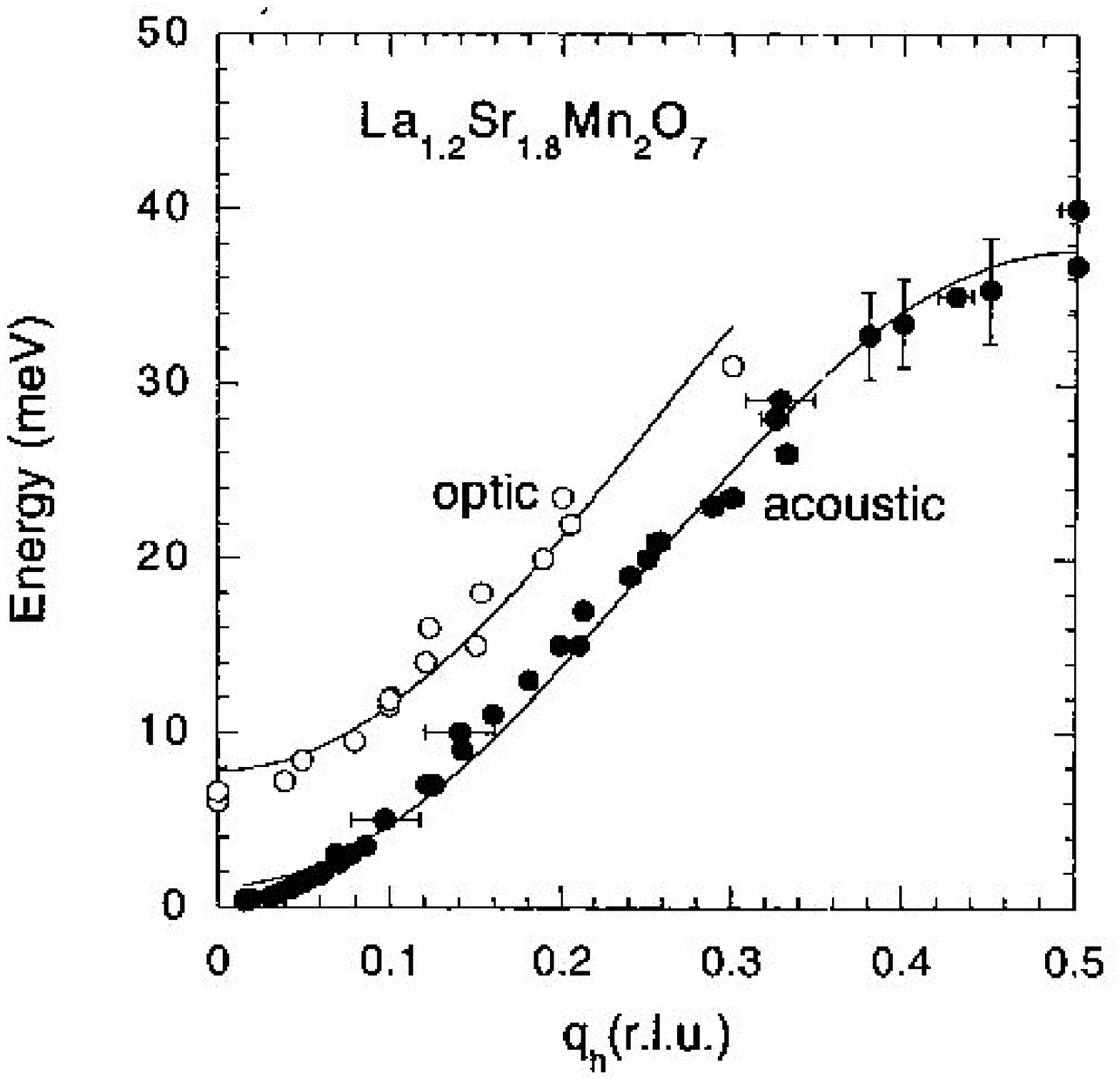}
\hfill
\raisebox{0.3cm}
{\includegraphics[clip,width=6.5cm]{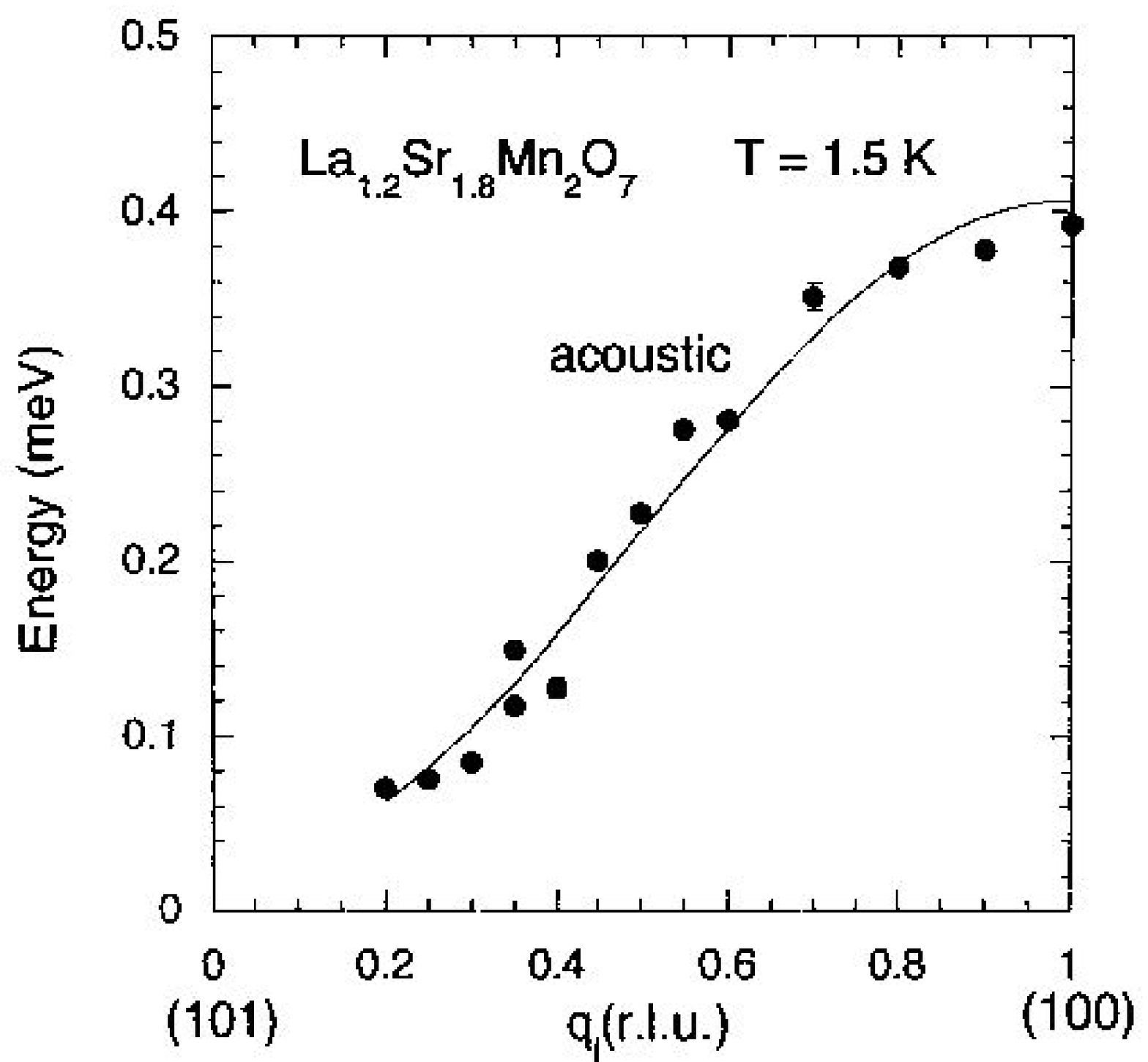}}
\end{center}
\vspace{0.5cm}
\caption{Left panel: Acoustic and optic spin wave branches along
[100] direction with a splitting $\Delta_{AO}$ = 6 meV and overall
dispersion of 40 meV. Full lines are obtained from the classical
dispersion expression in Eq.~(\ref{BIDISP}). Right panel: Acoustic spin wave
dispersion along [001]. An extrapolated anisotropy gap $\Delta_A(0)$ = 0.04 meV
is obtained. (After \protect\cite{Chatterji01}.)}
\label{fig:Bidispa}
\end{figure}
%%%%%%%%%%%%%%%%%%%%%%%%%%%%%%%%%%%%%%%%%%%%%%%%%%%%%%%%%%%%%%%%%%%%%%%%%%%%
%
%6.48
\begin{eqnarray}
\label{DEXPAR}
J^{DE} & = & -\frac{1}{2S^2}\frac{1}{2z} \left( \epsilon_0 + \epsilon_\pi
\right) \quad \mbox{with} \quad \epsilon_{0,\pi} = \int_{-W}^{\epsilon_F\pm
  t_\perp} N(\epsilon)\epsilon d\epsilon\nonumber\\
J^{DE}_{\perp} & = &\frac{1}{2S^2} \frac{t_\perp}{2} \left( n_0 - n_\pi \right)
\quad \mbox{with} \quad n_{0,\pi} = \int_{-W}^{\epsilon_F\pm t_\perp} N
(\epsilon) d\epsilon~~~. 
\end{eqnarray}
The effective t$_{2g}$ exchange parameters J$^{DE}$, J$_\perp^{DE}$
are therefore completely determined by the e$_g$ band parameters in
the limit $J_H\rightarrow\infty$.  This surprisingly simple result for
the spin waves in the double exchange ferromagnet is in good agreement
with inelastic neutron scattering experiments on the x = 0.4
\LSXM~. This is seen in the left panel of Fig.~\ref{fig:Bidispa}. Two
parallel A and O modes are observed with a maximum zone-boundary
energy of $\omega_A(q_x = \frac{\pi}{a})\sim$ 40 meV and an A-O
splitting of $\Delta_{AO}$= 6 meV. From these values one obtains
SJ$^{DE}$ = 10 meV and SJ$_\perp^{DE}$ = 3 meV. Using
Eqs.~(\ref{DEXPAR}) this leads to e$_g$ band parameters t = 0.175
meV and t$_\perp$ = 0.1 eV.

In the present theory there is no double exchange between adjacent
bilayers and hence no spinwave dispersion along the c-axis. But
experimentally a small dispersion along c was found
(Fig.~\ref{fig:Bidispa}, right panel) although it is two orders of
magnitude smaller than the one along the a-axis. This requires an
inter-bilayer DE constant J'$^{DE}_\perp$ with
J'$^{DE}_\perp$/J$^{DE}_\perp\simeq$ 1.5$\cdot$10$^{-2}$ \cite{Chatterji99a}
and shows that \LSXM~is an almost ideal two dimensional double exchange
ferromagnet. This conclusion was supported by an analysis of diffuse neutron
scattering which exhibits long range FM in-plane correlations far above the
Curie temperature ($\sim$ 2.3 T$_C$) \cite{Chatterji02}. Finally,
Fig.~\ref{fig:Bidispa} shows that a small extrapolated anisotropy gap
$\Delta_{A}(0)\sim$ 0.04 meV for the acoustic mode exists at the zone center
whose microscopic origin is not clear. 

The double exchange model based on the FM Kondo lattice Hamiltonian
is able to describe ferromagnetism and basic properties of spin wave
excitations quite well. However it has its limits. Firstly quantum
corrections of order 1/S$^2$ cannot be completely neglected as discussed in
Refs. \cite{Shannon02,Chatterji04} and references cited therein. They lead to
two effects: i) reduction of the overall dispersion of spin waves. In the
effective Heisenberg model this would necessitate a rescaling of the parameters
J$^{DE}$, J$^{DE}_\perp$ or possibly an inclusion of more parameters. ii) the
local spin moment is coupled to density fluctuations in the itinerant system
which leads to damping effects not present in the classical (1/S)
approximation. Comparison with experiment shows that (1/S$^2$) corrections
provide still insufficient damping and also cannot explain the observed
deviations from the classical \v q-dependence of spin waves
\cite{Shannon02}. It has also been proposed that a crossing with a phonon
branch might be involved in these anomalies. 
%6.27
%%%%%%%%%%%%%%%%%%%%%%%%%%%%%%%%%%%%%%%%%%%%%%%%%%%%%%%%%%%%%%%%%%%%%%%%%%%%
\begin{figure}[tb]
\begin{center}
\includegraphics[clip,width=8cm]{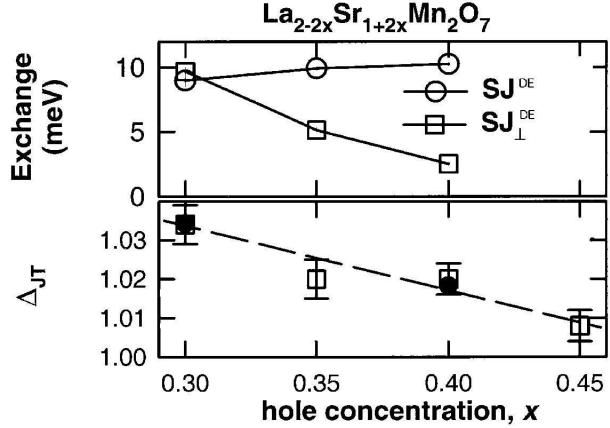}
\end{center}
\vspace{0.3cm}
\caption{Upper panel: Doping dependence of effective exchange parameters. Note
  that $J^{DE}_\perp$ is proportional to A-O spin wave splitting and $J^{DE}$
  to the overall dispersion. Lower panel: Doping dependence of JT distortion
  $\Delta_{JT}$ (elongation along c) of MnO$_6$ octahedra. (After
  \protect\cite{Perring01}.)} 
\label{fig:BiAOgap}
\end{figure}
%%%%%%%%%%%%%%%%%%%%%%%%%%%%%%%%%%%%%%%%%%%%%%%%%%%%%%%%%%%%%%%%%%%%%%%%%%%%
%
Another shortcoming of the model is the neglect of orbital degeneracy
of e$_g$ states. This has indeed dramatic effects on the doping
dependence of the A-O spin wave splitting and the resulting effective exchange
constants in the FM regime x = 0.3-0.4 as shown in Ref. \cite{Perring01}. When
x decreases from the present value x = 0.4 the effective J$^{DE}_\perp$
strongly increases while J$^{DE}$ stays almost constant. In the above double
exchange model the anisotropy ratio is given by 
%6.49
\begin{eqnarray}
%\label{ANIS}
\frac{J^{DE}_{\perp}}{J^{DE}} = -\left( \frac{t_{\perp}}{t} \right)
\frac{W \left( n_0-n_\pi \right)}{\epsilon_0 + \epsilon_\pi}~~~. 
\label{ANIS}
\end{eqnarray}
Assuming a doping independent t$_\perp$/t the above ratio changes at
most by $\sim$ 10 \% in the doping range x = 0.3 - 0.4, which is much too
small to explain experimental observations. Therefore another mechanism must be
invoked. It has been found \cite{Kubota00} that for decreasing x the MnO$_6$
octahedra elongate significantly along the c axis  due to the JT effect on
e$_g$ orbitals which leads to a lower energy and hence larger occupation for
d$_{3z^2-r^2}$ orbitals as compared to d$_{x^2-y^2}$ orbitals. Since the former
have larger overlap along the c axis, the effective t$_\perp$ strongly
increases with decreasing x. Thus it is really the prefactor (t$_\perp/t$) in
the above equation which leads to the dramatic increase of J$^{DE}_\perp$ and
the A-O spin wave splitting with decreasing x. This effect was described
phenomenologically in Ref. \cite{Shannon02} and within a microscopic model
in Ref. \cite{Jackeli02}.

In our discussion of bilayer manganites we have focused on models for
charge ordering which can be described as simple periodic
superstructures caused by inter-site but short range Coulomb
interactions supplemented by small polaron formation. We have also
discussed the importance of the double exchange mechanism for
explaining the spin excitations in the metallic ferromagnet away from
half-doping. We have mostly neglected the complications of orbital
order, JT distortions and the effect of longer range Coulomb
interactions in the low doping regime. This may lead to the important
possibility of an inhomogeneous state due to phase separation of
ferromagnetic metallic and charge ordered insulating regions. Such
states may consist of metallic droplets or stripes of holes in an
insulating environment. This possibility has been reviewed in Ref.
\cite{Moreo99}. These aspects may also be of great importance for
explaining the giant magnetoresistance of the manganites
\cite{Imada98}.

\section{Geometrically Frustrated Lattices}
\resetdoublenumb 
\resetdoublenumbf

\label{Sect:GeometricFrustration}

%%%%%%%%%%%%%%%%%%%%%%%%%%%%%%%%%%%%%%%%%%%%%%%%%%%%%%%%%%%%%%%%%%%%%%%%%%%
\begin{figure}[t b]
\includegraphics[clip,width=5.5cm]{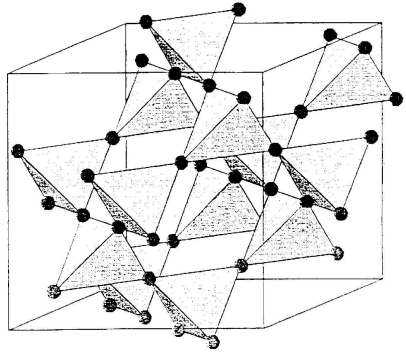}
\vspace{-0.5cm}
\caption{Pyrochlore lattice consisting of corner-sharing tetrahedra.}
\label{fig07.1}
\end{figure}
%%%%%%%%%%%%%%%%%%%%%%%%%%%%%%%%%%%%%%%%%%%%%%%%%%%%%%%%%%%%%%%%%%%%%%%%%%%

Usually the concept of frustration is used in connection with magnetic
systems. When Ising spins with an antiferromagnetic interaction are placed onto
certain lattices like a triangular one, the pair-wise interactions cannot be
satisfied simultaneously and therefore are frustrated. Here we will associate
the concept of frustration with lattice structures. We call a
lattice geometrically frustrated when in case that its sites are occupied by
antiferromagnetically coupled Ising spins the interactions are
frustrated. Examples are in two dimensions the just mentioned triangular, the
checkerboard or the kagomé lattice. In three dimensions the pyrochlore
lattice (see Fig. \ref{fig07.1}) is the one most frequently investigated. 
For those lattices we
want to study charge degrees of freedom, i.e., when the electron number at a
lattice site is fluctuating. A frustrated lattice structure can have a
degenerate ground state for special band fillings when electron correlations
are strong. In fact, in the limit of large on-site and nearest-neighbor 
electron repulsions there exists an exponentially large number of
configurations with minimal potential energy. This is particularly so when the
number of electrons equals half the number of lattice sites. It is not
surprising that this special feature which is closely related to a frustrated
geometry leads to new theoretical models and special effects when
the electrons are strongly correlated. In the center of our attention will be
the above mentioned pyrochlore lattice and a two-dimensional projection of it,
the checkerboard lattice. 

A pyrochlore lattice is a substructure of the spinels which have the
composition AB$_2$O$_4$. They can be considered as face-centered cubes of
O$^{2-}$ ions. The B ions are surrounded by an octahedron of O$^{2-}$ ions,
i.e., BO$_6$ and are positioned on corner-sharing tetrahedra which define the
pyrochlore lattice. Here we want to consider metallic spinels in which the
electrons are itinerant. They may undergo metal to insulator transitions which
are usually accompanied by a structural distortion. The most studied example
has been magnetite Fe$_3$O$_4$. A transition to an insulator already indicates
that electron correlations may be strong in spinels but an unambiguous proof is
the observed heavy-fermion behavior of LiV$_2$O$_4$ at low temperatures
\cite{Kondo99}. Another interesting case is AlV$_2$O$_4$. This material is
either a poor metal or a semiconductor at low temperatures. It undergoes a
structural phase transition at lower temperatures which apparently is caused by
strong electron correlations. Finally, LiTi$_2$O$_4$ is a metallic spinel
which becomes superconducting at a relatively high transition temperature of
T$_c$=13.7 K \cite{Johnston76,McCallum76}. Table \ref{tab:VII.I} summarizes
these materials with half-integer valency of the cations.

\begin{table}
\vspace{0.3cm}
{\centering \begin{tabular}{|lr|c|c|c|c|}
\hline 
& M =\ \  & Ti & V & V(Cr) & Mn\\
\hline 
\hline 
Li(Al)M$_2$O$_4$ && LiTi$_2$O$_4$ & LiV$_2$O$_4$ & AlV$_2$O$_4$ &
LiMn$_2$O$_4$\\ 
&&&& (LiCr$_2$O$_4$) &\\
\hline 
average d-electron count && d$^{0.5}$ & d$^{1.5}$ & d$^{2.5}$ & d$^{3.5}$\\
per M-atom &&&&&\\
\hline 
\hline 
\end{tabular}\par}\vspace{0.3cm}
\caption{Spinels with a half-integer valency of $d$ ions
\label{tab:VII.I}}
\end{table}

In the following we start out with a reminder on Fe$_3$O$_4$ for which a huge
literature does exist. For references see, e.g., Ref. \cite{Isoda00}. The
purpose is to merely recall some of the 
basic facts in order to understand better the special features of strong
electron correlations in LiV$_2$O$_4$, AlV$_2$O$_4$ and other spinels. This is
followed by a discussion of fractional charges. They are found when a model
Hamiltonian describing strongly correlated electrons with strong on-site and
nearest-neighbor repulsions is used and applied to a frustrated lattice.

Magnetite Fe$_3$O$_4$ has been much investigated because of the important role
it has played in the development of magnetism and magnetic materials. It
is a spinel of the form AB$_2$O$_4$ with A = Fe$^{3+}$ and B = Fe$^{2.5+}$
sites. We assume that 50 \% each of the B sites are in a Fe$^{2+}$ and
Fe$^{3+}$ configuration each, i.e., that electron correlations are so strong
that Fe$^+$ or Fe$^{4+}$ configurations are suppressed. Magnetite undergoes at
120 K a phase transition from a metallic high-temperature phase to an
insulating low-temperature phase. This transition was first observed by Verwey
and Haayman \cite{Verwey41} and is usually referred to as Verwey
transition. Verwey presented also a model for its description in terms of a
order-disorder transition, which is entropy driven. The implicit assumption
regarding B sites is thereby that the repulsions of electrons on neighboring
sites are so strong that the kinetic energy term of the electrons plays only a
minor role and may be neglected when the phase transition is considered. With
an average valence of the B sites of +2.5 this implies that two neighboring
Fe$^{2+}$ sites and also two Fe$^{3+}$ sites repel each other, while
Fe$^{2+}$-Fe$^{3+}$ sites attract. Let us denote by V$_{\alpha\beta}$ the
interaction between two neighbors Fe$^{\alpha +}$-Fe$^{\beta +}$. An
order-disorder phase transition will take place when

%7.1
\begin{equation}
\delta V = V_{33} + V_{22} - 2V_{23} > 0~~.
\label{DeltaV}
\end{equation} 

Verwey suggested a particular charge ordering for the insulating
low-temperature phase in which the Fe$^{2+}$ and Fe$^{3+}$ sites of the
pyrochlore lattice order in form of two families of chains pointing in the
[110] and [1-10] direction, respectively. However, the situation is more
complicated than that. First one should realize that in the absence of electron
hopping the ground state is highly degenerate. In order to minimize the Coulomb
interactions V$_{\alpha\beta}$, two of the sites of each tetrahedron of the
pyrochlore structure must be occupied by a Fe$^{2+}$ and two by a Fe$^{3+}$ ion
(tetrahedron rule). There is an exponentially large number of different
configurations which satisfy this rule \cite{Anderson56}. When a small hopping
of electrons is taken into account this degeneracy is partially lifted. How
that is taking place remains an unsolved problem. It is possible that the
electronic ground state would remain disordered, i.e., liquid like as long as
the lattice is unchanged. In that case
charge order could result from a structural distortion which is accompanying
the metal-insulator transition. Indeed, the experimental determination of the
electronic low-temperature phase has been a challenging and controversial
subject as has been a proper theoretical interpretation. A recent review
\cite{Walz02} describes the development of different theoretical models
starting from a Hubbard- like Hamiltonian including nearest-neighboring
repulsions \cite{Cullen70} up to inclusion of electron-phonon interactions
together with the tetrahedron rule \cite{Ihle80}. Calculations in the
frame of density functional theory based on LDA+U and the observed low
temperature lattice structure produce charge order which agrees with estimates
based on a valence bond analyses \cite{Leonov04}. They also conclude that the
tetrahedron rule is not strictly fulfilled. This is due to the kinetic energy,
i.e., electronic hopping terms which lead to violations of that
rule. Nevertheless, for a simple reason the tetrahedron rule must be satisfied
to a high degree. Without considerable short-range order it would be difficult
to understand the relatively low transition temperature of the Verwey
transition. For a conventional order-disorder phase transition the transition
temperature is given by $T_V=2\delta V/k_B$, where $k_B$ is Boltzmann's
constant. This would imply a T$_V$ of several thousands of Kelvin
\cite{Anderson56}. 

\subsection{Metallic Spinels: LiV$_2$O$_4$ - a Metal with Heavy Quasiparticles}
\label{Sect:Met_spin}

In order for a spinel oxide to be conducting, the electron count of the B ions
in AB$_2$O$_4$ should differ from an integer number. The compounds listed in
Table I are therefore of special interest. As mentioned before LiTi$_2$O$_4$
with $d^{0.5}$ per Ti ion is a superconductor with a transition temperature of
T$_c$ = 13.7 K. LiV$_2$O$_4$ with $d^{1.5}$ per V ion is a metal with heavy
quasiparticle excitations. At ambient pressure no spin- or charge order has
been observed down to the lowest temperature. The compound LiCr$_2$O$_4$ is not
stable and therefore AlV$_2$O$_4$ with $d^{2.5}$ per V ion is particularly
interesting. This system becomes a charge ordered insulator by so-called
valence skipping (see the next Section). Finally LiMn$_2$O$_4$ with $d^{3.5}$
per Mn ion is an antiferromagnetic insulator with a N\'eel temperature of
T$_N$=280 K. Charge ordering is taking place in that material which has been
used in batteries. In the following we will discuss LiV$_2$O$_4$ in more
detail. 

As pointed out above LiV$_2$O$_4$ shows at low temperatures heavy-fermion
behavior, i.e., it supports heavy quasiparticle excitations
\cite{Kondo97,Kondo99}. It has been the first system where the heavy
quasiparticles originate from $d$ electrons. Experiments show that the $\gamma$
coefficient of the low temperature specific heat C = $\gamma$T is strongly
enhanced and of order $\gamma\simeq$0.4 Jmol$^{-1}$K$^{-2}$. The spin
susceptibility is equally enhanced at low T. Over a large temperature regime it
shows a behavior 

%7.2
\begin{equation}
\chi_S = \chi_0 + \frac{C}{T+\Theta}~~,~~~\Theta = 63 K
\label{TTheta}
\end{equation}

\noindent i.e., of Curie-Weiss type. The Sommerfeld-Wilson ratio R$_W = \pi
k^2_B \chi_S(T = 0)/(3\mu^2_B\gamma)$ is found to be R$_W$ = 1.7. The
temperature independent term $\chi_0 = 0.4 \cdot 10^{-4}$ cm$^3$/mol and the
Curie constant is C = 0.47 cm$^3$K/(molV). The sign of $\Theta$ 
indicates antiferromagnetic interactions between V sites but no magnetic
ordering was found down to 4.2 K. In view of the frustrated lattice this is
understandable. The resistivity is found to be $\rho$(T) = $\rho_0$ + AT$^2$
with a large coefficient A = 2$\mu\Omega$ cmK$^{-2}$. The Kadowaki-Woods ratio
$A/\gamma^2$ \cite{Kadowaki86}, a hallmark of heavy quasiparticles is in the
range of other heavy-quasiparticle materials \cite{Urano00}. These findings are
typical signatures of heavy fermion systems. From the specific heat data one
may determine the entropy S(T). One finds that S(T = 60K) - S (T = 2K) =
10J/(mol$\cdot$K) which is close to 2R ln2 where R is the gas
constant. The implication is that at 60 K the system has almost one excitation
per V ion. This is inconceivable with a conventional band description of the
$d$ electrons. According to Pauli's principle only a small fraction of them is
participating in the excitations when a one-electron picture is
applied. Calculations based on the LDA show that the electrons near the Fermi
energy have $t_{2g}$ character \cite{Matsuno99,Singh99,Eyert99}. These states
are well separated from the $e_g$ 
states as well as from the oxygen states (see Fig. \ref{fig07.2}). The width of
the $t_{2g}$ bands is of order 2 eV and therefore at 60 K only a small fraction
of the electrons in $t_{2g}$ states would contribute to the excitations. In
fact, the calculated density of states must be multiplied by a factor of 25 in
order to account for the large $\gamma$ value. This provides convincing
evidence for strong electron correlations in LiV$_2$O$_4$. Further support is
given by the observation that the material undergoes a phase transition into a
charge ordered state at approximately 6 GPa
\cite{Takeda05,Fujiwara04}. Presumably this metal-insulator transition is again
accompanied by a structural distortion as in Fe$_3$O$_4$. 

%%%%%%%%%%%%%%%%%%%%%%%%%%%%%%%%%%%%%%%%%%%%%%%%%%%%%%%%%%%%%%%%%%%%%%%%%%%
\begin{figure}[t b]
\includegraphics[clip,width=6.0cm]{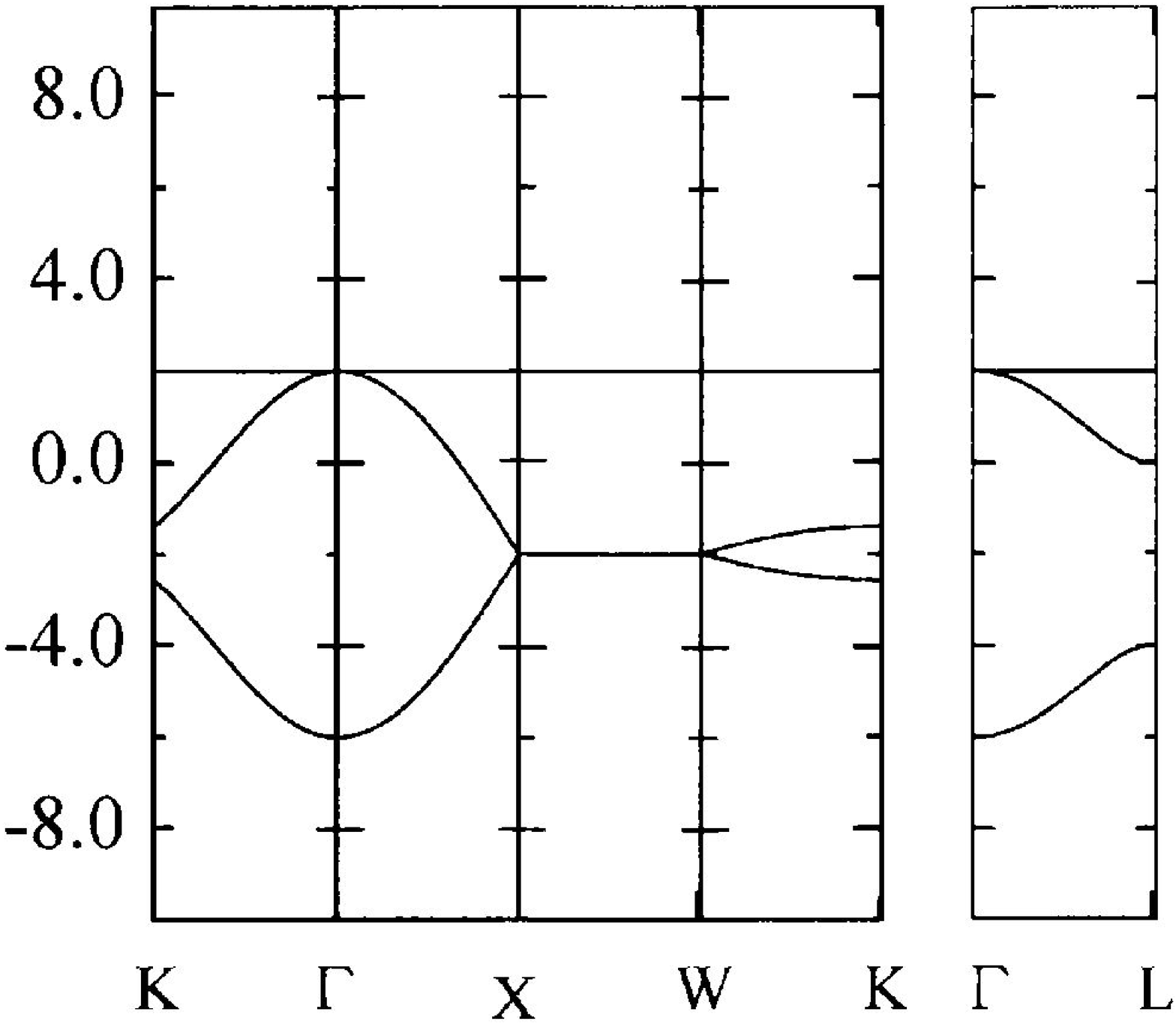}
%\vspace{0.5cm}
\caption{Energy bands for electrons in a pyrochlore lattice with one orbital
  per site in the presence of nearest neighbor hopping. The upper flat band is
  two-fold degenerate. (After \cite{Isoda00})} 
\label{fig07.3}
\end{figure}
%%%%%%%%%%%%%%%%%%%%%%%%%%%%%%%%%%%%%%%%%%%%%%%%%%%%%%%%%%%%%%%%%%%%%%%%%%%

It is worth pointing out that the band structure of a pyrochlore lattice
has interesting features, which are simple to derive when only nearest-neighbor
hopping is taken into account. The Hamiltonian of noninteracting electrons with
nearest-neighbor hopping $t$ is given in diagonal form by

%7.3
\begin{equation}
H_0 = \sum_{{\bf k} \alpha \sigma} \left[ \epsilon_\alpha \left( {\bf k}
  \right) - \mu \right] a^\dagger_{{\bf k} \alpha \sigma} a_{{\bf k} \alpha
  \sigma}  
\label{akalpha}
\end{equation}     

\noindent where $\alpha$ = 1, ..., 4 is a band index due to 4 atoms/unit cell
and one orbital per site is assumed. The band energies are

%7.4
\begin{eqnarray}
\epsilon_\alpha \left( {\bf k} \right) & = & \left\{
\begin{array}
{l@{\quad;\qquad\qquad}l}
2t & \alpha = 3,4\\ 
-2t \left[ 1 \pm \left( 1 + \eta_{\bf k} \right)^\frac{1}{2} \right] & \alpha =
1,2 
\end{array} \right.\nonumber \\
\eta_{\bf k} & = & {\rm cos} \left( 2k_x \right) {\rm cos} \left( 2k_y \right)
+ {\rm cos} \left( 2k_y \right) {\rm cos} \left( 2k_z \right) + {\rm cos}
\left( 2k_z \right) {\rm cos} \left( 2k_x \right)
\label{2t_2t}
\end{eqnarray}  

\noindent where $k_\nu$ is given in reciprocal lattice units $\frac{2
  \pi}{a}$. The bandstructure is shown in Fig. \ref{fig07.3}. One notices a
  two-fold degenerate flat band which is unoccupied provided $t>0$. Hopping
  processes beyond nearest neighbors do not give the flat band a
  dispersion. This is only the case when the hopping matrix elements differ for
  different orbitals. The essential features of the simplified bands are still
  visible in the bands calculated by LDA (see Fig. \ref{fig07.2}). 

%%%%%%%%%%%%%%%%%%%%%%%%%%%%%%%%%%%%%%%%%%%%%%%%%%%%%%%%%%%%%%%%%%%%%%%%%%%
\begin{figure}[t b]
\includegraphics[clip,width=7.0cm]{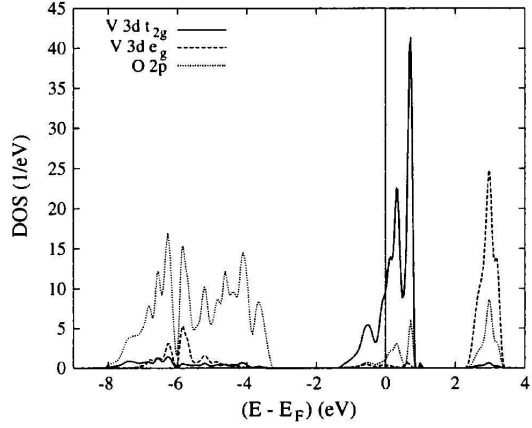}
%\vspace{0.5cm}
\caption{Partial densities of states (DOS) for LiV$_2$O$_4$ calculated in
  LDA. The flat band of Fig. \ref{fig07.3} corresponds to the spike at $\simeq$
  1 eV. (After \cite{Eyert99})}
\label{fig07.2}
\end{figure}
%%%%%%%%%%%%%%%%%%%%%%%%%%%%%%%%%%%%%%%%%%%%%%%%%%%%%%%%%%%%%%%%%%%%%%%%%%%

As regards heavy quasiparticles the crucial question is which degrees of
freedom are associated with their formation. Usually one relates spin degrees
of freedom with the low energy excitations giving rise to the heavy
quasiparticles. This has been discussed at length, e.g., in Sec.
\ref{sec:PartialLocalization} and \ref{sect:ChargeOrdering}. However, in a
frustrated lattice one might also think of charge degrees 
of freedom giving rise to a large number of low energy excitations. The high
degeneracy of the ground state in the absence of a kinetic energy term in the
Hamiltonian is lifted when hopping processes are included and the entropy can
be released over a small temperature range. This is discussed in
the next section for the strong correlation limit. Nonetheless, the finding
that the entropy at 60K is close to $2R \cdot ln2$ suggests that here too spin
degrees of freedom are responsible for the large low temperature specific heat
coefficient $\gamma$.

In the following we give an estimate which shows that spin degrees of freedom
in the pyrochlore structure are indeed able to explain the size of $\gamma$ in
LiV$_2$O$_4$. In setting up the Hamiltonian we include repulsive interactions
between electrons on neighboring sites. In view of the observed charge ordering
under pressure, the following Hamiltonian seems appropriate

%7.5
\begin{eqnarray}
H & = & -\sum_{\langle ij \rangle \nu} t_\nu \left( c^\dagger_{i \nu \sigma}
c_{j \nu \sigma} + h.c. \right) + U \sum_{i \nu} n_{i \nu \uparrow} n_{i \nu
  \downarrow} + U \sum_{i; \nu > \mu} n_{i \nu} n_{i \mu}\nonumber \\
&& + \tilde{J} \sum_{i \nu \mu} {\bf s}_{i \nu} {\bf s}_{i \mu} + V
\sum_{\langle ij \rangle} n_i n_j + \sum_{\langle ij \rangle} J_{ij} \left(
S_i, S_j \right) {\bf S}_i {\bf S}_j~~~. 
\label{Hsumij}
\end{eqnarray}  

\begin{figure}[t b]
\includegraphics[clip,width=5.5cm]{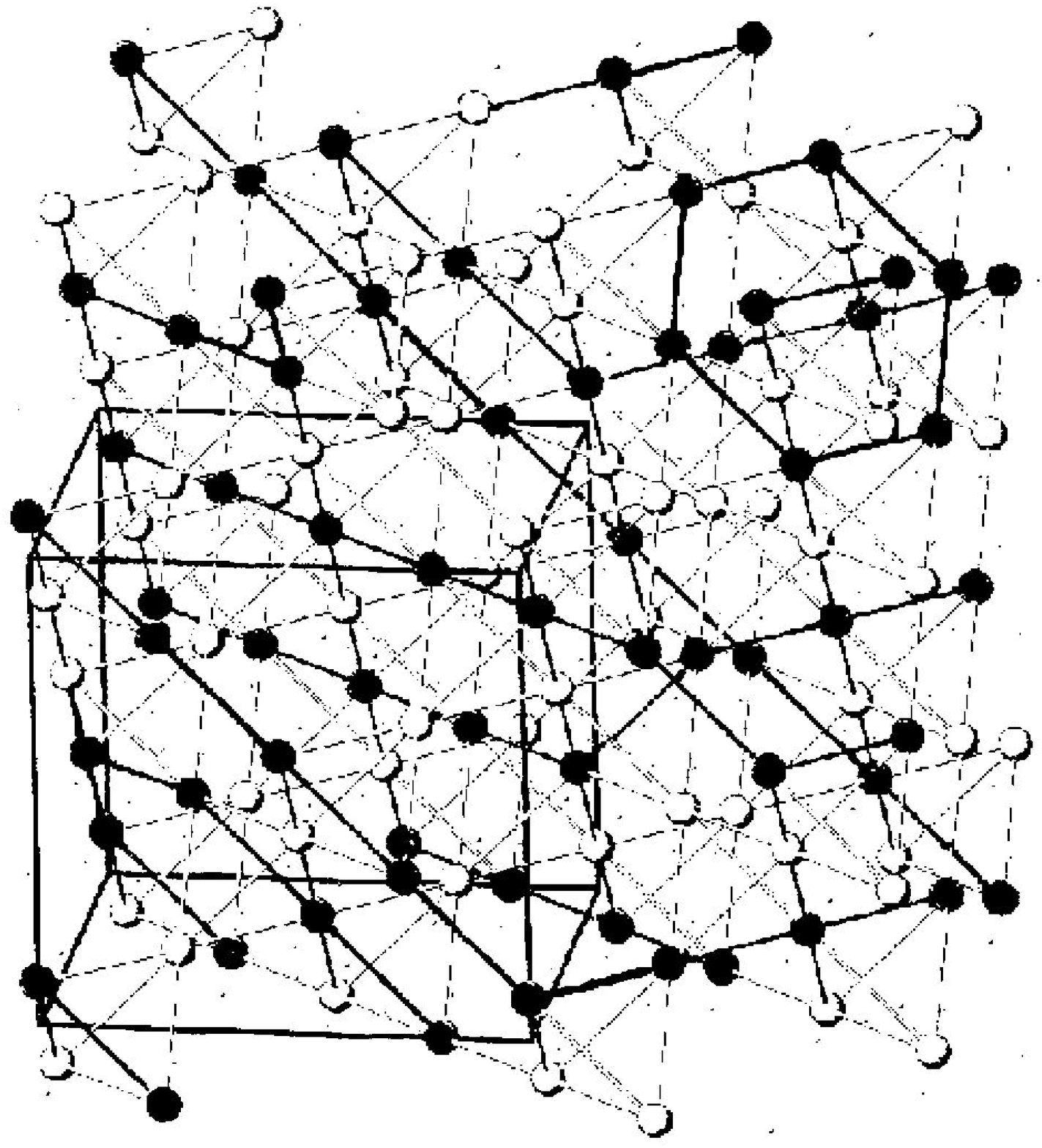}
%\vspace{0.5cm}
\caption{Pyrochlore lattice: Example of a configuration satisfying the
  tetrahedron rule. Occupied 
  sites with S = 1 (black dots) are connected by thick solid lines which form
  chains or rings. The same may be done for occupied sites with S = 1/2 (yellow
  dots).}  
\label{fig07.4}
\end{figure}

Here $i$ is a site and $\nu$ is an orbital index ($\nu$ = 1,2,3) denoting the
different $t_{2g}$ orbitals. The first term is the kinetic energy or electronic
hopping term. For the purpose of the intended estimate for $\gamma$ we will
later neglect it. The following three terms describe the intra-atomic Coulomb
repulsions and spin interactions. For simplicity we have neglected the
differences in the repulsions when different orbitals at site $i$ are
involved. Otherwise we would have to introduce an additional parameter U$'$
(compare with Eq. (\ref{Ht12c}) below). Finally, the last two terms are due to the Coulomb repulsions and antiferromagnetic spin-spin
interactions between neighboring sites. Hereby ${\bf S}_i = \sum_\nu {\bf s}_{i
\nu}$. For an estimate of the spin contributions to the $\gamma$ coefficient we
assume that the $t_\nu$ are very small so that they do not play a role. Then
the $d^2$ configurations have spin S = 1. The Coulomb repulsions are minimized
if on each tetrahedron two sites are in a $d^1$ configuration with S = 1/2 and
two are in a $d^2$ configuration with S = 1. Let us pick out one of the
exponentially large number of degenerate ground-state configurations (see
Fig. \ref{fig07.4}). One notices that all S = 1/2 
sites form chains and rings and the same holds true for the sites with spin
1. The smallest rings consist of six sites. These features are independent of
the chosen configuration. By means of constrained LDA+U calculations one can
determine the nearest-neighbor spin coupling constants J$_{ij}$(S$_i$S$_j$)
\cite{Fulde01}. One finds for J(1/2, 1/2) = 3 meV and J(1,1) = 24 meV implying
that the spin 1 sites are much stronger coupled to each other than the spin 1/2
sites. Note that spin 1 chains have a gap in the expectation spectrum, i.e.,
the Haldane gap $\Delta_H$ \cite{Haldane83,HaldaneD83}. Therefore spin 1/2
chains and rings are virtually uncoupled from each other. They can be coupled
only via spin 1 chains. But the coupling between the two is frustrated and it
takes a  considerable energy $\Delta_H \simeq$ 0.41 J(1,1) to excite the spin 1
chains and rings. Therefore the spin 
1/2 chains remain essentially uncoupled and we can determine directly the
$\gamma$ coefficient of the specific heat and the susceptibility from the
relations \cite{Bonner64}. 

%%%%%%%%%%%%%%%%%%%%%%%%%%%%%%%%%%%%%%%%%%%%%%%%%%%%%%%%%%%%%%%%%%%%%%%%%%%
%%%%%%%%%%%%%%%%%%%%%%%%%%%%%%%%%%%%%%%%%%%%%%%%%%%%%%%%%%%%%%%%%%%%%%%%%%%

%7.6
\begin{equation}
\gamma = \frac{2}{3} \frac{k_B R}{J \left( 1/2, 1/2 \right)}~~~,~~~ \chi_s =
\frac{4 \mu^2_{\rm eff} R}{\pi^2 J \left( 1/2, 1/2 \right)}~~~.
\label{gammakB}
\end{equation}  

Note that the Sommerfeld-Wilson ratio is $R_W$ = 2. An experimental fit of
$\gamma_{\rm exp}$ would require J(1/2, 1/2) = 1.2 meV instead of the
calculated 3 meV. It is known that spin interactions are overestimated by a
LDA+U calculation. But in any case, the improvement which is needed by applying
the above localized electron picture to determine the $\gamma$ coefficient is
much less than the factor of 25 which is missing when a band approach is
used. The above estimate suggests that a description of electrons in
LiV$_2$O$_4$ should start from the localized limit instead from the band limit
because in the former case the required corrections by including $t_\nu \neq 0$
are much less. 

There have been also a number of attempt to explain the heavy quasiparticle in
LiV$_2$O$_4$ with an on-site Hubbard interaction U only, i.e., without
including Coulomb repulsions between neighboring sites. Thereby one of the
$t_{2g}$ electrons is kept as localized while the remaining 0.5 electron per V
site is treated as delocalized. The following argument is used to justify this
distinction. Due to a slight distortion of the oxygen octahedra surrounding the
V sites the $t_{2g}$ states split into a lower $a_{1g}$ and two $e'_g$
states. The splitting is much smaller than the corresponding
bandwidths. Indeed, a LDA bandstructure calculation
\cite{Matsuno99,Eyert99,Singh99} finds the total occupancies $n(e'_g)$ = 1.1
and $n(a_{1g})$ = 0.4 implying a similar population of the different $t_{2g}$
orbitals. But when instead a LDA+U calculation is performed \cite{Anisimov99} 
the $a_{1g}$ state is singly occupied while the remaining 0.5 electrons per V
site are of $e'_g$ character. However, this seems to be a typical mean-field
result. From an atomic point of view there is no reason why an $a_{1g}$
electron should not be able to hop to a neighboring site like an $e'_g$
type electron. The following Hamiltonian based on the LDA+U findings has been
used and investigated beyond mean-field approximation \cite{Laad03}

%7.7
\begin{eqnarray}
H & = & -\sum_{\langle ij \rangle} t_{12} \left( c^\dagger_{i1\sigma}
c_{j2\sigma} + h.c. \right) + U \sum_{i \alpha} n_{i \alpha \uparrow} n_{i
  \alpha \downarrow} + U' \sum_i n_{i1} n_{i2}\nonumber \\ 
&& -\tilde{J} \sum_i {\bf S}_i \left( \sigma_{i1} + \sigma_{i2} \right) + J
\sum_{\langle ij \rangle} {\bf S}_i {\bf S}_j~~~.
\label{Ht12c}
\end{eqnarray}  

The indices 1 and 2 refer to the two $e'_g$ orbitals. Due to Hund's rule the
coupling at site $i$ between a localized $a_{1g}$ electron with spin ${\bf
  S}_i$ and an $e'_g$ electron with spin $\frac{1}{2}\sigma_{i \alpha}$
($\alpha$ = 1,2) is ferromagnetic. The spin-spin interactions between
neighboring sites $i$ and $j$ are antiferromagnetic. Otherwise the system would
order ferromagnetically. The on-site Coulomb repulsion of $e'_g$ electrons is
chosen to be different when the electrons are in the same orbital and when they
are not. The spin-spin interaction between the $e'_g$ electrons is neglected. A
strong Hund's rule coupling is assumed between the $a_{1g}$ and the 
$e'_g$ electrons by taking the limit $\tilde{J} \to \infty$. The nearest
neighbor spin correlations between the localized $a_{1g}$ electrons imply that
the effective hopping matrix element $t_{12}(S)$ of the $e'_g$ electrons
depends on the relative spin orientation of the $a_{1g}$ electrons, i.e., 

%7.8
\begin{equation}
t_{12} (S) = t_{12} \sqrt{\frac{1 + \langle {\bf S}_i {\bf S}_j
	\rangle}{2S^2}}~~~. 
\label{t12S}
\end{equation}

The determination of $\langle {\bf S}_i {\bf S}_j \rangle$ takes the frustrated
lattice, here the pyrochlore structure into account \cite{Canals98}. With the
above simplification the Hamiltonian (\ref{Ht12c}) reduces to 

%7.9
\begin{equation}
H = -\sum_{\langle ij \rangle} t_{12} (S) \left( c^\dagger_{i1} c_{j2} +
c^\dagger_{i2} c_{j1} + h.c. \right) + U' \sum_i n_{i1} n_{i2}~~~.
\label{Hijt12S}
\end{equation}

The $c_i$ operators have only one additional index which takes the values 1 and
2 and acts like a pseudospin. Therefore Eq. (\ref{Hijt12S}) has the form of a
Hubbard Hamiltonian with a spin dependent hopping matrix element. This
Hamiltonian has been treated for the 1/4 filled case by iterated perturbation
theory \cite{Laad03}. When U' is increased a Kondo-like sharp resonance is
obtained at the Fermi surface resulting in heavy quasiparticles at low temperatures. Thermodynamic as well as transport properties can be expressed in
terms of a cross-over temperature T$^*$, going from a heavy Fermi liquid to a
spin liquid at $T > T^*$.

There have been also a number of other attempts to explain the heavy
quasiparticles which we want to mention. One approach starts from a Hamiltonian
similar to (\ref{Ht12c}) but replacing the two $e'_g$ orbitals by a single one
\cite{Lacroix00}. Nearest-neighbor spin correlations are treated by a
mean-field ansatz $\langle {\bf S}_i {\bf S}_j \rangle = -\frac{3}{2}\Gamma^2$
and so are Hund's rule correlations $\langle {\bf S}_i {\bm{\sigma}}_i
\rangle = u^2$. The mean field u is determined from a pseudo-hybridization
between the $a_{1g}$ electron and the itinerant, i.e., $e'_g$ electron. A
subsidiary condition ensures that there is one $a_{1g}$ electron per site. Two
temperatures characterize that approach. Above T = T$_{\rm mag}$ the quantity
$\Gamma$ = 0 and the susceptibility is Curie-Weiss like because the intersite
correlations have vanished. Similarly, a vanishing mean field u marks the
second temperature T$_{\rm HF}$($\simeq$ 20 K) below which a narrow $a_{1g}$
band at E$_{\rm F}$ appears giving rise to heavy quasiparticles. More details
are found in Ref. \cite{Lacroix00}. 

The mean-field approach has been generalized by treating also the $a_{1g}$
electron as itinerant. The on-site Coulomb repulsion between the $e'_g$ and
$a_{1g}$ electrons is taken into account in a slave boson mean-field
approximation \cite{Kusunose00}. In effect the $a_{1g}$ bandwidth is strongly
renormalized and gives rise to a sharp resonance at E$_{\rm F}$.

There have been also weak coupling approaches to explain the heavy
quasiparticles. One suggestion is based on multicomponent fluctuations due to
the $t_{2g}$ and spin degrees of freedom \cite{Yamashita03}. One of the
consequences of the large orbital contributions to the $\gamma$ coefficient of
the specific heat is a small Sommerfeld-Wilson ratio of R$_w \simeq$
0.1. Another approach treats the pyrochlore structure of the V ions as a 
network of Hubbard chains \cite{Fujimoto02}. Due to this one-dimensional
feature electron correlations have strong effects on electron-hole excitations
and hence on the self-energy. 

\subsection{Structural Transition and Charge Disproportionation: AlV$_2$O$_4$}
\label{Sect:StrucTransChargDispAlV2O4}

\begin{figure}[t b]
\includegraphics[clip,width=10.0cm]{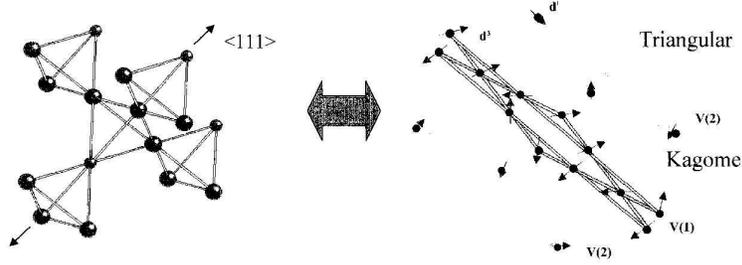}
%\vspace{0.5cm}
\caption{Distortion of a pyrochlore lattice by elongation along the [111]
  axis. The resulting rhombohedral lattice consists of Kagom\'e and triangular
  planes.} 
\label{fig07.5}
\end{figure}

The spinel AlV$_2$O$_4$ is of interest because the average $d$-electron
number per V ion is 2.5 and therefore the configurations are expected to
fluctuate between $3d^2$ and $3d^3$. However, what actually happens is that
the system undergoes a phase transition at approximately $T_c = 700 K$ to a
charge ordered state. It is associated with a change of the lattice structure
from pyrochlore to alternating Kagom\'e and triangular planes. This is shown
in Fig. \ref{fig07.5}. 

%%%%%%%%%%%%%%%%%%%%%%%%%%%%%%%%%%%%%%%%%%%%%%%%%%%%%%%%%%%%%%%%%%%%%%%%%%%
%%%%%%%%%%%%%%%%%%%%%%%%%%%%%%%%%%%%%%%%%%%%%%%%%%%%%%%%%%%%%%%%%%%%%%%%%%%

\begin{figure}[t b]
\includegraphics[clip,width=7.0cm]{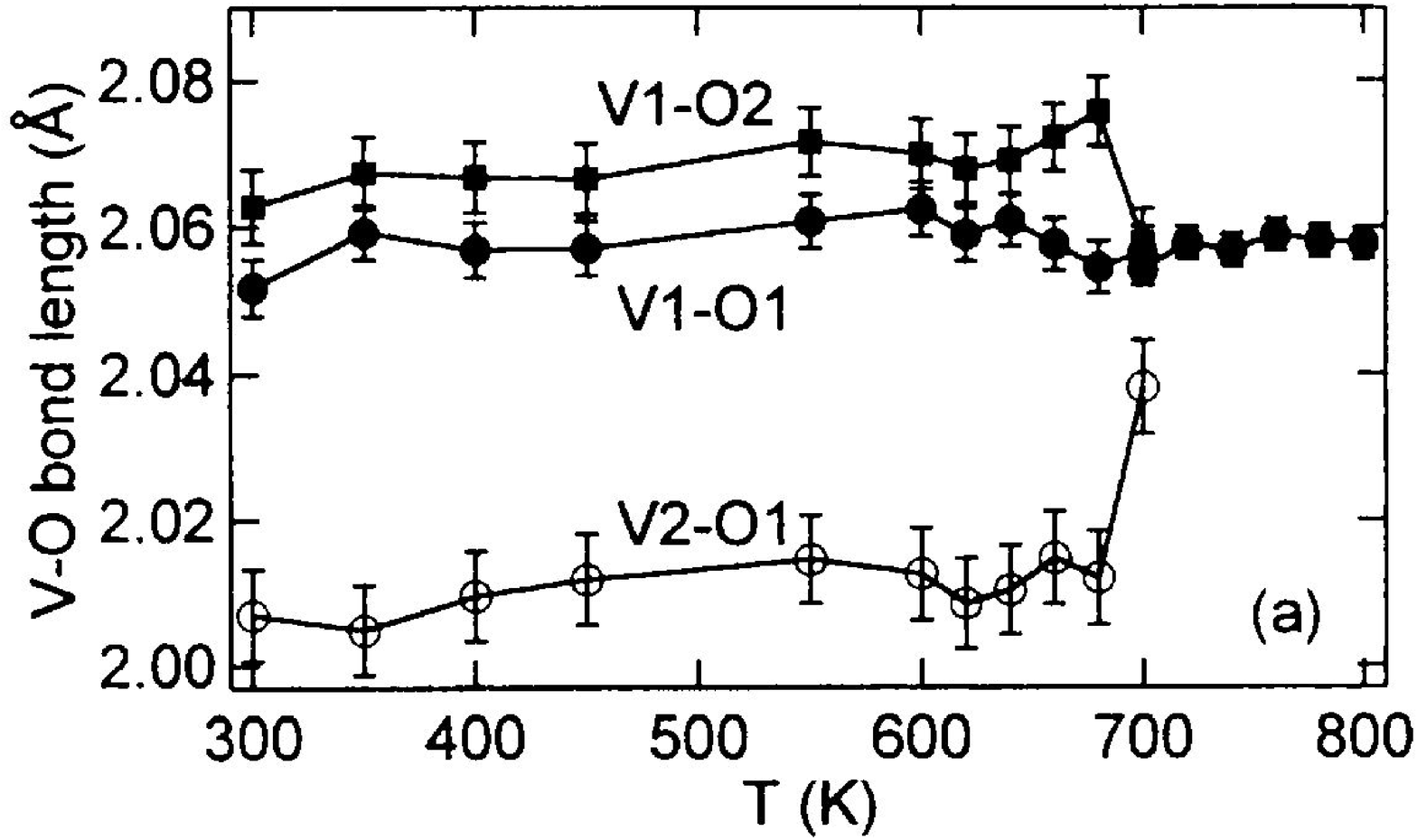}
%\vspace{0.5cm}
\caption{AlV$_2$O$_4$: Dependence of the V-O bond lengths on temperature. V1
  and V2 refer to V$^{2+}$ and V$^{4+}$ ions while O1 and O2 are different
  oxygen ions. (After \cite{Matsuno01})}    
\label{fig07.6}
\end{figure}

\noindent In the low temperature rhombohedral phase the [111] axis
is elongated while the perpendicular axes are shortened in order to keep the
volume of the unit cell nearly constant. Experimental results are shown in
Fig. \ref{fig07.6}. The figure contains also a plot of the observed changes of
the angle between two unit vectors of the rhombohedral lattice as the
temperature is lowered below the phase transition temperature. Note that there
are three times as many sites on the Kagom\'e lattice than there are on the
triangular one.
 
%%%%%%%%%%%%%%%%%%%%%%%%%%%%%%%%%%%%%%%%%%%%%%%%%%%%%%%%%%%%%%%%%%%%%%%%%%%
%%%%%%%%%%%%%%%%%%%%%%%%%%%%%%%%%%%%%%%%%%%%%%%%%%%%%%%%%%%%%%%%%%%%%%%%%%%

\begin{figure}[t b]
\includegraphics[clip,width=7.0cm]{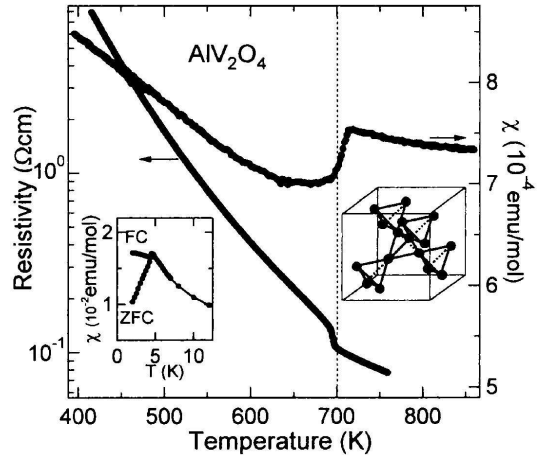}
%\vspace{0.5cm}
\caption{Resistivity and magnetic susceptibility at 1 Tesla as function of
  temperature for AlV$_2$O$_4$. The left inset shows the low temperature field
  cooled (FC) and zero field cooled (ZFC) susceptibility in a field of
  10$^{-2}$ Tesla. (After \cite{Matsuno01})} 
\label{fig07.7}
\end{figure}

\noindent In a simplified description the V ions have a valency of
V$^{4+}$ on the triangular and V$^{2+}$ on the Kagom\'e sites corresponding to
$d^1$ and $d^3$ configurations, respectively. The charge disproportionation is
often called valence skipping. Of course, this is a simplified view 
since in reality the disproportionation is considerably less than the
separation $4 \times d^{2.5} \rightarrow 3 \times d^3 + 1 \times d^1$ would
suggest. The system avoids frustration by distorting. Unfortunately the
experimental results are still sparse. It is not even clear whether
AlV$_2$O$_4$ is a poor metal or a small gap semiconductor at low
temperatures. Resistivity measurements as well as those of the susceptibility
are shown in Fig. \ref{fig07.7}. The structural phase transition shows up in
both quantities, i.e., by a small but steep increase in the resistivity and a
pronounced decrease of the magnetic susceptibility. 

%%%%%%%%%%%%%%%%%%%%%%%%%%%%%%%%%%%%%%%%%%%%%%%%%%%%%%%%%%%%%%%%%%%%%%%%%%%
%%%%%%%%%%%%%%%%%%%%%%%%%%%%%%%%%%%%%%%%%%%%%%%%%%%%%%%%%%%%%%%%%%%%%%%%%%%

For a description of the
phase transition one must set up a model Hamiltonian. It should also allow for
explaining differences of charge ordering in AlV$_2$O$_4$ and LiV$_2$O$_4$
under pressure. While AlV$_2$O$_4$ is a semiconductor or a metal in the
charge ordered state, LiV$_2$O$_4$ becomes an insulator
\cite{Takeda05}. Electronic structure calculations based on LDA+U provide a
realistic description of magnetic insulators but are not suitable for strongly
correlated paramagnetic metals. Therefore we exclude them here and start from a
microscopic model Hamiltonian at the price of having to introduce adjustable
parameters. Nevertheless we can uncover this way the processes which lead to
the observed structural transition and the accompanying charge order. From
standard band-structure calculations it is known that only $t_{2g}$ states are
near the Fermi energy (compare with Fig. \ref{fig07.2}). All other bands are
well above or below $E_{\rm F}$. Therefore it suffices to include only $t_{2g}$
electrons in the model Hamiltonian. We write it in the form 

%7.10
\begin{eqnarray}
H & = & H_0 + H_{\rm int} + H_{\rm e-p}~~,~~{\rm with} \nonumber \\
H_0 & = & \sum_{\langle l \mu, l'\mu' \rangle} t^{\nu \nu'}_{\mu \mu'} \left(
l,l' \right) c^\dagger_{l \mu \nu \sigma} c_{l' \mu' \nu' \sigma} \nonumber \\
H_{\rm int} & = & \sum_{l \mu} \left\{ \left( U + 2J \right) \sum_\nu
n_{l \mu \nu \uparrow} n_{l \mu \nu \downarrow} + U \sum_{\nu > \nu'} n_{l \mu
\nu \sigma} n_{l \mu \nu' \bar{\sigma}} \right. \nonumber \\
&& + \left. \left( U - J \right) \sum_{\nu > \nu'} n_{l \mu \nu \sigma} n_{l
    \mu \nu' \sigma} \right\} + \frac{V}{2} \sum_{\langle l \mu, l' \mu'
  \rangle \nu \nu' \sigma \sigma'}  n_{l \mu \nu \sigma} n_{l' \mu' \nu'
  \sigma'} \nonumber \\
H_{\rm e-p} & = & \epsilon \Delta \sum_{l \nu \sigma} \sum_\mu \left(
  \delta_{\mu, 1} - \frac{1}{3} \left( 1 - \delta_{\mu, 1} \right) \right)
n_{l \mu \nu \sigma} + K \sum_l \epsilon^2_l~~~.
\label{HepDelta}
\end{eqnarray}  

Here $H_0$ describes the kinetic energy. The electron creation and
annihilation operators are specified by four indices, i.e., for the unit cell
(denoted by $l$), the sublattice ($\mu = 1 - 4$), the $t_{2g}$ orbital ($\nu
= d_{xy}, d_{yz}, d_{zx}$), and the spin ($\sigma = \uparrow,
\downarrow$). The brackets $\langle ... \rangle$ indicate a summation over
nearest-neighbor sites. The term $H_{\rm int}$ describes the
on-site Coulomb and exchange interactions $U$ and $J$ among the $t_{2g}$
electrons. The last term contains the Coulomb repulsion $V$ of an electron with
those on the six neighboring sites. Finally $H_{\rm e-p}$ describes the
coupling to lattice distortions. The deformation potential is denoted by
$\Delta$. It is due to a shift in the orbital energies of the $V$ sites caused
by relative changes in the oxygen positions. While the energy shift is positive
for the $V$(1) sites it is negative for the $V$(2) sites. 

The elastic constant $K$ refers to the $c_{44}$ mode and describes the energy
due to the rhombohedral lattice deformation. It is reasonable to assume that
like in Fe$_3$O$_4$ \cite{Schwenk00} and Yb$_4$As$_3$ \cite{Goto99} only the
$c_{44}$ mode is strongly coupled with the charge disproportionation. One can
give at least approximate values for all parameters except for $V$ and
$\Delta$. Their ratio will be fixed by the charge-ordering transition
temperature while keeping the constraint $V \ll U$. For the on-site Coulomb-
and exchange integrals we set $U$ = 3.0 eV and $J$ = 1.0 eV which are values
commonly used for vanadium oxides \cite{Fujimori88}. Band structure
calculations which we have performed demonstrate that the hopping matrix
elements $t^{\nu \nu'}_{\mu \mu'}$ between different orbitals $\nu \neq \nu'$
are negligible. For simplicity we can therefore omit $\nu, \nu'$. Then $t_{\mu
  \mu'} (l,l') = -t$ when $l, l'$ are nearest neighbors. We will take into
account the orbital dependent hopping matrix elements coming from a
tight-binding fit of LDA calculations. Furthermore the $c_{44}$ elastic
constant is neither known for AlV$_2$O$_4$ nor for LiV$_2$O$_4$, while
computational methods for its ab initio calculation in the case of materials
with strong electronic correlations are not mature enough. Therefore a
representative value $c^{(0)}_{44}/\Omega = 6.1 \cdot 10^{11} {\rm erg/cm}^3$
is used for AlV$_2$O$_4$ where $\Omega$ is the volume of the cubic unit cell
with a lattice constant of a = 5.844 ${\AA}$. This value is close to the
experimental value for Fe$_3$O$_4$ which has also the spinel structure. This
leads to $K \simeq 1.1 \cdot 10^2$ eV. The deformation potential $\Delta$ is
not known but is commonly of the order of the band width. For convenience we
introduce the dimensionless coupling constant $\lambda = \Delta^2/Kt$ and
lattice distortion $\delta_L = \epsilon \Delta/t$. From LDA calculations the
bandwidth is 8$t$ = 2.7 eV, and therefore a reasonable value is $\lambda t =
\Delta^2/K$ = 1 eV. This means $\Delta$ = 10.5 eV which is twice the value of
Yb$_4$As$_3$ \cite{Fulde95}. 

Not contained in the Hamiltonian (\ref{HepDelta}) is a spin-spin interaction
term between the V ions. This might turn out a shortcoming since bandstructure
calculations based on a local spin-density approximation (LSDA) to density
functional theory find effective exchange constants which are strongly enhanced
in the low temperature phase. Estimates are J$_{kk}$ = J$_{kt}$ = 202 K for the
high temperature phase where the subscripts $k$ and $t$ refer to Kagom\'e and
triangular sites. For the low temperature phase the corresponding estimates are
J$_{kk}$ = 360 K and J$_{kt}$ = 167 K \cite{Yaresko05}.

Let us first consider $H_0$ which is easily diagonalized. With four $V$ ions
per unit cell and three $t_{2g}$ orbitals there are altogether 24 bands,
assuming that the spin symmetry is broken. Of those twelve are dispersionless
and degenerate. Furthermore, there are two sixfold degenerate dispersive bands
(compare with Fig. \ref{fig07.3}). The Fermi energy is in a region of high
density of states. More details are found in Ref. \cite{Isoda00}.

The next term we discuss is $H_{\rm e-p}$ which describes the deformation
potential coupling. When it is sufficiently strong it leads to charge
ordering. There is no opening of a gap though, but only a sharp decrease of
the density of states near $E_{\rm F}$, i.e., the system remains a metal. This
feature does not change when the interactions $U$ and $V$ are included in
mean-field approximation, provided we deal with a paramagnetic
state. Generally $U$ suppresses charge ordering while $V$ enhances it. Again,
no gap opens at $E_{\rm F}$ but the density of states decreases in its
neighborhood. In mean-field approximation the relation between the homogeneous
lattice distortion $\delta_L$ along [111] and charge disproportionation is
given by 

%7.11
\begin{equation}
\delta_L = \frac{\lambda \left( n_2 - n_1 \right)}{2}
\label{deltaL}
\end{equation}

\noindent where $n_1 = \sum_{\nu \sigma} \langle n_{l1\nu\sigma} \rangle$ is
the occupational number of the triangular sites while $n_2 = n_3 = n_4$ with
$n_2 = \sum_{\nu \sigma} \langle n_{l2\nu\sigma} \rangle$ is the one of
the Kagom\'e sites.

\begin{figure}[t b]
\includegraphics[clip,width=7.0cm]{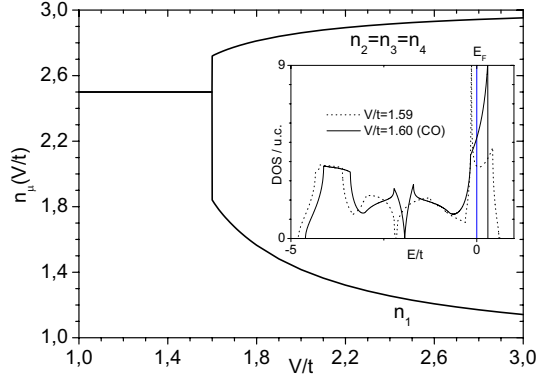}
%\vspace{0.5cm}
\caption{Charge disproportionation based in Eq. (\ref{HepDelta}) as function of
  V/t. The $n_\mu$ denote the 
  occupation numbers of the four sites of a tetrahedron. In the inset the
  changes in the density of states are shown when V/t is just below and above
  the critical value at which charge ordering sets in. For V/t = 1.67 one has $n_1$ = 2.5 - 3$\delta$ and $n_i$ = 2.5 + $\delta$ (i = 2 - 4) with  a charge
  disproportionation  $\delta \simeq$ 0.25. Here U = 3.0 eV, J = 1.0 eV,
  $\lambda$t = 1.0 eV and 8t = 2.7 eV. (After \cite{Zhang05}).} 
\label{fig07.8}
\end{figure}

A shortcoming of the mean-field analysis is that it leads to a second-order
phase transition instead of the observed first-order one. This is almost
certainly due to strong correlations which suppress charge fluctuations between
different vanadium sites. In order to incorporate them at least approximately,
one must allow for unrestricted mean-field solutions by breaking the spin
symmetry. But the constraint of zero total magnetic moment has to remain. This
is done by ascribing to the sites $\mu$ of a tetrahedron an occupational number
$n_\mu$ and a magnetization $m_\mu = \sum_{\nu \sigma} \langle
n_{l\mu\nu\sigma} \rangle \sigma$. The spins are assumed to be directed
towards the center of the tetrahedron in the undistorted phase. For them the
[111] direction is a convenient quantization axis. In the distorted phase the
spins of the V(1) sites are slightly tilted with respect to this axis so that
the net magnetization remains zero. The free energy is a function of the
different $n_\mu$ and $m_\mu$ and must be minimized with respect to both. The
energy bands do now depend on spin $\sigma$. The hopping matrix elements
between nearest neighbor sites must be transformed accordingly, so that the
different spin directions are accounted for. For details see \cite{Zhang05}. As
a result the charge disproportionation is found as function of the various
parameters. We show in Fig. \ref{fig07.8} the results for $T = 0$ as function
of the ratio $V/t$. As usual the calculated disproportionation is larger, here
by a factor of 2.5 than the one obtained from a valence band analysis of the
measured distorted structure. This is due to the simplified model Hamiltonian
which does not allow for screening by non-$d$ electrons. The same difficulty
was found for Yb$_4$As$_3$ \cite{Staub02,Fulde95}. Also shown in 
Fig. \ref{fig07.8} as an inset is the change in the density of states in the
vicinity of the critical ratio $(V/t)_{\rm crit} = 1.6$ at which the
transition to a charge ordered state does occur. The strong change which one
can notice might explain the observed small but steep increase in the
resistivity when charge ordering sets in. The phase transition is found to be
of first order when the free energy is evaluated. For $V/t = 1.67$ the
calculated transition temperature is $T_c = 660 K$ which is reasonably close to
the observed one of $T_{\rm exp} = 700 K$. Of course, there is some
arbitrariness in the particular choice of $V/t$. It should be pointed out that
the LSDA calculations yield a value of $\delta$ = 0.17 for the
disproportionation and a small hybridization gap.

%%%%%%%%%%%%%%%%%%%%%%%%%%%%%%%%%%%%%%%%%%%%%%%%%%%%%%%%%%%%%%%%%%%%%%%%%%%
%%%%%%%%%%%%%%%%%%%%%%%%%%%%%%%%%%%%%%%%%%%%%%%%%%%%%%%%%%%%%%%%%%%%%%%%%%%

In the undistorted phase the point symmetry of the vanadium ions is $D_{3d}$
and the $t_{2g}$ degeneracy is reduced to an $a_{1g}$ singlet and an $e'_g$
doublet. In the distorted, i.e., charge ordered phase the symmetry of the
triangular sites remains $D_{3d}$ while the one of the Kagom\'e sites is
lowered to $C_{2h}$. This lifts also the $e'_g$ degeneracy. It is due to an
elongation of the crystal in [111] direction and a contraction perpendicular to
it. While in the distorted phase of AlV$_2$O$_4$ the energies of the three
orbitals are nearly the same because of a very small distortion of the oxygen
octahedra one expects that in LiV$_2$O$_4$ the energy of the $a_{1g}$ orbital
is highest. There the distortion of the octahedra is fairly large. Since the
changes in orbital energies are small in AlV$_2$O$_4$ one expects that the
system remains gapless even in the charge ordered state in agreement with LDA
calculations. The LDA band structure can be reproduced by a proper choice of
the hopping parameters. By determining self-consistently the occupational
numbers of the three $V$(1) sites one finds that $n_2^{xy} = n_2^{zx} >
n_2^{yz}$, $n_3^{xy} = n_3^{yz} > n_3^{zx}$ and $n_4^{yz} = n_4^{zx} >
n_4^{xy}$ in the $t_{2g}$ basis. This shows that the ordered orbitals on
sublattices 2, 3 and 4 are perpendicular to each other. There is no orbital
ordering on the $V$(2) sites in the charge ordered state because of the small
energy difference between the $a_{1g}$ and $e'_g$ orbitals as compared with the
bandwidth. The sharp decrease of the density of states near $E_{\rm F}$ as well
as the order of the charge ordering phase transition are not influenced by
orbital order. The same holds true for the absence of an energy gap at $E_{\rm
  F}$. Therefore orbital ordering is of little importance for AlV$_2$O$_4$. 

These findings should be compared with charge ordering observed in LiV$_2$O$_4$
under pressure. A crucial difference is the average $d$-electron number per $V$
site which is 1.5 for LiV$_2$O$_4$ as compared with 2.5 in the case of
AlV$_2$O$_4$ and the distinct role of the $a_{1g}$ orbital. Consequently in the
limit $t \to 0$ the $a_{1g}$ orbital is empty at the $V$(1) sites of
LiV$_2$O$_4$ while the split $e'_g$ orbitals are singly occupied with $S =
1$. On the $V$(2) sites with $d^0$ also the split $e'_g$ orbitals
are unoccupied. Therefore a gap at $E_{\rm F}$ is expected in the charge
ordered state. This is what is found when the Hamiltonian (7.5) is treated in
mean-field approximation with a filling factor of 1/4 instead of 5/12 as in
the case of AlV$_2$O$_4$. The opening of a gap can also be inferred from
measurements of $\rho(T)$ \cite{Takeda05}. Orbital ordering is obviously
not relevant here. For further details we refer to \cite{Zhang05}. The
structure in the distorted phase on which the above theory has been based was
recently called into question in Ref. \cite{Horibe05}. Instead of a
low-temperature structure consisting of triangular and Kagom\'e planes those
authors interprete their data in terms of V$_7$ molecular clusters. The future
has to show which structure is the correct one.

\subsection{Fractional Charges due to Strong Correlations}

Consider the pyrochlore lattice occupied by half as many electrons than there
are number of sites. We take the limit U $\rightarrow \infty$ and neglect for
simplicity the electron spin so that we are dealing with spinless fermions. In
that case half of the lattice sites are singly occupied and half of them are
empty. Restricting the inter-site Coulomb repulsion to nearest neighbors the
Hamiltonian is 

%7.12
\begin{equation}
H = -t \sum_{\langle ij \rangle} \left( c^\dagger_i c_j + h.c. \right) + V
\sum_{\langle ij \rangle} n_i n_j~~~.
\label{HtVij}
\end{equation}

\begin{figure}[t b]
\includegraphics[clip,width=6.0cm]{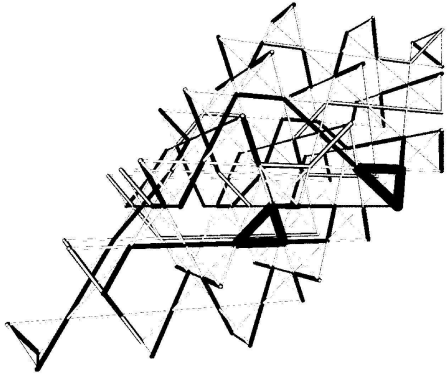}
%\vspace{0.5cm}
\caption{Separated tetrahedra on a pyrochlore lattice after an electron has
  been added to the system. Occupied sites are connected by solid red lines and
  empty by green lines (courtesy of F. Pollmann).}  
\label{fig07.9}
\end{figure}

The $c^\dagger_i$ create spinless fermions at sites $i$ and as usual $n_i =
c^\dagger_i c_j$. As discussed before, the repulsion $V$ is minimized when in
each tetrahedron  two sites are occupied and two sites are empty (tetrahedron
rule). Let us now add one particle to the system. Since each site belongs to
two tetrahedra, the above rule is violated for two neighboring tetrahedra which
now contain three electrons each. The interaction energy is increased by
4$V$. But when one of the four nearest neighbors of the added particle hops
onto an empty site the two tetrahedra violating the rule are
separated. This is shown in Fig. \ref{fig07.9}. The important point is that the
interaction energy remains unchanged by this separation , i.e., it is still
4$V$. When the charge of the added particle is $e$, each fragment must be
assigned a charge e/2. 

%%%%%%%%%%%%%%%%%%%%%%%%%%%%%%%%%%%%%%%%%%%%%%%%%%%%%%%%%%%%%%%%%%%%%%%%%%%
%%%%%%%%%%%%%%%%%%%%%%%%%%%%%%%%%%%%%%%%%%%%%%%%%%%%%%%%%%%%%%%%%%%%%%%%%%%

%%%%%%%%%%%%%%%%%%%%%%%%%%%%%%%%%%%%%%%%%%%%%%%%%%%%%%%%%%%%%%%%%%%%%%%%%%%
\begin{figure}[t b]
\includegraphics[clip,width=6.0cm]{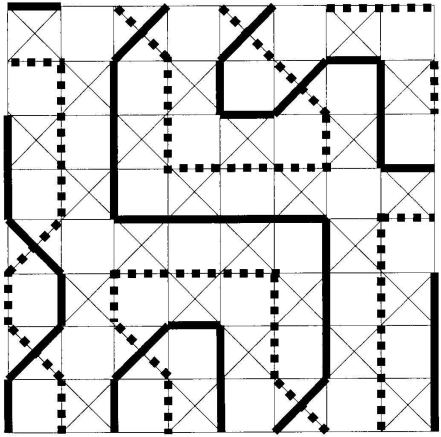}
%\vspace{0.5cm}
\caption{Ground-state configuration of a checkerboard lattice with half-filling
of spinless fermions. Occupied sites are connected by solid lines and empty
sites by dashed lines. Hopping takes place along thin lines with matrix element
$-t$.}   
\label{fig07.10}
\end{figure}
%%%%%%%%%%%%%%%%%%%%%%%%%%%%%%%%%%%%%%%%%%%%%%%%%%%%%%%%%%%%%%%%%%%%%%%%%%%

Taking into account the kinetic energy term of the Hamiltonian lifts the
exponentially large degeneracy of the ground state which is present when the
kinetic energy term is absent. Deconfinement of the charges e/2 remains
intact provided t/V is sufficiently small
\cite{Polyakovbook,FuldeP02,Betouras06}. Since a numerical study of a
pyrochlore lattice is difficult we shall investigate instead the simpler
checkerboard lattice. The latter can be considered a projection of the
pyrochlore lattice onto a plane. But one must keep in mind that the lower
dimension of the checkerboard lattice may result in different behaviors as
regards confinement or deconfinement of the fractional charges e/2. This is
known from lattice gauge theories which are closely related to the present
problem \cite{Polyakovbook,Fradkin91}. Also the statistics of fractionally
charged excitations may differ in two and three dimensions. In two dimensions
the wavefunction of the particles belongs to a representation of the braid
group while in three dimensions it is one of the permutation
group. Fig. \ref{fig07.10} shows one of the ground state configurations. They
have the form of string nets (see also 
Ref. \cite{Levin05}). In the absence of hopping the degeneracy is $N_{\rm deg}
= (4/3)^{3N/4}$ where $N = N_x N_y$ and $N_x, N_y$ are the number of lattice
sites in $x$ and $y$ direction. In order to study numerically the ground state
of the system for large ratios of $V/t$ it is advantageous to introduce an
effective Hamiltonian \cite{Runge04} which acts on the different configurations
obeying the tetrahedron rule. They form what will be called the {\it allowed
  subspace} of all configurations. An effective Hamiltonian allows for the
diagonalization of much larger clusters than does the full $H$. To leading
order in $t/V$ one finds that the energy is lowered for all allowed
configurations by an amount of 

%7.13
\begin{equation}
\Delta E = -\frac{4t^2}{V} \sum_i n_i~~~.
\label{HSigma}
\end{equation} 

\noindent Therefore it does not lift the
degeneracy and one has to go to the next higher order term. Different allowed
configurations are connected through ring hopping processes. The smallest
non-vanishing process is

%7.14
\begin{equation}
%{\begin{figure}[t b]
\includegraphics[width=7.0cm]{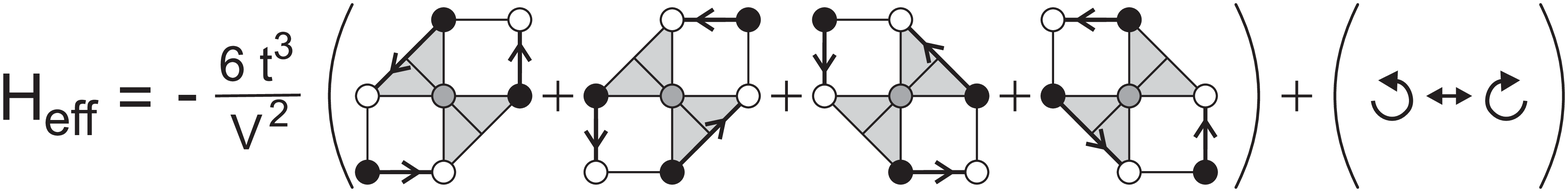}
%\end{figure}}
\label{Heff6tV}
\end{equation} 

\noindent where filled dots denote occupied and empty dots empty sites,
respectively. The site in between can be empty or occupied. The arrows stand
for particle hopping. With 

%7.15
\begin{equation}
t_{\rm ring} = \frac{12 t^3}{V^2}
\label{tring}
\end{equation} 

\noindent this Hamiltonian can be written as

%7.16
\begin{equation}
H_{\rm eff} = t_{\rm ring} \sum_{\hexagon} c^\dagger_{j_6} c^\dagger_{j_4}
c^\dagger_{j_2} c_{j_5} c_{j_3} c_{j_1} ~~~.
\label{Hefftring}
\end{equation} 

Hopping on larger rings implies higher orders in $t/V$. The matrix elements of
$H_{\rm eff}$ with respect to different allowed configurations $|i\rangle$ and
$|j\rangle$ are 

%7.17
\begin{equation}
\langle j \mid H_{\rm eff} \mid i \rangle = (-1)^{n_0} t_{\rm ring}
\label{jiring}
\end{equation} 

\noindent where $n_0$ is the occupation number of the site inside the ring. It
is worth realizing that the sign dependence of $\langle j | H_{\rm eff} | i
\rangle$ is absent when instead of a checkerboard lattice the pyrochlore
lattice is considered. In the latter structure there is no lattice site inside
a 6-ring loop. The partial lifting of the degeneracy of the ground-state
manifold by $H_{\rm eff}$ can be understood with the help of the height
representation. For that 
purpose one divides the criss-crossed squares of the checkerboard lattice into
sublattices $A$ and $B$ and assigns a clockwise and counter-clockwise
orientation to them. To each occupied site is attached a unit vector the
direction of which is in accordance with the orientation of the corresponding
criss-crossed squares. At each empty site the vector is pointing into opposite
direction. This defines a vector field ${\bf f}({\bf r})$ where ${\bf r}$ is
defined with respect to the uncrossed 
plaquettes. Because of the tetrahedron rule, i.e., two sites are occupied and
two are unoccupied on a criss-crossed square, the vector sum over
a closed loop vanishes, implying curl ${\bf f} = 0$. Therefore  ${\bf f}$ can
be written as the gradient of a scalar field $h({\bf r})$, i.e., a potential
which is called height field. It allows for the introduction of two topological
quantum numbers $\kappa_x$ and $\kappa_y$. They quantify the difference in the
potential at the upper and lower boundary and at the right and left one. Their
values cover the range $-N_x/2 \leq \kappa_x \leq N_x/2$ and $-N_y/2 \leq
\kappa_y \leq N_y/2$. 

%%%%%%%%%%%%%%%%%%%%%%%%%%%%%%%%%%%%%%%%%%%%%%%%%%%%%%%%%%%%%%%%%%%%%%%%%%%
\begin{figure}[t b]
\includegraphics[clip,width=7.0cm]{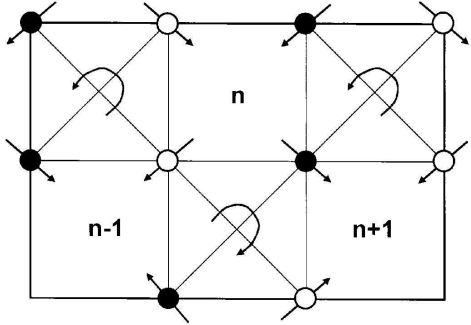}
%\vspace{0.5cm}
\caption{Values of the height field $h$ for a part of a given configuration:
  solid and empty dots mark occupied and empty sites, respectively.}  
\label{fig07.11}
\end{figure}
%%%%%%%%%%%%%%%%%%%%%%%%%%%%%%%%%%%%%%%%%%%%%%%%%%%%%%%%%%%%%%%%%%%%%%%%%%%

In Fig. \ref{fig07.11} an example is given of the change
of the height field $h({\bf r})$ defined on the uncrossed squares. On
a torus $\kappa_x$ and $\kappa_y$ are winding numbers. An important point is
that the application of $H_{\rm eff}$ does not change the topological quantum
numbers of a configuration. Therefore the degenerate ground-state
configurations are divided into different classes according to ($\kappa_x,
\kappa_y$). The matrix $\langle i| H_{\rm eff} | j \rangle$
 reduces into block matrice characterized by ($\kappa_x,\kappa_y$). It turns out that
 $H_{\rm eff}$ does not necessarily connect all configurations within a
 class. Therefore the matrix $(H_{\rm eff})_{ij}$ for the class
 ($\kappa_x,\kappa_y$) may reduce further to irreducible blocks
 ($\kappa_x^{(\alpha)},\kappa_y^{(\alpha)}$). This holds true in particular for
 the class (0, 0) which has the largest number of elements. It is found that
 for a system of size 8 x 8 with $N_f = 32$ spinless fermions the ground state
 has topological quantum numbers (0, 0). The subclass to which it belongs has
 more than $10^5$ elements. The ground state is two-fold degenerate and the two
 states are related by particle-hole symmetry. Note that $H_{\rm eff}$ has this
 symmetry while $H$ does not have it. The two states are charge ordered along
 one of the diagonals and can be transformed into each other by a rotation of
 90 degrees. This invalidates a supposition in Ref. \cite{FuldeP02}. The
 participation ratio (PR) of a wavefunction  is a measure of how extended that
 function is. With  

%7.18
\begin{equation}
\mid \psi \rangle = \sum_\nu \alpha_\nu \mid \nu \rangle
\label{psisumnu}
\end{equation} 

\noindent where $| \nu \rangle$ denotes different configurations the
participation ratio is 

%7.19
\begin{equation}
PR [\psi] = \frac{1}{\sum\limits_\nu \mid \alpha_\nu \mid^4}~~~.
\label{PRpsi}
\end{equation} 

For the ground state $| \psi_0 \rangle$ of the 8 x 8 system this ratio is very
large, i.e.,

%7.20
\begin{equation}
PR [\psi_0] = 46.9 \cdot 10^3
\label{PRpsi0}
\end{equation}

\noindent indicating that many configurations contribute to it. The largest
possible value of PR $[\psi_0]$ in a Hilbert space of dimension $N$ is PR
$[\psi_0] = N$. The total
density of state is

%7.21
\begin{equation}
\rho_{\rm tot} (E) = \sum_l \delta \left( E - E_l \right)
\label{rhotot}
\end{equation}

\noindent where $E_l$ are the eigenvalues of $H$. From it we can compute the
specific heat $C(T)$ according to

%7.22
\begin{equation}
C(T) = \frac{\partial}{\partial T} \frac{\int dEE \rho_{\rm tot} (E) e^{-
	\beta E}}{\int dE \rho_{\rm tot} (E) e^{-\beta E}}~~~.
\label{CTtot}
\end{equation}

\begin{figure}[t b]
\includegraphics[clip,width=7.0cm]{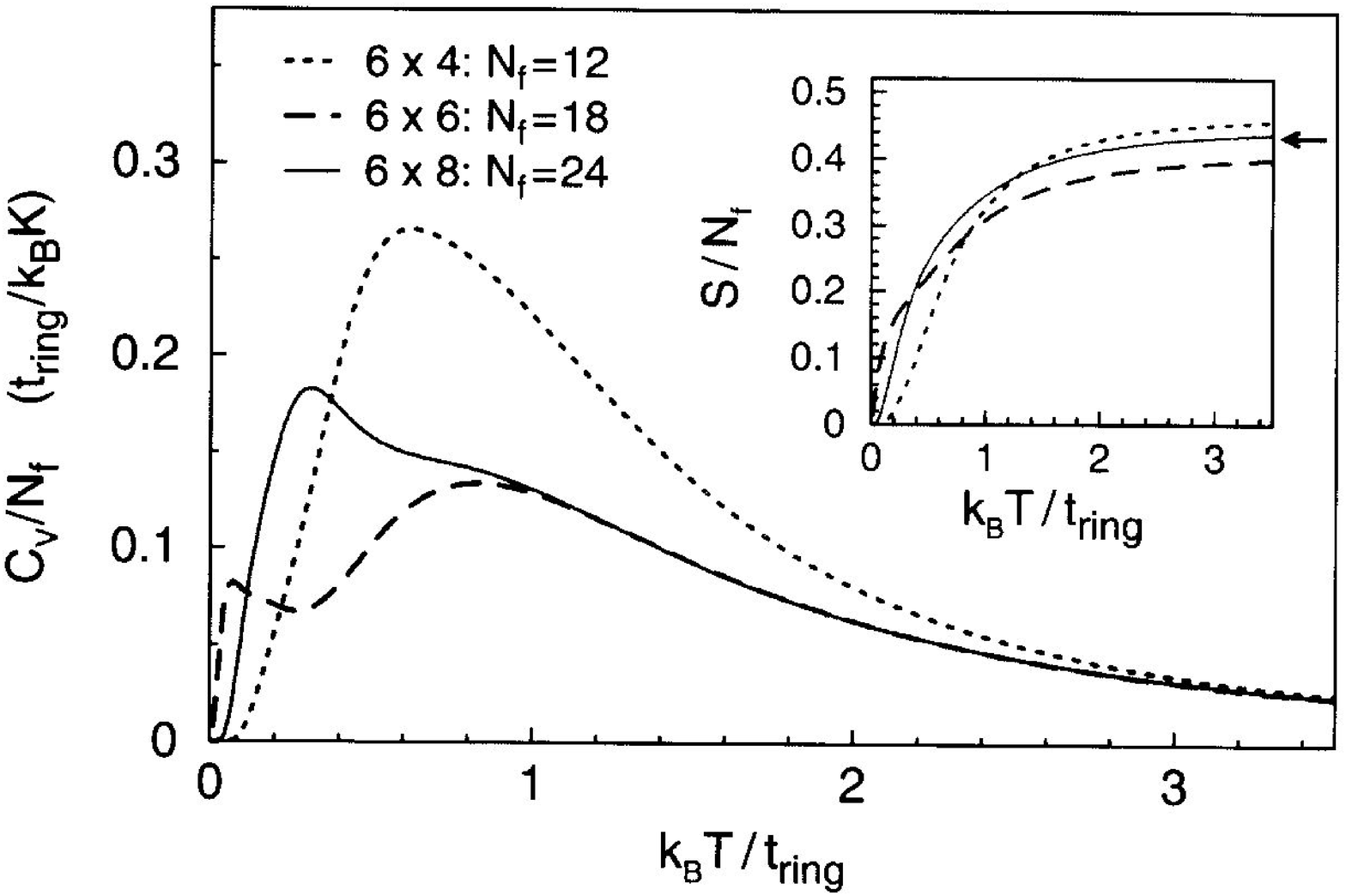}
%\vspace{0.5cm}
\caption{Specific heat per particle for a checkerboard lattice of spinless
  fermions at half filling for various system sizes. Inset: corresponding
  entropies, the arrow marks the value of 2(3/4) ln(4/3). (After
  \cite{Runge04})}   
\label{fig07.12}
\end{figure}

Numerical results for different system sizes are shown in
Fig. \ref{fig07.12}. One notices an almost linear in $T$ behavior. This
suggests that a large quasiparticle mass in geometrically frustrated lattices
can possibly be due to charge instead of spin degrees of freedom, the usual
source of heavy quasiparticles. The large number of low lying excitations
causing the steep linear increase of the specific heat with temperature is here
due to a release of entropy (3/4) ln (4/3) $\approx$ 0.22 per site over a
temperature range of $k_BT \leq 2t_{\rm ring}$.

%%%%%%%%%%%%%%%%%%%%%%%%%%%%%%%%%%%%%%%%%%%%%%%%%%%%%%%%%%%%%%%%%%%%%%%%%%%
%%%%%%%%%%%%%%%%%%%%%%%%%%%%%%%%%%%%%%%%%%%%%%%%%%%%%%%%%%%%%%%%%%%%%%%%%%%

\begin{figure}[t b]
\includegraphics[clip,width=3.8cm]{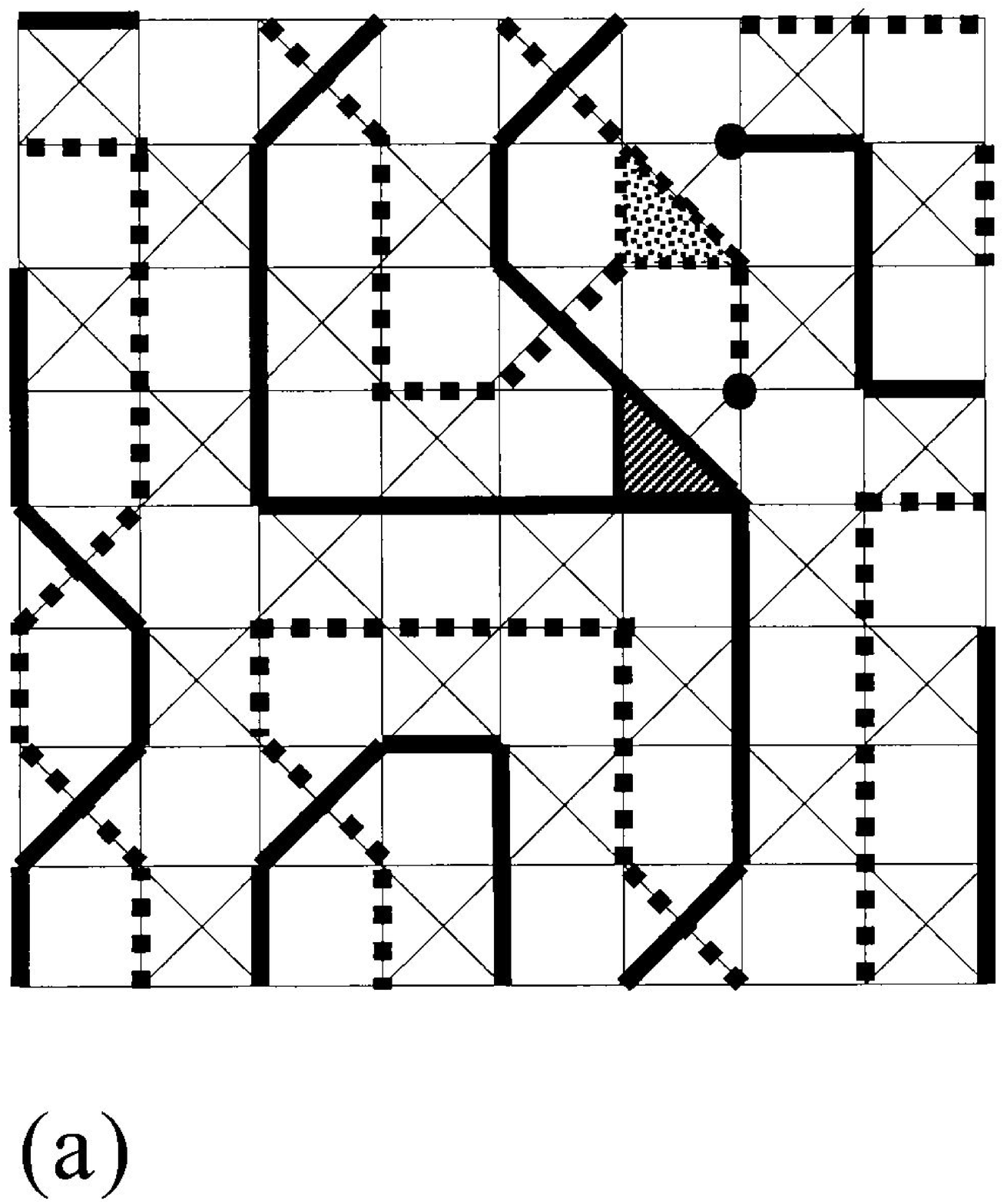}
\hspace{.5cm} \includegraphics[clip,width=4.0cm]{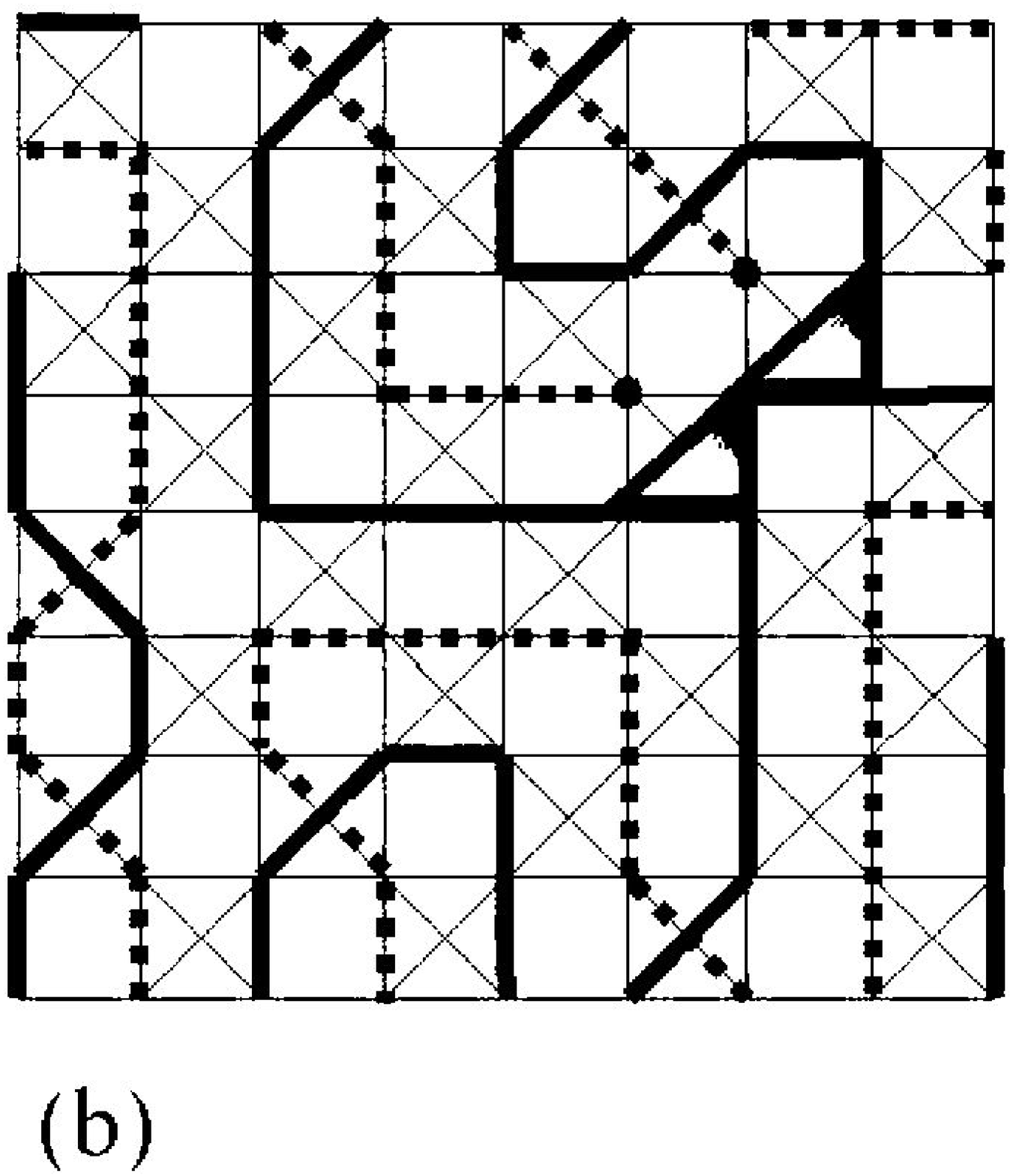}
\hspace{.5cm} \includegraphics[clip,width=4.0cm]{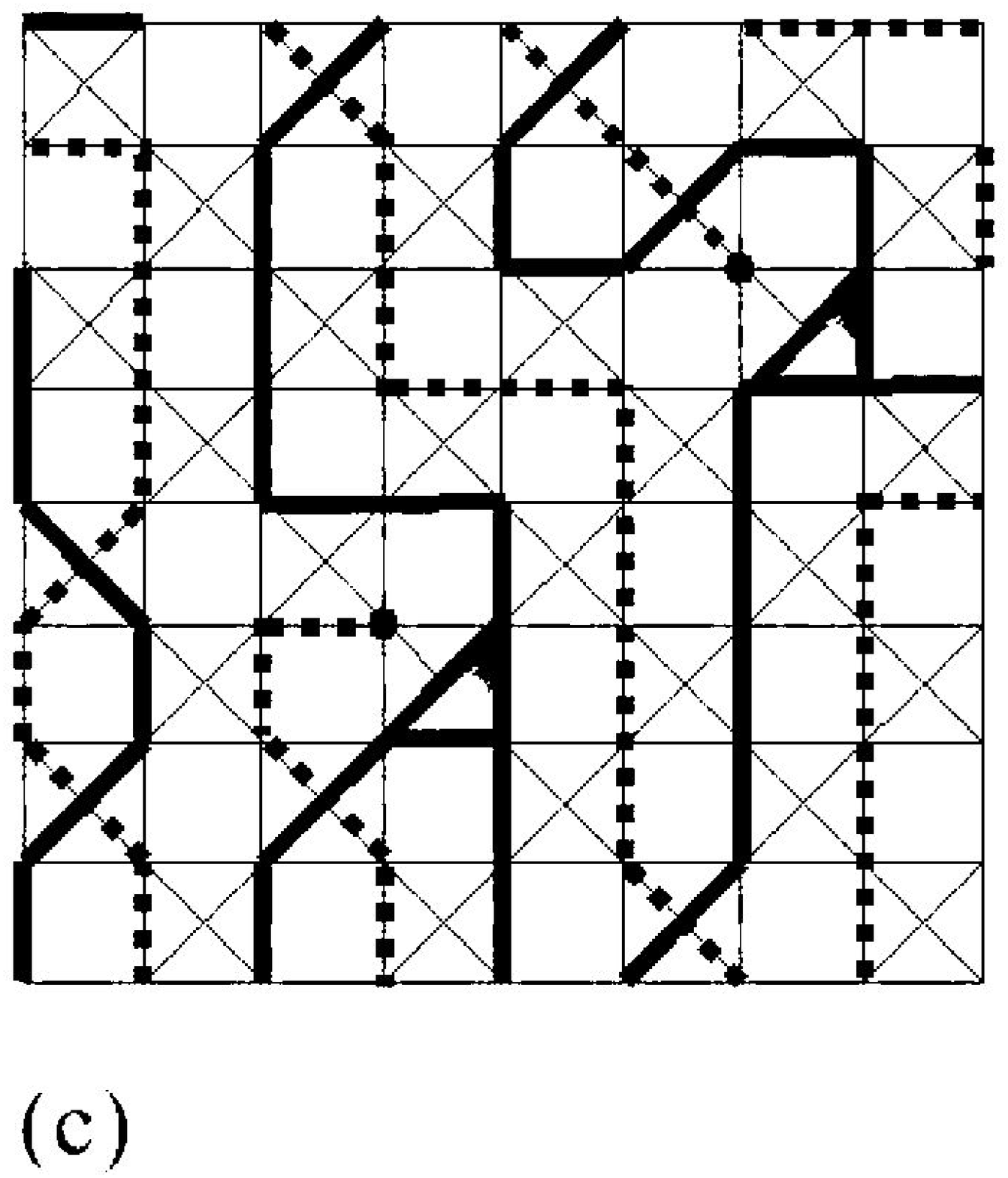}
%\vspace{0.5cm}
\caption{(a) Vacuum fluctuation. One criss-crossed square (tetrahedron)
  contains three particles while another one contains one particle only;
(b) particle added to the otherwise half-filled checkerboard
  lattice; (c) after a hop of one of the electrons the two squares with three
  particles each have separated. (After \cite{FuldeP02})}   
\label{fig07.13}
\end{figure}

In the following we want to go beyond the space of allowed configurations which
obey the tetrahedron rule. When a particle is hopping to a neighboring empty
site (vacuum fluctuation) the tetrahedron rule is broken for two tetrahedra:
now one tetrahedron contains three particles, while another contains one only 
(see Fig. \ref{fig07.13}a). For the corresponding uncrossed squares curl
${\bf f} \neq 0$. By subsequent hopping these two objects can separate carrying
a charge of $\pm e/2$ each. As a result of the vacuum fluctuation the vector
field ${\bf f}(n)$ contains a vortex-antivortex pair. An insulator to metal
transition may be viewed as a proliferation of such pairs and resembles a
Kosterlitz-Thouless transition. It is difficult to compute numerically the
ratio $V/t$ at which this transition takes place, but an estimate yields a
critical ratio of $(V/t)_{\rm cr} \simeq 7$. This is considerably larger than
the value obtained by conventional equations of motion methods in Hubbard
I-type approximation \cite{Zhang05a}.

%%%%%%%%%%%%%%%%%%%%%%%%%%%%%%%%%%%%%%%%%%%%%%%%%%%%%%%%%%%%%%%%%%%%%%%%%%%
%%%%%%%%%%%%%%%%%%%%%%%%%%%%%%%%%%%%%%%%%%%%%%%%%%%%%%%%%%%%%%%%%%%%%%%%%%%

When an extra particle is added to the system two neighboring criss-crossed
squares contain three occupied sites each (see Fig. \ref{fig07.13}b). These two
special squares separate from each other when one of the electrons is hopping
to a neighboring site (see Fig. \ref{fig07.13}c). There is no change in the
particle repulsions associated with this separation of the two. Since the
charge of the added particle is e, each of the separated squares must have
charge e/2. The reason for that is easily seen. When a square with three
particles moves in the checkerboard lattice plane, the squares left behind must
again satisfy the tetrahedron rule. Inspection shows that this requires the
backflow of a charge e/2, thus reducing the charge flow due to hopping of an
electron from e to e/2. For the checkerboard lattice it turns out that ring
hopping processes cause a (weak) restoring force on the two particles with
fractional charge e/2 \cite{Betouras06}. We have not discussed the issue of the
spin here but instead have considered spinless or fully polarized fermions. In
passing we point out that not only do we have spin-charge separation when the
spin is included but also is the spin distributed over parts of the sample. For
more details we refer to Ref. \cite{FuldeP02}. 

The spectral density is a quantity which is expected to show signatures of
fractional charges. For that reason we determine the integrated spectral
density 

%7.23
\begin{equation}
S (\omega) = \sum_{{\bf k} \nu} \left| \left< \left. \psi_{{\bf k} \nu}^{N + 1} \right| \right. c_{{\bf k} \nu}^\dagger \left| \left. \psi_0^N \right> \right.
\right|^2 \delta \left( \omega - E_{{\bf k} \nu}^{N + 1} + E_0^N \right)~~~.
\label{Domega}
\end{equation}  

Here $| \psi_{{\bf k} \nu}^{N + 1} \rangle$ are the different excited states
characterized by momentum ${\bf k}$ and band index $\nu$ of the $N + 1$
particle system and $E_{{\bf k} \nu}^{N+1}$ are the corresponding energies. The
energy of the $N$-particle ground state is $E_0^N$. The $c^\dagger_{{\bf
	k}\nu}$ create particles in quantum states ${\bf k}, \nu$. 

%%%%%%%%%%%%%%%%%%%%%%%%%%%%%%%%%%%%%%%%%%%%%%%%%%%%%%%%%%%%%%%%%%%%%%%%%%
\begin{figure}[t b]
\includegraphics[clip,width=6.0cm]{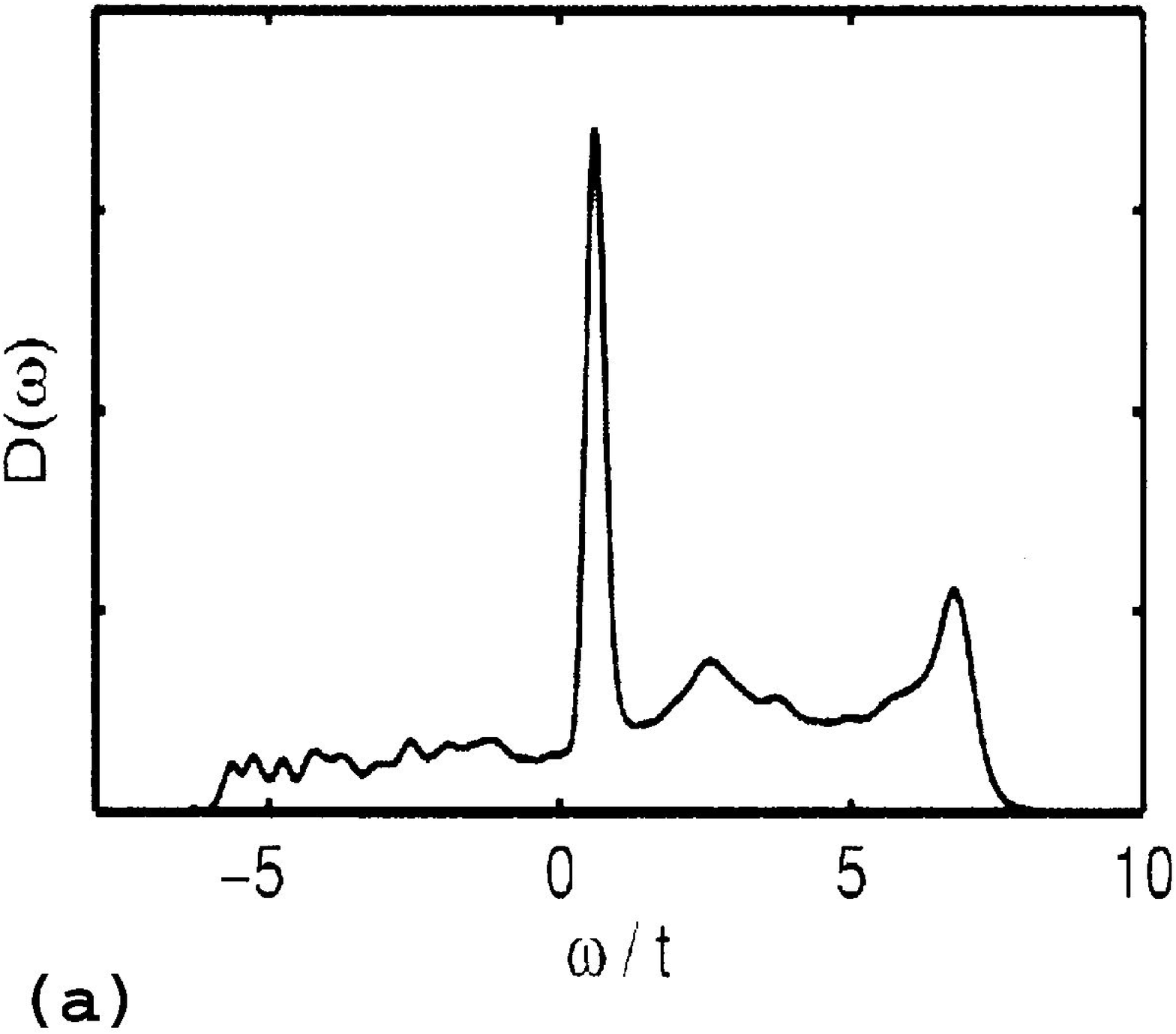}
\hspace{.5cm} \includegraphics[clip,width=6.0cm]{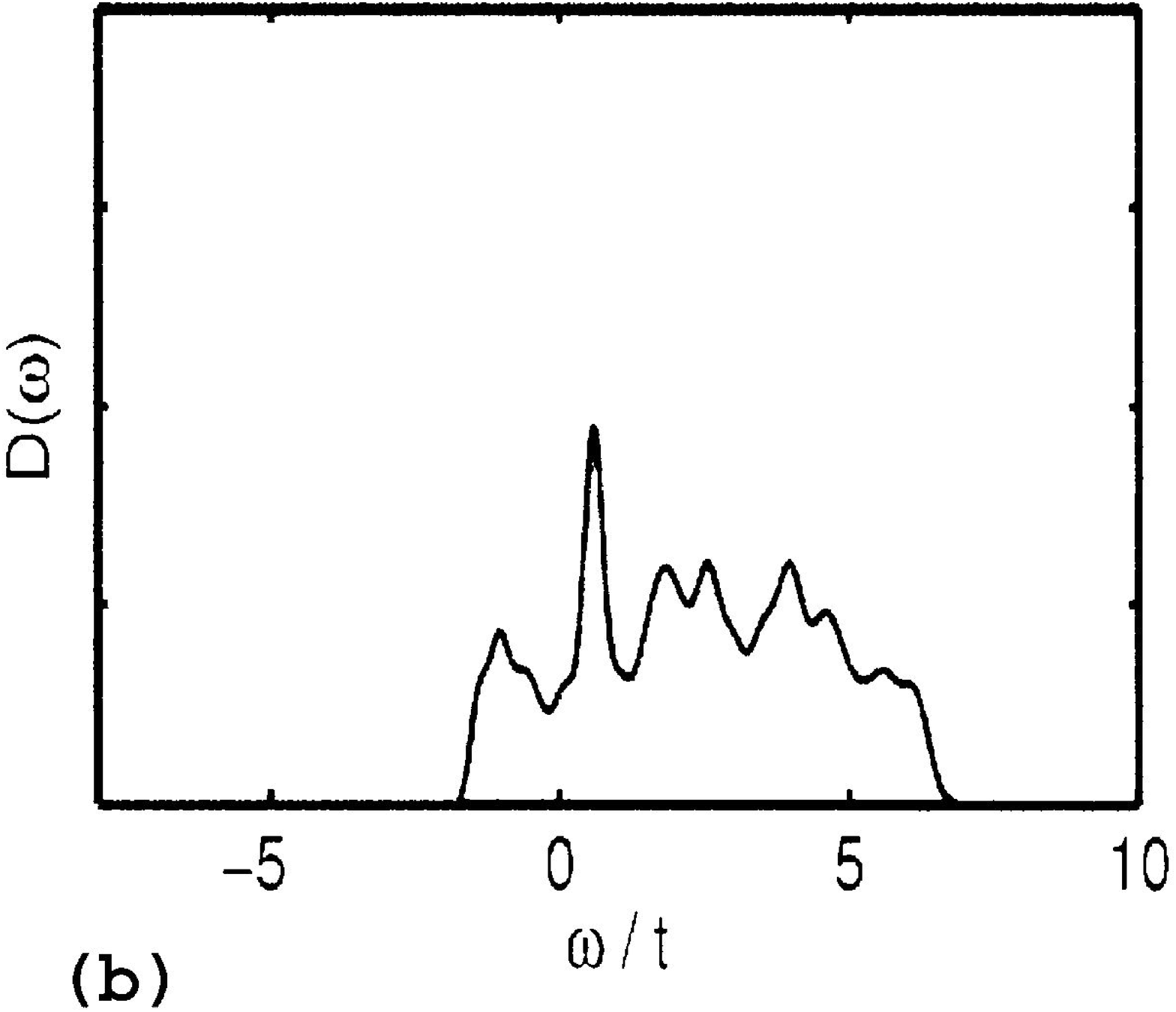}
%\vspace{0.5cm}
\caption{Integrated spectral density as function of energy for a 36 sites
 checkerboard system with 18 particles and V/t = 25. (a) When the 
 excited states are fully accounted for and (b) when only configurations are used in
 which the added particle remains an entity. (After \cite{PollmannF05})}   
\label{fig07.14}
\end{figure}
%%%%%%%%%%%%%%%%%%%%%%%%%%%%%%%%%%%%%%%%%%%%%%%%%%%%%%%%%%%%%%%%%%%%%%%%%%%

For a 32 sites checkerboard cluster with $N = 16$ spinless fermions and $V/t =
25$ the integrated spectral density is shown in Fig. \ref{fig07.14}a. When one
includes only configurations of the $N + 1$ particle system in which
the added particle is not disintegrated like in Fig. \ref{fig07.14}a one
obtains the spectral density shown in Fig. \ref{fig07.14}b. A comparison with
Fig. \ref{fig07.14}a shows a considerable reduction of the width of the
excitation spectrum and an absence of the low frequency part. Therefore we may
conclude that hopping processes leading to a separation of the charge e into
two parts have an important effect on the spectral density and constitute
further evidence for the appearance of fractional charges.  

The above considerations suggest that in 2D there are also other lattices than
the checkerboard one allowing for charge fractionalization. That holds
particularly true for the kagom\'e lattice which at 1/3 filling can support
excitations with charge e/3 and 2e/3 when a spinless fermion is added
\cite{PollmannF06}. The requirement hereby is a short-range repulsion $Vn_in_j$
when $i$ and $j$ are on the same hexagon. 

The fermionic character of the particles discussed here makes the present model
different from related spin models and models for hard core
bosons. Nevertheless, there exist also similarities between the problem of
fractional charges and resonating valence bonds (RVB). For example, Kalmeyer
and Laughlin \cite{Kalmeyer89} have shown that the ground state wavefunction of
an antiferromagnetic Heisenberg Hamiltonian with nearest-neighbor interactions
on a triangular lattice is practically the same as the fractional quantum Hall
effect (FQHE) wavefunction for bosons. As the ground state of a Heisenberg
Hamiltonian on a frustrated lattice is of the RVB type, the two phenomena,
i.e., FQHE effect and RVB's are closely related. RVB systems support spinons,
i.e., a spin flip with $\Delta s_z = 1$ breaks up into two spinons with spin
1/2 each \cite{Moessner01,Balents02,Motrunich02,Misguish04}. The situation
resembles the one in Fig. \ref{fig07.13} in particular when an Ising
Hamiltonian us used. Spins up and down on a checkerboard lattice correspond to
occupied and unoccupied sites and a state with total spin $S^z_{\rm tot} = 0$
has a half-filled lattice of hard-core bosons as analogue. While a large body
of work exists connecting RVB models with confined and deconfined phases of
compact gauge theories (see, e.g., Refs. \cite{Fradkin91,Moessner01,Hermele04})
corresponding work for fermionic systems is still missing.

\begin{figure}[t b]
\includegraphics[clip,width=6.0cm]{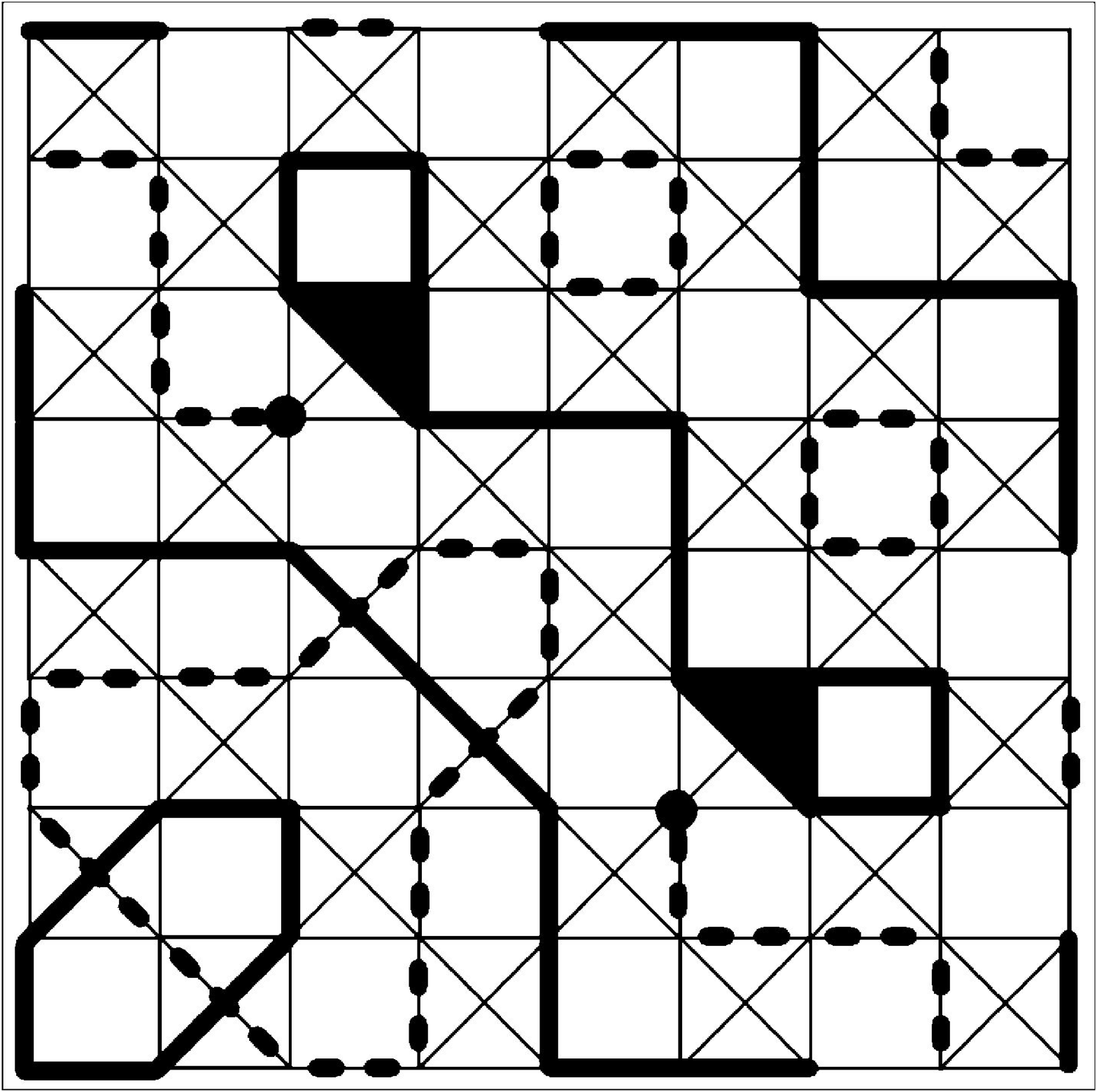}
%\vspace{0.5cm}
\caption{String (marked in red) of an odd number of sites (here five)
  connecting two particles with charge e/2.}   
\label{fig07.15}
\end{figure}

Finally we want to comment on some modifications which occur when the spin of
the fermions is included. The question may be asked where the spin of an added
electron goes when the excitation falls into two parts with charge e/2
each. The answer is found by looking at Fig. (\ref{fig07.15}) where
periodic boundary conditions are used. The two charges e/2 are connected by a
spin chain containing an odd number of sites. The ground state of that chain is
two-fold degenerate and represents the spin degrees of freedom. Thus one may
state that the spin is smeared over parts of the system as is the connected
spin chain. This is a rather novel feature which we have not met 
before in condensed matter physics to the best of our knowledge. It should be
also noticed that in the presence of spins the lowest order ring exchange
process on a checkerboard lattice involves four sites instead of six. This is
different for the pyrochlore lattice where the smallest possible ring involves
six sites (see Fig. \ref{fig07.4}). So the ground-state degeneracy is lifted in
order $t^2/V$. 

%%%%%%%%%%%%%%%%%%%%%%%%%%%%%%%%%%%%%%%%%%%%%%%%%%%%%%%%%%%%%%%%%%%%%%%%%%
%%%%%%%%%%%%%%%%%%%%%%%%%%%%%%%%%%%%%%%%%%%%%%%%%%%%%%%%%%%%%%%%%%%%%%%%%%%

\newpage

\section{High-Energy Excitations}
\resetdoublenumb 
\resetdoublenumbf

\label{Sect:HighEnergyExcitations}

In Sec. \ref{Sect:SignStrongCorr} we demonstrated that strong electron
correlations generate characteristic low energy scales. They are much smaller
than the Fermi energy, the typical energy scale in a metal. But it is well
known that strong correlations influence high-energy excitations as
well. Hereby the expression {\it high energy} is used quite flexible. For
example, it may include satellite structures or peaks as they are found in the
photoelectron spectra of Ni metal \cite{Huefner74,Eberhardt80} or of the high
temperature superconducting cuprates. But it also includes energies which are
larger than the one at which Fermi liquid behavior breaks down. In practice
this may be a rather low energy or temperature. These effects are best
studied by investigating the one-particle Green's function. In Dyson's
representation it reads 

%8.1
\begin{equation}
G_\nu ({\bf k}, z) = \frac{1}{z - \epsilon_\nu ({\bf k}) - \Sigma_\nu ({\bf k},
  z)}~~~. 
\label{Gnukz}
\end{equation}  

The energy $\epsilon_\nu ({\bf k})$ describes the dispersion of electrons in
band $\nu$ within an effective single-electron approximation while the
self-energy $\Sigma_\nu ({\bf k}, \omega)$ contains all effects beyond
that approximation. In case that $\epsilon_\nu ({\bf k})$ describes the
Hartree-Fock bands, $\Sigma_\nu ({\bf k}, \omega)$ contains the electronic
correlations. When Landau's Fermi liquid theory does apply, the low-energy
excitations near the Fermi energy are quasiparticles. They are obtained from

%8.2
\begin{equation}
G ({\bf k}, \omega) = \frac{Z}{\omega - \epsilon_{qp} ({\bf k}) - i \gamma_{\bf
	k} {\rm sgn}~\omega} + G_{\rm inc}({\bf k}, \omega) ~~~. 
\label{Gkomeg}
\end{equation}  

\noindent For convenience we have omitted the band index $\nu$. The first term
contains the quasiparticle pole at

%8.3
\begin{equation}
\omega = \epsilon_{qp} ({\bf k}) + i \gamma_{\bf k}
\label{omepsqp}
\end{equation} 

\noindent with the quasiparticle dispersion $\epsilon_{qp}({\bf k})$ and
lifetime $\gamma_{\bf k}$. The second term is the incoherent part which is of
special interest here. Usually it is not further discussed but we want to draw
attention to its importance and to significant features when electron
correlations are strong. We may think of it as being due to the internal
degrees of freedom of the correlation hole which is surrounding an
electron. The bare electron has weight $Z < 1$ in the quasiparticle. The
reduction is due to the interaction of the electron (or hole) with the
surroundings. The internal degrees of freedom of the modified surroundings,
i.e., the correlation hole shows up in the integrated spectral density in form
of quasiparticle damping and satellite structures. Those structures have
generally a ${\bf k}$ dependence which is much weaker than the one of the
quasiparticles. An example is the shadow band in the Hubbard model when the
Hubbard I approximation \cite{Hubbard63} is made. 

The internal degrees of freedom of the correlation hole are best described by
the projection operator method. The idea is to select those particular
operators which describe the most important microscopic processes in setting up
the correlation hole. Green's function is determined within that restricted
operator space. The Hubbard I approximation to the Hubbard model is the
simplest example. Here only one operator, i.e., $c^+_{i \sigma} n_{i - \sigma}$
where $i$ is the site index is used in order to describe the correlation
hole. More interesting cases are discussed in the following subsections. 

We want to draw attention to the zero-point fluctuations of internal excitation
modes of the correlation hole. A well known example is Gaskell's ansatz
\cite{Gaskell58} for the ground-state $|\psi_0 \rangle$ of a homogeneous
electron gas. It is of the form 

%8.4
\begin{equation}
\mid \psi_0 \rangle = {\rm exp} \left( \sum_{\bf q} \tau ({\bf q})
\rho^\dagger_{\bf q} \rho_{\bf q} \right) \left. \left| \Phi^{\rm SCF}_0
\right. \right>  
\label{psi0ranexp}
\end{equation} 

\noindent where $|\Phi^{\rm SCF}_0 \rangle$ is the Hartree-Fock ground state
and $\rho_{\bf q}$ are the density fluctuations of wavenumber $q$. Furthermore
$\tau (q) \sim q^{-2}$ when $q \rightarrow 0$. Here the zero-point fluctuations
of plasmons are taken into account which are the internal excitation modes of
the correlation hole when the latter is described within the random phase
approximation (RPA). It is well known that the RPA models very well the
long-range part of the correlation hole but fails to describe properly the
short-range part. Zero-point fluctuations of other modes, in particular of
those associated with the short-range part of the correlation hole enter in a
similar way the Jastrow prefactor in Eq. (\ref{psi0ranexp}). We start out by
describing the projection operator formalism before we discuss a number of
applications.  

\subsection{Projection Operators}

In the following we shall use the retarded Green function in order to determine
the excitations of the system. It is given by

%8.5
\begin{equation}
G^R_{\nu \sigma} \left( {\bf k}, t \right) = -i \Theta (t) \left< \psi_0 \left|
\left[ c_{\nu \sigma} ({\bf k}, t), c^\dagger_{\nu \sigma} ({\bf k}) \right]_+
\right| \psi_0 \right>  
\label{GRnusig}
\end{equation}

where $| \psi_0 \rangle$ is the exact ground state and $\Theta (t)$ is the step
function, i.e., $\Theta (t) = 1$ for $t \geq 0$ and zero otherwise. The
superscript R will be left out in the following discussion for simplicity. We
assume that the Fourier transforms of the most important microscopic processes
for the generation of the correlation hole are represented by a set of
operators 
$\{ A_\mu ({\bf k}) \}$, the dynamical variables. But we want to include in
this set also the original operators $c^\dagger_{\nu \sigma} ({\bf
  k})$. Explicit examples for proper choices of the $ A_\mu ({\bf k})$ will be
given when specific applications of this method are discussed. Within that
reduced operator space $\Re_0$ spanned by the $\{ A_\nu ({\bf k}) \}$ we define
the Green function matrix 

%8.6
\begin{equation}
G_{\mu \nu} \left( {\bf k}, t \right) = -i \Theta (t) \left< \psi_0 \left|
\left[ A^\dagger_\mu ({\bf k}, t), A_\nu ({\bf k}, 0) \right]_+ \right| \psi_0
\right>~~~.  
\label{Gmunukt}
\end{equation}

\noindent Note that the Green function (\ref{GRnusig}) is just a diagonal
element of that matrix. It is convenient to introduce the following notation

%8.7
\begin{equation}
(A|B)_+ = \left< \psi_0 \left| \left[ A^\dagger, B \right]_+ \right| \psi_0
  \right>~~~. 
\label{ABplus}
\end{equation}

\noindent This enables us to rewrite (\ref{Gmunukt}) after a Fourier
transformation in the condensed form

%8.8
\begin{equation}
G_{\mu \nu} ({\bf k}, z) = \left( A_\mu \left| \frac{1}{z - L} \right. A_\nu
\right)_+ 
\label{GmnkzA}
\end{equation}

\noindent where $z = \omega + i \eta$ and $\eta$ is a positive infinitesimal
number. The Liouvillean $L$ corresponds to $H$ and is defined by its
action on an arbitrary operator $A$ through

%8.9
\begin{equation}
LA = i \frac{dA}{dt} = \left[ H, A \right]_- 
\label{LAHA}
\end{equation}

so that

%8.10
\begin{equation}
A(t) = e^{iLt} A(0)~~~.
\label{Ate}
\end{equation}

By making use of the identity

%8.XX
\begin{equation}
\frac{1}{a + b} = \frac{1}{a} - \frac{1}{a} b \frac{1}{a + b}\nonumber
\label{abaab}
\end{equation}

we can rewrite (\ref{GmnkzA}) in the form of the following matrix equation

%8.11
\begin{equation}
\left[ z \openone - \left( \mathbb{L} + \mathbb{M} (z) \right)
\mathbb{\bbchi}^{-1} \right] \mathbb{G} (z) = \mathbb{\bbchi}
\label{openone}
\end{equation}

with matrix elements

%8.12
\begin{eqnarray}
L_{\mu \nu} & = & \left( A_\mu \left| LA_\nu \right)_+ \right. \nonumber \\
\chi_{\mu \nu} & = & \left( A_\mu \left| A_\nu \right)_+ \right. \nonumber \\
M_{\mu \nu} (z) & = & \left( A_\mu \left| LQ \frac{1}{z - QLQ} QLA_\nu
\right)_+ \right.~~~. 
\label{LchiM}
\end{eqnarray}

The matrix $\mathbb{M} (z)$ is called memory function
\cite{Mori65,ZwanzigBook61}. It couples the relevant operators $\left\{ A_\nu
\right\}$ to the remaining degrees of freedom. The operator $Q$

%8.13
\begin{equation}
Q = 1 - \sum_{ij} \left. \left| A_i \right. \right)_+ \chi^{-1}_{ij} \left(
\left. A_j \right| \right. 
\label{Q1ij}
\end{equation}

\noindent projects onto those remaining degrees, i.e., onto an operator space
perpendicular to the $\left\{ A_\nu \right\}$, i.e., $Q | A_\nu)_+ =
0$. Setting $\mathbb{M} (z) = 0$ implies that the dynamics of
the system is approximated by the  $\left\{ A_\nu \right\}$ and takes place
within $\Re_0$. One may either choose a large basis $\left\{ A_\nu
\right\}$ and set $\mathbb{M} (z) = 0$ or work with a small basis and keep
$\mathbb{M} (z)$. Examples are given below. If not stated otherwise we shall
set $\mathbb{M} (z) = 0$.

The dimension of the matrix equation (\ref{openone}) equals the number of
dynamical variables $A_\nu ({\bf k})$. The energy resolution of the excitation
spectrum depends on the size of the set $\{A_\nu ({\bf k})\}$. For high-energy
excitations a relatively small number of dynamical variables is sufficient. By
increasing their number one can increase the energy resolution of the spectral
density calculated from Eq. (\ref{GmnkzA}).

\subsection{The Hubbard Model: Appearance of Shadow Bands}  

The Hubbard model shows particularly well a number of generic features caused
by strong electron correlations. Among them are the appearance of satellite
structures, shadow bands and a marginal Fermi liquid like behavior close to
half filling. The Hamiltonian is of the well known form

%8.14
\begin{eqnarray}
H & = & -t \sum_{\langle ij \rangle} \left( c^\dagger_{i \sigma} c_{j \sigma} +
h.c. \right) + U \sum_i n_{i \uparrow} n_{i \downarrow}\nonumber\\
& = & H_0 + H_I
\label{H_tij}
\end{eqnarray}

\noindent in standard notation. In the following we discuss solely for
pedagogical and illustrative reasons the spin-density wave (SDW) and Hubbard I
approximation by applying projection operators. The simplest case in which a
shadow band does appear, is a square lattice at half filling when a
spin-density wave (SDW) approximation is made. This simple approximation leads
to an antiferromagnetic ground state $| \Phi_{AF} \rangle$. Charge fluctuations
are suppressed here on a mean-field level by symmetry breaking. Breaking a
symmetry can reduce intersite charge fluctuations similarly as strong
correlations do without symmetry breaking. Therefore features of strong
correlations show up already in this simple mean-field scheme.

From the kinetic energy term in (\ref{H_tij}) one obtains the dispersion
$\epsilon ({\bf k}) = -2t ({\rm cos} k_x + {\rm cos} k_y)$. The lattice
constant has been set equal to unity. Therefore, at half filling the Fermi
surface is nested and the ground state in mean-field approximation is a spin
density wave. One finds

%8.15
\begin{equation}
\left| \left. \Phi_{AF} \right> \right. = \prod_{{\bf k} \sigma} \left[
  u_{\bf k} c^\dagger_{{\bf k} \sigma} + \sigma v_{\bf k} c^\dagger_{{\bf k} +
  {\bf Q} \sigma} \right] \mid 0 \rangle  
\label{PhiAFpi}
\end{equation}

\noindent where ${\bf Q}$ is a reciprocal lattice vector and $c^\dagger_{{\bf
  k} \sigma}$ is the Fourier transform of $c^\dagger_{i \sigma}$. Furthermore
  $u^2_{\bf k} + v^2_{\bf k} = 1$ with

%8.16
\begin{eqnarray}
u^2_{\bf k} & = & \frac{1}{2} \left( 1 - \frac{\epsilon({\bf k})}{E({\bf k})}
  \right)~~~,~~v^2_{\bf k} = \frac{1}{2} \left( 1 + \frac{\epsilon({\bf
  k})}{E({\bf k})}\right) \nonumber\\ 
E({\bf k}) & = & \left( \epsilon({\bf k})^2 + \frac{m^2_0~ U^2}{4} \right)
  ^\frac{1}{2}~~~ .  
\label{u2k12}
\end{eqnarray}

\noindent Here $m_0$ is the staggered magnetization. The latter has to be
calculated self-consistently. In order to calculate the excitations we choose
for the relevant dynamical variables $\left\{ A_\nu \right\}$

%8.17
\begin{equation}
A_1 ({\bf k}) = c^\dagger_{{\bf k} \sigma}~~~, ~~A_2 ({\bf k}) =
c^\dagger_{{\bf k} + {\bf Q} \sigma}~~~.
\label{A1kc}
\end{equation}

When the 2 x 2 matrix (\ref{openone}) with $\mathbb{M} (z) = 0$ is evaluated
one finds 

%8.18
\begin{equation}
G_{11} \left( {\bf k}, z \right) = \frac{u^2_k}{z - \left( \frac{U}{2} - E
  ({\bf k}) \right)} + \frac{v^2_k}{z - \left( \frac{U}{2} + E ({\bf k})
  \right)} 
\label{G11kz}
\end{equation}

\noindent and for the spectral density
%8.19
\begin{equation}
D \left( {\bf k}, \omega \right) = u^2_{\bf k} \delta \left( \omega -
\frac{U}{2} + E ({\bf k}) \right) + v^2_{\bf k} \delta \left( \omega -
\frac{U}{2} - E ({\bf k}) \right)~~~.  
\label{Dk}
\end{equation}

For each ${\bf k}$ point there are two contributions to $D( {\bf k}, \omega)$,
i.e., one $\delta$-function peak with a large weight and one with a small
one. The peaks with the smaller weight form a shadow band $E_{sb} ({\bf k}) =
-E ({\bf k})$ which complements the band resulting from the peaks with large
weight \cite{Kampf90,Kampf94}. Because of the mean-field level the shadow band
disappears for temperatures higher than the N\'eel temperature. This is in
reality not the case and therefore one would like to reproduce a shadow band
also for arbitrary filling factors and for the paramagnetic state. This can be
done, of course, only by accounting for the strong correlations. They are
described in the simplest way by the Hubbard I \cite{Hubbard63}
approximation. In that case the following choice is made for the dynamical
variables $\{ A_\nu \}$

%8.20
\begin{equation}
A_1 ({\bf k}) = c^\dagger_{{\bf k} \sigma}~~~, ~~A_2 ({\bf k}) =
\frac{1}{\sqrt{N_0}} \sum_i e^{i{\bf k R}_i} c^\dagger_{i \sigma} \delta n_{i -
  \sigma} 
\label{A1kck}
\end{equation}

\noindent with $\delta n_{i - \sigma} = n_{i - \sigma} - \langle n_{i - \sigma}
\rangle$. The ${\bf R}_i$ denote the positions of the $N_0$ lattice
sites. Again, we want to determine $G_{11} ({\bf k}, \omega)$ from
(\ref{GmnkzA}) by setting $\mathbb{M} (z) = 0$. The 2 x 2 matrix equation can
be easily solved and one finds

%8.21
\begin{equation}
G_{11} ({\bf k}, z) = \left[ z - \epsilon ({\bf k}) - \frac{U}{2} n \left( 1 +
  \frac{\hat{U}}{z - \hat{U}} \right) \right]^{-1}
\label{G11kzz}
\end{equation}

\noindent where $n$ is the number of electrons per site and $\hat{U} = U (1 -
n/2)$. $G_{11} ({\bf k}, \omega)$ has two poles centered at $z = 0$ and $z =
U$. Neglecting terms of order $U^{-1}$ we can rewrite (\ref{G11kzz}) as

%8.22
\begin{equation}
G_{11} ({\bf k}, z) = \frac{1 - \frac{n}{2}}{z - \epsilon ({\bf k}) \left( 1 -
  \frac{n}{2} \right)} + \frac{\frac{n}{2}}{z - U - \epsilon ({\bf k})
  \frac{n}{2}}~~~.  
\label{G11kzn}
\end{equation} 

\begin{figure}[t b]
\includegraphics[clip,width=7.0cm]{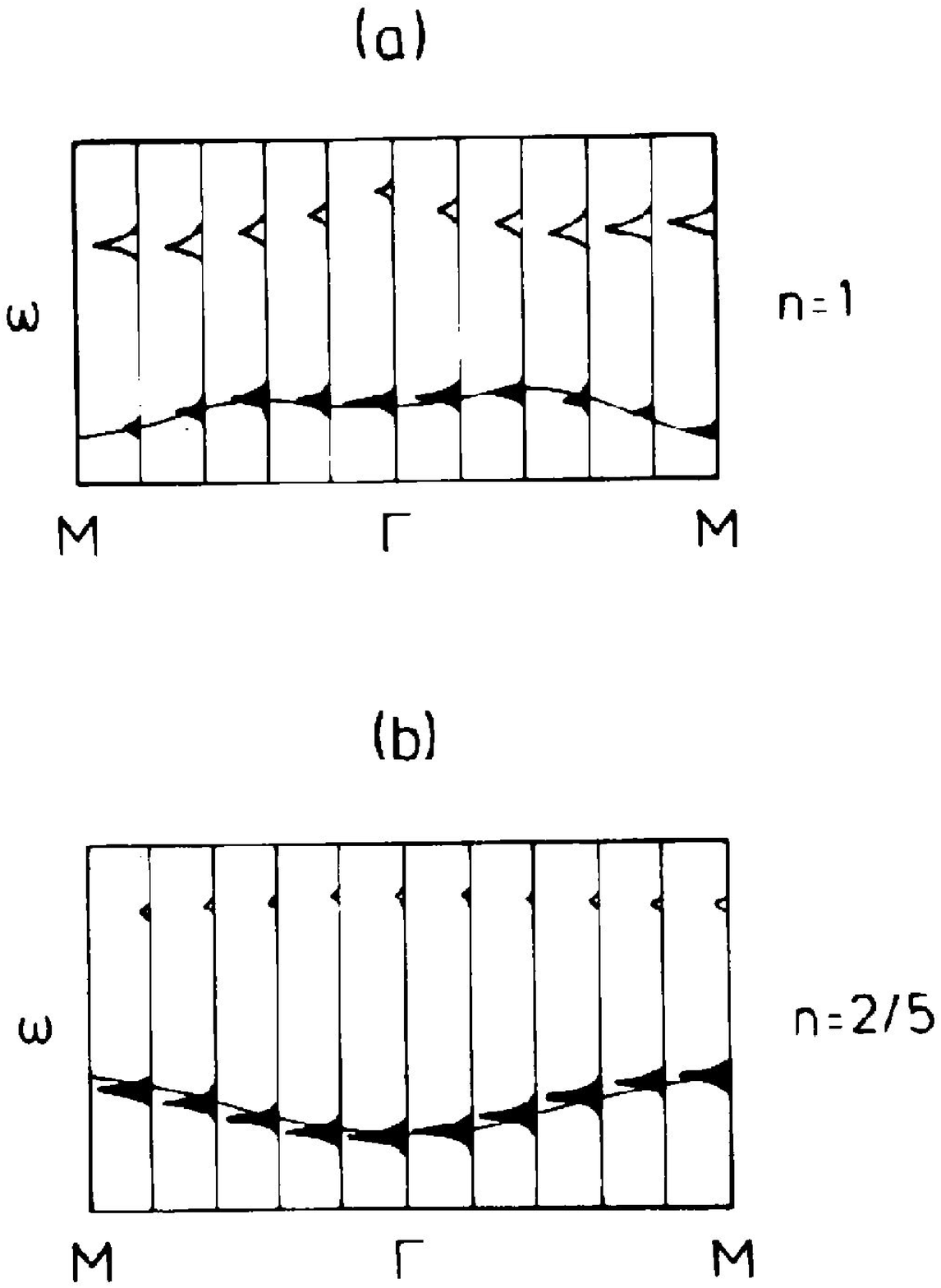}
%\vspace{0.5cm}
\caption{Schematic representation of the two bands resulting from the two
  variables of Eq. (\ref{A1kck}), i.e., the upper and the lower Hubbard band.
  The electron number per site is $n$, the broadening is artificial.}
\label{fig08.1}
\end{figure}

The poles give raise to two bands, i.e., the upper and the lower Hubbard
band. Their widths differ except for $n = 1$. For $n = 0.25$ the upper Hubbard
band is reduced to a satellite structure with a ${\bf k}$-dependent
dispersion. This is illustrated in Fig. \ref{fig08.1} and is well
known. The reason for repeating these facts here is that we want to proceed
similarly when we discuss the satellite structure in Ni or the spectral density
of electrons in Cu-O planes. We want to  emphasize the point of view that the
incoherent part of a Green function is the superposition of satellite peaks
which have a small ${\bf k}$-dependence each. But before, we demonstrate that
for special band fillings a Hubbard model can show marginal Fermi liquid
behavior. This is shown explicitely for a square lattice.

%%%%%%%%%%%%%%%%%%%%%%%%%%%%%%%%%%%%%%%%%%%%%%%%%%%%%%%%%%%%%%%%%%%%%%%%%%%
%1
%%%%%%%%%%%%%%%%%%%%%%%%%%%%%%%%%%%%%%%%%%%%%%%%%%%%%%%%%%%%%%%%%%%%%%%%%%%

\subsection{Marginal Fermi Liquid Behavior and Kink Structure}  

The Coherent Potential Approximation (CPA) has been widely used in electronic
structure calculations of disordered systems
\cite{Soven67,Taylor67,Velicky68,Yonezawa68,Elliott74,Ehrenreich76}. Here we
want to combine it with the projection operator technique and apply it to the
Hubbard model. We remind the reader that a CPA was introduced first by
Hubbard when treating his Hamiltonian \cite{Hubbard64}. In the
following we want to treat the many-electron problem as accurately as possible
for a number of sites which need not be connected. This cluster is embedded in
a medium with a coherent potential $\tilde{\Sigma}(\omega)$. The potential is
determined self-consistently so that it agrees with the momentum integrated
self-energy $\Sigma ({\bf k}, \omega)$ of the cluster. It is known
\cite{Kakehashi04,KakehashiYo02} that the local projection operator method
combined with the CPA \cite{KakehashiYos04} is equivalent to the Many-Body CPA
\cite{Hirooka77}, the Dynamical CPA \cite{Kakehashi92,KakehashiY92} as well as
to the Dynamical Mean Field Theory (DMFT) which is based on many-body physics
in infinite dimensions \cite{Metzner89,MuellerHartmann89}. For reviews of the
latter see \cite{Pruschke95,Georges96}. Here we want to discuss the more
general nonlocal version of the projection operator method \cite{KakehashiY04}
combined with the CPA. It is based on a decomposition of the scattering matrix
of the system into one-site, two-site etc. scattering matrices. This way the
scattering matrix can be calculated successively in terms of increments, a
method successfully applied in the theory of wavefunction based electronic
structure calculations of solids. For a review of that method see Refs.
\cite{Stoll05,Fulde02}. We apply the theory to the Hubbard model on a square
lattice near half filling \cite{Hubbard63,Hubbard64}. The results are
interesting. A marginal Fermi liquid behavior is found for large $U$ values and
small hole doping and a corresponding phase diagram is worked out
\cite{Kakehashi05}. It is also found that for similar parameters the excitation
spectrum (or real part of the self-energy) has a kink near $E_F$
\cite{KakehashiY05}. Such a structure was observed in a number of underdoped
cuprates like Bi$_2$Sr$_2$CaCuO$_{8 + \delta}$ or La$_{2 -
  \delta}$Sr$_\delta$CuO$_4$ \cite{Bogdanov00,Cuk05}. It has been attributed to
electron-phonon interactions \cite{Shen02,Verga03,Ishihara04,Cuk05} and also to
the interaction with a magnetic resonance mode observed by inelastic neutron
scattering \cite{Eschrig00,Johnson01,Eschrig03,Schachinger03}. But as it turns
out, also strong electron correlations in a 2D Hubbard model can produce it.  

To reach our goal we start out by choosing for the $\{ A_\nu \}$ simply the
operators $c^\dagger_{i \sigma}$ but we keep this time the memory matrix, i.e.,
$\mathbb{M} (z) \neq 0$. In that case we write for a paramagnetic system

%8.23
\begin{equation}
G ({\bf k}, z) = \frac{1}{z - \epsilon ({\bf k}) - U^2 M ({\bf k}, \omega)}~~~.
\label{Gkz1z}
\end{equation}  

The matrix $\mathbb{L}$ in (\ref{openone}) gives only a constant energy shift
and is neglected here. For convenience a factor $U^2$ has been extracted from
$M ({\bf k}, z)$. We decompose

%8.24
\begin{equation}
M ({\bf k}, z) = \sum_j M_{j_0} (z) e^{i{\bf kR}_j} 
\label{Mkzj}
\end{equation}  

\noindent and write for the reduced memory matrix according to (\ref{LchiM})

%8.25
\begin{equation}
M_{ij} (z) = \left( c^\dagger_{i \sigma} \delta n_{i - \sigma} \left|
\frac{1}{z - \bar{L}} \right. c^\dagger_{j \sigma} \delta n_{j - \sigma}
\right)_+~~~, 
\label{MijzC}
\end{equation}

\noindent where $\bar{L} = QLQ$ with the projector $Q = 1 - \sum_{i \sigma} |
c^\dagger_{i \sigma})_+ (c^\dagger_{i \sigma}|$. Stopping at this stage, i.e.,
neglecting the new memory function in the denominator of (\ref{MijzC}) would
bring us back to the Hubbard I approximation. In order to establish a
connection to the CPA we define the Hamiltonian of an effective medium through

%8.26
\begin{equation}
\tilde{H} (z) = H_0 + \tilde{\Sigma} (z) \sum_i n_i
\label{HzH0z}
\end{equation}

\noindent with the corresponding Liouvillean $\tilde{L}$, i.e., $\tilde{L}B =
[ \tilde{H} (z), B ]_-$ for arbitrary operators $B$. The aim is to account
efficiently for local as well as nonlocal correlations by means of an effective
medium characterized by $\tilde{\Sigma} (z)$. 

The coherent potential $\tilde{\Sigma} (z)$ is determined self-consistently
from 

%8.27
\begin{equation}
\tilde{\Sigma} (z) = \frac{U^2}{N} \sum_{\bf k} M ({\bf k}, z)
\label{sumzu2}
\end{equation} 

\noindent where $N$ is the number of sites. With this in mind we decompose 

%8.28
\begin{eqnarray}
\bar{L} & = & Q\tilde{L}Q + \sum_i Q \left( U \delta n_{i \uparrow} \delta n_{i
  \downarrow} - \tilde{\Sigma} (z)n_i \right) Q \nonumber \\ 
& = & L_0 (z) + \sum_i L^{(i)}_I (z) \nonumber \\
& = & L_0 (z) + L_I (z)~~~. 
\label{LiQU}
\end{eqnarray}

Next we express $(z - \bar{L})^{-1}$ in terms of a scattering (super)operator
$T$ as 

%8.29
\begin{eqnarray}
\frac{1}{z - \bar{L}} & = & \frac{1}{z - L_0} + \frac{1}{z - L_0} T \frac{1}{z
  - L_0} \nonumber \\  
& = & g_0 (z) + g_0 (z) T g_0 (z)
\label{frac1zL}
\end{eqnarray}

\noindent with 

%8.XX
\begin{equation}
T = L_I + L_I g_0 L_I + ...~~~.\nonumber
\label{}
\end{equation}

\noindent The $T$-operator can be decomposed into a sequence of many-sites
increments

%8.30
\begin{equation}
T = \sum_i T_i + \sum_{\langle ij \rangle} \delta T_{ij} + \sum_{\langle ijk \rangle} \delta T_{ijk} + ...~~~.
\label{TiTiij}
\end{equation}

Those increments are closely related to the $T$ operators $T_i, T_{ij},
T_{ijk}, ...$ of single-site, 2-sites, 3-sites etc. clusters. It is

%8.31
\begin{eqnarray}
\delta T_{ij} & = & T_{ij} - T_i - T_j \nonumber \\  
\delta T_{ijk} & = & T_{ijk} - \delta T_{ij} - \delta T_{ik} - \delta T_{jk} -
T_i - T_j - T_k \nonumber \\ 
& \vdots &
\label{TijTij}
\end{eqnarray}

\noindent This enables us to introduce retarded memory functions for clusters

%8.32
\begin{equation}
M^{(c)}_{ij} (z) = \left( c^\dagger_{i \sigma} \delta n_{i - \sigma} \left|
\frac{1}{z - \bar{L}^{(c)}} \right. c^\dagger_{j \sigma} \delta n_{j - \sigma}
\right)_+~~~. 
\label{Mcijzci}
\end{equation}

\noindent These matrices have dimensions 1 x 1 when $c = i$ (one-site cluster)
and 2 x 2 when $c = (i, j)$, i.e., in the 2-site cluster approximation. The
Liouvillean $\bar{L}^{(c)} (z)$ is given by

%8.33
\begin{equation}
\bar{L}^{(c)} = L_0 (z) + \sum^{N_c}_{n \in c} L^{(n)}_I (z)~~~.
\label{LcL0}
\end{equation}

The sum over $n$ involves all $N_c$ sites belonging to a given cluster
$c$. Within zeroth-order renormalized perturbation theory (RPT-0) only that
part of $L^{(n)}_I (z)$ is used which projects onto the operators $c^\dagger_{i
  \sigma} \delta n_{i - \sigma}$, i.e., $\bar{P} L^{(n)}_I (z) \bar{P}$ with 
the projector.

%8.34
\begin{equation}
\bar{P} = \sum_{i \sigma} \left. \left| c^\dagger_{i \sigma} \delta n_{i -
  \sigma} \right. \right)_+ \chi^{-1}_i \left( \left. c^\dagger_{i \sigma}
  \delta n_{i - \sigma} \right| \right. 
\label{barPi}
\end{equation}

\noindent and $\chi_i = \langle n_{i - \sigma} \rangle (1 - \langle n_{i -
  \sigma} \rangle )$. With this simplification the memory matrix can be
  expressed in terms of a ''screened'' one as

%8.35
\begin{equation}
M^{(c)}_{ij} (z) = \left[ \mathbbm{g} \cdot \left( 1 - \mathbb{L}^{(c)}_I
  \mathbbm{g} \right)^{-1} \right]_{ij}~~~. 
\label{Mcgg}
\end{equation}

\noindent The screened memory matrix is given by

%8.36
\begin{equation}
g_{ij} (z) = \left( c^\dagger_{i \sigma} \delta n_{i - \sigma} \left|
\frac{1}{z - L_0 (z)} \right. c^\dagger_{j \sigma} \delta n_{j - \sigma}
\right)_+~~~. 
\label{gijzZ}
\end{equation}

\begin{figure}[t b]
\includegraphics[clip,width=8.0cm]{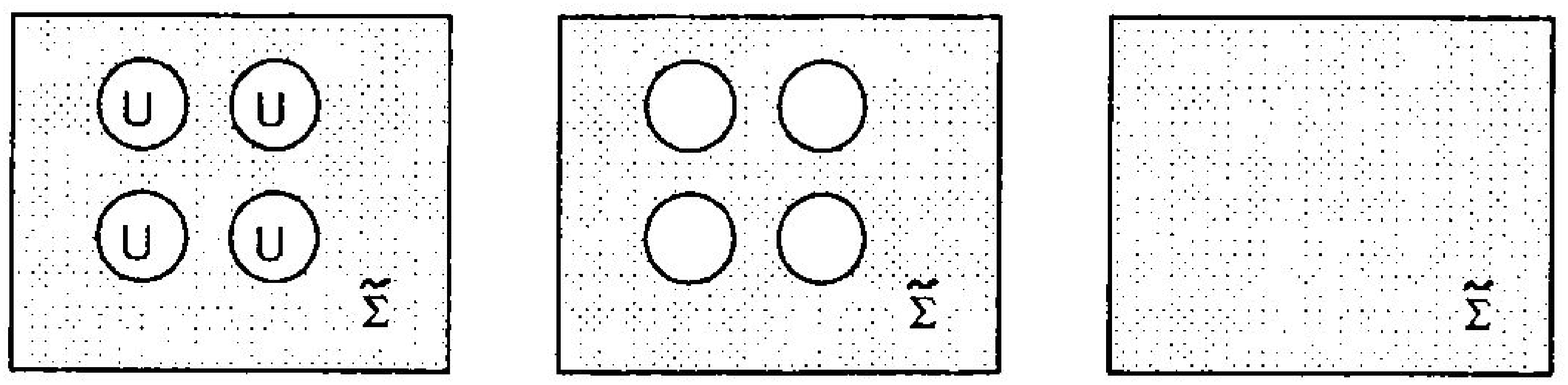}
%\vspace{0.5cm}
\caption{Left-site: Cluster with on-site Coulomb repulsion U embedded in a
  medium. Middle: Cavities replacing the sites of the cluster. Right side:
  Uniform medium with self-energy $\tilde{\Sigma} (\omega)$. (After
  \cite{KakehashiY04})} 
\label{fig08.2}
\end{figure}

The cluster memory matrix $M^{(c)}_{ij} (z)$ describes a Hubbard cluster
embedded in a uniform medium with a Hamiltonian $\tilde{H}(z)$ (see left side
in Fig. (\ref{fig08.2})). The interactions in the embedded cluster are given by
$\sum_{i \epsilon c} (U \delta n_{i \uparrow} \delta n_{i \downarrow} -
\tilde{\Sigma} (z) n_i)$. We start from a uniform medium
described by $\tilde{H}(z)$ (see Eq. (\ref{LiQU}) and the right side of
Fig. (\ref{fig08.2})). An alternative would have been to start from the
same medium but with cavities at the sites of the cluster (see the middle of
Fig. (\ref{fig08.2})), i.e., from a Hamiltonian $\tilde{H}^{(c)}$ with

%8.37
\begin{equation}
\tilde{H}^{(c)} = \tilde{H} (z) - \tilde{\Sigma} (z) \sum^{N_c}_{n \in c}
n_i~~~. 
\label{HcHzz}
\end{equation} 

\noindent In this case the interaction in the cluster would be $\sum_{n \in c}
U \delta n_{i \uparrow} \delta n_{i \downarrow}$.

\begin{figure}[t b]
\includegraphics[clip,width=9.0cm]{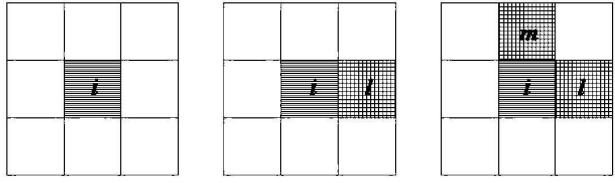}
%\vspace{0.5cm}
\caption{Schematic drawing of the multisite cluster expansion of Green's
  function. From left to right: single-site, two-sites and three-sites
  contributions. Note that the sites of a cluster need not be nearest
  neighbors.} 
\label{fig08.3}
\end{figure}

The diagonal matrix $\mathbb{L}^{(c)}_I$ in (\ref{Mcgg}) describes the atomic
  excitations. The matrix elements are given by $[L^{(i)}_I, L^{(j)}_I,
  ...]$. It has one element $L^{(i)}_I = U (1 - 2 \langle n_{i - \sigma}
  \rangle )/ [\langle n_{i - \sigma} \rangle ( 1 - \langle n_{i - \sigma}
  \rangle)]$ when $c = i$ and a second one $L^{(j)}_I$ when $c = (i,j)$. The
  incremental cluster expansion is depicted in Fig. \ref{fig08.3}. With the
  above approximations the theory reproduces two limiting cases exactly, i.e.,
  the limit of small $U$ when perturbation theory is applicable as well as the
  atomic limit. This is an important feature of the present theory. 

%%%%%%%%%%%%%%%%%%%%%%%%%%%%%%%%%%%%%%%%%%%%%%%%%%%%%%%%%%%%%%%%%%%%%%%%%%%
%2
%%%%%%%%%%%%%%%%%%%%%%%%%%%%%%%%%%%%%%%%%%%%%%%%%%%%%%%%%%%%%%%%%%%%%%%%%%%

In the limit of large clusters the memory functions are independent of the
medium. But since the cluster expansion must be truncated in practical
applications there is a dependence of the memory functions on the medium and we
must make an optimal choice for it. When correlation effects on the {\it
  static} matrix elements are neglected one can write down an explicit
expression for the screened memory function of the form 

%8.38
\begin{equation}
g^{(c)}_{ij} (z) = A_{ij} \int \frac{d \epsilon d \epsilon' d \epsilon''
  \rho^{(c)}_{ij} (\epsilon) \rho^{(c)}_{ij} (\epsilon') \rho^{(c)}_{ji}
  (\epsilon'') \chi (\epsilon, \epsilon', \epsilon'')}{z - \epsilon - \epsilon'
  + \epsilon''} 
\label{gcijzA}
\end{equation} 

\noindent with $A_{ii} = [\langle n_{i - \sigma} \rangle ( 1 - \langle n_{i -
	\sigma} \rangle )]/ [\langle n_{i - \sigma} \rangle_c ( 1 - \langle n_{i -
	\sigma} \rangle_c )]$ and $A_{i \neq j} = 1$. Here $\rho^{(c)}_{ij}
	(\epsilon)$ is the density of states of a system with one or two empty
	sites (depending on the cluster $c$) embedded in a medium with a coherent
	potential $\tilde{\Sigma} (z)$. More specifically

%8.39
\begin{equation}
\rho^{(c)}_{ij} (z) = - \frac{1}{\pi} Im [(z - \tilde{H}^{(c)})^{-1}]_{ij}~~~.
\label{rhocijz}
\end{equation} 

%%%%%%%%%%%%%%%%%%%%%%%%%%%%%%%%%%%%%%%%%%%%%%%%%%%%%%%%%%%%%%%%%%%%%%%%%%%
%3
%%%%%%%%%%%%%%%%%%%%%%%%%%%%%%%%%%%%%%%%%%%%%%%%%%%%%%%%%%%%%%%%%%%%%%%%%%%

Furthermore $\langle n_{i \sigma}\rangle_c = \int d \epsilon \rho^{(c)}_{ii}
(\epsilon) f (\epsilon)$ where $f (\epsilon)$ is the Fermi function. Moreover
$\chi (\epsilon, \epsilon', \epsilon'') = f (-\epsilon) f (-\epsilon') f
(-\epsilon'') + f (\epsilon) f (\epsilon') f (-\epsilon'')$. These
approximations are used in order so solve Eq. (\ref{sumzu2}) self-consistently
whereby $M({\bf k}, z)$ is replaced by Eq. (\ref{Mcgg}).

We reemphasize that in the two-site cluster approximation for fixed site $i$,
all sites $j$ are taken into account until convergence is achieved. For the
Hubbard model on a square lattice this includes sites which are more than ten
lattice vectors apart! In this respect the present theory resembles incremental
schemes which have been applied in the treatment of solids by quantum chemistry
methods \cite{Stoll92}. For a review see Ref. \cite{Fulde02}. The self
consistent projection method (SCPM) provides
an interesting link between treatments of model Hamiltonians in solid state
theory and true ab initio calculations with controlled approximations based on
methods used in quantum chemistry. Our theory differs from extensions of the
DMFT such as the Dynamical Cluster Approximations \cite{Hettler00} or the
Cellular Dynamical Mean-Field Theory \cite{Kotliar01} where clusters of {\it
  connected} sites are treated. They require a truly impressive amount of
numerical work and have been carefully reviewed in Ref. \cite{Maier05}.

Determining $M ({\bf k}, \omega)$ might look like a very demanding
computational problem too, but that is not really the case. In fact, we can
calculate $M ({\bf k}, \omega)$ directly without a numerical analytic
continuation from the imaginary axis as is the case in other schemes
\cite{Maier05}. Therefore a high numerical resolution as regards energy and
momentum can be obtained.

\begin{figure}[t b]
\includegraphics[clip,width=10.0cm]{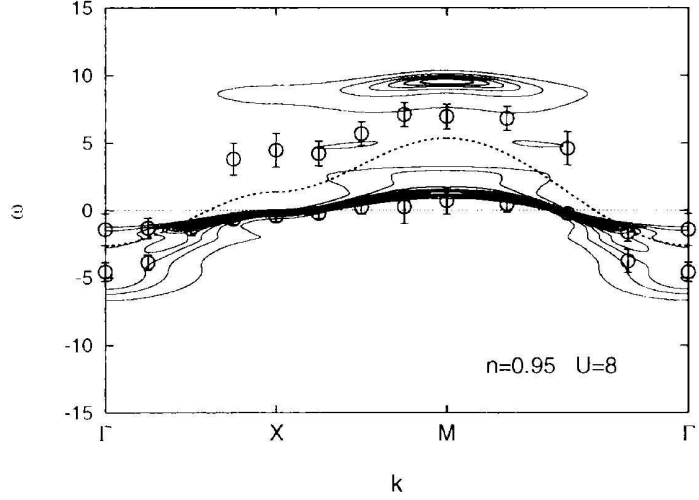}
%\vspace{0.5cm}
\caption{Single-particle excitations at zero temperatures for the Hubbard model
on a square lattice in units of the nearest-neighbor transfer integral
$t$. Dashed line: Hartree-Fock result. Open circles with error bars: QMC
results for $T = 0.33$ \cite{Groeber00}. (After \cite{Kakehashi05})} 
\label{fig08.4}
\end{figure}

The following results are obtained for $U = 8|t|$ and $T = 0$ in the underdoped
regime, i.e., for $n \leq 1$. At half-filling, the system will always be an
antiferromagnetic insulator due to nesting. But in the effective medium
approach described here one finds in addition also a paramagnetic metallic
solution. We use the latter here, being aware of the fact that an arbitrarily
small temperature or doping concentration will destroy antiferromagnetic order
in two dimensions (Mermin-Wagner theorem). A metallic solution does exist only
for $U < U_{\rm crit}$ where in single-site approximation $U_{\rm crit} \simeq
14|t|$ is obtained. Therefore with a value of $U = 8|t|$ we are still in the
metallic regime. We find that a flat quasiparticle band is crossing the Fermi
energy $E_F$. There is also an empty upper Hubbard band found which is
centered around the $M$ point and there is incoherent spectral density near the
$\Gamma$ point resulting from the lower Hubbard band. The results compare well
with those of Quantum Monte Carlo (QMC) calculations for finite temperatures
\cite{Bulut94,Preuss94,Groeber00}. This is seen in Fig. \ref{fig08.4}. The
Fermi surface for $n = 0.95$ (underdoped regime) is found to be hole like. This
is due to a collapse of the lower Hubbard band which causes a portion of the
flat quasiparticle band near the $X$ point to sink below $E_F$.

%%%%%%%%%%%%%%%%%%%%%%%%%%%%%%%%%%%%%%%%%%%%%%%%%%%%%%%%%%%%%%%%%%%%%%%%%%%
%4
%%%%%%%%%%%%%%%%%%%%%%%%%%%%%%%%%%%%%%%%%%%%%%%%%%%%%%%%%%%%%%%%%%%%%%%%%%%

These findings are contrary to expections from the Hubbard I
approximation. From (\ref{G11kzn}) it is seen that within that approximation
the spectral weight of the lower Hubbard band increases while that of the upper
Hubbard band decreases in case of hole doping $n < 1$. In the present,
improved approximation this is not the case for small hole doping and implies
that Luttinger's Theorem \cite{Luttinger60} does not apply here. It would
require a reduction of the volume enclosed by the Fermi surface with increasing
hole concentration. For $n = 0.8$ (overdoped regime) an electron-like Fermi
surface is found. These results are in agreement with the Dynamical Cluster
Approximation \cite{Maier02} and with QMC calculations
\cite{Bulut94,Preuss94,Groeber00}. 

An important finding is that for doping less than 2 \% the effective mass
$m_{\bf k}$ changes strongly between the $M$ point, i.e., $[\pi, \pi]$ where it
is minimal and the $X$ point, i.e., $[\pi, 0]$ where it has its maximum. It is
computed numerically from $m_{\bf k} = 1 - U^2 Re \partial M ({\bf k}, \omega)
/ \partial \omega |_{\omega = 0_+}$. Near the $X$ point one finds that the
smaller the $\delta \omega$ steps are in computing the derivative, the larger
becomes $m_{{\bf k} = X}$. One finds that approximately $m_{\bf k} \sim {\rm
  ln} \delta \omega$, indicating marginal Fermi liquid behavior. The marginal
Fermi liquid theory \cite{Varma89,Varma90,Littlewood91,Zimanyi93} had been
designed in order to explain a number of features, observed in the normal
state of the high-$T_c$ cuprates where electron correlations are strong. They
deviate from normal Fermi liquid behavior. Most noticeable is a linear
temperature dependence of the resistivity $\rho (T) \sim T$ which has been
observed, e.g., in doped La$_2$CuO$_4$ materials. It is suggestive to associate
this dependence with a quasiparticle scattering rate $1/\tau (T) \sim \rho
(T)$, i.e., $\tau^{-1} \sim Im \Sigma(T) \sim T$. Then it is plausible to
assume that for $T < |\omega|$ the corresponding expression is $Im \Sigma
(\omega) \sim |\omega|$. Indeed, optical reflectivity measurements on
YBa$_2$Cu$_3$O$_4$ are in accordance with this form \cite{Collins89}. But also
the frequency-dependent spin susceptibility at low $T$ shows marginal Fermi
liquid behavior. Real and imaginary part of $\Sigma ({\bf k}, \omega)$ are
related to each other via a Kramers-Kroning relation from which is follows that
$Re \Sigma (\omega) \sim \omega {\rm ln} (\omega/\omega_0)$ where $\omega_0$ is
a cut-off parameter. From this form of $Re \Sigma (\omega)$ we conclude that
the residuum of the Green's function pole (see Eq. (\ref{Gkomeg})) is
$Z (\omega) \sim 1 / |{\rm ln} (\omega/\omega_0) |$ and that the quasiparticle
mass $m_{\bf k}$ diverges at the Fermi energy like $m_{\bf k} \sim |{\rm ln}
(\omega/\omega_0) |$.

\begin{figure}[t b]
\includegraphics[clip,width=8.0cm]{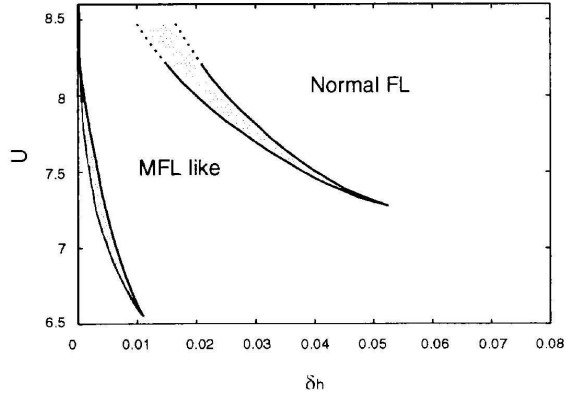}
%\vspace{0.5cm}
\caption{Phase diagram for the Hubbard model on a square lattice as function of
hole doping $\delta n$. In the upper shaded region two self-consistent
solutions are found one of which corresponds to a marginal and the other to a
normal Fermi liquid. In the lower shaded region two slightly different marginal
Fermi liquid solutions are found. (After \cite{Kakehashi05})} 
\label{fig08.5}
\end{figure}

The above treatment shows that a marginal Fermi liquid type of behavior can be
obtained from a Hubbard Hamiltonian on a square lattice, provided one is in a
certain $U$ dependent doping regime. We expect that this region will be
enlarged when long range antiferromagnetic correlations are taken into account
which have been neglected here. The phase diagram in the $U$ vs. $\delta n$
plane is shown in Fig. \ref{fig08.5}, where $\delta n$ denotes hole
doping. There are two small regions in which two self-consistent solutions are
found. They separate the marginal Fermi liquid regime from the normal Fermi
liquid one. For smaller $U$ values the cross-over between the two regimes is
continuous.  

%%%%%%%%%%%%%%%%%%%%%%%%%%%%%%%%%%%%%%%%%%%%%%%%%%%%%%%%%%%%%%%%%%%%%%%%%%%
%5
%%%%%%%%%%%%%%%%%%%%%%%%%%%%%%%%%%%%%%%%%%%%%%%%%%%%%%%%%%%%%%%%%%%%%%%%%%% 

It is instructive to consider the origin of the marginal Fermi liquid like
behavior of the self-energy. For half-filling a van Hove singularity shows up
in the density of states at $E_F$ which leads, as is well known, to a
self-energy of marginal Fermi liquid type. With hole doping this singularity
moves rather fast away from $E_F$ provided correlation effects are small
\cite{Virosztek90,Schweitzer91,Kastrinakis00}. But when $U$ is large as assumed
here, and for small hole doping spectral density moves to high
energies and therefore $E_F$ remains virtually pinned to the van Hove
singularity. This changes at higher doping concentrations where $E_F$ moves to
lower energies and normal Fermi liquid behavior is recovered.

\begin{figure}[t b]
\includegraphics[clip,width=9.0cm]{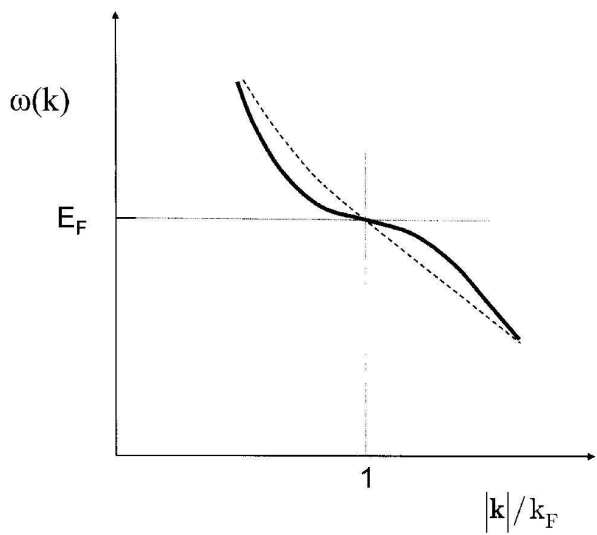}
%\vspace{0.5cm}
\caption{Schematic drawing of the dispersion $\omega({\bf k})$ near $k_F$ in
  the presence of a kink.}  
\label{fig08.6}
\end{figure}

It is interesting that for small hole doping the calculations for $U = 8 |t|$
show a kink in the excitation spectrum at approximately $\omega_{\rm kink} = -
0.8 |t|$. A kink has been observed in high-resolution photoemission experiments
in the quasiparticle band dispersion of high-$T_c$ cuprates
\cite{Bogdanov00,Cuk05}. It was found that along the nodal direction $(0,
0)-(\pi, \pi)$ the effective Fermi velocity $v_F$ as defined by the form
$\omega ({\bf k}) = v_F ({\bf k}) (|{\bf k}| - k_F)$ is neither sensitive to
the type of cuprate, nor to doping concentration or isotope substitution
\cite{Gweon04} as long as $\omega < \omega_{\rm kink}$ (= 60 - 70 meV). But
when $\omega > \omega_{\rm kink}$ a strong dependence of $v_F$ on those
quantities as well as on ${\bf k}$ is observed. The form of the dispersion is
schematically shown in Fig. \ref{fig08.6}. 

%%%%%%%%%%%%%%%%%%%%%%%%%%%%%%%%%%%%%%%%%%%%%%%%%%%%%%%%%%%%%%%%%%%%%%%%%%%
%6
%%%%%%%%%%%%%%%%%%%%%%%%%%%%%%%%%%%%%%%%%%%%%%%%%%%%%%%%%%%%%%%%%%%%%%%%%%%

\begin{figure}[t b]
\includegraphics[clip,width=8.0cm]{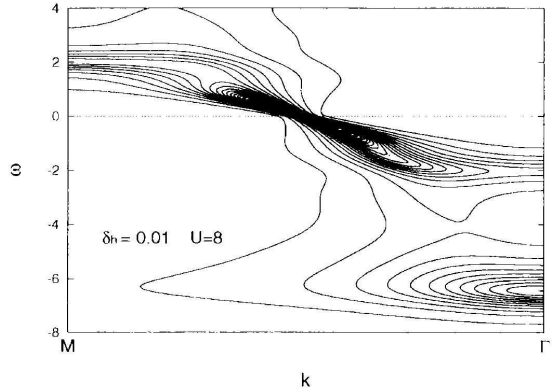}
%\vspace{0.5cm}
\caption{Hubbard model on a square lattice near half-filling. Contours of
  spectral function of excitations along $M - \Gamma$ for 
  hole doping $\delta n = 0.01$ showing a kink. (After \cite{KakehashiY05})} 
\label{fig08.7}
\end{figure}

\noindent There have been two possible explanations advanced for this
experimental observation. One is based on a coupling of electrons to the
longitudinal optical phonon mode found in inelastic neutron scattering
experiments \cite{Cuk05}. The other is based on a coupling of the electronic
quasiparticles to spin fluctuations, in particular to an observed resonance
mode \cite{Eschrig00,Johnson01,Chubukov04}. No consensus has been reached
yet. So it is interesting to note that a kink is also obtained from the Hubbard
model on a square lattice for small doping concentrations. Here it is based on
long-ranged electron correlations. In Fig. \ref{fig08.7} we show the excitation
spectrum near $E_F$ for a small hole concentration of $\delta n = 0.01$ and $U
= 8 |t|$ as before \cite{KakehashiY05} (compare with Fig. \ref{fig08.4}). 

%%%%%%%%%%%%%%%%%%%%%%%%%%%%%%%%%%%%%%%%%%%%%%%%%%%%%%%%%%%%%%%%%%%%%%%%%%%
%7
%%%%%%%%%%%%%%%%%%%%%%%%%%%%%%%%%%%%%%%%%%%%%%%%%%%%%%%%%%%%%%%%%%%%%%%%%%%

\noindent As before, the calculations assume a paramagnetic ground state. The
kink structure becomes weaker as the doping concentration increases while the
kink position changes little 
with $\delta n$. For $\delta n = 0.05$ the kink has disappeared together with
the remnants of the lower Hubbard band. The ratio of the Fermi velocities
$v'_F/v_F$ above and below $\omega_{\rm kink}$ is found to be $v'_F/v_F = 1.8$
for $\delta n = 0.01$ and $1.5$ for $\delta n = 0.02$. The kink is caused by a
hybridization of the quasiparticle excitation with short-range
antiferromagnetic fluctuations. With decreasing antiferromagnetic correlations,
e.g., by hole doping the kink structure does also decrease and eventually
disappears. Whether or not it provides an explanation for the observed kink in
the underdoped regime of La$_{2-x}$Sr$_x$CuO$_4$ remains an open question.

\subsection{Nickel and its Satellite}

Photoelectron spectroscopy has revealed a pronounced satellite structure in Ni
metal which is approximately 6 eV below the Fermi energy $E_F$
\cite{Huefner74}. It is due to strong electron correlations and a number of
different theoretical model calculations have been performed to explain it. The
simplest way of including correlations is by calculating the self-energy in
second-order perturbation theory \cite{Treglia80,Treglia82}. This results in a
$d$-band narrowing effect as well as in a satellite structure. However, for a
quantitative comparison with experiments this is not sufficient. Another well
known method is the $t$-matrix approach of Kanamori \cite{Kanamori63}. It
accounts for multiple scattering of two $d$-holes, but it is strictly valid
only in the cases of small or almost complete band filling. It has been used to
explain the satellite structure \cite{Penn79,Liebsch79} and was subsequently
extended to include also multiple electron-hole scattering
\cite{Liebsch81,Igarashi83,Igarashi85}. These calculations establish also a
link to calculations based on couplings of a hole to magnons
\cite{Roth69,Hertz73,HertzJA73,Matsumoto78}. There have been also numerical
diagonalizations based on a 4-sites tetrahedral cluster with five $d$-orbitals
per site \cite{Victora85}. A correctly positioned multiplet structure was
obtained for ferromagnetic Ni but more detailed information is beyond the
scheme of that approach. Other calculations were based on a LDA+DMFT approach
\cite{Lichtenstein01} and on a Gutzwiller projected wavefunction
\cite{Buenemann03}. In the latter case band narrowing but no satellite
structure was obtained.

Here we want to start from a five-band Hubbard model and to calculate the
spectral density for paramagnetic Ni with the help of the projection operator
method \cite{Unger94}. It gives good insight into the relevant microscopic
processes which contribute to the satellite and are described by the set of
operators $\{ A_\nu \}$ (see Eq. (\ref{Gmunukt})). The results are found to be
similar to those obtained when the two-hole one-electron $t$ matrix is solved
by employing Faddeev's equations \cite{Igarashi94}.

We write the five-band Hubbard Hamiltonian for the $d$ electrons of Ni in the
following form

%8.40
\begin{eqnarray}
H & = & H_0 + \sum_l H_1 (l) \nonumber \\
H_0 & = & \sum_{{\bf k} \nu \sigma} \epsilon_\nu ({\bf k}) n_{\nu \sigma} ({\bf
  k}) \nonumber \\ 
H_1 (l) & = & \frac{1}{2} \sum_{ijmn} \sum_{\sigma \sigma'} V_{ijmn}
a^\dagger_{i \sigma} (l) a^\dagger_{m \sigma'} (l) a_{n \sigma'} (l) a_{j
  \sigma} (l)~~~. 
\label{HH0lH1}
\end{eqnarray} 

\noindent Here $l$ is a site index and $i, j, m, n$ denote different $d$
orbitals. The $\epsilon_\nu ({\bf k})$ with $\nu = 1, ..., 5$ are the energy
dispersions of canonical $d$-bands obtained by solving a one-particle
equation. We use for them the LDA bands being aware of the fact that they
contain already some correlation effects. The $n_{\nu \sigma} ({\bf k}) =
c^\dagger_{{\bf k} \nu \sigma} c_{{\bf k} \nu \sigma}$ are number operators for
Bloch states with quantum numbers ${\bf k}, \nu$ and $\sigma$. Their creation
operators are expressed in terms of the basis operators $a^\dagger_{i \sigma}
(l)$ through 

%8.41
\begin{equation}
c^\dagger_{{\bf k} \nu \sigma} = \frac{1}{\sqrt{N_0}} \sum_{ln} \alpha_{\nu n}
({\bf k}) e^{i{\bf kR}_l} a^\dagger_{n \sigma} (l)
\label{cknusig}
\end{equation}  

\noindent where N$_0$ is the number of sites.

The interaction matrix elements in (\ref{HH0lH1}) are of the form

%8.42
\begin{eqnarray}
V_{ijmn} & = & U_{im} \delta_{ij} \delta_{mn} + J_{ij} (\delta_{in} \delta_{jm}
+ \delta_{im} \delta_{jn}) \nonumber \\
U_{im} & = & U + 2J - 2J_{im}
\label{Vijmn}
\end{eqnarray}  

\noindent where $U$ and $J$ are average values of the Coulomb- and exchange
interaction, respectively. The matrix $J_{ij}$ can be expressed in terms of $J$
and a single anisotropic parameter $\Delta J$ provided we deal with a cubic
lattice as is the case here. For more details we refer, e.g., to
Refs. \cite{Kleinmann81,Oles84}. When electronic correlations are neglected the
ground state is of the form

%8.43
\begin{equation}
\mid \Phi_0 \rangle = \prod_{|{\bf k}| < k_F \atop \nu \sigma} c^\dagger_{{\bf
	k} \nu \sigma} \mid 0 \rangle~~~. 
\label{Phi0pro}
\end{equation}  

The ground state with the inclusion of correlations $|\psi_0\rangle$ can be
determined  within the Hubbard model (\ref{H_tij}) by means of local
correlation operators \cite{Stollhoff81,Fulde87}. They are applied on $\mid
\Phi_0 \rangle$ and generate a correlation hole around each $d$ electron. We do
not describe this process here in detail but refer to the original
literature. Instead, we concentrate on identifying those operators $\{ A_\nu\}$
which are needed in order to describe the satellite structure in Ni as well as
possible. They can be obtained by simply calculating $[H_1 (l), a^\dagger_{i
	\uparrow} (l)]_-$ and considering the operators which are generated by this
commutator. They will be important to generate a local, i.e., on-site
correlation hole for an electron in orbital $i$ at site $l$. We find this way

%8.44
\begin{eqnarray}
A^{(1)}_{ij}(l) & = & \left\{
\begin{array}{l}
           2a^\dagger_{i \uparrow} (l) \delta n_{i \downarrow} (l),~~~ i = j \\
           a^\dagger_{i \uparrow} (l) \delta n_j (l), ~~~~~ i \neq j \\
\end{array}~~~\right. \nonumber\\
A^{(2)}_{ij}(l) & = & \frac{1}{2} \Big( a^\dagger_{i \uparrow} (l) s^z_j (l) +
a^\dagger_{i \downarrow} s^+_j (l) \Big)~~~~~{\rm and} \nonumber \\ 
A^{(3)}_{ij}(l) & = & \frac{1}{2}  a^\dagger_{j \downarrow} (l) a^\dagger_{j
  \uparrow} (l) a_{i \downarrow} (l) ~~~. 
\label{A1A2A3ija}
\end{eqnarray}   

Thereby the notation $\delta n_{i \sigma} (l) = n_{i \sigma} (l) - \langle n_{i
\sigma} (l) \rangle$ with $n_{i \sigma} (l) = a^\dagger_{i \sigma} (l) a_{i
  \sigma} (l)$ and ${\bf s}_i (l) = \frac{1}{2} \sum_{\alpha \beta}
a^\dagger_{i \alpha} (l) \boldsigma_{\alpha \beta} a_{i \beta} (l)$ has been
used. The operators $A^{(1)}_{ij} (l)$ describe density correlations while the
$A^{(2)}_{ij} (l)$ and $A^{(3)}_{ij} (l)$ describe spin correlations of the
added electron on site $l$. The relevant operators $\{ A_\nu \}$ consist
therefore of $A^{(0)}_\nu ({\bf k}) = c^\dagger_{{\bf k} \nu \uparrow}$ and

%8.45
\begin{equation}
A^{(\alpha)}_{ij} ({\bf k}) = \frac{1}{\sqrt{N_0}} \sum_l A^{(\alpha)}_{ij}
e^{i {\bf kR}_l}~~~,~~~~~\alpha = 1, 2, 3~~~.
\label{AalphaijA}
\end{equation}

One checks easily that for each ${\bf k}$ point there are 66 relevant operators
$\{ A_m ({\bf k}) \}$, i.e., 1 + 25 + 20 + 20. Thus the (66 x 66) matrix

%8.46
\begin{equation}
\mathbb{G} (z) = \mathbb{\bbchi} (z \mathbb{\bbchi} - \mathbb{L})^{-1}
\mathbb{\bbchi} 
\label{mbbGz}
\end{equation}

\noindent (see (\ref{GmnkzA})) has to be diagonalized for each ${\bf k}$
point. The roots of the secular equation yield the $d$ bands as well as the
satellite structures.  

%%%%%%%%%%%%%%%%%%%%%%%%%%%%%%%%%%%%%%%%%%%%%%%%%%%%%%%%%%%%%%%%%%%%%%%%%%%
%8
\begin{figure}[t b]
\includegraphics[clip,width=7.0cm]{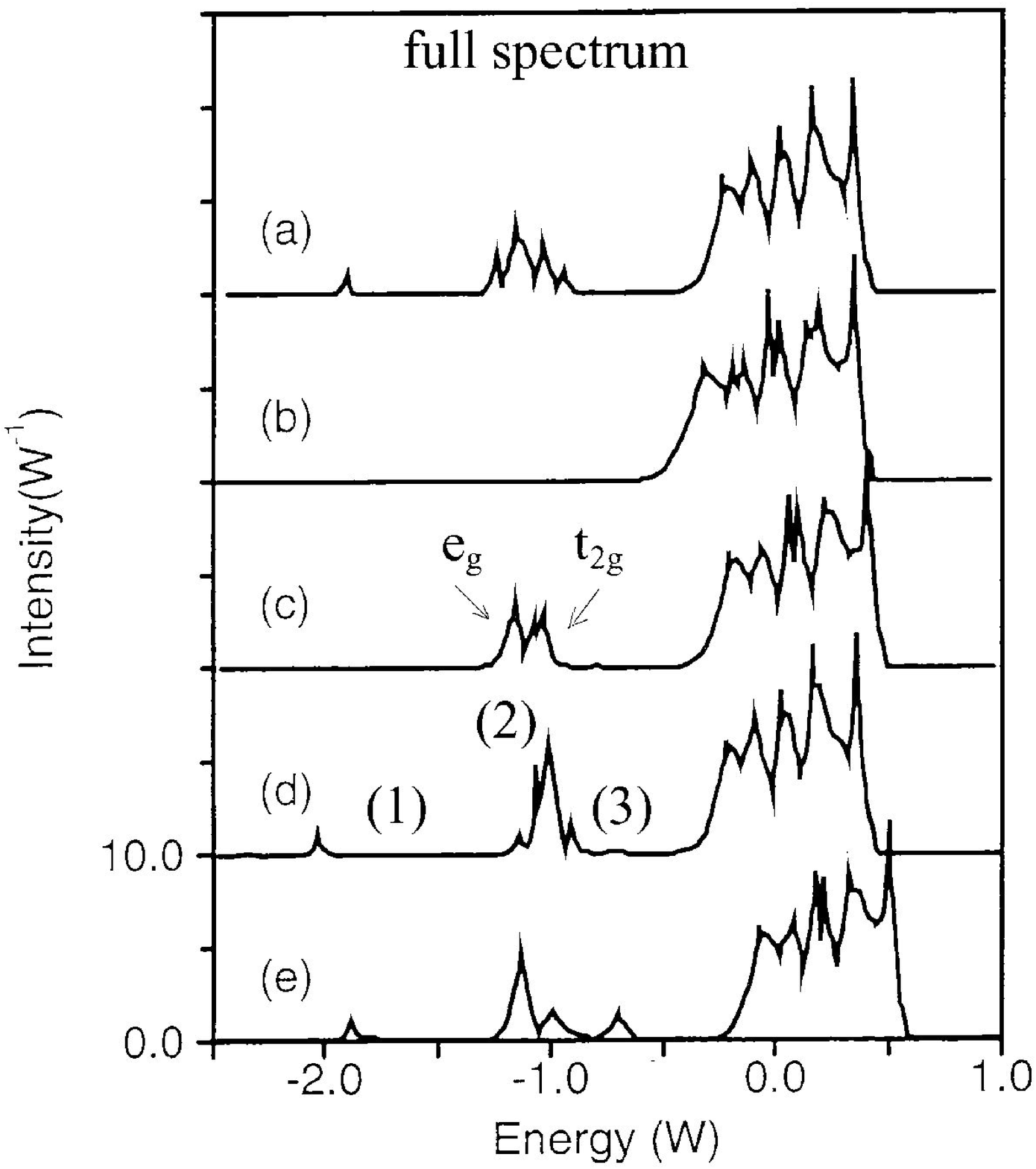}
%\vspace{0.5cm}
\caption{(a) full spectrum of paramagnetic Ni for $U = 0.56$, $J = 0.22$ and
  $\Delta J = 0.031$ in units of $W$; (b) LDA results; (c) spectrum when $J =
  \Delta J = 0$; (d) spectrum when $\Delta J = 0$. The peaks (1) - (3)
  correspond to $^1S$, $^1G$ and $^1 D$, $^3P$ and $^3 F$ atomic $d^8$
  configurations; (e) spectrum when $| \Phi_0 \rangle$ instead of $|
  \psi_0\rangle$ is used. (After \cite{Unger94})}  
\label{fig08.8}
\end{figure}
%%%%%%%%%%%%%%%%%%%%%%%%%%%%%%%%%%%%%%%%%%%%%%%%%%%%%%%%%%%%%%%%%%%%%%%%%%%

\begin{figure}[t b]
\includegraphics[clip,width=8.0cm]{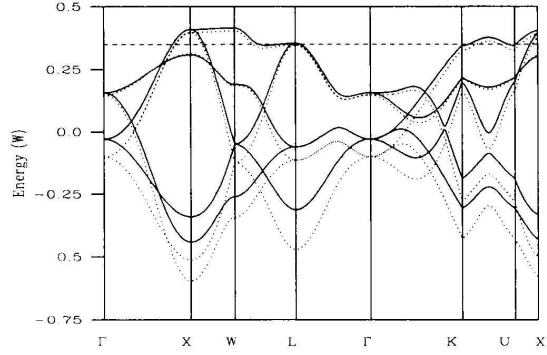}
%\vspace{0.5cm}
\caption{Narrowing of the quasiparticle bands of fcc paramagnetic Ni due to
  electron correlations (solid lines). The bands without correlations are
  identified here with the LDA bands (dotted lines). (After \cite{Unger94})} 
\label{fig08.9}
\end{figure}

The spectral density obtained this way for paramagnetic Ni is shown in
Fig. \ref{fig08.8}. The parameters have been chosen as $U/W = 0.56$, $J/W =
0.22$ and $\Delta J/W = 0.031$ in units of the bandwidth $W$. Those values
were obtained from a fit of experiments which measure the multiplet structure
of Ni ions put into an Ag matrix \cite{vanderMarel88,Igarashi94}. A $d$
electron number of $n_d = 9.4$ has been taken. A comparison with the
calculations in SCF approximation, here identified with the LDA results show a
narrowing of the bandwidth and a pronounced satellite structure. With a spin
averaged bandwidth of $W = 4.3$ eV obtained from a local spin-density
approximation this satellite is peaked at 6.8 eV below the top of the $d$
bands (Fig. \ref{fig08.8}). This has to be compared with an experimental value
of 6.3 eV. The shape of the satellite reflects the atomic $d^8$ multiplet. When
the anisotropy parameter $\Delta J = 0$ is set to zero, it is split into three
substructures representing a $^1S$ state, two degenerate singlets $^1G$ and
$^1D$ and two degenerate triplets $^3P$ and $^3F$. The energy difference
between $^1G$ and $^3F$ is $2J$ and between $^1S$ and $^1G$ is $5J$. The three
structures at $-1.9 W$, $-1.1 W$ and $-0.7 W$ 
correspond to the three atomic levels. The main peak at $-1.1 W$ is split by
the anisotropy parameter into finer structures. But also the $e_g - t_{2g}$
splitting as well as the quasiparticle dispersion have an effect on the
satellite structure. Also shown are the modifications in the satellite
structure which arise when the ground state $| \psi_0 \rangle$ is replaced by
the one without correlations $| \Phi_0 \rangle$. For completeness we show also
the reductions in the widths of the different quasiparticle bands when
correlations are taken into account (see Fig. \ref{fig08.9}). We want to
mention that a satellite below the $d$ bands is also found for fcc Co as well
as for bcc Fe while for fcc Sc a satellite above the $d$ bands is obtained. We
repeat that our theoretical findings are based on a 5-band Hubbard model. It
might not be sufficient for a satisfactory description in all cases because $s$
electrons have been left out altogether and because intersite correlations are
not taken into account in the choice for the set $\{A_\nu \}$. 

%%%%%%%%%%%%%%%%%%%%%%%%%%%%%%%%%%%%%%%%%%%%%%%%%%%%%%%%%%%%%%%%%%%%%%%%%%%
%9
%%%%%%%%%%%%%%%%%%%%%%%%%%%%%%%%%%%%%%%%%%%%%%%%%%%%%%%%%%%%%%%%%%%%%%%%%%%

\subsection{Multiplet Effects in $5f$ Systems}
\label{sub:Multieffin5fsystems}

The high-energy effects due to strong correlations discussed so far result from
the overall suppression of charge fluctuations. They reflect the presence of
atomic configurations with well-defined occupations of the partially filled
inner valence shells. In a free atom or ion, the degeneracies of these
configurations are (partially) lifted by the electron-electron
interaction. This leads to the formation of atomic multiplets where the scale
for the excitation energies is set by the exchange constant. The latter
involves the anisotropic part of the Coulomb interaction which remains (almost)
unscreened in a metal. As a consequence we expect pronounced multiplet
effects in the single-particle spectra whenever the the exchange constant
exceeds the gain in kinetic energy as measured by the corresponding
effective band width. This is the case for the high-energy satellites
in Ni which were discussed in Sec. \ref{sub:Multieffin5fsystems}.
Multiplet effects are also strongly evident in the actinide compounds
where the $5f$ exchange constant is of order 1 eV and exceeds the bare
effective band width. Here we focus on U-based heavy-fermion
compounds where strong intra-atomic correlations are responsible
for the dual character of the $5f$ electrons. We restrict ourselves
to qualitative features of the 5$f$-spectral function 

%8.47
\begin{eqnarray}
%\lefteqn
A_{j_{z}}({\bf k},\omega ) =
\begin{cases}
 \sum_{n} \left| \left< \Psi_{n}^{(N + 1)} \left| c_{j_{z}}^{\dagger} ({\bf k})
 \right| \Psi_{0}^{(N)} \right> \right|^{2} \delta \left( \omega -
 \omega_{n0}^{(+)} \right) & ;\, \omega > 0\\
 \sum _{n} \left| \left< \Psi_{n}^{(N - 1)} \left| c_{j_{z}} ({\bf k}) \right|
 \Psi_{0}^{(N)} \right> \right|^{2} \delta \left( \omega + \omega_{n0}^{(-)}
 \right) & ;\, \omega < 0 
\end{cases}
\label{eq:DefSpectralFunction}
\end{eqnarray}
 
with 

%8.48
\begin{equation}
c_{j_{z}} ({\bf k}) = \sum_{\bf a}e^{i{\bf ka}} c_{j_{z}} (a)~~; ~~~~~
\omega_{n0}^{(\pm )} = E_{n}^{(N\pm 1)} - E_{0}^{(N)}~~~.
\label{eq:cjzka}
\end{equation}

It yields the probability for adding ($\omega >0$) or removing
($\omega <0$) a $5f$ electron with energy $\omega $ in a state
characterized by momentum ${\bf k}$ and angular momentum projection $j_{z}$
to the $N$-particle ground state $|\Psi _{0}^{(N)}\rangle $
with energy $E_{0}^{(N)}$. The states with $N\pm 1$ and their energies
are denoted by $|\Psi _{n}^{(N\pm 1)}\rangle $ and $E_{n}^{(N\pm 1)}$,
respectively. The notation was introduced in
Sec. \ref{sec:PartialLocalization}.  

\begin{figure}
\includegraphics[width=0.60\columnwidth]{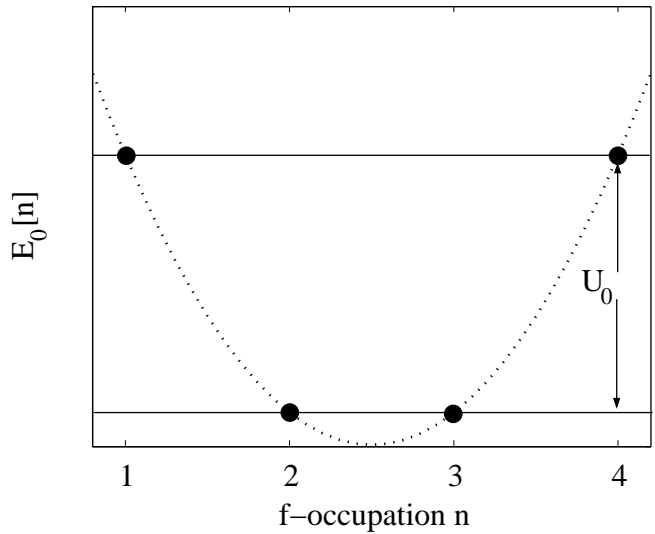}
\caption{Configurational energies vs. $5f$-valence for the microscopic model
adopted to describe the U sites. For a qualitative discussion, the
value U$_{0}$ is treated as a parameter since it is screened in a
metal and concomitantly will depend upon the crystallographic environment
of the U site.
\label{fig:Evsn}}
\end{figure}

The structure of the spectra in the strong-coupling limit can be understood
by considering the atomic limit where the system is modeled as a
statistical ensemble of isolated $5f$ atoms. The absence of pronounced
valence peaks in the photoemission and inverse photoemission spectra
suggests that the energies of the $f^{2}$- and $f^{3}$-configurations,
are (almost) degenerate i.e., $E(5f^{2})=E(5f^{3})$. The variation
with $5f$-valence of the configurational energies $E(f^{n})$ is
shown schematically in Fig. \ref{fig:Evsn}.

%%%%%%%%%%%%%%%%%%%%%%%%%%%%%%%%%%%%%%%%%%%%%%%%%%%%%%%%%%%%%%%%%%%
%10
%%%%%%%%%%%%%%%%%%%%%%%%%%%%%%%%%%%%%%%%%%%%%%%%%%%%%%%%%%%%%%%%%%%%%

The ground state will mainly involve $5f^{2}$ and $5f^{3}$ configurations.
The corresponding spectral functions are obtained in close analogy
to the classical work of Hubbard \cite{Hubbard63}. Let us first consider
the zero-configuration width approximation which neglects intra-atomic
correlations. The valence transitions $f^{2}\rightarrow f^{1}$ and
$f^{3}\rightarrow f^{4}$ occur at high energies set by the isotropic
average of the Coulomb interaction and, concomitantly, do not affect
the low-temperature behavior. The latter is determined by the low-energy
peak resulting from the transitions $f^{2}\leftrightarrow f^{3}$
within the $f^{2}$- and $f^{3}$-configurations. This peak is a direct
consequence of the intermediate-valent ground state. The distribution
among the peaks can be estimated from combinatorial considerations.
The weight $Z(f^{2}\rightarrow f^{1})$ of the transition $f^{2}\rightarrow
f^{1}$ equals the probability that a state with a given $j_{z}$ is occupied
in that $f^{2}$ contribution of the mixed-valent ground-state. Following
these lines one finds

%8.49
\begin{eqnarray}
Z \left( f^{2} \rightarrow f^{1} \right) = \frac{1}{6} & \quad & Z \left( f^{2}
\rightarrow f^{3} \right) = \frac{1}{3} \nonumber \\
Z \left( f^{3} \rightarrow f^{2} \right) = \frac{1}{4} & \quad & Z \left( f^{3}
\rightarrow f^{4} \right) = \frac{1}{4}
\label{eq:WeightsForValenceTransitions}
\end{eqnarray}

%%%%%%%%%%%%%%%%%%%%%%%%%%%%%%%%%%%%%%%%%%%%%%%%%%%%%%%%%%%%%%%%%%%%%%%%%
%11
\begin{figure}
\raisebox{1cm}{\includegraphics[width=0.40\columnwidth]{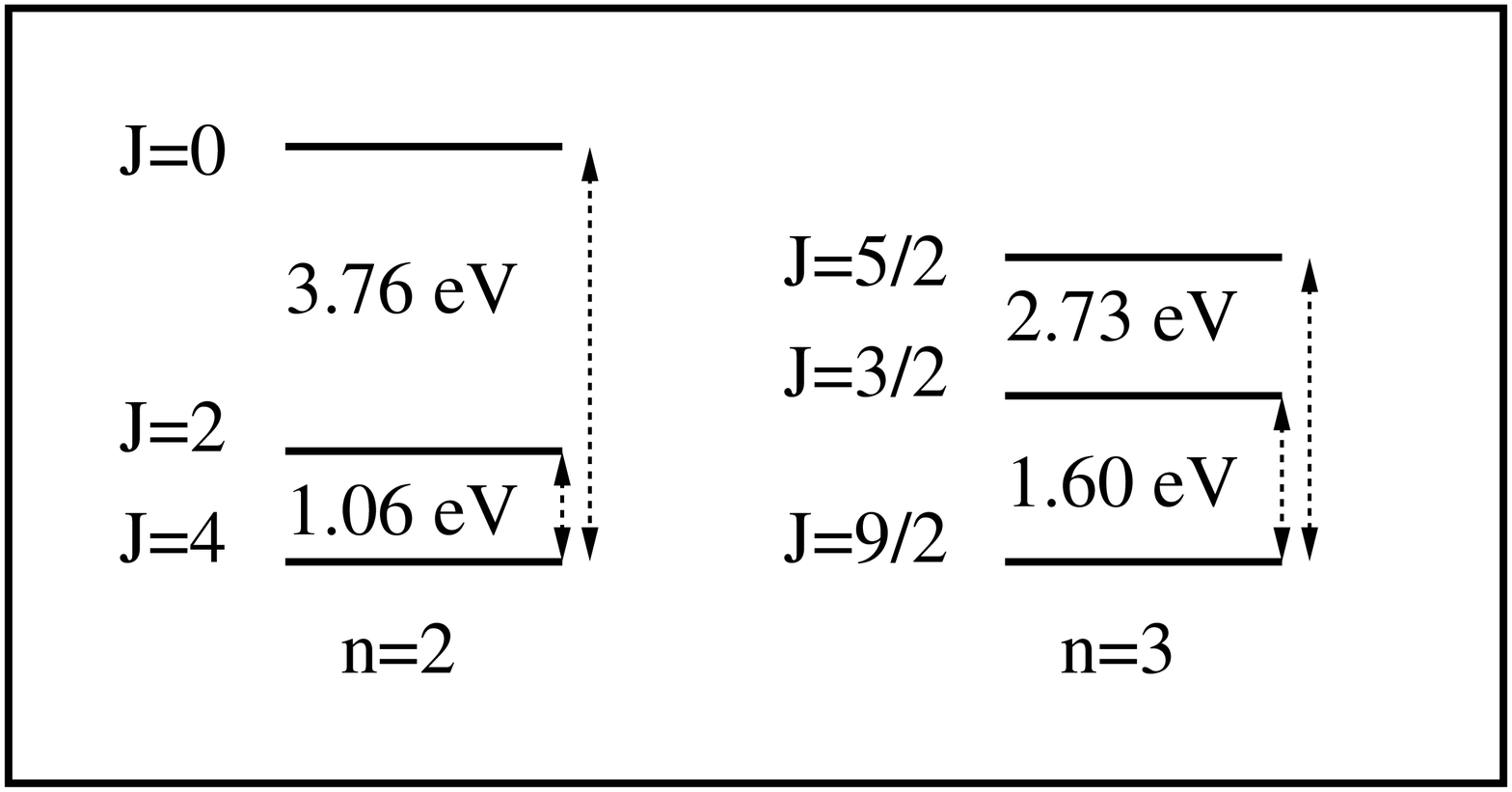}}
\hspace*{1cm}
\includegraphics[width=0.40\columnwidth]{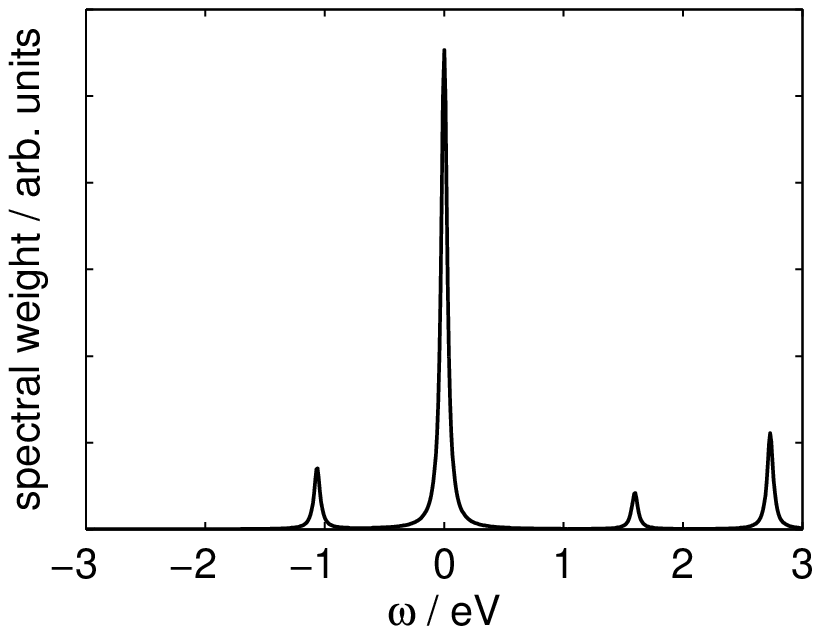}
\caption{Left panel: Multiplets for $f^{2}$ and $f^{3}$ configurations of the U
model in the atomic limit. The levels are obtained by diagonalizing
the Coulomb matrix using j-j coupling and Coulomb parameters of UPt$_{3}$.
Right panel:  Spectral function for the model in the atomic limit.
\label{fig:Multiplets}}
\end{figure}
%%%%%%%%%%%%%%%%%%%%%%%%%%%%%%%%%%%%%%%%%%%%%%%%%%%%%%%%%%%%%%%%%%%%%%%%%%%

The central focus is the evolution of the low-energy peak whose spectral
weight sums up to 7/12 (= 1/3 + 1/4). Intra-atomic correlations which are
usually described by Hund's rules yield the multiplets displayed schematically
in Figure \ref{fig:Multiplets}. The values for the excitation energies
are calculated from the Coulomb parameters of UPt$_{3}$ adopting
$j-j-$coupling. In the atomic limit the model spectral function has
a low-energy peak resulting from transitions between the Hund's rule
ground state manifolds $\left|f^{2},J=4,J_{z}\right\rangle $ and
$\left|f^{3},J=9/2,J_{z}\right\rangle $ as well as peaks corresponding
to transitions into excited multiplets. Due to the rotational invariance
of the Coulomb interaction, the spectral functions do not depend upon
the magnetic quantum number $j_{z}$. The spectral weights for the
photoemission and inverse photoemission parts, $\frac{1}{10}\sum_{J_{z}}|
\langle f^{2}, J', J_{z}-j_{z} | c_{j_{z}} | f^{3}, J = 9/2, J_{z} \rangle
|^{2}$ and $\frac{1}{9} \sum_{J_{z}} | \langle f^{3}, J',J_{z} + j_{z} |
c_{j_{z}}^{\dagger} | f^{2}, J = 4, J_{z} \rangle |^{2}$ are easily obtained by
expressing the matrix elements in terms of reduced matrix elements and
Clebsch-Gordan coefficients.The excitation energies as well as the
corresponding spectral weights are listed in Table
\ref{tab:PositionsWeightsAtomicLimit}. We should like to mention that there is
no transition into the excited multiplet $| f^{2}, J = 0 \rangle$. 

The model therefore predicts $5f$ spectral weight at $\simeq$ 1 eV
binding energy in the photoemission spectra from U-based heavy fermion
compounds. The position of these $5f$ structures should not depend
sensitively on the actual crystalline environment since they result
from intra-atomic excitations. The weights, however, will be altered
by the reconstruction of the ground state and the low-energy excitations
in an extended solid and may vary with chemical composition and temperature.
These predictions seem to be in agreement with recent experiments
\cite{Eloirdi05}. Recent photoemission studies by Fujimori et
al. \cite{Fujimori05} seem to indicate the presence of features in the proper
energy range. The fact that they cannot be attributed to LDA energy bands
further supports our interpretation in terms of multiplet side bands. An
unambiguous identification, however, will require resonant photoemission
experiments. 

%%%%%%%%%%%%%%%%%%%%%%%%%%%%%%%%%%%%%%%%%%%%%%%%%%%%%%%%%%%%
\begin{table}
\begin{tabular}{llll|lll}
\hline 
& \multicolumn{2}{c}{PES} && \multicolumn{3}{c}{BIS}\\
\hline
Position / [eV] \quad & \quad -1.06 \quad & \quad 0.00 & \quad & \quad 0.00
\quad & \quad 1.60 \quad & \quad 2.73\\
Spectral weight \quad & \quad 0.054 \quad & \quad 0.196 & \quad & \quad 0.218
\quad & \quad 0.032 \quad & \quad 0.083 \\
\hline
\end{tabular}
\caption{Positions and spectral weights of the atomic transitions in
  uranium. When added the weight at energy position 0.00 of the particle and
  hole propagation is 0.414.
\label{tab:PositionsWeightsAtomicLimit}}
\end{table}
%%%%%%%%%%%%%%%%%%%%%%%%%%%%%%%%%%%%%%%%%%%%%%%%%%%%%%%%%%%%%% 

\subsection{Excitations in Copper-Oxide Planes}

It is well known that the copper-oxide based perovskites which play a major
role in high-temperature superconductivity are strongly correlated electron
systems. This is immediately obvious by realizing that, e.g., La$_2$CuO$_4$ is
an insulator and not a metal despite the fact that with one hole per unit cell
one would expect a half-filled conduction band. Note that insulating behavior
is found also above the antiferromagnetic transition temperature and therefore
is unrelated to the doubling of the unit cell when the material becomes an
antiferromagnet. Indeed, LDA calculations give an effective Coulomb repulsion
of two $d$ holes on a Cu site of $U_d = 10.5$ eV and a hopping matrix element
of a hole from a Cu $3d_{x^2 - y^2}$ orbital to an O $2p_{x(y)}$ orbital of
$t_{pd} = 1.3$ eV. Therefore the bandwidth is small as compared with $U_d$.

The strong correlations lead to a single-particle spectral density which is
quite distinct from the one of weakly correlated electrons. In fact, the
excitations are dominated by the internal degrees of freedom of the correlation
hole so that the one-particle or coherent part of Green's function plays a
secondary role here. This will become clear towards the end of this section.

%%%%%%%%%%%%%%%%%%%%%%%%%%%%%%%%%%%%%%%%%%%%%%%%%%%%%%%%%%%%%%%%%%%%%%%%%%%
%12
\begin{figure}[t b]
\includegraphics[clip,width=8.0cm]{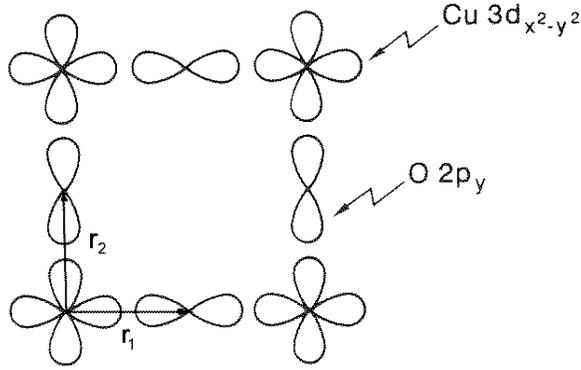}
%\vspace{0.5cm}
\caption{Cu and O orbitals accounted for in the 3-band Hubbard Hamiltonian
  (\ref{Hsummksi})}  
\label{fig08.10}
\end{figure}
%%%%%%%%%%%%%%%%%%%%%%%%%%%%%%%%%%%%%%%%%%%%%%%%%%%%%%%%%%%%%%%%%%%%%%%%%%%

Again we shall use a model Hamiltonian
although an ab initio calculation would be highly desirable. We want to write
it in hole representation in the form of a three-band Hubbard or Emery
model. The unit cell consists of one Cu $3d_{x^2 - y^2}$ orbital and two O
$2p_{x(y)}$ orbitals (see Fig. \ref{fig08.10}). The orbital energies are
$\epsilon_d$ and $\epsilon_p$. Two holes on a Cu site repel each other with
$U_d$ and on a O site with $U_p$. The hopping matrix element between a Cu
orbital and an O orbital is $t_{pd}$ and $t_{pp}$ between the orbitals of the
two O atoms. The parameter values can be derived from a constrained LDA
calculation \cite{Hybertsen89} with the following results: $U_d = 10.5$ eV,
$U_p = 4.0$ eV, $t_{pd} = 1.3$ eV, $t_{pp} = 0.65$ eV, $\epsilon_p - \epsilon_d
= 3.6$ eV. It is helpful to first introduce a basis of oxygen orbitals which is
diagonal with respect to oxygen-oxygen hopping processes $t_{pp}$. Let us
denote the corresponding creation operators by $c^\dagger_{m {\bf k} \sigma}$
where $m$ is an oxygen band index. Those bands have a dispersion

%8.50
\begin{equation}
\epsilon_m ({\bf k}) = \epsilon_p \pm 2t_{pp} \left[ {\rm cos} \left[ {\bf k}
	\left( {\bf r}_1 + {\bf r}_2 \right) \right] - {\rm cos} \left[ {\bf k}
	\left( {\bf r}_2 - {\bf r}_1 \right) \right] \right] ~~~~~~~~~~,(m =
	1,2) 
\label{epsilmkp}
\end{equation}

\noindent where the vectors ${\bf r}_1$ and ${\bf r}_2$ are shown in
Fig. \ref{fig08.10}. Due to the different orientations of the O $2p_{x(y)}$
orbitals, the sign of $t_{pp}$ depends on the direction. It is positive in the
direction ${\bf r}_1 + {\bf r}_2,~ - ({\bf r}_1 + {\bf r}_2)$ and negative in
the direction ${\bf r}_1 - {\bf r}_2,~ {\bf r}_2 - {\bf r}_1$.

Next we introduce a linear combination of oxygen orbitals which possesses the
same symmetry on a CuO$_4$ plaquette as the $t_{pd}$-matrix elements, i.e.,

%8.51
\begin{equation}
p^\dagger_{{\bf k} \sigma} = \sum_m \phi_{m {\bf k}} c^\dagger_{m {\bf k}
  \sigma}~~~, 
\label{pksigmm}
\end{equation}

\noindent with

%8.52
\begin{equation}
\phi_{m {\bf k}} = -\frac{i}{\sqrt{2}} \left[ {\rm sin} k{\bf r}_1 \pm {\rm
  sin} k{\bf r}_2 \right] ~~~. 
\label{phimki}
\end{equation}

\noindent The model Hamiltonian we shall be using can thus be written as

%8.53
\begin{eqnarray}
H & = & \sum_{m {\bf k} \sigma} \epsilon_m ({\bf k}) c^\dagger_{m {\bf k}
  \sigma} c_{m {\bf k} \sigma} + U_p \sum_J n_{p \uparrow} (J) n_{p\downarrow}
  (J)\\ 
\nonumber 
&& + \epsilon_d \sum_{{\bf k} \sigma} d^\dagger_{{\bf k} \sigma} d_{{\bf k}
  \sigma} + U_d \sum_I n_{d \uparrow} (I) n_{d \downarrow} (I)\\ \nonumber 
&& + 2t_{pd} \sum_{{\bf k} \sigma} \left( p^\dagger_{{\bf k} \sigma} d_{{\bf
	k}\sigma} + h.c. \right)~~~. 
\label{Hsummksi}
\end{eqnarray}

\noindent As usual $n_{p \sigma} (J)$ and $n_{d \sigma} (I)$ are the occupation
number operator of the O $2p_{x(y)}$ orbital on site $J$ and of the Cu $3d_{x^2
- y^2}$ orbital on site $I$, respectively. 

With the help of the above Hamiltonian we calculate the spectral density of the
Cu-O planes, with and without hole doping. For that purpose we have to choose
the right set of relevant operators $\{ A_n ({\bf k})\}$. They must include the
most important microscopic processes in the strongly correlated system which
generate the correlation hole of an electron. First of all, the hole operators 

%8.54
\begin{equation}
A_p (m, {\bf k}) = p^\dagger_{m {\bf k}}~~~~~~~, A_d ({\bf k}) =
d^\dagger_{{\bf k} \uparrow}~~,~~~~~~ m = 1,2  
\label{apmkp}
\end{equation}

must be part of the set. They are supplemented by a number of {\it local}
operators. In order to suppress double occupancies of the $d$ and $p$ orbitals
due to the large values of $U_d$ and $U_p$, we must include
$\bar{d}^\dagger_\uparrow = d^\dagger_\uparrow (I) n_{d \downarrow} (I)$ and
$\bar{p}^\dagger_\uparrow = p^\dagger_{J \uparrow} n_{p \downarrow} (J)$, i.e.,
their Fourier transforms 

%8.55
\begin{eqnarray}
A_d({\bf k})  & = & \frac{1}{\sqrt{N}} \sum_J e^{-i {\bf kR}_I}
\bar{d}^\dagger_{I \uparrow} \nonumber \\
A_p({\bf k})  & = & \frac{1}{\sqrt{2N}} \sum_J e^{-i {\bf kR}_J}
\bar{p}^\dagger_{J \uparrow} 
\label{AdkN}
\end{eqnarray}

\noindent where $N$ is the number of Cu sites and there are two O sites per Cu
site. In addition we want to account for spin flips of $d$ holes at Cu sites in
combination with spin flips at neighboring O sites. Those processes are
important for the formation of a Zhang-Rice singlet \cite{ZhangF88}. This is a
singlet state formed by a hole at a Cu site and another one at a
nearest-neighbor O site. 

The corresponding microscopic operator is

%8.56
\begin{equation}
A_f ({\bf k}) = \frac{1}{\sqrt{N}} \sum_I e^{-i {\bf kR}_I}
\tilde{p}^\dagger_{I \uparrow} S^+_I 
\label{afk1N}
\end{equation}

\noindent where

%8.57
\begin{equation}
\tilde{p}^\dagger_{I \sigma} = \frac{1}{2} \left( p^\dagger_{1 \sigma} +
p^\dagger_{2 \sigma} - p^\dagger_{3 \sigma} - p^\dagger_{4 \sigma} \right) 
\label{pfrac12p}
\end{equation}

\noindent is a superposition of the four O $2p$ orbitals which surround the
  $3d$ orbital of Cu site $I$. The operator $S^+_I = d^+_{I \uparrow} d_{I
  \downarrow}$. For a possible formation of triplet states also the operator

%8.58
\begin{equation}
A_a({\bf k}) = \frac{1}{\sqrt{N}} \sum_I e^{-i {\bf kR}_I} \tilde{p}^\dagger_{I
\uparrow} n_{d \downarrow} (I)
\label{Aak1N}
\end{equation}

\noindent is needed. In order to describe charge transfer in the vicinity of an
added hole we also include

%8.59
\begin{equation}
A_c({\bf k}) = \frac{1}{\sqrt{N}} \sum_I e^{-i {\bf kR}_I} p^\dagger_{I
\uparrow} p^\dagger_{I \downarrow} d_{I \downarrow}~~~.
\label{Ack1N}
\end{equation}

\noindent This completes the choice of the set $\{ A_\nu ({\bf k}) \}$. With
9 operators $A_\nu ({\bf k})$ we have to diagonalize for each ${\bf k}$ point
(9 x 9) matrices $L_{\mu \nu}$ and $\chi_{\mu \nu}$ (see (\ref{LchiM})). The
static expectation values which enter the matrix elements are determined from
the spectral functions to which they are related via

%8.60
\begin{equation}
\left< \psi_0 \left| A^\dagger_m ({\bf k}) A_n ({\bf k}) \right| \psi_0 \right>
= \int\limits^{+\infty}_{-\infty} d \omega f (\omega)  A_{mn} ({\bf k},
\omega)  
\label{psi0AmkAn}
\end{equation}

%%%%%%%%%%%%%%%%%%%%%%%%%%%%%%%%%%%%%%%%%%%%%%%%%%%%%%%%%%%%%%%%%%%%%%%%%%%
%13
\begin{figure}[t b]
\includegraphics[clip,width=8.0cm]{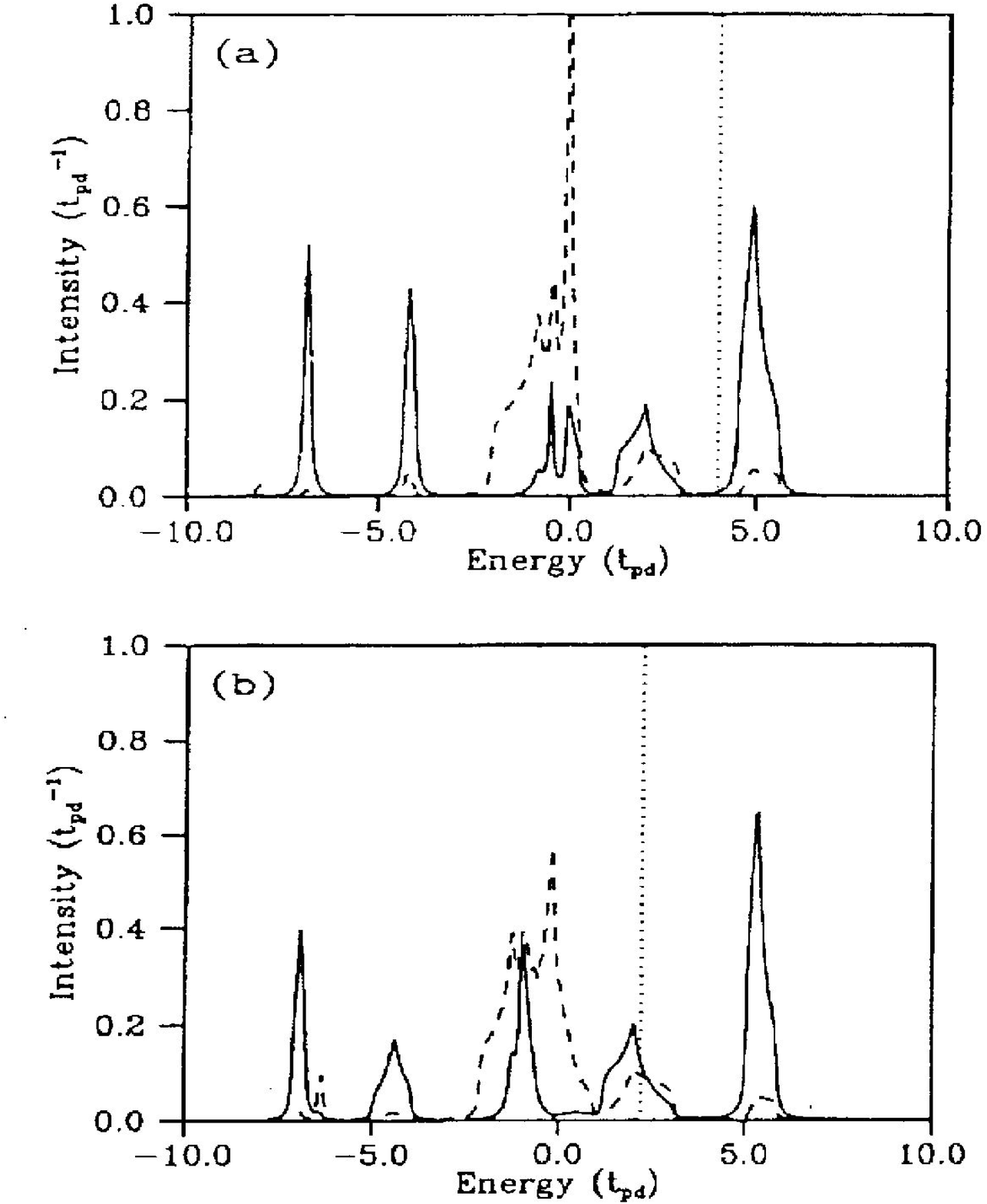}
%\vspace{0.5cm}
\caption{Spectral density of Cu-O planes described by the Hamiltonian
  (\ref{Hsummksi}); (a) at half-filling; (b) for 25 \% hole doping. Solid and
  dashed lines show the Cu and O contributions. In the hole doped case spectral
  density has been shifted from the upper Hubbard band to the region near $E_F$
  (dotted line). Parameters are $U_d$ = 8, $U_p$ = 3, $t_{pp}$ = 0.5, $t_{pd}$
  = 1, $\epsilon_p - \epsilon_d$ = 4. Compare also with
  Fig. \ref{fig08.12}. (After \cite{Unger93})}  
\label{fig08.11}
\end{figure}
%%%%%%%%%%%%%%%%%%%%%%%%%%%%%%%%%%%%%%%%%%%%%%%%%%%%%%%%%%%%%%%%%%%%%%%%%%%

\noindent where $f (\omega) = [e^{\beta \omega} + 1]^{-1}$ with $\beta =
(k_BT)^{-1}$ is the Fermi distribution function. It can be replaced by a step
function. By solving Eq. (\ref{psi0AmkAn}) self-consistently the static
correlation functions are obtained. For more details we refer to the original
literature \cite{Unger93,UngerP93}. The resulting density of states are shown
in Figs. \ref{fig08.11} for the half-filled case as well as for $25 \%$ of hole
doping. The two cases are supposed to simulate La$_2$CuO$_4$ and
La$_{2-x}$Sr$_x$CuO$_4$, respectively. The agreement with results of exact
diagonalization of a small cluster containing four CuO$_2$ units
\cite{Tohyama92} is very good. It is noticed that at half filling the system is
insulating and a gap is present. Around $\omega = 0$ one notices the O $2p$
band and a $3d$ component due to hybridization. The upper Hubbard band is
centered at $5t_{pd}$. The structure near $-7t_{pd}$ results mainly from $A_d
({\bf k})$ and is interpreted as the lower Hubbard band. The peak near
$-4t_{pd}$ comes mainly from $A_c ({\bf k})$, i.e., from charge
fluctuations. The structure close to $2.5 t_{pd}$ marks the Zhang-Rice singlet
state. It is important for the interpretation of spectroscopic experiments
\cite{Romberg90}. When the system is doped with holes the Fermi energy moves
into the Zhang-Rice singlet band. Simultaneously there is a transfer of
spectral density  taking place from the upper Hubbard band to energies below
$E_F$. 

%%%%%%%%%%%%%%%%%%%%%%%%%%%%%%%%%%%%%%%%%%%%%%%%%%%%%%%%%%%%%%%%%%%%%%%%%%%
%14
\begin{figure}[t b]
\includegraphics[clip,width=6.0cm]{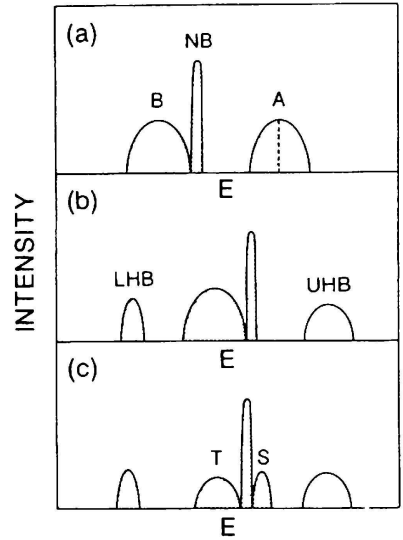}
%\vspace{0.5cm}
\caption{Schematic representation of the density of states of a Cu-O plane
  described within a 3-band Hubbard model. (a) independent electron
  approximation. A, B, NB denote antibonding, bonding and nonbonding part; (b)
  splitting of the d band into a lower (LHB) and upper (UHB) Hubbard band; (c)
  singlet-triplet splitting (S, T) S = Zhang-Rice singlet} 
\label{fig08.12}
\end{figure}
%%%%%%%%%%%%%%%%%%%%%%%%%%%%%%%%%%%%%%%%%%%%%%%%%%%%%%%%%%%%%%%%%%%%%%%%%%%

As mentioned at the beginning of this section we are dealing here with a
situation where the internal degrees of freedom of the correlation hole
dominate the spectral density. This is most clearly seen by comparing the
density of states for the half-filled case with the one obtained in the
independent electron approximation. The latter is shown in
Fig. \ref{fig08.12}a. It consists of a bonding, nonbonding and antibonding
part. The splitting of the $d$ electron contributions into a lower and upper
Hubbard band is schematically shown in Fig. \ref{fig08.12}b and the
singlet-triplet splitting in Fig. \ref{fig08.12}c. The difference between
Figs. \ref{fig08.11}a and b demonstrates nicely the effect of the excitations
of the correlation hole.

\newpage

\section{Summary and Outlook}
\resetdoublenumb 
\resetdoublenumbf

\label{Sect:Summary}

The different topics we have discussed in this review concern mainly
low energy effects of strongly correlated electrons, though not
exclusively. In Sec. \ref{Sect:HighEnergyExcitations} we have also given
examples of high energy features like shadow bands, satellites and kink effects
which originate from strong electron correlations. Low energy effects
are mainly due to heavy quasiparticles consisting of an electron with
its (rigid) correlation hole. The latter is very pronounced when correlations
are strong and therefore an electron moves with a reduced Fermi velocity
through the system dragging the correlation hole with it. Low temperature
properties are strongly influenced by the effective mass of the
quasiparticles. Important examples are Ce-based heavy fermion metals where an
interplay of strong on-site 4f Coulomb correlations and hybridization with
conduction electrons leads to the heavy quasiparticle
mass. These features may be described within Renormalized Band
Theory by assuming periodic scattering with a resonant phase shift. In
this way one obtains the proper heavy quasiparticle bands and the Fermi
surface. In recent years it has been realized that in 
U-based heavy fermion compounds like, e.g., UPd$_2$Al$_3$ heavy quasiparticles
originate from a special mechanism: the 5f electrons form a two-component
system with partly itinerant and partly localized orbitals and
the renormalization of the former via intra-atomic excitations of the
localized 5f electrons leads to strong mass enhancements. In fact, this
mechanism has also been identified as the origin of superconductivity in
UPd$_2$Al$_3$ which is not based on electron-phonon interactions.

We have also given examples where the low energy excitations cannot
simply be described by a Fermi liquid of heavy quasiparticles. In
general, even in weakly correlated metals, one expects non-Fermi
liquid behavior close to quantum phase transitions. The earliest
examples are the logarithmic corrections to the linear specific heat
term coming from electrons dressed by paramagnon excitations close to
a quantum critical point of ferromagnetism. By now numerous examples
of mostly Ce-based heavy fermion metals have been found where the
vicinity to an AF quantum phase transition leads to pronounced
non-Fermi liquid anomalies. Their microscopic understanding is still
at an early stage. 

A different type of deviation from the Fermi liquid picture emerges in
strongly correlated electron compounds which exhibit Wigner-lattice
type charge order as is the case e.g., in Yb$_4$As$_3$. Here spin and
charge degrees of freedom which are responsible for low energy
excitations may belong to different types of electrons, i.e., they
appear to be separated. In the example of Yb$_4$As$_3$ the spin
degrees of freedom involve 4$f$ holes of Yb while the charge degrees
of freedom were found to be due to As 4$p$ holes. The spins of 4$f$
holes are aligned along chains and then the 1D neutral spinon
excitations cause the large linear specific heat term while light 4$p$
holes carry the current.

Quite generally charge ordering in 3d or 4f mixed valent systems is a
promising route to obtain low-dimensional spin structures. An
example is NaV$_2$O$_5$. There the 1D spin excitations that emerge via
charge ordering interact with the lattice leading to a spin-Peierls
type dimerization and spin-gap formation. An even stronger kind of
lattice coupling occurs in the manganites which leads to polaron
formation. The intersite Coulomb correlations then drive charge order
of polaronic quasiparticles as observed in the bilayer-manganites. The
non-stoichiometric compounds of this class are examples of strong
Hund's rule correlations of itinerant (e$_g$) and localized (t$_{2g}$)
electrons. This leads to almost ideal 2D ferromagnetic order driven by
kinetic or double exchange mechanisms.

Another deviation from a simple Fermi liquid heavy-quasiparticle
description was found for the Hubbard model on a square lattice at a
particular range of doping and for U above the critical value of the
metal-insulator transition for half filling. Here the excitations were of a
form previously termed marginal Fermi liquid and suggested to explain some
of the properties of layered cuprates.

High-energy features of strongly correlated electrons result on the
other hand from excitations of internal degrees of freedom of the
correlation hole. While the slowly moving quasiparticle has a low
excitation energy the internal excitations of its cloud correspond to
high energies and show up as satellite structures in photoelectron
spectroscopy. Examples were given in Sec. \ref{Sect:HighEnergyExcitations}. 

In which direction will research on the theory of strongly correlated electrons
develop in the future? Here one can only speculate at the risk of being
completely wrong. Firstly there is the field of electronic structure
calculations.  After LDA based density functional calculations turned out to
describe insufficiently strongly correlated electrons they were combined
successfully with other approaches to improve on the shortcomings. As discussed
in this review Renormalized Band Structure calculations showed considerable
predictive power. But also LDA+U and LDA+GW schemes have been applied and
yielded insight into strongly correlated electrons. However the more
corrections are added to the original approximation to density functional
theory like LDA, the less controlled the results are. Double counting of
correlation contributions or screening is one of the uncertainties, accounting
for the spatial extend of the correlation hole is another. The latter is a
problem which also dynamical mean field theory (DMFT) is facing in its present
form. Extension of that theory to clusters \cite{Maier05} are desirable but
difficult to realize in practice. This suggests that over the long run
wavefunction-based approaches using quantum chemical methods may be a possible
solution to the problem. They are technically very demanding but they are
well controlled and allow for a detailed understanding of different correlation
processes. They may have a great future \cite{Fulde02}.

Exact diagonalizations will certainly also play their role in the
future. Usually they use model Hamiltonians with model parameters
obtained from single particle LDA(+U) calculations. With increasing
possibilities to treat larger clusters they will lead to much new insight into
the nature of the correlated ground state as well as the excitations
responsible for low temperature thermodynamic anomalies.

Analyses of the excitation spectrum of correlation holes seem to be a very
promising field. By projecting the calculations onto a few selected variables
which represent the most important degrees of freedom, insight may be gained
about the relative importance of different correlations. We expect considerable
extensions of our understandings of elementary excitations here. As was shown
in Sec. \ref{Sect:GeometricFrustration} in special lattices (here
geometrically frustrated ones) it may happen that an electron added to the
system  may separate into two parts when correlations are strong. This may
give raise to charge fractionalization, a phenomenon known from the fractional
quantum Hall effect (FQHE) \cite{Laughlin83}. Even if there is a weak restoring
force between the two parts we would have ''quasiparticles'' of considerable
spatial extent. They would have little in common with the usual picture of an
electron surrounded by a correlation hole. Interestingly enough, certain
analogies to field theories in elementary particle physics do appear here when
dealing with confinement and deconfinement \cite{Polyakovbook,Fradkin91}. After
all, why should nature restrict itself to realize certain basic properties only
in one particular field of physics and not in others too.

\begin{table}[pthb]
{\em List of acronyms}\\[0.5cm]
\begin{tabular}{lp{10cm}}
AF     &   antiferromagnet\\
ARPES  &   angle resolved photoemission spectroscopy\\
BEC    &   Bose Einstein condensation\\
BZ     &   Brillouin zone\\
CDW    &   charge density wave\\
CEF    &   crystalline electric field\\
CMR    &   colossal magnetoresistance\\
CO     &   charge order\\
CPA    &   coherent potential approximation\\
DE     &   double exchange\\
dHvA   &   de Haas-van Alphen\\
DOS    &   density of states\\
DM     &   Dzyaloshinsky-Moriya\\
DMFT   &   dynamical mean-field theory\\
EHM    &   extended Hubbard model\\
e-p    &   electron-phonon\\
FL     &   Fermi liquid\\
FM     &   ferromagnet\\
FQHE   &   fractional quantum Hall effect\\
FS     &   Fermi surface\\
HF     &   heavy fermion or Hartree-Fock\\
ICM    &   incommensurate magnet\\
INS    &   inelastic neutron scattering\\
ISSP   &   Ising-spin-Peierls\\
ITF    &   Ising model in transverse field\\
JT     &   Jahn-Teller\\
JW     &   Jordan-Wigner\\
KS     &   Kondo singlet\\
\end{tabular}
\end{table}
\begin{table}[pthb]
\begin{tabular}{lp{10cm}}
LDA    &   local density approximation\\
LSDA   &   local spin density approximation\\
n.n.   &   nearest neighbor\\ 
n.n.n. &   next nearest neighbor\\
NCA    &   non-crossing approximation\\
NFL    &   non-Fermi liquid\\
NMR    &   nuclear magnetic resonance\\
ODLRO  &   off-diagonal long range order\\
PM     &   paramagnet\\
QCP    &   quantum critical point\\
QMC    &   quantum Monte Carlo\\
QPT    &   quantum phase transition\\
RPA    &   random phase approximation\\
RKKY   &   Ruderman-Kittel-Kasuya-Yoshida\\
SCF    &   self consistent field\\
SCPM   &   self consistent projection method\\
SDW    &   spin density wave\\
TB     &   tight binding\\
WFM    &   weak ferromagnet
\end{tabular}
\end{table}

\section*{Acknowledgment}

\resetdoublenumb 
\resetdoublenumbf

We thank Prof. Y. Kakehashi for a critical reading of
Sec. \ref{Sect:HighEnergyExcitations} and for helpful comments. Last not least
we thank Mrs. Regine Schuppe for carefully preparing the manuscript.

\bibliography{ReferencesPT}

\end{document}